\immediate\write18{makeindex \jobname.nlo -s nomencl.ist -o \jobname.nls}

\documentclass[
	pdfa,
	oneside,
	saychapapp,
	fancyheaders]{ualberta}
	\settoclevel{3}

	\addlatexfiles{./00_LaTeX_Files/}
	\input{\@addlatexfilesincludePackages} 

	\addprefatory{./01_Prefatory/}

	\addchapters{./02_Chapters/}

	\addappendices{./04_Appendices/}

	\addmedia{./99_Inclusions/}
	\addimages{Images/}
	\addtables{Tables/}
	\addcode{Code/}
	\adddata{Data/}
	\addpdf{PDFs/}

	\input{\@addlatexfilesincludeTheorems}
	\input{\@addlatexfilesincludeMacros}
	\input{\@addlatexfileslistingCodeFormatting}

	\title{Physical Effects in the Vicinity of Black Holes}                         
	\author{Alexander Koek}                   
	\degree{\input{\@addlatexfilesselectDegree}}      
	\specialization{}                            
	\deptfac{\input{\@addlatexfilesselectDepartment}} 
	\convocationdate{\the\year}                  

	\input{\@addprefatoryAbstract}

	\input{\@addprefatoryPreface}

	\input{\@addprefatoryQuotes_Dedications}

	\input{\@addprefatoryAcknowledgements}

\begin{document}

	\maketitle                   
	\makeabstract                
	\makepreface                 
		\makededicationandquote      
	\acknowledgements                

	\onehalfspacing              

	\tableofcontents             
	\listoffigures               
	\listofsymbols               
	\glsaddall                   

	\bodyoftext 

	\truedoublespacing     

%

	\chapter{Introduction}\label{ch1}

In 1784, English astronomer John Michell first formulated the idea of a "dark star," a Newtonian object so dense that at its surface its escape velocity exceeded even the speed of light \cite{michell1784}.  Although these objects do not have a true "event horizon," in the sense that there is no one surface that determines an object's "point of no return," this description provided the first known example of a massive astrophysical object that is invisible to observations in the electromagnetic spectrum. Over a hundred years later, just months after Einstein's revolutionary discovery of the theory of general relativity (GR), it was shown by Karl Schwarzschild that within the framework of GR, very similar objects still theoretically existed, although these objects possessed true "event horizons," where passing light rays would be trapped for eternity \cite{Schwarzschild:1916uq}.  

These Schwarzschild black holes\footnote{It was not until 1967 that these objects were termed "black holes" by John Wheeler, and up until around this point in time, there was disagreement in the scientific community if these solutions actually corresponded to true astrophysical objects, or if they were merely a subject of pure mathematical interest with no physical backing.} were intensely studied for the decades that followed, with a specific focus on the meaning and interpretation of their event horizons.  Specifically, it was an object of debate whether or not this horizon was a defect of the spherically symmetric black hole solution to Einstein's equations or a true physical feature of the solutions.  These arguments were eventually resolved, and it was determined that this event horizon was indeed physical and corresponded to a surface from which light and massive particles could not escape.

Around the same time as these developments, a rotating black hole solution to Einstein's equations was discovered by Roy Kerr and was published in 1963 \cite{Kerr1963}.  These black holes were no longer spherically symmetric, and possessed many properties that differentiated them from Schwarzschild black holes.  The theoretical study of these features and their applications to astrophysics is still under immense investigation to this day.

Although black holes were studied from a theoretical point of view for the better part of the twentieth century, it was not until the second decade of the twenty-first century that the existence of these objects was confirmed by gravitational wave detection of binary black hole mergers via the Laser Interferometer Gravitational-Wave Observatory (LIGO) and by interferometry via the Event Horizon Telescope (EHT)\cite{EHT,LIGO}.

The goal of this thesis is to study classical interactions of black holes with their environments, and to conclude that their implications are far-reaching, with the hopes of demonstrating that although classical GR is a theory that already spans centuries, it is one whose fruits are still ripe to bear, and can help us better orient ourselves in the landscape of modern black hole physics.

In the entirety of this thesis, unless otherwise noted, we will adopt geometrized units of $G=c=1$.  When studying Maxwell's equations, we also use Gauss units where $\epsilon_0=1/(4\pi)$ and $\mu_0=4\pi$. We will also use standard sign conventions consistent with those of \emph{Gravitation} by Misner, Thorne, and Wheeler, such as a spacetime signature of $(-,+,+,+)$ \cite{MTW}.

\section{Einstein's Field Equations and Black Hole Solutions} \label{s11}
The equations of general relativity, which were published by Albert Einstein in late November of 1915 \cite{Einstein:1916vd}, are
\begin{equation} \label{EFE}
R_{\mu\nu}-\frac{1}{2}Rg_{\mu\nu} = 8\pi T_{\mu\nu}\,.
\end{equation}
Here, $g_{\mu\nu}$ is the spacetime metric, $R_{\mu\nu}$ is the Ricci tensor, $R$ the Ricci scalar, and $T_{\mu\nu}$ is the stress-energy tensor describing the matter content of the spacetime.

An arguably more natural way of determining Einstein's equations is via a gravitational action, which was proposed by David Hilbert in 1915 \cite{Hilbert1915}, five days before Einstein's formulation of GR was sent for publishing\footnote{Although Einstein's formulation of general relativity was published after Hilbert's formulation, it is widely recognized that Einstein was the founder of the theory, and had been laying its groundwork for many years before the release of both his and Hilbert's publications.}.  The action
\begin{equation} \label{EH}
S = \int \Big[\frac{1}{16\pi}R+\mathcal{L}_M\Big]\sqrt{-g}\,d^4x\,,
\end{equation}
where $g$ is the determinant of the metric and $\mathcal{L}_M$ is the matter Lagrangian density, successfully reproduces Einstein's field equations if we vary it with respect to the metric.  This action is called the Einstein-Hilbert action.  In the rest of this section we will give a historical overview of classical black hole solutions to Einstein's equations.  We will attempt to summarize these results in historical order, but we may jump around every now and then to preserve logical continuity.

\subsection{Schwarzschild Black Holes} \label{s111}
The first non-Minkowski solution to the Einstein field equations was discovered by Karl Schwarzschild in December of 1915, and formally published in January of 1916 \cite{Schwarzschild:1916uq}.  He searched for a spherically symmetric metric of the form
\begin{equation}
ds^2 =-e^{\alpha(r)}dt^2+e^{\beta(r)}dr^2+r^2(d\theta^2+\sin^2\theta d\phi^2)\,,
\end{equation}
where $\alpha$ and $\beta$ must be functions of solely $r$ to preserve the aforementioned spherical symmetry.  Note that if we set $\alpha=\beta=0$, then we retrieve flat Minkowski space in spherical coordinates.  For a more general solution, the one which Schwarzschild found, we can substitute this ansatz into Einstein's equations \eqref{EFE} and ensure that they are satisfied.  This leads to the following metric
\begin{equation} \label{Schw}
ds^2=-\bigg(1-\frac{2M}{r}\bigg)dt^2+\bigg(1-\frac{2M}{r}\bigg)^{-1}dr^2+r^2(d\theta^2+\sin^2\theta d\phi^2)\,.
\end{equation}
Here, $M$ is a constant, and one can show that it corresponds to the mass of the black hole by taking the asymptotic limit of the metric to the order of $1/r$, and identifying the first-order perturbation of the $g_{00}$ term as proportional to the Newtonian gravitational potential.

One might worry about the two visible singularities of the Schwarzschild solution \eqref{Schw}, $r=0$ and $r=2M$. It was shown by Arthur Eddington in 1924 that the latter of the two singularities can be eliminated via a specific transformation of the time coordinate \cite{Eddington1924}. About ten years later, the first acknowledgment that this singularity was not a physical singularity, but a "fictitious" one, was made by Georges Lema\^{\i}tre in 1933 when he introduced his own coordinate transformation that removed the singular $r=2M$ surface completely \cite{Lemaitre:1933gd}. Eddington's original coordinate transformation was later rediscovered and expounded upon by David Finkelstein in 1958, who understood this surface to be a "unidimensional membrane" for passing light rays \cite{Finkelstein1958}.  This "singularity" is known as the "event horizon" of the black hole, and is determined by the equation $g^{rr}(r)=0$, which gives the solution $r\equiv r_S=2M$. This radius is deemed the "Schwarzschild radius." 

The corresponding metric that eliminates this "singularity," which can be achieved by a coordinate transformation, is called the Eddington-Finkelstein metric.  As for the former singularity, $r=0$, no such coordinate transformation has been found.  It is thought that this singularity is a "true" singularity of the spacetime, and one hint of this can be seen if we calculate a curvature invariant of the metric \eqref{Schw}, $K \equiv R^{\mu\nu\alpha\beta}R_{\mu\nu\alpha\beta} = 48/r^6$, which is called the Kretschmann scalar.  This scalar is coordinate-independent, and formally approaches infinity as $r\to 0$.

Returning to the surface $r=2M$, it was first contextualized by Penrose in 1965 to be the outer boundary of "trapped surfaces" that is a natural consequence of gravitational collapse, where both ingoing and outgoing rays at any radius $r<2M$ are destined to head towards the singularity $r=0$ \cite{Penrose1965}. This paper was a precursor to the famous Penrose-Hawking singularity theorems, which showed that this spacetime is geodesically incomplete (see \cite{HawkingPenrose1970} and references therein).

Schwarzschild black holes make for great objects of theoretical study due to their simplicity and our deep understanding of them. For example, they allow for a natural generalization to spherically-symmetric black holes that contain a central charge, which are described by the Reissner-Nordstr\"{o}m metric.  This metric was independently discovered by Reissner in 1916 and Nordstr\"{o}m in 1918 \cite{Reissner1916,Nordstrom1918}. The metric is given by
\begin{equation} \label{RN}
ds^2=-\bigg(1-\frac{2M}{r}+\frac{Q^2}{r^2}\bigg)dt^2+\bigg(1-\frac{2M}{r}+\frac{Q^2}{r^2}\bigg)^{-1}dr^2+r^2(d\theta^2+\sin^2\theta d\phi^2)\,.
\end{equation}
and one can see that to recover Schwarzschild, we simply set $Q=0$.

Furthermore, in classical general relativity, one could consider the Vaidya metric, which models a typical Schwarzschild black hole in the case where it is allowed to absorb or emit null pressureless fluid \cite{Vaidya1943}.  This metric can be used to study black hole radiation, gravitational collapse, and many other phenomena of theoretical and observational interest.

\subsection{The Kerr Spacetime and its Geometry} \label{s113}

A natural next step when searching for more nontrivial solutions of Einstein's equations is to look for a rotating solution.  This was achieved by Roy Kerr in 1963 \cite{Kerr1963}.  The most common coordinate representation of his solution was found by Robert H. Boyer and Richard W. Lindquist in 1967 \cite{Boyer1967}, which is written as follows
\begin{align} \label{Kerr}
ds^2=&-\Big(1-\frac{2Mr}{\Sigma}\Big)dt^2+\frac{\Sigma}{\Delta}dr^2+\Sigma d\theta^2\\\nonumber
&+\Big(r^2+a^2+\frac{2Mra^2}{\Sigma}\sin^2\theta\Big)\sin^2\theta d\phi^2-\frac{4Mra\sin^2\theta}{\Sigma}dtd\phi\,,
\end{align}
where $\Sigma=r^2+a^2\cos^2\theta$ and $\Delta=r^2-2Mr+a^2$.  Here, $a$ is a new parameter called the rotation parameter, and it is related to the angular momentum of the black hole via $a=J/M$.

The Kerr metric is of particular interest in the domain of astrophysics, as it is the best black hole model in general relativity to describe the rotating black holes that we see in the universe, such as the one at the center of the galaxy.  There are many features of Kerr black holes that grant us a greater understanding of the universe in the astrophysical setting.  We will briefly name a few of them, which are especially important for our study.

Note that, like the Schwarzschild metric, the Kerr metric also contains a coordinate singularity defined by $g^{rr}=0$.  This equation can be recast as a quadratic equation in $r$ which determines the horizons of the black hole. It is easy to see that it will have not one, but two horizons, called the "inner" and "outer" horizons
\begin{equation}
    r_H = M \pm \sqrt{M^2-a^2}\,,
\end{equation}
where the outer horizon corresponds to a "plus" sign and the inner horizon corresponds to a "minus" sign. The outer horizon is the one which we still refer to as the "event horizon." In the case where the rotation parameter $a$ is set to zero, we recover the familiar Schwarzschild radius corresponding to the event horizon of a Schwarzschild black hole for the "plus" solution, and the "minus" solution goes to $r=0$.  

If we look carefully once more at the Schwarzschild metric, we notice that at the Schwarzschild radius $r_S$, the $g_{tt}$-component of the metric also vanishes.  However, this no longer happens when $r=r_H$ in the Kerr metric.  What happens if we search for the radius where $g_{tt}=0$ in the Kerr metric in Boyer-Lindquist coordinates?  We arrive at a new radius, which corresponds to a three-dimensional hypersurface known as the
"ergosurface," which is the outer boundary of a four-dimensional region known as the "ergoregion." Once more, we see that the equation describing the ergosurface is quadratic, and hence we will have an "inner ergoregion" and an "outer ergoregion."  We define the outer (inner) ergoregion as the region sandwiched between the outer (inner) ergosurface and  the outer (inner) event horizon\footnote{In the literature, the term "ergosphere" is sometimes used to refer to the "ergosurface," and in other instances is used to refer to the "ergoregion."  To avoid this confusion, we stick with the clearly defined terms "ergosurface" and "ergoregion."}. The defining radii of the outer and inner ergosurfaces are 
\begin{equation}
    r_E = M \pm \sqrt{M^2-a^2\cos^2\theta}\,.
\end{equation}
What exactly are these ergosurfaces?  Since they correspond to a change in sign of the $g_{tt}$-component of the metric, we can make two conclusions about them.  First, they are surfaces of infinite redshift for an observer at radial infinity that sees a ray penetrate these surfaces.  Second, one can check that any particle trajectory that is within the ergoregion must co-rotate with the black hole.  One can think of circular particle orbits in this region as rotating so fast that it is not possible to overcome the rotation of the spacetime and counter-rotate against the black hole.  Again, we restate that in the Schwarzschild metric, the ergosurface and the event horizon are the same surface, but in the Kerr metric they split into four unique inner and outer ergosurfaces and horizons.  The Kerr spacetime is also different from the Schwarzschild spacetime in that it has a "ring" singularity at $r=0$ instead of a point singularity.  This can be seen by noting that Eq.~\eqref{Kerr} is written in oblate spheroidal coordinates.

Another distinct feature of the Kerr metric, which is not present for non-rotating black holes, is that of "frame-dragging."  Due to the presence of the off-diagonal element in Eq.~\eqref{Kerr}, the spherical symmetry of the spacetime is broken, and we get a spacetime with axial symmetry instead.  Furthermore, this off-diagonal component suggests an influence on the clocks of observers that travel around the black hole, depending on if they are co-rotating with the black hole, or counter-rotating against it.  

Similar to the Reissner-Nordstr\"{o}m solution for Schwarzschild black holes, there also exists a charged version of Kerr black holes, derived by Newman and collaborators shortly after Kerr's discovery in 1965 \cite{Newman:1965my}.  In Boyer-Lindquist form, it is given by
\begin{align} \label{KerrNewman}
ds^2=&-\Big(1-\frac{2Mr-Q^2}{\Sigma}\Big)dt^2+\frac{\Sigma}{\Delta}dr^2+\Sigma d\theta^2\\\nonumber
&+\Big(r^2+a^2+\frac{(2Mr-Q^2)a^2}{\Sigma}\sin^2\theta\Big)\sin^2\theta d\phi^2\\\nonumber
&-\frac{2(2Mr-Q^2)a\sin^2\theta}{\Sigma}dtd\phi\,.
\end{align}

\subsection{Uniqueness Theorems and Complete Spacetimes}

When discussing the nature of Schwarzschild and Kerr black holes, an important question to address is that of uniqueness. It was first discovered by J{\o}rg Tofte Jebsen in 1921 and later independently rediscovered by George David Birkhoff in 1923 that any spherically-symmetric vacuum solution to the Einstein field equations is always static and asymptotically flat\footnote{Historically, this theorem is referred to as Birkhoff's theorem, due to the fact that Birkhoff's solution was initially much more well-known.  However, in hindsight, given that Jebsen provided the first proof of this theorem, it seems only fair to call this theorem the Jebsen-Birkhoff theorem.} \cite{Jebsen1921,Birkhoff1923}.  This uniquely relates the exterior metric (think Gauss' law) for any spherically-symmetric mass distribution to the Schwarzschild metric.  We note that there is no equivalent of the Jebsen-Birkhoff theorem for rotating black holes, meaning that the exterior vacuum solution of a rotating, axially-symmetric mass is not necessarily given by the Kerr metric.

A natural question to ask would be, is the converse of the Jebsen-Birkhoff theorem true?  In other words, does staticity and asymptotic-flatness of a spacetime guarantee that it is Schwarzschild?  The answer is yes, as long as such a spacetime has a regular event horizon. This was discovered first by Werner Israel in 1967 \cite{Israel:1967nh}.  This is considered the first of many "no hair" and uniqueness theorems, that suggest black holes are uniquely determined solely by their mass, charge, and angular momentum.  The extension of the original Schwarzschild case to Reissner-Nordstr\"{o}m black holes was later given by Israel in 1967 \cite{Israel:1967za}.  For the Kerr case, a proof of its uniqueness theorem was first found by Carter in 1971, which was improved upon by Robinson in 1975 \cite{Carter:1971bh,Robinson1975}.  Lastly, two uniqueness proofs for Kerr-Newman black holes were discovered concurrently by Mazur and Bunting in the early 1980's \cite{Mazur:1982,Bunting1983}.

Another important question regarding black hole spacetimes is, are they able to be analytically extended? The first maximal extension of the Schwarzschild metric using hyperbolic functions and separate coordinate patches was provided by J.L. Synge in 1950 \cite{Synge:1950gk}.  We call this extension "maximal" because it is the largest analytical extension of the coordinate system where every geodesic path is complete excluding the paths that terminate at singularities.  This discovery heavily inspired the development of the Kruskal-Szekeres coordinate system, which was discovered independently by both Martin Kruskal and George Szekeres in 1960 \cite{Kruskal:1959vx,Szekeres1960}.  These coordinates consist of four distinct regions of spacetime, which correspond to a black hole, a white hole, and their exterior regions.  

Shortly after the discovery of these coordinates, Penrose published his landmark paper defining and laying out the construction of "conformal diagrams" (also known as "Penrose diagrams" or "Carter-Penrose diagrams\footnote{Brandon Carter is often credited when naming these diagrams due to the fact that he was the first person to use Penrose's techniques to derive a maximal analytical extension of the Kerr metric and study properties of causality and geodesic completeness in such a framework \cite{Carter:66,Carter:68}}"), which allow one to visualize an entire spacetime as a series of connected boxes via conformal transformations where infinities are represented as conformal boundaries \cite{Penrose:1964dga}.  These maps are crucial for the study of black hole and cosmological spacetimes, as they preserve causality and easily allow one to detail the lightcone structure of such spacetimes.

\section{Tests of General Relativity} \label{s12}

As we have previously asserted, there are a wealth of weak-field and strong-field tests and effects within classical general relativity that offer simple and elegant explanations for more complicated and realistic phenomena that we see when we turn our telescopes to the stars.  We will now cover some of these effects as examples, and as an argument for why it is always a good thing to have more of them.  For an in-depth overview of these tests, one should consult \cite{Will1986}.

\subsection{Altering of Trajectories} \label{s121}

The first technical confirmation of general relativity actually didn't come from an experimental observation of an event predicted by the theory, but from the ability of the theory to confirm an observational anomaly that Newtonian gravity was unable to remedy.  Namely, it was observed by the French astronomer Urbain Le Verrier in 1859 that the orbit of Mercury did not follow a perfect elliptical pattern as was predicted by Newton.  Instead, he found that this ellipse was precessing at a slow but detectable rate \cite{LeVerrier1859}. It wasn't until 1915, over a half-century later, that Einstein discovered this precession could be accounted for by his theory of general relativity \cite{Einstein1915}.  

Although this calculation for Mercury's orbit is a bit more complicated due to its highly elliptical nature, an approximate calculation for almost-circular orbits can be made. Take a particle orbiting around an object of mass $M$ and gravitational radius $r_S=2M$, and assume that the periapsis $r_2$ (distance from particle to object at closest approach) and apoapsis $r_1$ (distance from particle to object at farthest approach) are approximately equal. Then the following series expansion in terms of $(\zeta_2-\zeta_1)^2$ is valid, where $\zeta_i=r_s/r_i$:
\begin{equation}
    \Delta\phi = \frac{2\pi}{\sqrt{1-\frac{3}{2}(\zeta_1+\zeta_2)}}+O\big((\zeta_2-\zeta_1)^2\big)\,.
\end{equation}
Here, $\Delta\phi$ represents the total angle traversed by a particle in an elliptical orbit around $M$ as it travels from periapsis to apoapsis and then back to periapsis.  A value of $\Delta\phi >2\pi$ demonstrates that the angular position of the particle after a full orbit around $M$ is different from its initial angular position, indicating a precession of the orbit. We can see that as the mass of the object $M$ becomes increasingly small, we recover $\Delta\phi=2\pi$ at leading order, as we would expect from Newtonian gravity. For a more detailed calculation of this phenomenon, see \cite{FroZel2011}.

The next confirmation of general relativity came from Arthur Eddington and collaborators during a total solar eclipse in 1919.  They showed that, as predicted by general relativity, the deflection angle of light rays by a massive object should be twice as much as the value that one would expect from Newtonian gravity \cite{Dyson1920}.  Specifically, for an object with mass $M$, in Newtonian gravity we expect a ray with impact parameter $b$ to be deflected by a total angle of 
\begin{equation}
    \alpha_{N} = \frac{2GM}{c^2b} = \frac{r_S}{b}\,,
\end{equation}
where $r_S$ is the Schwarzschild radius of the black hole, and we have temporarily restored parameters $G$ and $c$.  However, calculations in general relativity give that
\begin{equation}
    \alpha_{GR} = \frac{4GM}{c^2b} = \frac{2r_S}{b}=2\alpha_N\,.
\end{equation}
This is what Eddington's 1919 test confirmed.  The phenomenon of gravitational lensing has a long history, and a wide variety of applications for observational effects derived from classical relativistic principles.  For an in-depth overview on this topic, see \cite{Schneider:1992bmb}.

\subsection{Gravitational Redshift} \label{s122}

One of the first proposed predictions of general relativity, which is known as gravitational redshift, was calculated by Einstein in 1907 \cite{Einstein1907}. This effect can be described heuristically as the stretching of wavelengths of waves traveling at the speed of light due to their need to shed energy in order to escape a gravitational potential well.  In practical terms, this stretching comes from the difference between a far-away observer's clock $t$ and the clock of an observer at the emitted ray $t'$ within the gravitational potential well.  The relation between these two clocks is given by the time-component of the Schwarzschild metric \eqref{Schw}
\begin{equation}
\frac{dt'}{dt} = \sqrt{1-\frac{2M}{r}}\,.
\end{equation}
Therefore, one can imagine that the ratio of wavelengths between an emitter that records wavelength $\lambda_e$ at radius $r_e$ and an observer that records wavelength $\lambda_o>\lambda_e$ at radius $r_o>r_e$ is given by
\begin{equation} \label{lole}
    \frac{\lambda_o}{\lambda_e} = \sqrt{\frac{1-2M/{r_o}}{1-2M/{r_e}}}\,.
\end{equation}

For an observer at radial infinity, we find that
\begin{equation}
    1+z_{\infty} = \frac{\lambda_\infty}{\lambda_e} = \bigg(1-\frac{2M}{r_e}\bigg)^{-1/2}\,,
\end{equation}
where $z_\infty$ is called the redshift factor.  The first serious observational confirmation of gravitational redshift emitted by a distant star was made by Daniel Popper in 1954\cite{Popper1954}.

The next landmark test of gravitational redshift of the decade was the famous Pound-Rebka experiment, which was performed in 1959 \cite{PoundRebka1960}.  They measured the difference in wavelength between an observer $r_0$ and an emitter $r_e$ at the top and bottom of a 74-foot tall building, respectively.  Using Eq.~\eqref{lole}, and temporarily restoring $G$ and $c$ in these equations, we can simplify this equation to obtain the approximate result
\begin{equation}
    \frac{\Delta\lambda}{\lambda_e} = \frac{\lambda_o-\lambda_e}{\lambda_e} \approx \frac{gh}{c^2}\,,
\end{equation}
where $g\approx 9.81 \,m/s^2$ is Earth's gravitational acceleration at its surface, and $h$ is the height of the building.  Their observational results were in very good agreement with the theoretical predictions given by the above formula.

For a more complete breakdown of the history and physics behind tests of general relativity, see \cite{MTW} or \cite{Will1986}.

\subsection{Gravitational Waves}

Arguably the most important general relativistic effect for past, current, and future tests and observations is that of gravitational wave propagation.  For this reason, we dedicate a standalone subsection to this phenomenon. In general relativity, gravitational waves are propagating perturbations of spacetime, which are sourced by time-varying quadrupole moments of stress-energy tensors.  Due to the fact that this propagation is reliant on quadrupole moments, an object like a pulsating star will not emit gravitational waves, but for example a system of two massive compact objects rotating around their shared center of mass will propagate gravitational waves.

Although he was not the first to postulate the existence of gravitational waves, Albert Einstein was the first to show that they existed within his general theory of relativity, and that they necessarily propagate at the speed of light \cite{Einstein:1918btx}.  The existence and interpretation of gravitational waves was a subject of hot debate for over half a century after Einstein's theoretical framework was developed.  The first indirect detection of gravitational waves came shortly after the first observation of a binary pulsar system by Hulse and Taylor in 1974 \cite{Hulse_Taylor_1975}.  A few years after this detection, it was shown by Taylor and collaborators that the orbits of the objects in this system were experiencing a gradual inspiral pattern, which could be attributed to the loss of energy within the system due to the emission of gravitational waves \cite{Weisberg1981}.

A few years before this, efforts to observe gravitational waves directly via laser interferometers began to surface.  The most notable of these efforts came from the construction and design of the Laser Interferometer Gravitational-Wave Observatory, or LIGO for short.  This detector came online in 2002, but it wasn't until after many trials and improvements that the first direct detection of gravitational waves from the merger of two black holes was made by LIGO in 2015 \cite{LIGO}.  For their significant contributions to this project, Rainer Weiss, Barry Barish, and Kip Thorne were awarded the Nobel Prize in Physics in 2017 \cite{nobelphx2017}.  Today, there exist three more gravitational wave detectors that are in operation, which are used to confirm and cross-check detections from LIGO.  They are called VIRGO, KAGRA, and GEO600.

To date, LIGO has made over three-hundred detections of mergers of two massive compact objects, including black hole-black hole mergers, black hole-neutron star mergers, and even neutron star-neutron star mergers \cite{LIGOScientific:2026sit}.  The detector continues to provide results that are in excellent agreement with the results of general relativity, and offers a unique way to observationally analyze the inspiral, merger, and ringdown of binary black hole systems.

The study of this "ringdown," the final stage of a black hole merger in which the final object resulting from the collision of a black hole "rings" like a bell and emits specific gravitational wave signatures, is crucial for our current and future understanding of gravitational physics.  There is a whole field of gravitational physics known as "ringdown spectroscopy" that is devoted to the study of the gravitational waves emitted during the ringdown \cite{Dreyer:2003bv}, which are called "quasi-normal modes."  Ringdown spectroscopy can be used as a test of the no-hair theorem for rotating black holes \cite{Isi:2019aib}, as well as a tool for searching for theories of gravity beyond general relativity \cite{Berti:2025hly}.

\subsection{Other Modern Tests and Future Prospects} \label{s123}

When discussing tests of general relativity in past, present, and future landscapes, it is crucial to mention the post-Newtonian (PN) approximation method.  The idea behind this method is to expand solutions to Einstein's equations in orders of Newton's gravitational constant $G$, so that one derives weak-field approximations of such solutions at varying orders of precision.  Technically, the first use of this approximation was done by Einstein in his 1915 paper calculating the precession of the perihelion of Mercury \cite{Einstein1915}.

Modern observational tests of general relativity use a modified version of this formalism, called the parameterized post-Newtonian (PPN) formalism, which introduces two adjustable parameters $\beta$ and $\gamma$ into the post-Newtonian framework in order to better characterize how a theory deviates from both special relativity and general relativity.  Many texts have been written about the PPN formalism, but we offer the following two texts as references \cite{will1993theory,poisson2014gravity}.

We will now dive into a few effects arising from general relativity that are impactful on modern observational techniques, and which will likely come into play for future gravitational observations.

Frame-dragging offers a robust test of general relativity\footnote{This phenomenon was even detected from the Earth's rotating gravitational field by an orbiting satellite called Gravity Probe B \cite{Ciufolini:2004rq}!} and opens up possible observational signatures for modified theories of gravity. This frame-dragging effect also alters the image of what is known as the black hole's "shadow."  This shadow is defined by the boundary of gravitational capture for light rays traveling around the black hole that start off at radial infinity.  If these light rays are scattered by the black hole, then they can travel away from its grasp and eventually be observed by telescopes.  However, if they are captured by the black hole and fall in, we will never be able to observe them, and therefore in this boundary between the two cases, we see the edges of the black hole's "shadow."  

Furthermore, the frame-dragging effect warps the image of this shadow compared to that of a sphere.  This effect eluded resolution by the Event Horizon Telescope during the EHT collaboration's extraction of the first ever image of a black hole shadow \cite{EHT}.  However, it is argued that the effect should be visible to the Event Horizon Telescope in its upgraded and improved form, which is called the "next-generation Event Horizon Telescope" (ngEHT) \cite{Ricarte:2022wpd}.  The ngEHT is currently operational and is expected to complete its second full run in 2030.  For a somewhat accurate depiction of the frame-dragging effect that can be visually understood (with some creative liberties taken), one can consult the paper written on the methods used to generate the black hole in the movie Interstellar \cite{James:2015yla}.  We discuss the mathematical framework behind black hole shadows in a little more depth in Appendix~\ref{B42}.

Another important aspect of Kerr black holes which is thought to have major astrophysical relevance is something which is called the Penrose process, which was first laid-out by Roger Penrose in 1969 \cite{Penrose:1969pc}.  This effect relies on the unique properties of the ergoregion that is only found within rotating black holes. The process works as follows: say that we have a particle moving through the ergoregion.  Due to the fact that the Killing vector that describes time-translational invariance becomes spacelike within the ergoregion (see Appendix~\ref{apxA} for a definition of Killing vectors), we can have particle or photon trajectories that take on negative energy within the ergoregion.  Therefore, if one of these particles decays in such a manner that one of the decay particles is ejected from the ergoregion while the other particle remains within the ergoregion, we have a process that generates positive overall energy.

Although the Penrose process itself does not give enough energy to power astrophysical phenomena surrounding a black hole \cite{Bardeen_2}, there is a phenomenon that follows the same line of reasoning, called the Blandford-Znajek process, which does provide a significant amount of energy due to processes that occur in the ergoregion \cite{Blandford:1977ds}.  A similar effect, known as the Blandford-Payne mechanism, was shown to also extract energy from accretion disks surrounding black holes via surrounding magnetic fields \cite{Blandford:1982xxl}.  Combining these two ideas shows us that from processes that occur due to the geometry of rotating black holes, there are important astrophysical phenomena that arise which are of great observational interest.

Another modern test of general relativity, which is bound to play an important role in future tests of the theory, is the Shapiro time delay effect \cite{Shapiro64}.  For more details on the physics of the effect itself, see Appendix~\ref{B41}.  In short, the effect leads to delays in arrival time of incoming radiation due to the influence of gravitational fields that it passes along its trajectory.  Since the delay itself is sensitive to the characteristics of the gravitational field that it is altered by, this effect can play a major role in testing modified theories of gravity (see e.g. \cite{Moffat:2006gz}).

\section{Overview of Thesis} \label{s13}

The body of this thesis contains four principal chapters, each of which covers a specific work regarding the interaction of a black hole with its surrounding environment. The content of each chapter is briefly summarized below:

Chapter~\ref{ch2} describes what happens when one boosts a compact spinning matter distribution in the limit that approaches the speed of light.  We call such a boost an "ultrarelativistic" boost, and we also refer to this limit as the "ultrarelativistic" limit.  Within this limit, the problem of scattering of light rays by this compact object becomes one of a light ray being perturbed by a two-dimensional impulse plane.  We see that this consideration greatly simplifies calculations, and we provide some specific examples of what one can calculate in this framework and compare these examples with known calculations in black hole spacetimes.

Chapter~\ref{ch3} focuses on the evolution of a rotating Kerr black hole in the background of a homogeneous scalar field with a preferred gradient magnitude and direction.  Specifically, we allow for the field gradient vector to have arbitrary strength, and consider cases where it is timelike, spacelike, and null.  We consider specific cases of these three vector fields where we choose its orientation, along with the orientation of the black hole's spin and velocity vectors.  The evolution of the black hole's spin and mass parameters are analyzed, and we use the evolution of these quantities to prove that the second law of black hole thermodynamics is still valid in any of these cases.  We also derive and analyze the equations of motion for a rotating black hole in this scalar field background. We note that this analysis is done for a weak-field approximation of the homogeneous scalar field, in that we assume its stress-energy does not contribute to the overall black hole spacetime geometry, and furthermore we assume that the evolution of the black hole parameters are slow enough so that its initial geometry is preserved throughout the process.

The problem at hand in Chapter~\ref{ch4} is very similar to that of Chapter~\ref{ch3}.  Instead of considering a rotating black hole in a homogeneous scalar field background, we turn our attention to a rotating black hole with electric and magnetic monopole charges in a homogeneous electromagnetic field background.  Similar to our previous analysis, we select various orientations for the electric and magnetic fields with respect to the black hole's velocity and spin directions, and analyze the evolution of the black hole's spin, mass, and velocity.  We discuss similarities and differences between the scalar field and electromagnetic field setups.  This calculation is also considered to be a weak-field approximation, in that the charges and homogeneous fields do not contribute at all to the black hole's spacetime geometry.

In Chapter~\ref{ch5}, we turn our attention to the spinoptics effects in the Kerr spacetime.  This effect describes the spin-gravitational interaction between a rotating black hole, and waves propagating at the speed of light that carry spin, such as electromagnetic and gravitational waves.  We first discuss the electromagnetic Faraday effect, and draw comparisons between this effect and the spinoptics effects.  We then derive the formalism necessary to study this phenomenon, which we consider as a perturbative approximation for waves with high frequency.  Their motion can be analyzed as the motion of individual null rays, which turn out to be slightly non-geodesic due to the spinoptics interaction.  We use the spinoptics framework to study the effect for the scattering of light rays by rotating Kerr black holes, and we derive approximate analytical results for rays that are initially in the equatorial plane of the Kerr spacetime.  Future research questions regarding this effect are outlined and discussed.

Lastly, in Chapter~\ref{ch6}, we summarize the results given throughout this dissertation and discuss their meaning and relevance.

	\chapter{A Nifty Trick: Boosted Black Holes} \label{ch2}

In this chapter we will show that, using a Lorentz boost that approaches the speed of light, we can derive a novel plane wave metric for the gravitational field of any compact mass with arbitrary mass and angular momentum distributions.  With this plane wave solution, we can greatly simplify the problem of light ray scattering by a gravitating object and expand its scope, such that we will obtain the deflection angle for a ray that is scattered by any arbitrary mass distribution with arbitrary internal angular momentum.

This "trick" can be qualitatively described as follows:  A light ray starts out on its initial trajectory, is scattered by some arbitrary but compact mass distribution, and then it travels off to infinity.  Therefore, what one can do is perform an arbitrarily-directed Lorentz boost of constant velocity of the frame where the center of mass of the mass distribution is at rest.  If this boost is chosen with velocity $\text{v}\to 1$, that is, a velocity that approaches the speed of light, then the problem simplifies greatly; due to Lorentz length contraction, the three-dimensional mass distribution reduces to that of a two-dimensional plane distribution.  Therefore, in this "ultrarelativistically" boosted frame, a null ray which starts out at ingoing infinity is deflected by some infinitely-thin impulse upon reaching the two-dimensional plane distribution, and then is scattered off to outgoing infinity.

The first known implementation of this "ultrarelativistic" boosting procedure on a metric was done to the Schwarzschild metric by Aichelburg and Sexl in 1971 \cite{AichelburgSexl}.  Not long after, this procedure was generalized to both the Kerr and Kerr-Newman metrics \cite{Balasin:1995tj,Lousto:1992th}.  The basis of all of these transformations was originally outlined by Penrose in 1976, who suggested that any spacetime could be ultrarelativistically boosted to some metric that resembles a plane wave \cite{PenroseLimit} (more on this later).  This general formalism is also closely related to the idea of gyratons; spinning ultrarelativistic objects whose gravitational fields are exact solutions to Einstein's equations.  These were discovered and formulated by Frolov and collaborators in 2005, and are a useful tool for describing the dynamics of spinning particles moving near the speed of light in the realm of general relativity \cite{Frolov:2005gy,Frolov:2005zq,Frolov:2005ja}.  These gyratons have attracted considerable interest in relation to the idea of microscopic higher-dimensional black hole formation from high-energy particle collisions \cite{Giddings:2001bu,Eardley:2002re,Kanti:2004nr}.

From the black hole uniqueness theorems, it is known that the exterior of the Kerr spacetime is uniquely determined solely by the mass and the angular momentum of the black hole.  This, along with other symmetry properties of the spacetime, allows one to exactly solve for scattering trajectories of massive and massless particles.  Generally, the metric that arises from some arbitrary rotating object, such as a rotating star, does not coincide with the Kerr metric. In this chapter, we set out to find a plane wave metric solution for compact masses with arbitrary mass distributions and angular momentum distributions that may exhibit differential rotation. To achieve this, we take an ultrarelativistic boost of the spacetime of the compact source, and we find the form of the source metric after this boost, which corresponds to a two-dimensional distribution within a null plane.  Studying the trajectories of null rays passing through this plane, we conclude that they can be separated into freely-moving ingoing and outgoing trajectories, which are connected in a simple manner by a finite set of parameters related to the two-dimensional impulse plane.

After this, we show that applying an inverse boost transformation to the scattering trajectories allows us to re-derive classical scattering angle and time delay formulas in the rest frame of a gravitating object in a very simple fashion. These formulas are shown to correspond exactly to the weak-field limit of scattering angles and time delays for black holes in general relativity without any boosting effects.  Lastly, we identify our maps relating inward and outward trajectories as a specific case of the well-known gravitational memory effect, and we remark on the importance of this effect.

\section{The Weak-Field Limit: Linearized Einstein's Equations} \label{s21}

\subsection{Perturbative Expansion and Gauge Freedoms} \label{s211}

In the ultrarelativistic limit where $\text{v}\to 1$, we must take the mass of the compact object $M\to 0$, so that the energy of the system remains fixed and the geometry describing such an object remains well-defined.  Therefore, since the mass of the compact source vanishes in this limit, we pose our problem within the weak-field approximation of general relativity, meaning that we will expand the metric as a perturbative approximation of the flat Minkowski metric.  To properly study light ray propagation within this geometry, we must impose the following restrictions:
\begin{itemize}
    \item The radial distance $r$ from a passing null ray to the center of mass of the compact massive object must always be much greater than some characteristic length scale that describes the gravitational strength of the object.  Naturally, we choose this length to be the Schwarzschild radius of the object, $r_S=2M$, and so $r \gg 2M$ at any point on the null ray.
    \item For the same null ray, we also require it to pass by the compact object at a distance larger than $L$, the maximum distance from the center of the mass of the object to any point within its mass distribution.  In other words, $r > L$.
\end{itemize}
We can then assume that the metric has the form
\begin{equation} \label{pert}
    g_{\mu\nu} = \eta_{\mu\nu}+h_{\mu\nu}\,,
\end{equation}
where we enforce the condition that any component of the perturbation metric should be much smaller in magnitude than any component of the background metric, so that in a Minkowski background in Cartesian coordinates, we have $|h_{\mu\nu}|\ll 1$.

Cutting off this perturbation after first order means that raising and lowering of indices of any tensor can be done simply by using $\eta_{\mu\nu}$ and $\eta^{\mu\nu}$ instead of $g_{\mu\nu}$ and $g^{\mu\nu}$.  This means that 
\begin{equation}
    g^{\mu\nu} = \eta^{\mu\nu} - h^{\mu\nu}\,.
\end{equation}
One can easily verify that at first order, $g_{\mu\nu}$ and $g^{\mu\nu}$ are indeed inverses, so that $g_{\mu\lambda}g^{\lambda\nu}=\delta_{\mu}^{\nu}$.

One last detail that we must be careful to preserve when linearizing gravity is that of diffeomorphism invariance.  One can imagine that the preservation of Einstein's equations under coordinate transformations will have direct consequences on how $h_{\mu\nu}$ is allowed to transform.  The question becomes, how does the gauge freedom of full diffeomorphism invariance translate to the perturbed metric?  The answer is quite simple and satisfying.  

Say that we have a vector field $\alpha^{\mu}(x)$, and let the coordinates transform as $x^{\mu}\to\tilde{x}^{\mu}=x^{\mu}+\alpha^{\mu}(x)$.  The only assumption we need to make is that $\boldsymbol{\alpha}(x)$ is small in magnitude everywhere in order to preserve the smallness of $h_{\mu\nu}$, and the rest will follow.  Note that in the following calculations, we assume that the background is the Minkowski metric in Cartesian coordinates. Taylor expanding the metric, one has
\begin{equation}
    \tilde{g}_{\mu\nu}(\tilde{x}) = \tilde{g}_{\mu\nu}(x)+\alpha^{\lambda}(x)g_{\mu\nu,\lambda}(x)\,.
\end{equation}
From the perspective of a coordinate transformation, we also have the following equation
\begin{equation}
    \tilde{g}_{\mu\nu}(\tilde{x}) = \frac{\partial x^{\lambda}}{\partial\tilde{x}^{\mu}}\frac{\partial x^{\kappa}}{\partial\tilde{x}^{\nu}}g_{\lambda\kappa}(x) = g_{\mu\nu}(x)-\big(g_{\mu\lambda}(x)\alpha^{\lambda}_{\;,\nu}+g_{\nu\lambda}(x)\alpha^{\lambda}_{\;,\mu}\big)\,.
\end{equation}
Matching these equations, and identifying the first-order terms as the Lie derivative of $\boldsymbol{g}$ along $\boldsymbol{\alpha}$, we have
\begin{equation}
    \tilde{g}_{\mu\nu} = g_{\mu\nu} - \mathcal{L}_{\boldsymbol{\alpha}}g_{\mu\nu}\,.
\end{equation}
Then, simply plugging in the metric perturbation equation \eqref{pert} gives us the transformation law
\begin{equation} \label{egauge}
    h_{\mu\nu}\to\tilde{h}_{\mu\nu}=h_{\mu\nu}-\mathcal{L}_{\alpha}g_{\mu\nu} = h_{\mu\nu}-2\alpha_{(\mu,\nu)}\,.
\end{equation}

This is an answer that one may have guessed from the beginning: The change in the metric perturbation under the flow of an infinitesimal vector field compensates exactly for the overall change in the metric along such a flow. Furthermore, the above equation dictates the gauge freedom that we have for $\boldsymbol{h}$ under diffeomorphisms.  It is no longer full diffeomorphism invariance, but it still leaves us with a vast number of degrees of freedom to play around with.

Lastly, we should note that we were rather "laissez-faire" in our treatment of maps, transformations, pullbacks, etc.  For a more mathematically rigorous derivation and discussion of these gauge freedoms, see e.g. \cite{Carroll2019-di}.

\subsection{Linearized Einstein Field Equations} \label{s212}

We will not go through the precise derivation, but in order to obtain the linearized version of the Einstein tensor $G_{\mu\nu}\equiv R_{\mu\nu}-\frac{1}{2}Rg_{\mu\nu}$, one starts by inserting Eq.~\eqref{pert} into the definition of Christoffel symbols up to and including first order terms in the expansion.  From there, one can calculate the Riemann tensor, and therefore both the Ricci tensor and Ricci scalar.  Such a calculation gives
\begin{equation} \label{linG}
    G_{\mu\nu}=h^{\lambda}_{\;\,(\mu,\nu)\lambda}-\frac{1}{2}\big((h^{\lambda}_{\,\,\lambda})_{,\mu\nu}+\Box{h_{\mu\nu}}+\eta_{\mu\nu}h^{\alpha\beta}_{\;\;\,,\alpha\beta}-\eta_{\mu\nu}\Box{(h^{\lambda}_{\,\,\lambda})}\big)\,.
\end{equation}
Here, $\Box\equiv\partial^{\mu}\partial_{\mu}$ is the d'Alembertian operator in the Minkowski spacetime.

To clean things up, we can utilize the gauge freedom from Eq.~\eqref{egauge}.  We choose to impose the de Donder gauge condition (sometimes called the harmonic gauge or the Lorentz gauge):
\begin{equation}
    \bar{h}^{\mu\nu}_{\;\;,\nu}=0\,,\hspace{15pt}\bar{h}_{\mu\nu}=h_{\mu\nu}-\frac{1}{2}\eta_{\mu\nu}h^{\lambda}_{\,\,\lambda}\,.
\end{equation}

After some tensor algebra, one can check that Eq.~\eqref{linG} becomes
\begin{equation}
    G_{\mu\nu} = -\frac{1}{2}\Box\,\bar{h}_{\mu\nu}\,.
\end{equation}
Therefore, in the absence of a cosmological constant, the linearized version of Eq.~\eqref{EFE} is
\begin{equation} \label{linBarEFE}
    \Box\,\bar{h}_{\mu\nu} = -16\pi T_{\mu\nu}\,,
\end{equation}
or alternatively,
\begin{equation} \label{linEFE}
    \Box\,{h}_{\mu\nu} = -16\pi \bar{T}_{\mu\nu}\,,\hspace{15pt}\bar{T}_{\mu\nu}=T_{\mu\nu}-\frac{1}{2}\eta_{\mu\nu}T^{\lambda}_{\,\,\lambda}\,.
\end{equation}

\subsection{Multipole Expansion for the Gravitational Field of a Compact Source} \label{s213}

We now consider a compact stationary gravitating object. We assume that the center of mass of this compact source is at rest, and therefore, for a background Minkowski metric in Cartesian coordinates given by
\begin{equation} \label{Mink}
    \eta_{\mu\nu}dX^{\mu}dX^{\nu} = -dT^2+dX^2+dY^2+dZ^2\,,
\end{equation}
we have the following values for the four-momentum of the center of mass of the object $P^{\mu}=(P^0,P^i)$:
\begin{equation} \label{poincare1}
    P^0\equiv M = \int T^{00}\,d^3X\,,\hspace{15pt}P^{i} = \int T^{0i}\,d^3X = 0\,.
\end{equation}
We also have the following angular momentum tensor
\begin{equation} \label{poincare2}
    J^{\mu\nu}=\int\big(X^{\mu}T^{\nu0} - X^{\nu}T^{\mu0}\big)\,d^3X\,.
\end{equation}

It is important to note that $\boldsymbol{P}$ and $\boldsymbol{J}$ are conserved in this spacetime, at least at the lowest order of the perturbation.  Since $\boldsymbol{J}$ is by definition antisymmetric, we see that both tensors give a total of ten independent conserved quantities, as one would expect from the Poincar\'{e} invariance of the Minkowski spacetime. Moreover, out of all these lowest-order symmetries, we assume that one of them holds for all orders of the perturbation.  Specifically, for the Killing vector $\boldsymbol{\xi}=\partial_T$, we say that the Lie derivative $\mathcal{L}_{\boldsymbol{\xi}}\boldsymbol{T}=0$ in the full spacetime.  This allows us to write Eq.~\eqref{linBarEFE} as
\begin{equation} \label{poisson}
    \Delta\bar{h}_{\mu\nu} = -16\pi T_{\mu\nu}\,,
\end{equation}
where $\Delta=\partial^{i}\partial_{i}$ is the Laplacian operator in flat Minkowski space.

Using conservation of stress-energy, and once again enforcing that the center of mass of the compact object coincides with the origin of the coordinate system, we write down two additional formulas
\begin{equation} \label{poincare3}
    J^{ij}=2\int X^{i}T^{j0}\,d^3X\,,\hspace{15pt}\int X^{i}T^{00} d^3X = 0\,.
\end{equation}

Now we arrive at the multipole expansion of the tensorial Poisson equation \eqref{poisson}.  For more intricate details about this expansion, see \cite{FroZel2011}, but the gist of it is that we solve this equation via Green's functions, and use the weak-field condition for objects at far distance $r\gg r_S$ to obtain the expanded solution
\begin{equation} \label{multipole}
    \bar h_{\mu\nu}(X) = \frac{4}{r}\int T_{\mu\nu}(X')\,d^3X'+\frac{4X^{i}}{r^3}\int X'_i \,T_{\mu\nu}(X')\,d^3X'+...\,,
\end{equation}
where the ellipsis at the end of the equation refers to higher-order terms that are disregarded in the expansion, and in Cartesian coordinates $r$ is defined as $r^2=X_iX^i$.  Each successive term in this expansion describes a multipole moment, giving us higher angular resolution of the spacetime that comes from our stress-energy tensor.  The first term in the expansion corresponds to the monopole moment of the stress-energy tensor, while the second term corresponds to its dipole moment, and so on and so forth. Using Eqs.~\eqref{poincare1}-\eqref{poincare3}, we arrive at the following solutions to the perturbed metric
\begin{equation}
    h_{00} = \frac{2M}{r}\,,\hspace{10pt}h_{ij}=\frac{2M}{r}\delta_{ij}\,,\hspace{10pt}h_{0i}=\frac{2J_{ij}X^{j}}{r^3}\,.
\end{equation}
Briefly switching to a spherical Minkowski background, as well as gauge-fixing rotation in the spacetime so that the axial angular momentum vector aligns with the spherical north pole, allows us to write the full weak-field metric as
\begin{equation}
    ds^2 = -\Big(1-\frac{2M}{r}\Big)dT^2+\Big(1+\frac{2M}{r}\Big)(dr^2+r^2 d\Omega^2)-\frac{4J\sin^2\theta}{r}dTd\phi\,,
\end{equation}
where $d\Omega^2=d\theta^2+\sin^2\theta d\phi^2$ is the metric for a unit 2-sphere.  This metric is the well-known Lense-Thirring metric, which (to nobody's surprise) was discovered and studied by Lense and Thirring in 1918 \cite{LT1918}.  

If one were to study the trajectory of initially radially infalling geodesics within this metric, they would find that it has a near-exact correspondence with the Coriolis effect for a body falling radially inward towards a gravitational source in Newtonian gravity within a rotating, non-inertial coordinate system.  The place where the "exactness" of this analogy falls apart comes from the fact that the angular velocity at which the system is rotating is now a function of radius in the general-relativistic case.  This is another clear-cut example of the aforementioned frame-dragging effect.

\section{The Ultrarelativistic Limit} \label{s22}

It was shown by Roger Penrose in 1976 that "any space-time has a plane wave as a limit."  More specifically, the procedure goes as follows:

\begin{enumerate}
    \item Write down the metric $g_{\mu\nu}(x)$ in terms of lightcone coordinates for some null geodesic path within the metric.  In four dimensions we will have $x^{\mu}=(u,v,y,z)$.
    \item Rescale the coordinates as $x^{\mu}\to X^{\mu}=(U,V,Y,Z)=(u,e^{2\sigma}v,e^{\sigma}y,e^{\sigma}z)$.  This gives us a new metric $\tilde{g}_{\mu\nu}(X)$.
    \item One can see that $\tilde{g}_{\mu\nu}(X)=e^{-2\sigma}\hat{g}_{\mu\nu}(X)$ for some metric $\hat{g}_{\mu\nu}$.  Make this conformal transformation to $\hat{g}_{\mu\nu}(X)$.
    \item Take the limit $\sigma\to\infty$ for $\hat{g}_{\mu\nu}(X)$.  This is called the "Penrose limit", and we arrive at the metric $\boldsymbol{g}_p\equiv\lim_{\sigma\to\infty}\hat{\boldsymbol{g}}$.  This metric corresponds to a plane wave!
\end{enumerate}

The physical interpretation of this process is that we have a limit of wordlines of observers approaching the worldline of a null observer as they are boosted towards the speed of light.  This is what is happening as we make the conformal transformation, and the reason for the rescaling of the coordinates before this conformal limit is to recalibrate their clocks to run faster as they approach this limit to account for the time dilation of such a transformation.  For the original paper detailing this "Penrose limit," see \cite{PenroseLimit}.

We will be using a procedure that is akin to the Penrose limit in spirit, but different in implementation\footnote{The authors note that the procedure in this chapter, albeit similar to the Penrose limit in idea, is not the same as taking the Penrose limit of a spacetime.  There is very likely a direct connection between these two ideas, but we have not found this connection and so we make the distinction between the two different approaches.  For this reason, we will call this procedure the "ultrarelativistic limit" as opposed to the "Penrose limit."}.  The energy of an object of mass $m$ recorded by an observer approaching the limit $\text{v}\to 1$ goes as $E=\gamma m$, where $\gamma=(1-\text{v}^2)^{-1/2}$ is the usual Lorentz factor, meaning that the energy formally approaches infinity.  If we want the energy to remain fixed at a constant value, in the same way that we readjust the observers' clocks when performing the Penrose limit procedure above, then we must also require the mass $m\to 0$ as $\text{v} \to 1$ to compensate.

\subsection{Boosting the Source} \label{s221}

To understand how our boosted metric differs from that of the Aichelburg-Sexl solution, we must first see how its mass distribution transforms under this "ultrarelativistic" boost.  When we say "ultrarelativistic," we mean a boost that formally approaches the speed of light, so that $\text{v} \to 1$.

As one does for relativistic boost transformations, we define two reference frames.  The first frame, $\mathcal{K}$, corresponds to the $X^{\mu}$ coordinate system (the "rest" frame), and the second frame, $\mathcal{K}'$, will correspond to some $x^{\mu}$ coordinate system after a relativistic boost (the "boosted" frame).  We need to fix a boost-direction, and so we choose the boost transformation to go along the $X$-direction of the background metric described by Eq.~\eqref{Mink}.

How should the stress-energy tensor look in the rest frame $\mathcal{K}$?  We choose the matter to correspond to dust particles, so that any pressure may be neglected.  With this in mind, we write the following ansatz
\begin{equation} \label{Tdef}
    T_{\mu\nu} = \xi_{(\mu}K_{\nu)}\,,\hspace{15pt} K_{\nu} = \rho\xi_{\nu}+\mathcal{Z}_{\nu}\,,\hspace{15pt} \xi^{\nu}\mathcal{Z}_{\nu} = 0\,.
\end{equation}
For now, $\rho$ and $\mathcal{Z}_{\nu}$ are arbitrary functions related to the matter and linear momentum distributions of the source, and as we'll recall, $\boldsymbol{\xi}=\partial_T$ is the timelike Killing vector for the unboosted metric.

For $\mathcal{Z}_{\nu}=(0,\mathcal{Z}_i)$, we also make the educated guess that
\begin{equation} \label{zj}
    \mathcal{Z}^i = j^{ik}_{\;\;,k}\,,\hspace{15pt} j^{ik} = j^{[ik]}\,.
\end{equation}
One can see that this guess was indeed educated, since the conditions from Eqs.~\eqref{poincare1} and \eqref{poincare3} will now be automatically satisfied
\begin{align} \label{conservations}
   &M = \int \rho\,d^3X\,,\\\nonumber
   &P^{i} = \frac{1}{2}\int\mathcal{Z}^{i}\,d^3X = \frac{1}{2}\oint_{\Sigma} j^{ik}\,d\sigma_{ik} = 0\,,\\\nonumber
   &J^{ik} = \int X^{i}\mathcal{Z}^{k}\,d^3X = \oint_{\Sigma} X^{i}j^{kl}\,d\sigma_{l} - \int j^{ki}\,d^3X = \int j^{ik}\,d^3X\,.
\end{align}
Here we have used the divergence theorem, and taken the boundary surface $\Sigma$ to be outside of the compact mass distribution.

In the Cartesian Minkowski spacetime given by Eq.~\eqref{Mink}, we can see that any spatial direction also admits a Killing vector due to translational invariance in that spatial direction.  If we choose to boost the spacetime in the $X$-direction, then a natural choice for a second Killing vector would be $\boldsymbol{\xi}_{X}\equiv\partial_X$.  In addition to this, we will retroactively redefine the first Killing vector as $\boldsymbol{\xi}\equiv\boldsymbol{\xi}_T=\partial_T$, and will refer to it as such for the rest of this chapter for clarity.

We will refer to the spacetime that is spanned by $(\boldsymbol{\xi}_T,\boldsymbol{\xi}_X)$ as $\Pi_1$, and the remaining orthogonal spacetime as $\Pi_2$.  Note that while $\Pi_1$ is affected by our Lorentz boost, $\Pi_2$ is not.  Therefore, it will best serve us to break up all of the vectors and tensors we use into their $\Pi_1$ and $\Pi_2$ components, where the former components will be denoted by capital indices from the start of the Latin alphabet, and the latter components will be denoted by lowercase indices from the start of the Latin alphabet.  For example, we can write
\begin{equation}
    \mathcal{Z}_{\mu} = \mathcal{Z}_A\delta^A_{\mu}+\mathcal{Z}_a\delta^a_{\mu}\,,\hspace{15pt}\mathcal{Z}_{A}=(0,\mathcal{Z}_X)\,,\hspace{10pt}\mathcal{Z}_a=(\mathcal{Z}_Y,\mathcal{Z}_Z)\,.
\end{equation}
Furthermore, using Eq.~\eqref{zj}, we can write
\begin{equation}
    j_a\equiv j_{a\mu}\xi^{\mu}_X\,,\hspace{10pt}j_{ab}=e_{ab}j\,,\hspace{15pt}\mathcal{Z}_X = j^{a}_{\;,a}\,,\hspace{10pt}\mathcal{Z}_a=-\partial_{X}j_a+e_{ab}j^{,b}\,,
\end{equation}
where $e_{ab}$ is the two-dimensional Levi-Civita symbol with $e_{23}=1$.  From Eq.~\eqref{Tdef}, the stress-energy tensor then becomes
\begin{equation}
 T^{\mu\nu}=\rho\xi^{(\mu}_T\xi^{\nu)}_T+\mathcal{Z}_X\xi^{(\mu}_T\xi^{\nu)}_X+\mathcal{Z}^a\xi_T^{(\mu}\delta^{\nu)}_a\,.
\end{equation}

Since we will eventually be solving Eq.~\eqref{linEFE}, we also want to calculate $\bar{\boldsymbol{T}}$, which we find to be
\begin{equation}
    \bar{T}_{\mu\nu} = \frac{1}{2}\rho \mathcal{Q}_{\mu\nu}+\frac{1}{2}\mathcal{Z}_{X}\mathcal{P}_{\mu\nu}+\mathcal{S}_{\mu\nu}+\mathcal{N}_{\mu\nu}\,,
\end{equation}
where
\begin{align}
    &\mathcal{Q}^{\mu\nu}=\xi^{\mu}_{T}\xi^{\nu}_{T}+\xi^{\mu}_{X}\xi^{\nu}_{X}\,,\hspace{15pt}\mathcal{P}^{\mu\nu}=\xi^{\mu}_{T}\xi^{\nu}_{X}+\xi^{\mu}_{X}\xi^{\nu}_{T}\,,\\\nonumber
    &\mathcal{S}^{\mu\nu}=\mathcal{Z}^{a}\xi^{(\mu}_{T}\delta^{\nu)}_{a}\,,\hspace{32pt}\mathcal{N}^{\mu\nu}=\frac{1}{2}\rho\delta^{\mu}_{a}\delta^{\nu}_{b}\delta^{ab}\,.
\end{align}
Note that only $\boldsymbol{\mathcal{N}}$ will remain invariant under Lorentz boosts, as it fully lies in $\Pi_2$.

Next, as per the ultrarelativistic limit prescription, we will transform the coordinates to lightcone coordinates.  For a standard Lorentz-boost
\begin{equation} \label{boost}
    X=\gamma(x-vt)\,,\hspace{10pt}T=\gamma(t-vx)\,,\hspace{10pt}Y=y\,,\hspace{10pt}Z=z\,,
\end{equation}
we have the following set of lightcone coordinates in both $\mathcal{K}'$ and $\mathcal{K}$, respectively:
\begin{align} \label{lightcone}
    &U=\frac{1}{\sqrt{2}}(T-X)\,,\hspace{15pt}V=\frac{1}{\sqrt{2}}(T+X)\,,\\\nonumber
    &u=\frac{1}{\sqrt{2}}(t-x)\,,\hspace{15pt}v=\frac{1}{\sqrt{2}}(t+x)\,.
\end{align}
We will also find it useful to define the following relational quantities
\begin{align} \label{lightcone2}
    &U=\alpha u\,,\hspace{15pt}V=\alpha^{-1}v\,,\hspace{15pt}\alpha=\sqrt{\frac{1+\text{v}}{1-\text{v}}}\,,\\\nonumber
    &T=\frac{1}{\sqrt{2}}(\alpha u+\alpha^{-1}v)\,,\hspace{15pt}X=\frac{1}{\sqrt{2}}(-\alpha u+\alpha^{-1}v)\,.
\end{align}

In lightcone coordinates $X^{\mu}=(U,V,Y,Z)$, we can write
\begin{equation}
    \xi_{(T)A} = -\frac{1}{\sqrt{2}}(1,\;1)\,,\hspace{15pt}\xi_{(X)A} = -\frac{1}{\sqrt{2}}(1,\;-1)\,,
\end{equation}
meaning that in these same coordinates
\begin{align}
    &\mathcal{Q}_{AB} = 
    \begin{pmatrix}
        1 & 0\\ 0 & 1
    \end{pmatrix}\,,\hspace{52pt}
    \mathcal{P}_{AB} = 
    \begin{pmatrix}
        1 & 0\\ 0 & -1
    \end{pmatrix}\,,\\[10pt]\nonumber
    &\mathcal{S}_{aB} = 
    -\frac{1}{2\sqrt{2}}\begin{pmatrix}
        \mathcal{Z}_Y & 0 \\
        \mathcal{Z}_Z & 0
    \end{pmatrix}\,,\hspace{15pt}
    \mathcal{N}_{ab} = 
    \frac{1}{2}\rho\begin{pmatrix}
        1 & 0\\ 0 & 1
    \end{pmatrix}\,.
\end{align}

Via Eq.~\eqref{lightcone2}, it is easy to see that in the boosted $\mathcal{K}'$ frame, we will have
\begin{align}
    &\mathcal{Q}_{A'B'} = 
    \begin{pmatrix}
        \alpha^2 & 0\\ 0 & \alpha^{-2}
    \end{pmatrix}\,,\hspace{51pt}
    \mathcal{P}_{A'B'} = 
    \begin{pmatrix}
        \alpha^2 & 0\\ 0 & -\alpha^{-2}
    \end{pmatrix}\,,\\[10pt]\nonumber
    &\mathcal{S}_{a'B'} = 
    -\frac{1}{2\sqrt{2}}\begin{pmatrix}
        \alpha\mathcal{Z}_Y & 0 \\
        \alpha^{-1}\mathcal{Z}_Z & 0
    \end{pmatrix}\,,\hspace{15pt}
    \mathcal{N}_{a'b'} = 
    \frac{1}{2}\rho\begin{pmatrix}
        1 & 0\\ 0 & 1
    \end{pmatrix}\,.
\end{align}
Doing the proper matrix addition, the non-zero components of the $\bar{\boldsymbol{T}}$ tensor will be
\begin{align} \label{TbarPrime}
    &\bar{T}_{u'u'} = \frac{1}{2}(\rho + \mathcal{Z}_X)\alpha^2\,,\hspace{15pt}\bar{T}_{v'v'} = \frac{1}{2}(\rho - \mathcal{Z}_X)\alpha^{-2}\,,\\\nonumber
    &\bar{T}_{u'a'} = -\frac{1}{2\sqrt{2}}\alpha\mathcal{Z}_{a}\,,\hspace{15pt}\bar{T}_{v'a'} = -\frac{1}{2\sqrt{2}}\alpha^{-1}\mathcal{Z}_a\,,\hspace{15pt}\bar{T}_{a'b'} = \frac{1}{2}\rho\delta_{a'b'}\,.
\end{align}

\subsection{Ultrarelativistic Limit of the Source} \label{s222}

It is yet to be seen how the source quantities will transform under the ultrarelativistic limit as we take $\text{v}\to 1$, and correspondingly $\alpha\to\infty$. As previously mentioned, we invoke the guiding principle that the energy of the source should remain fixed in this limit.  Using this principle, one can immediately write down the rescaling
\begin{equation} \label{sourceLim}
    \hat{\rho} = \alpha\rho\,,\hspace{15pt}\hat{j}_a = \alpha j_{a}\,,\hspace{15pt}\hat{j}=j\,,
\end{equation}
and require that these "hatted" functions remain constant under this rescaling.  We can see that the first condition in Eq.~\eqref{sourceLim} corresponds exactly to fixing the energy, since
\begin{equation}
    E=\lim_{\text{v}\to 1}\frac{1}{2}\int\hat{\rho}\,d^3X = \lim_{\text{v}\to 1}\frac{1}{2}\alpha M=\gamma M\,,
\end{equation}
where we use the following asymptotic expansion as $\text{v}\to1$ to identify $\alpha\approx2\gamma$ in the leading order
\begin{equation}
    \alpha = 2\gamma\bigg(1-\frac{1}{4}\gamma^{-2}-\frac{1}{16}\gamma^{-4}+O(\gamma^{-6})\bigg)\,.
\end{equation}
As for the rescaling of the axial angular momentum density vector components that are orthogonal to the $X$-direction, $j_a$, one can imagine that a similar line of reasoning holds.  For completeness, we write $j$ as $\hat{j}$ to denote that it is also a fixed variable.

Next, we will want to show that any remaining $O(\alpha)$ terms in Eq.~\eqref{TbarPrime} will correspond to an impulse that any passing null ray experiences as it transverses the Lorentz-contracted plane of the source.  Heuristically, this makes sense, but we still need to develop a sound mathematical mechanism for this, as we don't yet know what prefactors might come along with this impulse (for example the delta function of a normalized Gaussian comes with a prefactor of $(2\pi)^{-1/2}$).

To show this, let's consider a function $f(X,Y,Z)$ with compact support in a sphere of radius $L$ from its center of mass, and a separate function $B(u)$ that changes slowly as a function of $u$, so that we can write
\begin{equation}
    \text{supp}(f)=\{r\equiv\sqrt{X^2+Y^2+Z^2}\leq L\}\,,\hspace{15pt} B(u) = B(0) + \frac{dB}{du}\Big|_{u=0} \;u+\,...\,\,.
\end{equation}
Then, if we consider the following integral along $v=0$
\begin{equation}
    I(Y,Z) = \int \alpha f(X,Y,Z) B(u)\,du\,,
\end{equation}
and use Eq.~\eqref{lightcone2} to make a coordinate transformation from $u$ to $X$, we have in the leading order
\begin{equation}
    I(Y,Z) \approx -\sqrt{2}\,\tilde{f}(Y,Z)\, B(0)\,,\hspace{15pt}\tilde{f}(Y,Z) \equiv \int f(X,Y,Z)\,dX\,.
\end{equation}
Therefore, we have the following identification
\begin{equation} \label{deltaSub}
    \alpha f(X,Y,Z) \longrightarrow -\sqrt{2}\,\delta(u)\,\tilde{f}(Y,Z)\,.
\end{equation}

Denoting
\begin{equation} \label{ul}
    \hat{\mu} = \int (\hat{\rho}+\hat{\mathcal{Z}}_X)\,dX\,,\hspace{15pt}\lambda_a = e_{ab}\int j^{,b}\,dX\,,
\end{equation}
and utilizing Eqs.~\eqref{sourceLim} and \eqref{deltaSub}, we obtain the following ultrarelativistic limits on the components of the stress-energy tensor in the $\mathcal{K}'$ frame
\begin{equation}
    \bar{T}_{u'u'}\approx -\frac{1}{\sqrt{2}}\delta(u)\,\hat{\mu}(Y,Z)\,,\hspace{15pt}\bar{T}_{u'a'} \approx \frac{1}{2}\delta(u)\,\lambda_{a}(Y,Z)\,.
\end{equation}
All other components should vanish under this limit.  One quick note to make is that we will continue to interchange $a'$ and $a$ as indices, as they both belong to $\Pi_2$, which is not affected by our chosen Lorentz boost.

Now that we know how the traceless stress-energy tensor $\bar{\boldsymbol{T}}$ of an arbitrary compact source transforms under the ultrarelativistic limit, we can make an Ansatz for our metric under such a limit.

\subsection{Ultrarelativistic Limit of the Metric} \label{s223}

We know that the background metric in lightcone coordinates in the $\mathcal{K}'$ frame should be
\begin{equation} \label{lightBackground}
    \eta'_{\mu\nu}dx'^{\mu}dx'^{\nu} = -2\,du\,dv + dy^2 + dz^2\,.
\end{equation}
Then, rewriting Eq.~\eqref{poisson} to shift the "bar" back onto the stress-energy tensor, and noting that the only dependence of the functions $u$ and $v$ in this tensor comes from a $\delta(u)$, we can simplify the equation to
\begin{equation}
    ^{(2)}\Delta h'_{\mu\nu} = -16\pi\bar{T}'_{\mu\nu}\,,
\end{equation}
where $^{(2)}\Delta$ is the Laplacian in the $\Pi_2$ spacetime, given by
\begin{equation}
    ^{(2)}\Delta \equiv \partial^2_y + \partial^2_z\,.
\end{equation}
Therefore, we can guess the form of the total metric in the boosted $\mathcal{K}'$ frame to be
\begin{equation} \label{PenMetric}
    ds'^2 = -2 \,du\,dv + dw_a\,dw^{a} + \Phi(u,w)\,du^2 + 2A_a(u,w)\,du\,dw^{a}\,,
\end{equation}
where for the rest of this chapter we employ the shorthand that $dw^a=(dy,\,dz)$, and $w=(y,\,z)$.

Looking at the form of $\bar{\boldsymbol{T}}'$, the perturbed metric functions can be broken down further into
\begin{equation}
    \Phi(u,w) = \delta(u)\,F(w)\,,\hspace{15pt}A_{a}(u,w) = \delta(u)\,\mathcal{A}_a(w)\,,
\end{equation}
leading to the set of two equations that determines the system
\begin{equation}
    ^{(2)}\Delta F = 8\sqrt{2}\pi\hat{\mu}\,,\hspace{15pt}^{(2)}\Delta \mathcal{A}_a = -8\pi\lambda_a\,.
\end{equation}
These equations can be readily solved by the method of Green's functions.  For the two-dimensional flat Laplacian, we have the following equation
\begin{equation}
    ^{(2)}\Delta G(w,w') = \delta^2(w-w')\,,
\end{equation}
which gives the Green's function
\begin{equation}
    G(w,w') = \frac{1}{4\pi}\ln(|w-w'|^2/C^2)\,,
\end{equation}
where $|w-w'|^2=|y-y'|^2+|z-z'|^2$, $d^2w=dy\,dz$, and $C$ is an arbitrary constant of integration.  Therefore,
\begin{align} \label{FA}
    &F(w) = 2\sqrt{2}\int\ln(|w-w'|^2/C^2)\hat{\mu}(w')\,d^2w'\,,\\\nonumber
    &\mathcal{A}_a(w) = -2\int\ln(|w-w'|^2/C^2)\lambda_a(w')\,d^2w'\,.
\end{align}
These relations fully define the metric for the "squeezed" two-dimensional compact mass with arbitrary mass distribution and angular momentum distribution that lies in a null plane.

In the next section, we will calculate the scattering of null rays by the infinitely-thin impulse plane in the boosted frame $\mathcal{K}'$. In order to extract certain values related to this scattering and compare them with already-known values, we will need to "undo" the ultrarelativistic limit on the boosted variables in order to bring them back to the rest frame $\mathcal{K}$.  For this purpose, we will need to define the notion of an "inverse ultrarelativistic transform."  If we define the ultrarelativistic limit as the limit of the map
\begin{equation} \label{Penrose}
    \mathcal{P}: (T,X)\to (u,v)\,,\hspace{10pt}\hat{\rho}=\alpha\rho\,,\hspace{10pt}\hat{j}_a = \alpha j_a\,,\hspace{10pt}\hat{j}=j\,,
\end{equation}
as $\alpha\to\infty$, then the inverse ultrarelativistic transform is just the inverse of this map, $\mathcal{P}^{-1}$.  We will return to this inverse map later on.

\section{Null Ray Scattering Via the Ultrarelativistic Limit} \label{s23}

We describe the problem at hand as follows:
\begin{itemize}
    \item We consider null geodesic rays that live in the spacetime defined by the metric given in Eq.~\eqref{PenMetric}.
    \item For any ray that passes through the null plane fixed at $u=0$, it will be scattered as it passes through the three-dimensional null hypersurface corresponding to the gravitational influence of the infinitely Lorentz-contracted massive object.  We denote this plane by $\Gamma$.
    \item If the spacetime in question is denoted by $M$, then $\Gamma$ splits $M$ into a "past" domain $M_-$ and a "future" domain $M_+$.  For a geodesic ray denoted by $\gamma$, this means that $\gamma$ is also split into $\gamma_-$ and $\gamma_+$.
    \item The wordline of $\gamma$ can be described by $x'^{\mu}=(u(\lambda),\,v(\lambda),\,w^{a}(\lambda))$, where $\lambda$ is the affine parameter of the ray.
    \item Inspecting Eq.~\eqref{PenMetric}, one can see that $v$ is a cyclic variable, and so $\partial_v$ is a Killing vector of the metric.  This means that $\mathcal{U}\equiv du/d\lambda$ is a conserved quantity.  We choose the affine parameterization such that $u=\lambda$, and we define a "dot" to be a derivative with respect to $u$ (\emph{i.e.} $\dot{w}^{a}=dw^{a}/du$).
    \item For such a null ray in the given spacetime, it can be defined by 5 constant independent parameters.  This comes from the fact that our ray moves in six-dimensional phase space corresponding to three position variables and three momentum variables, and moreover our trajectories do not depend on the energy of rays, thus reducing the number of variables that each ray depends on by one. 
\end{itemize}

\begin{figure}[!hbt]
    \centering
      \includegraphics[width=0.7\textwidth]{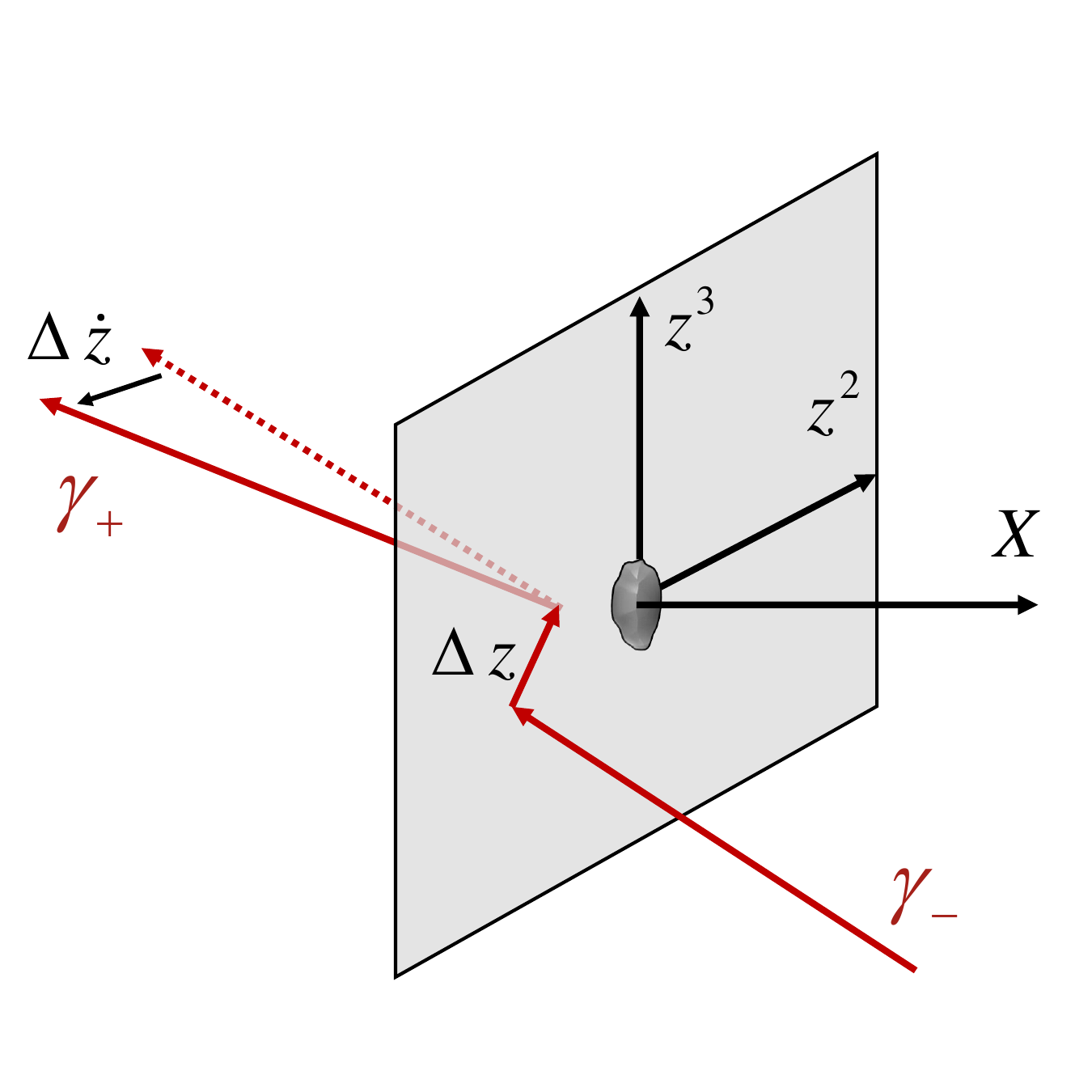}
    \caption{Schematic drawing of null ray scattering by a two-dimensional impulse plane.}
    \label{Fig_2_1}
\end{figure}

A schematic diagram of the above setup is shown in Fig.~\ref{Fig_2_1}.  We will first find the geodesic equations for the spacetime given by Eq.~\eqref{PenMetric}.  We will then substitute our desired setup into these geodesic equations.  Lastly, we will solve for some specific scenarios.

\subsection{Geodesic Equations in the Ultrarelativistic-Limited Metric} \label{s231}

We now derive the geodesic equations for any null ray in the spacetime given by Eq.~\eqref{PenMetric}.  As a reminder, we have
\begin{gather} \label{PenMetricRedux}
    ds'^2 = -2 \,du\,dv + dw_a\,dw^{a} + \Phi(u,w)\,du^2 + 2A_a(u,w)\,du\,dw^{a}\,,\\\nonumber
    \Phi(u,w) = \delta(u)\,F(w)\,,\hspace{15pt}A_a(u,w) = \delta(u)\,\mathcal{A}_a(w)\,.
\end{gather}
It has five independent non-vanishing Christoffel symbols
\begin{align}
    &\Gamma'^{v}_{uu} = -\frac{1}{2}\dot{\delta}F + \delta\dot{\delta}\mathcal{A}_a\mathcal{A}^a-\frac{1}{2}\delta^2\mathcal{A}_a F^{,a}\,,\\\nonumber
    &\Gamma'^{a}_{uu} = \dot{\delta}\mathcal{A}^a - \frac{1}{2}\delta F^{,a}\,,\\\nonumber
    &\Gamma'^{v}_{ua} = -\frac{1}{2}\delta F_{,a}-\frac{1}{2}\delta^2\mathcal{F}\mathcal{A}_b\,e_a^{\;b}\,,\\\nonumber
    &\Gamma'^{b}_{ua} = -\frac{1}{2}\delta\mathcal{F}e_a^{\;b}\,,\\\nonumber
    &\Gamma^{v}_{ab} = -\frac{1}{2}\delta(\mathcal{A}_{a,b} - \mathcal{A}_{b,a})\,,
\end{align}
where $e_{ab}$ is the two-dimensional Levi-Civita symbol with $e_{23}=1$, and
\begin{equation}
    \mathcal{F} = e^{ab}\mathcal{A}_{a ,b}\,.
\end{equation}

We now turn to the geodesic equations
\begin{equation} \label{geodesic}
    \frac{d^2x'^{\mu}}{d\lambda^2}+\Gamma'^{\mu}_{\alpha\beta}\frac{dx'^{\alpha}}{d\lambda}\frac{dx'^{\beta}}{d\lambda}=0\,,
\end{equation}
with the null constraint condition
\begin{equation} \label{nullray}
    g'_{\mu\nu}\frac{dx'^{\mu}}{d\lambda}\frac{dx'^{\nu}}{d\lambda} = 0\,.
\end{equation}

As we previously stated, since $\partial_v$ is a Killing vector of our metric, $du/d\lambda=const.$, and we can use affine parameterization to set $u=\lambda$ and denote derivatives as $\dot{(\,\,)}=d(\,\,)/du$.  This means that out of the four geodesic equations from Eq.~\eqref{geodesic}, the equation corresponding to $u$ is trivial.  This leaves us with the following three equations
\begin{align} \label{PenroseGeo}
    &\ddot{w}_a + \delta\mathcal{F}e_{ab}\dot{w}^b+\dot{\delta}\mathcal{A}_a-\frac{1}{2}\delta F_{,a} = 0\,,\\\nonumber
    & 2\dot{v} = \dot{w}_a\dot{w}^a + \delta F + 2\delta\mathcal{A}^{a}\dot{w}_a\,.
\end{align}
One can see that the vector equation does not depend on the scalar equation, so after solving the vector equation, one can easily find the solution to the scalar equation, giving a consistent and well-defined set of differential equations.

\subsection{General Solution to the Scattering Problem} \label{s232}

As previously discussed at the beginning of this section, our solutions should be of the form
\begin{equation}
    w_{\pm}^a(u) = c_{\pm}^a + \dot{c}_{\pm}^a\,u\,,\hspace{15pt} v(u) = b_{\pm} + \dot{b}_{\pm}\,u\,,
\end{equation}
where $\{b_{\pm},\,\dot{b}_{\pm},\,c_{\pm}^a,\,\dot{c}_{\pm}^a\}$ is a set of six constants that fully determine the system, along with the constraint given by the scalar equation in Eq.~\eqref{PenroseGeo}, which reduces the rank of this solution set to five, enforcing that the constraints are related by
\begin{equation} \label{constraint}
    \dot{b}_{\pm} = \frac{1}{2}c_{\pm}^a c_{\pm a}\,.
\end{equation}

One can imagine that for rays obeying Eq.~\eqref{geodesic} which are freely traveling through spacetime in $M_-$ and $M_+$, but are affected by an impulse at $\Gamma$, their equations of motion will be of the form
\begin{equation} \label{wv}
    w_+^a(u) = w_-^a(u) + \Delta c^a\,\theta(u) + \Delta\dot{c}^{a}\,u\,\theta(u)\,,\hspace{15pt}v_+(u) = v_-(u) + \Delta b\,\theta(u)\,,
\end{equation}
where $\theta(u)$ is the Heaviside function, and $\Delta c^a = c_+^a - c_-^a$, etc.  Since the Dirac delta function is known to be the derivative of the Heaviside function (ignoring the technicalities of defining distributions and the meaning of their derivatives), we readily derive that
\begin{align}
    &\dot{w}_+^a(u) = \dot{w}_-^{a}(u) + \Delta c^{a}\,\delta(u)+\Delta\dot{c}^{a}\,\theta(u)\,,\\\nonumber
    &\ddot{w}_+^{a}(u) = \ddot{w}_-^a(u) + \Delta c^{a}\,\dot{\delta}(u)+\Delta\dot{c}^{a}\,\delta(u)\,,\\\nonumber
    &\dot{v}_+(u) = \dot{v}_-(u) + \Delta b\,\delta(u)\,.
\end{align}

Moreover, since we are working in the weak-field approximation, we can enforce that the constants $\Delta b$, $\Delta c^a$, $\Delta\dot{c}^a$, as well as the source functions $F(w)$ and $\mathcal{A}_a(w)$, are small parameters.  Therefore, substituting Eq.~\eqref{wv} into Eq.~\eqref{geodesic}, we can ignore any terms where the product of two small parameters appears.  With this in mind, we arrive at the following equations of motion at leading order in the small parameters
\begin{equation} \label{PenroseEOM}
    \Delta\dot{c}^a = \frac{1}{2}F^{,a}-\mathcal{F}e^a_{\;b} c_-^b\,,\hspace{15pt}\Delta{c}^a = -\mathcal{A}^a\,,\hspace{15pt}\Delta{b}=\frac{1}{2}F + \mathcal{A}_a\dot{c}_-^a\,.
\end{equation}
These three equations tell us that, after passing through $\Gamma$, the geodesic rays experience the following changes in their trajectories:
\begin{itemize}
    \item On $\Gamma$ we see a shift in the $w^a$ coordinates given by $\Delta{c}^a$.
    \item There will be a change in direction of motion within $\Pi_2$ after passing through $\Gamma$, which is given by $\Delta\dot{c}^a$.
    \item There is a shift in $v$ given by $\Delta{b}$, which corresponds to a time-delay effect that one would anticipate for null rays passing by gravitating bodies.
\end{itemize}

We can further simplify the problem by fixing the initial conditions of the rays.  If we take them to be parallel to the $X$-direction when they are in $M_-$, then we should have $w_-^a-=c_-^a=const.$  With this in mind, we can also define the notion of an impact parameter $\ell\equiv\sqrt{c_{-a}c_-^a}$.  In the weak-field approximation, we enforce that $\ell\gg M$, \emph{i.e.} that the ray is passing by the compact massive object at a distance that is much larger than some characteristic distance describing the gravitational strength of the object.

Looking at Eq.~\eqref{wv}, we also have that
\begin{equation}
    w_+^a = c_-^a + \Delta{c}^a\,\theta(u)+\Delta\dot{c}^a\,u\,\theta(u)\,.
\end{equation}

If we want to calculate the ultrarelativistic-transformed scattering angles in the rest frame $\mathcal{K}$, we can use simple trigonometry:
\begin{equation}
    \Delta\hat{\theta}^a = -\lim_{u\to\infty}\frac{w_+^a(u)}{X} = \frac{\sqrt{2}}{\alpha}\Delta\dot{c}^a\,.
\end{equation}
Note that this angle tends to zero in the ultrarelativistic Limit $\alpha\to\infty$, but if we perform the inverse ultrarelativistic transform on this function that is defined by the inverse of the map \eqref{Penrose}, we have
\begin{align}
    &\Delta\theta_a = 2\frac{\partial}{\partial w^a} H(w)\Big|_{w^a=w^a_-}\,,\\\nonumber
    &H(w)\equiv\int \ln(|w-w'|^2)\Big[\rho(X',w')+j^a_{\;,a}(X',w')\Big]\,dX'd^2w'\,,
\end{align}
where we have inserted the definitions of $F(w)$ that was given in Eqs.~\eqref{ul} and \eqref{FA}\,, as well as the equation of motion defining $\Delta\dot{c}^a$ given in Eq.~\eqref{PenroseEOM}.  Note that the constant factor $C$ also disappears from the definition of $H(w)$, since it will not contribute to the overall scattering angle, as one would expect.

\subsection{Specific Case 1: Scattering Angle} \label{s233}

The first rational check to see if the ultrarelativistic limit really can be used to describe the scattering of null rays by compact massive objects should be to calculate the scattering angle of these rays.  In addition to the previous weak-field condition $\ell\gg M$, we also require $\ell\gg L$ so that the multipole expansion \eqref{multipole} is valid.  This allows us to write down the following series expansion
\begin{equation}
    |w-w'|^2 = |w|^2\bigg(1-2\frac{w\cdot w'}{|w|^2}+\frac{|w'|^2}{|w|^2}\bigg) \approx |w|^2\bigg(1-2\frac{w\cdot w'}{|w|^2}\bigg)\,,
\end{equation}
since $|w|\sim\ell$ and $|w'|\leq L$.  Applying this to the natural logarithm,
\begin{equation} \label{lnExpansion}
    \ln(|w-w'|^2)\approx \ln(|w|^2) - 2\frac{w\cdot w'}{|w|^2}\,.
\end{equation}
Differentiating this with respect to $w^a$,
\begin{align}
    &\frac{\partial}{\partial w^a}\ln(|w-w'|^2)\approx 2\frac{w_a}{|w|^2} - \frac{2}{|w|^4}q_{ab}w'^b\,,\\\nonumber
    &q_{ab} = |w|^2\delta_{ab} - 2w_aw_b\,.
\end{align}

Recall that the Poincar\'{e} symmetry of the background metric gives us Eqs.~\eqref{poincare1} and \eqref{poincare3}.  In the formalism that we have developed thus far, they can be rewritten as
\begin{equation}
    \int\rho\,d^3V=M\,,\hspace{15pt}\int j^{a'}_{\;,a'}\,d^3V = 0\,,\hspace{15pt} \int w'_{a}\,j^{b'}_{\;,b'}\, d^3V = -J_{Xa}\,,
\end{equation}
where $d^3V=dX'd^2w'$, and the last equation is obtained via integration by parts.  Putting this all together, we arrive at the following equation for the scattering angle
\begin{equation} \label{scattering}
    \Delta\theta_a = 4\frac{Mw_a}{|w|^2}+4q_{ab}\frac{J_{Xb}}{|w|^4}\,.
\end{equation}

How does this compare to the scattering angle of null rays in the Kerr metric?  In this case, we let $w^3=Z=0$, and we fix the orientation of the spacetime so that the axial angular momentum is pointing along the $Z$-axis. For this choice, $J_{Xb}=J\delta_b^Y$.  Inserting these two choices into Eq.~\eqref{scattering}, and defining $\Delta\theta_{2}=\Delta\theta_{Y}\equiv\Delta\phi-\pi$, we arrive at the following familiar scattering angle
\begin{equation} \label{DeltaPhi}
    \Delta\phi = \pi + \frac{4M}{\ell} - \frac{4J}{\ell^2}\,.
\end{equation}
This is the weak-field expansion for the scattering angle of a null ray deflected by a Kerr black hole in the equatorial plane\footnote{Let us note that there is an additional term $\sim(M/\ell)^2$ which is not present in in the linearized approximation used above, \eqref{DeltaPhi}, due to its proportionality to $G^2$ in restored SI units. For the Kerr black hole, this term is always larger than the $J/\ell^2$ term, but they can be on the same order for a near-extremal black hole. Nevertheless, the effect of the rotation can be distinguished from others by measuring the difference of the scattering angles for two opposite values of the impact factor $\ell$ and $-\ell$ with the same value for $J$. For extended objects with size $L\gg M$, the term proportional to $J$ can be larger than the $\sim (M/\ell)^2$ contribution.}!  At this point it is insightful to look at \cite{FroZel2011} for the complete derivation of this equation via standard methods for the Kerr metric.  One can see how useful this method of using the ultrarelativistic limit is, as the math breaks down into easily-solvable differential equations for an impulse force, without having to delve into the mess that is elliptical functions.

\subsection{Specific Case 2: Shift Effect} \label{s234}

Another instructive exercise is to think about what happens to a ray that starts off at radial ingoing infinity in the direction parallel to the axial angular momentum vector, passes the compact massive object, and then goes off to radial outgoing infinity.  One can imagine that, due to the frame-dragging of such a rotating object, there will be an azimuthal shift that displaces the ray with respect to its original azimuthal coordinate.  We will proceed with calculating this shift, and then compare it to the known shift effect in the case of a Kerr rotating black hole.

As previously discussed in Sec.~\ref{s232}, the information detailing the magnitude of this shift should be contained in the $\Delta{c}^a$ vector, and therefore can be calculated by looking at the source function $\mathcal{A}_a$. Assuming once more that we are in the weak-field approximation $\ell \gg M$, and that the multipole expansion is valid $\ell \gg L$, we utilize Eq.~\eqref{lnExpansion} to solve for $\mathcal{A}_a$ in Eq.~\eqref{FA}.  We have
\begin{equation}
    \mathcal{A}_a \approx -2e_{ab}\int\Big(\ln(|w|^2)-2\frac{w\cdot w'}{|w|^2}\Big)j^{,b}\,d^3V\,.
\end{equation}
Due to conservation of linear momentum given by Eq.~\eqref{conservations}, the first term in the integral will not contribute, and we calculate the changes in the ray's trajectory purely due to the shift effect to be
\begin{equation}
    \Delta{w}_a = \Delta{c}_a = \frac{4Je_{ab}w^{b}}{|w|^2}\,.
\end{equation}
Note that this shift occurs purely in the $\Gamma$-plane, since it is orthogonal to $w^a$.  Its magnitude is given by
\begin{equation}
    |\Delta{w}^a| = \frac{4J}{b}\,.
\end{equation}
Comparing once more with \cite{FroZel2011}, one can see that this exactly matches the shift effect seen in the Kerr metric.

\subsection{Specific Case 3: Shapiro Time Delay} \label{s235}

Lastly, we will show that in the case of a spherically-symmetric compact matter distribution, \emph{i.e.} one without intrinsic angular momentum, we can reproduce a weak-field limit of the Shapiro time delay given by Eq.~\eqref{shapiro}.

The setup of such a problem is as follows:  we start with a black hole of mass $M$, and assume that a null ray is emitted at some radius $r_0$ from the black hole.  This null ray is then scattered in the presence of the black hole's gravitational field, and arrives at a mirror placed at radius $r_1$.  The ray completes a return journey that is exactly opposite to its initial path, and is absorbed at $r_0$.  We assume that its radius of closest approach is given by $\ell$.  The overall travel time of the ray from the point of view of an observer at $r_0$ is measured, and it is concluded that this travel time is longer than the travel time of a ray that would complete the same path in the absence of the gravitational field of a black hole.  This time delay is attributed to gravitational time dilation from the gravitational field of the massive black hole.

Now, we simplify this setup, and find the weak-field limit of this setup for the Schwarzschild metric.  To this effect, we take $r_0=r_1\equiv R$ for some fixed radius $R$, and we assume that $R\gg \ell$.  Furthermore, we require that $\ell \gg M$. What does this mean in plain terms?
\begin{itemize}
    \item Since $r_0=r_1\equiv R$, the observer and reflecting mirror are at equal radial distance $R$ from the center of the black hole.
    \item By taking $\ell\gg M$, we are requiring that we are in the weak-field limit.  In this limit, the impact parameter coincides with the ray's radius of closest approach, so that $\ell \sim r_m$.
    \item By taking $R\gg\ell$, we are enforcing that both the observer and reflecting mirror are much farther away from the black hole than the radius of the ray's closest approach.
\end{itemize}
With these guiding principles in mind, the time delay equation reduces to the following first-order limit
\begin{equation} \label{ShapiroApprox}
    \Delta\tau\approx 4\sqrt{1-\frac{r_S}{R}}\bigg[R+r_S\ln\Big(\frac{R}{\ell}\Big)\bigg]\,.
\end{equation}
We recall that $r_S=2M$, and note that although it may not be proper in the true sense of an expansion, we maintain the time dilation factor un-expanded in front of the equation so that we preserve our understanding of why it's there.

Now that we have a final formula to compare to, we attempt to derive a first-order approximation of Shapiro time delay using ultrarelativistic limit methods.  As previously noted, the time delay effect is related to the change in the $v$-coordinate.  Looking to Eq.~\eqref{wv}, we have
\begin{equation}
    \Delta{v} = \Delta{b}\,.
\end{equation}
In the case of a source with zero angular momentum, $\mathcal{A}_a=\mathcal{Z}_X=0$, and our equations after applying the inverse ultrarelativistic transform are reduced to
\begin{equation}
    \Delta V= \sqrt{2}\int\rho\ln(|w-w'|^2/C^2)\,d^3V\,.
\end{equation}
Note that taking the ultrarelativistic Limit here also implies that $v\to V$ via Eq.~\eqref{lightcone2}. Now turning towards the above integral, we will see that taking the lowest-order term in the expansion from Eq.~\eqref{lnExpansion} will suffice.  This gives us the very simple answer
\begin{equation} \label{DV}
    \Delta{V}\approx \frac{r_S}{\sqrt{2}}\ln(|w|^2/C^2)\,.
\end{equation}

How do we solve for the time elapsed during this scattering event?  Using Eq.~\eqref{lightcone}, we can write
\begin{equation}
    \Delta T = -\Delta X + \sqrt{2}\Delta{V}\,,
\end{equation}
Now, what should $\Delta{X}$ be?  Since the equations we obtained using ultrarelativistic limit methods are only meant to be used for a single scattering event (and not the reflected path after bouncing off of a mirror), we choose that if the observer lies at $X_0=R$, and the mirror lies at $X_1=-R$, so that $\Delta{X}=-2R$.  This also implies that $|w|=R$ for our specific calculation.  Putting these facts together, Eq.~\eqref{DV} reduces to
\begin{equation}
    \Delta{T} = 2R + 2r_S\ln(R/|C|)\,.
\end{equation}

Only two points remain.  First, if we are considering a scattering event where the initial and final radii are the same, then the only sensible value to choose for the constant of motion $C$ is $C=r_m\sim\ell$, \emph{i.e.} the radius of closest approach.  Second, if we want to find the proper time over the entire event, we must include the gravitational time dilation prefactor that is also present in Eq.~\eqref{ShapiroApprox}, and we need to multiply the entire time duration by two in order to account for the return journey.

All in all, we have
\begin{equation} \label{ShapiroApprox2}
    \Delta\tau\approx 4\sqrt{1-\frac{r_S}{R}}\bigg[R+r_S\ln\Big(\frac{R}{\ell}\Big)\bigg]\,,
\end{equation}
which, as expected, is in exact agreement with Eq.~\eqref{ShapiroApprox}.

As a brief side note, one might go back to Eq.~\eqref{lnExpansion} and ask themself why only the lowest-order term in the expansion is necessary.  The answer lies in the form of Eq.~\eqref{ShapiroApprox2}.  If we were to factor an $R$ out of the bracketed term, then the second term is already small in the given approximation, and any extra terms from Eq.~\eqref{lnExpansion} can be neglected, as we would hope.

\section{The Gravitational Memory Effect} \label{s24}

Say that we have the following two sets of constants that fully constrain the motion of our ingoing null rays as they travel freely through $M_-$, are perturbed by an impulse at $\Gamma$, and then travel freely once more in the outgoing direction through $M_+$:
\begin{equation}
    \psi_{\pm} = \{c^{a}_{\pm},\,\dot{c}^a_{\pm},\,b_{\pm}\}\,.
\end{equation}
The equations in \eqref{PenroseEOM} determine the relation between these "in" and "out" states in such a way, that we could say they define a map
\begin{equation}
    \Psi: \psi_-\to\psi_+\,.
\end{equation}
This map not only contains information about the ultrarelativistic null rays, but also about the compact massive object.  Specifically, it contains nuggets of information about the mass density and angular momentum density of the massive object.

The question becomes: How much can we really learn about the internal structure of this compact massive object simply by sending rays past it and tracing their initial and final states?  In principle, at any fixed order of the multiple expansion of the metric, there should exist a setup of congruences of rays whose input and output information is sufficient to completely and unequivocally define all of the multipole moments of the stress-energy tensor of our compact massive object.  The configuration of null ray congruences necessary to obtain this information is left as an open question for study. 

The above phenomenon is a special case of what is known as the gravitational memory effect, which was discovered and formulated in the following two landmark papers \cite{Zeldovich:1974gvh,Christodoulou:1991me}, and later discussed extensively in \cite{Memory1,Memory2,Memory3,Memory4}. This effect states that gravitational waves passing through spacetime can perturb freely-floating test masses within that spacetime in such a way that they receive a permanent displacement.  In this sense, the gravitational wave imprints a "memory" of its passage onto these test particles.

In general, and as highlighted in the articles cited above, the gravitational memory effect could serve as an important check for verifying Einstein's equations.  It has been proposed as a mechanism for the search of gravitational waves, and even possibly a way of detecting low-energy gravitons \cite{Memory5,Memory6,Memory7}.

\section{Concluding Remarks} \label{s25}

In this chapter, we studied the scattering of light by the gravitational field of an ultrarelativistic massive spinning object. The corresponding metric of such an object was obtained by taking the ultrarelativistic limit of the object’s field in the weak-field approximation. We demonstrated that for spinning matter, it is not sufficient to keep the energy density fixed. One
also needs to perform a proper rescaling of the angular
momentum density components in the directions orthogonal to the direction of motion. We studied the scattering of light rays by the gravitational fields of the ultrarelativistic spinning objects, and constructed the map $\Psi$ relating the parameters of ingoing and outgoing trajectories in terms of the matter and spin distributions. 

We showed that the constructed map $\Psi$ can be used to calculate scattering angles for light propagating in the vicinity of a massive spinning object in its own rest frame. The latter result allows one to recover the standard results for gravitational lensing and Shapiro time delay, and obtain their generalization in the case where the matter producing this effect can have local differential rotation.  This framework gives a fresh new look at an old problem, which may be extended to more complicated cases that cannot be solved in the original framework.  It also highlights a specific case of the gravitational memory effect, which is an important object of study in the realm of tests of general relativity and its relation to a quantum theory of gravity.

	\chapter{Black Holes in Homogeneous Scalar Fields}\label{ch3}

The Kerr metric describes a rotating black hole in vacuum. However, this model is quite "bare-bones" if one wishes to study astrophysical black holes.  Astrophysical black holes that are observed in nature are known to be surrounded by matter, such as accretion disks, which can be absorbed by the black hole.  This is typically a very involved problem that requires complicated models and powerful computers to solve.  However, under certain assumptions, one can construct models of black hole evolution within both scalar and electromagnetic fields that have relatively simple analytical solutions.  In the next two chapters, this is what we set out to do.

We first turn our attention to the study of the motion of rotating black holes in homogeneous scalar fields.  The study of scalar fields may be important when it comes to both describing cosmological models of the universe and proposing modified theories of general relativity.  In the case of cosmology, a scalar field known as the "inflaton field" is often used as a mechanism to drive inflation in the early universe.  Scalar fields can also be used to describe phase transitions via spontaneous symmetry breaking.  This has far-reaching implications, as was seen by the discovery of the Higgs boson, the scalar particle which endows particles in the standard model with mass \cite{HiggsTh,HiggsEx}.  The importance of this particle cannot be understated, and is easily one of the most important discoveries of the twenty-first century.

An interesting and widely-discussed class of models of scalar fields are theories in which the gradient of the scalar field attains a nonzero constant value. Some pertinent examples of these recently-proposed scalar field models that lead to new and exciting physics are:
\begin{itemize}
    \item \textbf{Ghost condensate models:} These models propose a mechanism for inflation in the early universe via a shift-invariant scalar field that gives rise to spontaneous breaking of Lorentz-invariance when coupled with matter \cite{ArkaniHamed1,ArkaniHamed2}. There are very strict limits imposed by high-precision experiments on how large this symmetry-breaking could be, but nevertheless it is still theoretically possible at very small scales (see e.g. \cite{Mattingly2005}).
    \item \textbf{Solid inflaton model:} This model uses a "solid" scalar field with a spacelike gradient to describe early-universe inflation.  It is unique in that it preserves time-translation invariance \cite{Endlich2012}.
    \item \textbf{Emergent time models in Euclidean space:} These models work in Euclidean space and introduce a scalar "clock field," which under the right conditions can produce emergent Lorentzian spacetime \cite{Moffat:1992bf,Babichev:2007dw,Mukohyama:2013ls,Mukohyama:2013em}. Within this framework, it was recently shown that the Big Bang could be thought of as nothing more than Lorentzian "pockets" emerging from an overall Euclidean space \cite{Feng:2026em}.
    \item \textbf{Khrono-metric model:} This model can be thought of as a low-energy limit of Ho\v{r}ava gravity, which foliates spacetime and indicates a preferred time direction via the introduction of a scalar field known as the "khronon" \cite{Blas:2010hb,Kovachik:2023fos}. This theory also violates Lorentz invariance, and is closely related to Einstein-Aether gravity, which denotes a preferred direction of time via a timelike vector field $\boldsymbol{u}$, called the "aether" field \cite{Jacobson:2000xp,Jacobson:2007veq,Oost:2018ea}.
\end{itemize}

In the above models and other models similar to them, one often considers a scalar field $\Psi$ with a constant homogeneous field gradient. Looking at the above examples, it is also clear that the choice of such a gradient $\nabla\Psi$ plays a crucial role in determining what new physics can arise from its inclusion into well-established physical models.  Although the motion of a black hole in a homogeneous scalar field with a constant \emph{timelike} gradient was recently studied in \cite{Frolov:2024sf}, the above discussion motivates us to look at all three possible cases in our examination of how a rotating black hole interacts with a homogeneous scalar field.  Namely, we allow the constant gradient of our homogeneous scalar field to be timelike (which we will call a "$T$-field"), spacelike (which we will call an "$S$-field"), and null (which, you guessed it, we will call an "$N$-field").

In accordance with the famous uniqueness theorems for idealized black holes \cite{Israel:1967nh,Israel:1967za,Carter:1971bh}, we assume that a neutrally-charged black hole can be fully described by its mass $M$ and overall angular momentum $J$ as it travels through spacetime in a homogeneous scalar field.  We now look to study how these quantities evolve under the interaction of the black hole with the homogeneous scalar field, and we search for the black hole's equations of motion under such a setup.

\section{Homogeneous Scalar Fields in Flat Spacetime} \label{s31}

In what follows, we make the following simplifying assumption: We consider a solution of the source-free equation for a massless scalar field with a constant gradient in flat spacetime. We first need to discuss how a scalar field of constant gradient behaves in flat Minkowski spacetime, and see how it transforms under Lorentz transformations.  We start with Minkowski space in Cartesian coordinates

\begin{equation} \label{MinkCh3}
    \eta_{\mu\nu}d\tilde{X}^{\mu}d\tilde{X}^{\nu} = -d\tilde{T}^2 + d\tilde{X}^2 + d\tilde{Y}^2 + d\tilde{Z}^2\,.
\end{equation}

We consider a scalar field $\Psi$ with the following action
\begin{equation} \label{Sphi}
    S_{\Psi} = -\frac{1}{2}\int \sqrt{-\eta}\, \eta^{\mu\nu}\partial_{\mu}\Psi\partial_{\nu}\Psi\,d^4\tilde{X}\,,
\end{equation}
which can be varied to give us the massless Klein-Gordon equation
\begin{equation}
    \Box{\Psi} = 0\,.
\end{equation}
A scalar field with a constant field gradient is a special solution of this equation.

Scalar field models that are given by the action \eqref{Sphi} but have constant non-zero field gradients define a specific class of solutions with certain unique properties.  First, a constant non-zero solution for the field gradient picks out a preferred direction in the spacetime, which spontaneously breaks the spacetime's translational invariance. Second, this field will naturally have a non-zero field energy that can carry fluxes of energy and pressure which influence its surrounding environment.

We can calculate the stress-energy tensor of the scalar field via the variational definition
\begin{equation}
    T_{\mu\nu}\equiv \frac{-2}{\sqrt{-g}}\frac{\delta S_{\Psi}}{\delta g^{\mu\nu}}\,.
\end{equation}
Note that we replace $\boldsymbol{\eta}$ with $\boldsymbol{g}$ to signify that our variation with respect to the metric is off-shell. For Minkowski space, this gives
\begin{equation} \label{TPsi}
    T_{\mu\nu} = \Psi_{,\mu}\Psi_{,\nu} - \frac{1}{2}\eta_{\mu\nu}\Psi_{,\alpha}\Psi^{,\alpha}\,.
\end{equation}

As previously stated, we want to consider all three possible cases of its field gradient.  To this effect, we define
\begin{equation} \label{dPsi}
    \nabla_{\mu}\Psi = \Psi_0\begin{cases}
        \,(1,\,0,\,0,\,0) &\text{for $T$-field}\,,\\
        \,(0,\,N_X,\,N_Y,\,N_Z) &\text{for $S$-field}\,,\\
        \,(1,\,N_X,\,N_Y,\,N_Z) &\text{for $N$-field}\,,
    \end{cases}
\end{equation}
where $\vec{N}=(N_X,\,N_Y,\,N_Z)$ is a three-dimensional unit-vector in Minkowski space.  We will call the frame in which the gradient of the scalar field has the above form the "field frame" $\mathcal{K}'$.  We also have the set of Cartesian coordinates in the $\mathcal{K}'$-frame $x^{\mu}=(\tilde{T},\,\tilde{X},\,\tilde{Y},\,\tilde{Z})$.  One can check that the gradient of the homogeneous field has the norm
\begin{equation}
    \nabla_{\mu}\Psi\nabla^{\mu}\Psi = \epsilon\Psi_0^2\,,
\end{equation}
where $\epsilon=\{-1,\,0,\,1\}$ for the $T$-, $N$-, and $S$-fields, respectively.  One can imagine that by a rigid rotation of the frame, we can achieve the following definition for $\Psi$
\begin{equation} \label{Psi}
    \Psi = \begin{cases}
        \Psi_0\tilde{T} &\text{for $T$-field}\,,\\
        \Psi_0\tilde{X} &\text{for $S$-field}\,,\\
        \Psi_0(\tilde{T}+\tilde{X}) &\text{for $N$-field}\,.
    \end{cases}
\end{equation}
Therefore, the stress-energy tensor after this rigid rotation can be calculated as
\begin{equation} \label{SE}
    T_{\mu\nu} = \Psi_0^2\begin{cases}
        \,\frac{1}{2}\text{diag}(1,\,1,\,1,\,1) &\text{for $T$-field}\,,\\
        \,\frac{1}{2}\text{diag}(1,\,1,\,-1,\,-1) &\text{for $S$-field}\,,\\
        \,\delta^{\tilde{T}}_{\mu}\delta^{\tilde{T}}_{\nu}+\delta^{\tilde{X}}_{\mu}\delta^{\tilde{X}}_{\nu}+2\delta^{\tilde{T}}_{(\mu}\delta^{\tilde{X}}_{\nu)} &\text{for $N$-field}\,.
    \end{cases}
\end{equation}

One quick note that we will make is that a canonical version of the $\mathcal{K}'$-frame is not defined for the $N$-field.  To see this, we shift to lightcone coordinates
\begin{equation}
    \tilde{u} = \frac{1}{\sqrt{2}}(\tilde{T}-\tilde{X})\,,\hspace{15pt} \tilde{v} = \frac{1}{\sqrt{2}}(\tilde{T}+\tilde{X})\,,
\end{equation}
which gives us the following representation of the stress-energy tensor
\begin{equation}
    T_{\mu\nu} = 2\Psi_0^2\delta_{\mu}^{\tilde{v}}\delta_{\nu}^{\tilde{v}}\,.
\end{equation}
Imagine now that we decide to make a boost in the $\tilde{X}$-direction, as we did in Eqs.~\eqref{boost}-\eqref{lightcone2}.  Then, for
\begin{equation}
    v=\alpha^{-1}\tilde{v}\,,\hspace{15pt}u=\alpha \tilde{u}\,,\hspace{15pt}\alpha=\sqrt{\frac{1+\text{v}}{1-\text{v}}}\,,
\end{equation}
the stress-energy tensor transforms as
\begin{equation} \label{Talpha}
    T_{\mu\nu} = 2\alpha^2\Psi^{2}_0\delta_{\mu}^{v}\delta_{\nu}^{v}\,.
\end{equation}
This is to be expected, as null fluids should not "experience" a Lorentz boost like a $T$-field or $S$-field would. We should just see a change in its energy density because a boosted observer measuring this energy density sees the spacetime as stretching or shrinking.

Note that these $N$-fields have another interesting property.  In Eq.~\eqref{Psi}, we chose the null fluid to be "left-moving," \emph{i.e.} it travels in the negative $\tilde{X}$-direction assuming a positive arrow of time.  However, we could have just as easily defined a "right-moving" fluid.  Hence, we could encapsulate both of these scenarios by writing
\begin{equation}
    \Psi_N^{\pm} = \Psi_0(\tilde{T}\pm \tilde{X})\,,
\end{equation}
where $\Psi_N^-$ corresponds to a "right-moving" field and $\Psi_N^+$ corresponds to a "left-moving" field.  We can also write the $T$-field solution as a superposition of these two fields
\begin{equation}
    \Psi_T = \frac{1}{2}(\Psi_N^+ + \Psi_N^-)\,.
\end{equation}
This makes physical sense, as we want the energy fluxes of the "left-movers" and "right-movers" to cancel each other out, so that the stress-energy tensor of the $T$-field has no energy fluxes, as seen in the first equation of Eq.~\eqref{SE}.  To the same effect, one could also just as easily write
\begin{equation}
    \Psi_S = \frac{1}{2}(\Psi_N^+ - \Psi_N^-)\,.
\end{equation}

Let us now more explicitly discuss the transformation of the scalar field $\Psi$ under Lorentz transformations.  For the case where $\Psi$ takes on its most general form in Eq.~\eqref{dPsi}, a general Lorentz transformation from the field frame $\mathcal{K}'$ with coordinates $\tilde{X}^{\mu}=(\tilde{T},\,\tilde{X},\,\tilde{Y},\,\tilde{Z})$ to a corresponding frame $\mathcal{K}$ with coordinates $X^{\mu}=(T,\,X,\,Y,\,Z)$ that travels at velocity $\text{v}$ with respect to $\mathcal{K}'$ is given by
\begin{equation} \label{GenLorentz}
    \tilde{T} = \gamma(T + \vec{\text{v}}\cdot\vec{r})\,,\hspace{15pt}\vec{\tilde{r}} = \vec{r}+(\gamma-1)\frac{\vec{\text{v}}\cdot \vec{r}}{\text{v}^2}\vec{\text{v}}+\gamma T\vec{\text{v}}\,.
\end{equation}
Here, $\vec{\tilde{r}}=(\tilde{X},\tilde{Y},\tilde{Z})$, $\vec{r}=(X,Y,Z)$, and $\gamma=(1-\text{v}^2)^{-1/2}$.  We will eventually identify the $\mathcal{K}$-frame as the rest frame of the rotating black hole, and so to this effect, we denote it as the "BH frame."  In the $\mathcal{K}$-frame in the absence of a black hole, the scalar field achieves the form
\begin{equation} \label{PsiBoost}
    \Psi = \Psi_0(\Upsilon_T T + \Upsilon_X X + \Upsilon_Y Y + \Upsilon_Z Z)\,.
\end{equation}
Then, using Eq.~\eqref{GenLorentz} to compare Eqs.~\eqref{dPsi} and \eqref{PsiBoost}, we can find the values of $\{\Upsilon_T,\vec{\Upsilon}\} = \{\Upsilon_T,\Upsilon_X,\Upsilon_Y,\Upsilon_Z\}$ for each field configuration:
\begin{itemize}
    \item $T$-field
    \begin{equation} \label{UT}
        \Upsilon_T = \gamma\,,\hspace{15pt}\vec{\Upsilon} = \gamma\vec{\text{v}}\,.
    \end{equation}
    \item $S$-field
    \begin{equation} \label{US}
    \Upsilon_T = \gamma(\vec{N}\cdot\vec{\text{v}})\,,\hspace{15pt}\vec{\Upsilon} = \vec{N} + (\gamma-1)(\vec{N}\cdot\vec{\text{v}})\frac{\vec{\text{v}}}{\text{v}^2}\,.
    \end{equation}
    \item $N$-field
    \begin{equation} \label{UN}
        \Upsilon_T = \gamma(1+\vec{N}\cdot\vec{\text{v}})\,,\hspace{15pt}\vec{\Upsilon} = \vec{N}+\gamma\vec{\text{v}}+(\gamma-1)(\vec{N}\cdot\vec{\text{v}})\frac{\vec{\text{v}}}{\text{v}^2}\,.
    \end{equation}
\end{itemize}

This section was devoted to the study of a homogeneous scalar field in \emph{flat} spacetime, but what will happen if we place this scalar field in the rest frame of a rotating black hole?  How will it naturally accommodate itself?

\section{Scalar Field in the Black Hole Frame} \label{s32}

To discuss the form of the scalar field in the BH frame, we will first need to talk a bit about the Kerr metric itself, and its asymptotic symmetries in the regime at far radial distances from the black hole.  As a quick refresher of Sec.~\ref{s113}, we recall that the Kerr metric from Eq.~\eqref{Kerr} has the form
\begin{equation} \label{Kerr2}
    ds^2=-\Big(1-\frac{2Mr}{\Sigma}\Big)dt^2+\frac{\Sigma}{\Delta}dr^2+\Sigma d\theta^2+\frac{A\sin^2\theta}{\Sigma} d\phi^2-\frac{4Mra\sin^2\theta}{\Sigma}dtd\phi\,,
\end{equation}
where $A=\Sigma(r^2+a^2)+2Ma^{2}r\sin^2\theta$.  This metric has two commuting Killing vectors $\boldsymbol{\xi}=\partial_t$ and $\boldsymbol{\zeta}=\partial_{\phi}$, and the rotation parameter $a$ is related to the black hole's spin $J$ via $J=Ma$.  This metric harbors an inner and outer horizon, given by
\begin{equation}
    r_{\pm} = M\pm\sqrt{M^2-a^2}\,.
\end{equation}

With that refresher out of the way, and in support of our goal of extracting the asymptotic symmetries of the spacetime, we can write down the form of this metric in the absence of mass.  When $M/r\to 0$, we have
\begin{equation} \label{OblateMink}
    ds^2 = -dt^2 + \frac{\Sigma}{r^2+a^2}dr^2 + \Sigma d\theta^2 + (r^2+a^2)\sin^2\theta d\phi^2\,.
\end{equation}
This is nothing but the Minkowski metric in disguise, the disguise being that we are in oblate spheroidal coordinates. The transformation to ordinary flat Cartesian coordinates is 
\begin{align}
    &T = t\,,\\\nonumber
    &X = \sqrt{r^2+a^2}\sin\theta\cos\phi\,,\\\nonumber
    &Y = \sqrt{r^2+a^2}\sin\theta\sin\phi\,,\\\nonumber
    &Z = r\cos\theta\,.
\end{align}

As noted in Sec.~\ref{s213}, Minkowski space enjoys a global Poincare symmetry, given by 10 independent conserved quantities. Via Noether's theorem: 
\begin{itemize}
    \item 4 quantities correspond to invariance under coordinate translations.
    \item 3 quantities correspond to invariance under spatial axis rotations.
    \item 3 quantities correspond to invariance under Lorentz boosts.
\end{itemize}
In order to discuss changes in energy, linear momentum, and angular momentum of the black hole, we will want to write down the Killing vectors corresponding to translations and rotations in the spacetime given by Eq.~\eqref{OblateMink}.  In flat spacetime, the translational Killing vectors are given simply by $\boldsymbol{\xi}_{(T)}=\partial_T$, $\boldsymbol{\xi}_{(X)}=\partial_X$, etc.  Therefore, after converting to oblate spheroidal coordinates, we get
\begin{align} \label{transKilling}
    &\xi_{(T)}^{\mu}\partial_{\mu} = \partial_t\,,\\\nonumber
    &\xi_{(X)}^{\mu}\partial_{\mu} = \frac{\sqrt{r^2+a^2}}{\Sigma}\cos\phi\big(r\sin\theta\partial_r+\cos\theta\partial_{\theta}\big)-\frac{\sin\phi}{\sqrt{r^2+a^2}\sin\theta}\partial_{\phi}\,,\\\nonumber
    &\xi_{(Y)}^{\mu}\partial_{\mu} = \frac{\sqrt{r^2+a^2}}{\Sigma}\sin\phi\big(r\sin\theta\partial_r+\cos\theta\partial_{\theta}\big)+\frac{\cos\phi}{\sqrt{r^2+a^2}\sin\theta}\partial_{\phi}\,,\\\nonumber
    &\xi_{(Z)}^{\mu}\partial_{\mu} = \frac{(r^2+a^2)\cos\theta}{\Sigma}\partial_r-\frac{r\sin\theta}{\Sigma}\partial_{\phi}\,.
\end{align}
The rotational Killing vectors in flat Cartesian spacetime can be obtained by the following relation
\begin{equation} 
    \zeta_{(X^i)}^{\mu}\partial_{\mu} = e_{ijk}X^{j}\xi_{(X^k)}^{\mu}\partial_{\mu}\,,
\end{equation}
where $e_{ijk}$ is the three-dimensional Levi-Civita tensor in flat Euclidean space with $e_{123}=1$.  Once more, transforming these vectors to oblate spheroidal coordinates, we arrive at
\begin{align} \label{rotKilling}
    &\zeta_{(X)}^{\mu}\partial_{\mu} = \frac{\sqrt{r^2+a^2}}{\Sigma}\sin\phi\big(a^2\sin\theta\cos\theta\partial_r - r\partial_{\theta}\big)-\frac{r\cos\theta\cos\phi}{\sqrt{r^2+a^2}\sin\theta}\partial_{\phi}\,,\\\nonumber
    &\zeta_{(Y)}^{\mu}\partial_{\mu} = -\frac{\sqrt{r^2+a^2}}{\Sigma}\cos\phi\big(a^2\sin\theta\cos\theta\partial_r - r\partial_{\theta}\big)-\frac{r\cos\theta\sin\phi}{\sqrt{r^2+a^2}\sin\theta}\partial_{\phi}\,,\\\nonumber
    &\zeta_{(Z)}^{\mu}\partial_{\mu} = \partial_{\phi}\,.
\end{align}
Lastly, for completeness, we note that there are four more Killing vectors associated with Lorentz boosts in flat Cartesian spacetime, but for our purposes we will not need them.

Note that the Kerr metric in Boyer-Lindquist coordinates given by Eq.~\eqref{Kerr2} is singular at the event horizon $r=r_+$.  To avoid this problem, one can always switch to what are called "incoming Kerr coordinates," which are given by the following coordinate transformation $(t,\phi)\to(v,\psi)$:
\begin{align}
    &v = t + f(r)\,,\\\nonumber
    &f(r) = \int\frac{r^2+a^2}{\Delta}\,dr = r+\frac{2M}{r_+-r_-}\big[r_+\ln(r-r_+)-r_-\ln(r-r_-)\big]\,,\\[5pt]\nonumber
    &\psi = \phi + \phi_0(r)\,,\\\nonumber
    &\phi_0(r) = a\int\frac{dr}{\Delta} = \frac{a}{2\sqrt{M^2-a^2}}\ln\Big(\frac{r-r_+}{r-r_-}\Big)\,.
\end{align}

Now back to the previous question that was stated just before the beginning of this section: How does a scalar field situate itself within the Kerr spacetime?  In more mathematical terms: How do we solve the Klein-Gordon equation for a massless scalar field within this spacetime?  As was detailed in \cite{Frolov2017-wa}, the Kerr spacetime is quite special in that it automatically guarantees the separability of the Klein-Gordon equation for massless scalar fields. We won't get into the details of this, but if we know that we can perform a separation of variables on the massless scalar field $\Psi=\Psi(v,r,\theta,\psi)$, then we we can go ahead and solve four decoupled differential equations, similar in idea to the way that one would solve the Schrodinger equation for the Hydrogen atom by tackling the radial equation and spherical harmonics separately.  

We take this to be true, and propose the following solution which is achieved by our separation of variables
\begin{align} \label{PsiKerr}
    &\Psi = \Psi_0\big(\Upsilon_{T}\mathcal{U}_{T}+\Upsilon_{X}\mathcal{U}_{X}+\Upsilon_{Y}\mathcal{U}_{Y}+\Upsilon_{Z}\mathcal{U}_{Z}\big)\,,\\\nonumber
    &\mathcal{U}_{T} = t-r+f(r)-2M\ln(r-r_-)\\\nonumber
    &\hspace{12pt} = v - r - 2M\ln(r-r_-)\,,\\\nonumber
    &\mathcal{U}_{X} = \sin\theta\big((r-M)\cos\psi-a\sin\psi\big)\,,\\\nonumber
    &\mathcal{U}_{Y} = \sin\theta\big((r-M)\sin\psi+a\cos\psi\big)\,,\\\nonumber
    &\mathcal{U}_{Z} = (r-M)\cos\theta\,.
\end{align}
This solution has the following three properties, which one can verify if they so desire:
\begin{enumerate}
    \item It solves the massless Klein-Gordon equation in incoming Kerr coordinates.
    \item It is regular at the outer (or in this case the "future") event horizon.
    \item At large $r$, it presents the following asymptotics,
    \begin{equation}
        \mathcal{U}_T\approx T\,,\hspace{15pt}\mathcal{U}_X\approx X\,,\hspace{15pt}\mathcal{U}_Y\approx Y\,,\hspace{15pt}\mathcal{U}_Z\approx Z\,.
    \end{equation}
    That is to say, this solution asymptotically solves the Klein-Gordon equation in the flat Minkowski spacetime as we move far away from the black hole, and in this regime the scalar field has a constant gradient.
\end{enumerate}

This leads us to the interpretation that the above solution is simply the deformation of a scalar field with constant gradient in the presence of a massive rotating black hole.

\subsection{Fluxes} \label{s321}

Now that we have established a framework in which we can describe a scalar field with asymptotically constant gradient in the presence of a rotating black hole, we can now calculate its energy, linear momentum, and angular momentum fluxes due to this deformation. The trick for this lies in the utilization of the asymptotic Killing vectors that were defined in Eqs.~\eqref{transKilling} and \eqref{rotKilling}.  If we take the fluxes to be coming out of a surface that encapsulates the black hole, but is very far away from the hole itself, then we can use these asymptotic Killing vectors to obtain precise (but not exact) values for the fluxes.  

Since we are indeed not in the Minkowski spacetime, but instead in the Kerr spacetime in the presence of a scalar field, we stress the following two crucial reminders:
\begin{enumerate}
    \item The asymptotic Killing vectors that would be true Killing vectors in Minkowski space no longer lead to conserved quantities, because we are no longer in Minkowski space.  They are simply used to calculate energy-momentum fluxes.
    \item Even the two Killing vectors that the Kerr spacetime possess do not lead to conservation of energy and axial angular momentum of the black hole, because we also have to take into consideration the presence of the scalar field. The fact that they are Killing vectors in the Kerr spacetime, however, will still lead to some technical advantages in our calculations, as we will see shortly.
\end{enumerate}

To define flux, we must have some notion of currents.  Take $\boldsymbol{\eta}$ to be a vector in the Kerr spacetime in Boyer-Lindquist coordinates.  Then, a contraction with the stress-energy tensor 
\begin{equation}
    K_{\mu} = T_{\mu\nu}\eta^{\nu}\,,
\end{equation}
gives us the current of energy-momentum flow in the $\boldsymbol{\eta}$-direction.  For a two-dimensional surface $S_0$ defined by $t=const.$ and $r=r_0=const.$, we can also define a three-dimensional surface by shifting $S_0$ over a time interval $t\in(t_1,t_2)$.  We call this surface $\sigma$, and the flux of the current $K_{\mu}$ through this surface is given by
\begin{equation}
    K = \int_{\sigma} K_{\mu}\,d\sigma^{\mu}\,,
\end{equation}
where $d\sigma^{\mu}$ denotes the vector normal to $\sigma$, and can be calculated as
\begin{equation} \label{dsigma}
    d\sigma_{\mu} = \frac{1}{6}\sqrt{-g}\,\epsilon_{\mu\beta_1\beta_2\beta_3}\det\bigg(\frac{\partial x^{\beta_i}}{\partial y^j}\bigg)dtd\theta d\phi\,.
\end{equation}
For a derivation of this formula, see e.g. \cite{Nakahara1990-dr}.  Here, $\sqrt{-g}=\Sigma\sin\theta$, $\epsilon_{\mu\beta_1\beta_2\beta_3}$ is the Levi-Civita symbol with $\epsilon_{0123}=1$, and $y^{i}=(t,\theta,\phi)$ are the coordinates on $\sigma$. After explicit calculation, we find that
\begin{equation}
    d\sigma^{\mu} = -\Delta\delta^{\mu}_r\,dt\,d\Omega^2\,,\hspace{15pt}d\Omega^2 = \sin\theta\,d\theta\,d\phi\,.
\end{equation}
Note that the sign of the flux is chosen so that it represents radially-inflowing energy-momentum.

We assume that the Lie derivative $\mathcal{L}_{\boldsymbol{\xi}}K=0$, so that the flux is stationary, and we have
\begin{equation} \label{kappa}
    K = (t_2-t_1)\kappa\,,\hspace{15pt} \kappa = -\Delta\int_{S_0} T_{r\nu}\eta^{\nu}\,d\Omega^2\,.
\end{equation}
Therefore $\kappa$ represents the energy-momentum flux $K$ over some unit interval of time.  Note that if $\boldsymbol{\eta}$ is a Killing vector in the full Kerr spacetime, then $\kappa$ does not depend on the choice of the surface that envelops the black hole, and therefore it does not depend on the choice of $r_0$.  

If we wish to calculate energy flux, then we should recognize that its information is contained in the $T^{i 0}$ components of the stress-energy tensor.  Therefore, we define
\begin{equation} \label{Eflux}
    \mathcal{E} = \Delta\int_{S_0} T_{r\nu}\xi^{\nu}\,d\Omega^2\,,
\end{equation}
where we have obtained an extra negative sign compared to Eq.~\eqref{kappa} due to the lowering of the indices in $T^{i0}$.  Using the definition Eq.~\eqref{TPsi} for the field Eq.~\eqref{PsiKerr}, one obtains
\begin{equation}
    T_{tr} = \partial_t\Psi\partial_r\Psi\,.
\end{equation}
Furthermore, due to the form of the spatial components in Eq.~\eqref{PsiKerr}, only the $\mathcal{U}_T$ component of the field will contribute to the integral.  A quick calculation gives
\begin{equation}
    \partial_t\mathcal{U}_T = 1\,,\hspace{15pt}\partial_r\mathcal{U}_T = \frac{2Mr_+}{\Delta}\,.
\end{equation}
Hence,
\begin{equation}
    \mathcal{E} = 8\pi\Psi_0^2Mr_+\Upsilon_T^2\,.
\end{equation}
As expected, there is no $r_0$-dependence.

For momentum and angular momentum fluxes, we will make use of the asymptotic Killing vectors in Eqs.~\eqref{transKilling} and \eqref{rotKilling} to define
\begin{align} \label{PJ}
    &\mathcal{P}_{(i)} = -\lim_{r_0\to\infty}\bigg(\Delta\int_{S_0} T_{r\nu}\xi^{\nu}_{(i)} \,d\Omega^2\bigg)\,,\\\nonumber
    &\mathcal{J}_{(i)} = -\lim_{r_0\to\infty}\bigg(\Delta\int_{S_0} T_{r\nu}\zeta^{\nu}_{(i)} \,d\Omega^2\bigg)\,,
\end{align}
where we must take the above limit in order to be able to use these asymptotic Killing vectors, with the exception of $\mathcal{J}_{(Z)}$, which contains the Killing vector $\boldsymbol{\zeta}_{(Z)}=\boldsymbol{\zeta}$.

Since these calculations are rather long and dull, one should use a computer algebra system to carry them out.  In our calculations, we use the GRTensor package in Maple \cite{grtensorii}.  To make our lives simpler, it turns out that we can expand the integrand of Eq.~\eqref{PJ} as an asymptotic series in $1/r$, cutting off terms in the first equation after $O(r^{-2})$, and cutting off terms in the second equation after $O(r^{-3})$.  After this is done, we arrive at
\begin{align} \label{Ps}
    &\mathcal{P}_{(X)} = -8\pi\Psi_0^2 M\Upsilon_T\Big(r_+\Upsilon_X + \frac{1}{3}a\Upsilon_Y\Big)\,,\\\nonumber
    &\mathcal{P}_{(Y)} = -8\pi\Psi_0^2 M\Upsilon_T\Big(r_+\Upsilon_Y - \frac{1}{3}a\Upsilon_X\Big)\,,\\\nonumber
    &\mathcal{P}_{(Z)} = -8\pi\Psi_0^2 Mr_+\Upsilon_T\Upsilon_Z\,,
\end{align}
for the momentum fluxes, and
\begin{align} \label{Js}
    &\mathcal{J}_{(X)} = \frac{4}{3}\pi\Psi_0^2 Ma\Upsilon_Z\Big(M\Upsilon_X+\frac{8}{5}a\Upsilon_Y\Big)\,,\\\nonumber
    &\mathcal{J}_{(Y)} = \frac{4}{3}\pi\Psi_0^2 Ma\Upsilon_Z\Big(M\Upsilon_Y-\frac{8}{5}a\Upsilon_X\Big)\,,\\\nonumber
    &\mathcal{J}_{(Z)} = -\frac{4}{3}\pi\Psi_0^2 M^2a\big(\Upsilon_X^2+\Upsilon_Y^2\big)\,,
\end{align}
for the angular momentum fluxes.

\section{Evolution of Black Hole Parameters} \label{s33}

We will now discuss how the two parameters fully describing the black hole, $M$ and $J$ change and evolve in this system, and we will solve for the motion of the black hole as we switch from the BH frame $\mathcal{K}$ to the field frame $\mathcal{K}'$.

\subsection{Black Hole Frame} \label{s331}

Due to the fluxes of energy-momentum, the black hole can be thought of as acted on by a combination of forces and torques.  In the BH frame $\mathcal{K}$, the four-force and torque acting on the black hole are simply given by
\begin{equation}
    \mathcal{F}^{\mu} = (\mathcal{E},\,\mathcal{P}_{(X)},\,\mathcal{P}_{(Y)},\,\mathcal{P}_{(Z)})\,,\hspace{15pt} \vec{\mathcal{T}} = (\mathcal{J}_{(X)},\,\mathcal{J}_{(Y)},\,\mathcal{J}_{(Z)})\,.
\end{equation}

Note that the vector $\vec{N}$ which describes the orientation of the $S$- and $N$-fields has yet to be chosen.  In order to more-properly orient ourselves with respect to these vectors, we wish to write Eqs.~\eqref{Ps} and \eqref{Js} in a more convenient form.  To this effect, we proceed with the typical choice to orient the axial angular momentum vector of the Kerr black hole along the $Z$-axis of the coordinates, so that the spin vector of the black hole $\vec{J}$ is given by
\begin{equation}
    \vec{J} = J\vec{n}\,,\hspace{15pt}J=Ma\,,\hspace{15pt}\vec{n}=(0,\,0,\,1)\,.
\end{equation}
We also employ the following shorthand
\begin{equation}
    \beta = 8\pi\Psi_0^2\,.
\end{equation}

Now, writing the four-force and torque in a more palatable form, we have
\begin{align} \label{ForTor}
    &\mathcal{F}^0 = \beta Mr_+\Upsilon_T^2\,,\hspace{15pt}\vec{\mathcal{F}}=-\beta\Upsilon_T\Big(Mr_+\vec{\Upsilon}-\frac{1}{3}\vec{J}\times\vec{\Upsilon}\Big)\,,\\\nonumber
    &\vec{\mathcal{T}} = \frac{1}{6}\beta\Big[M\vec{\Upsilon}\times(\vec{\Upsilon}\times\vec{J})+\frac{8}{5M}(\vec{\Upsilon}\cdot\vec{J})(\vec{\Upsilon}\times\vec{J})\Big]\,.
\end{align}

With this tidy form of our equations, we can apply them directly to specific $T$-, $S$-, and $N$-field cases for the four-force via Eqs.~\eqref{UT}-\eqref{UN}:
\begin{itemize}
    \item $T$-field
    \begin{align} \label{FT}
        &\mathcal{F}_T^0 = \beta\gamma^2Mr_+\,,\\\nonumber
        &\vec{\mathcal{F}}_T = -\beta\gamma^2\Big[Mr_+\vec{\text{v}}-\frac{1}{3}(\vec{J}\times\vec{\text{v}})\Big]\,.
    \end{align}
    \item $S$-field
    \begin{align} \label{FS}
        &\mathcal{F}_S^0 = \beta\gamma^2Mr_+(\vec{N}\cdot\vec{\text{v}})^2\,,\\\nonumber
        &\vec{\mathcal{F}}_S = -\beta\gamma(\vec{N}\cdot\vec{\text{v}})\bigg[Mr_+\Big(\vec{N}+(\gamma-1)(\vec{N}\cdot\vec{\text{v}})\frac{\vec{\text{v}}}{\text{v}^2}\Big)\\\nonumber
        &\hspace{28pt}-\frac{1}{3}\Big((\vec{J}\times\vec{N})+(\gamma-1)(\vec{N}\cdot\vec{\text{v}})\frac{\vec{J}\times\vec{\text{v}}}{\text{v}^2}\Big)\bigg]\,.
    \end{align}
    \item $N$-field
    \begin{equation} \label{FN}
        \vec{\mathcal{F}}_N = (1+\vec{N}\cdot\vec{\text{v}})\Big(\mathcal{F}_T^{\mu}+\frac{1}{\vec{N}\cdot\vec{\text{v}}}\vec{\mathcal{F}}_S\Big)\,.
    \end{equation}
\end{itemize}

Since the black hole is at rest in the BH frame $\mathcal{K}$, we cannot know how its velocity evolves until we boost to the field frame $\mathcal{K}'$.  However, we can look at how its mass $M$ and spin $\vec{J}$ evolve in the presence of the scalar field.  Specifically, for a black hole with proper time $\tau$,
\begin{equation} \label{dMJ}
    \frac{dM}{d\tau} = \mathcal{F}^0\,,\hspace{15pt}\frac{d\vec{J}}{d\tau} = \vec{\mathcal{T}}\,.
\end{equation}
Via Eq.~\eqref{ForTor}, one can see that for any of the three given field configurations, $\mathcal{F}^0 = \beta Mr_+\Upsilon_T^2\geq 0$, meaning that the mass of the black hole will never decrease.  Moreover, it will only remain constant when $\Upsilon_T=0$. Can we make a similar statement about the magnitude of its spin, $J$?  Taking the following equation
\begin{equation}
    J\frac{dJ}{d\tau} = \frac{1}{2}\frac{dJ^2}{d\tau} = \frac{1}{2}\frac{d\vec{J}^2}{d\tau} = \vec{J}\cdot\frac{d\vec{J}}{d\tau}\,,
\end{equation}
and using Eq.~\eqref{ForTor} on Eq.~\eqref{dMJ}, we see that
\begin{equation} \label{dJdt}
    \frac{dJ}{d\tau} = -\frac{1}{6}\beta M J\big[\vec{\Upsilon}^2-(\vec{\Upsilon}\cdot\vec{n})^2\big]\,,
\end{equation}
meaning that the magnitude of the black hole's spin will never \emph{increase}, and that it will only remain the same when $\vec{\Upsilon}=\Upsilon\vec{n}$, where $\Upsilon=|\vec{\Upsilon}|$.

At this point, we need to make an important clarification.  We are choosing to describe the spacetime via the Kerr metric, but we are also claiming that the black hole's mass and angular momentum are changing with respect to time.  Can both of these things be true?  As long as we are thinking in a perturbative sense, such that $\dot{M}\ll 1$, and $\dot{J}/J\ll 1/M$, then there is nothing wrong with our analysis; so long as its time frame isn't many orders of magnitude larger than the characteristic "relaxation" time $\tau_{relax}\sim M$.  In the above equations, by a "dot" above a variable, we simply mean its derivative with respect to $\tau$.

Let's discuss a bit more in depth how these changes in mass and spin apply to specific field configurations:
\begin{itemize}
    \item $T$-field
    
    The mass of the black hole must always increase, since $\mathcal{F}_T^0>0$ becomes a strict inequality.  The spin of the black hole can remain constant, but only if $\vec{\text{v}}=\text{v}\vec{n}$.

    \item $S$-field

    The mass of the black hole now has the option to remain constant, under the specific condition that $\vec{\text{v}}\cdot\vec{n}=0$. If we uphold the above condition, and in addition require that spin remains constant, then we must also have $\vec{N}=\vec{n}$.

    \item $N$-field

    Similar to the $T$-field case, the mass of the black hole cannot remain the same and must always increase (unless we allowed it to travel at the speed of light, which of course we do not).  For the spin to remain constant, we must require that $\vec{\Upsilon}\times\vec{n}=0$.
\end{itemize}

Before moving on to the field frame $\mathcal{K}'$, we will do a brief thermodynamic analysis of the system at hand.  As we previously mentioned, the black hole's mass can never decrease.  Similarly, since the spin parameter $J$ can never increase, this also means that the event horizon radius $r_+$ can also never decrease (think of the relation between the event horizon radius and the spin parameter $a$).  With these two facts in mind, we conclude that the area of the black hole, $\mathcal{A}=8\pi Mr_{+}$, will also never decrease. 

Since a black hole's entropy is directly proportional to its area, we know that its entropy will never decrease.  Let's prove this via direct calculation.  The second law of black hole thermodynamics is
\begin{equation} \label{Therm2}
    TdS = dM - \omega dJ\,,
\end{equation}
where $\omega = a/2mr_+$ is the black hole's angular velocity at its event horizon, and its entropy is given by $S=\mathcal{A}/4$.  Note that in this definition we also take a dimensionless definition of $S$, where it is divided by Boltzmann's constant $k_B$, and similarly we take temperature $T$ to be in energy units, where it is multiplied by $k_B$.

The temperature of the black hole is proportional to its surface gravity, such that
\begin{equation}
    T = \frac{2\sqrt{M^2-a^2}}{\mathcal{A}}\,.
\end{equation}
We can also rewrite Eq.~\eqref{Therm2} to be an evolution equation of the thermodynamic variables with respect to the black hole's proper time, such that
\begin{equation}
    T\frac{dS}{d\tau} = \frac{dM}{d\tau} - \omega\frac{dJ}{d\tau}\,,
\end{equation}
and substituting in Eqs.~\eqref{dMJ} and \eqref{dJdt} gives
\begin{equation}
    T\frac{dS}{d\tau} = \frac{1}{2}\beta\bigg[2Mr_+\Upsilon_T^2+\frac{1}{6}\frac{Ma^2}{r_+}\Big(\vec{\Upsilon}^2-(\vec{\Upsilon}\cdot\vec{n})^2\Big)\bigg]\,.
\end{equation}
Taking our previous considerations of area of the black hole into account, we see that $dS/d\tau\geq 0$ always.  The only situation in which the entropy can remain the same is if $\Upsilon_T=0$ and $\vec{\Upsilon}\times\vec{n}=0$.  This can only happen for $S$-fields and $N$-fields, meaning that Kerr black holes in the presence of $T$-fields will always have increasing entropy.

For more information on black hole thermodynamics, and a reference from which the above formulas are taken, see e.g. \cite{FroZel2011}.

\subsection{Field Frame} \label{s332}

In order to trace the actual motion of the black hole within the homogeneous scalar field, we will need to go to the $\mathcal{K}'$ frame in order to see how its velocity evolves over time.  If we consider an observer at a distance $L\gg M$ from the black hole, we can think of the black hole as a point-like rotating massive object in this frame, with mass $M$, spin $\vec{J}$, and velocity $\vec{\text{v}}$ with respect to the frame.  Due to its motion with respect to the field frame, it now has four-momentum $P^{\mu}=(M\gamma,\,M\gamma\vec{\text{v}})$, and therefore a corresponding four-force
\begin{equation}
    f^{\mu} = \frac{dP^{\mu}}{d\tau}\,.
\end{equation}
This four-force can be related to our original four-force in the $\mathcal{K}$ frame, $\boldsymbol{\mathcal{F}}$, via the following Lorentz transformation
\begin{equation} \label{fF}
    f^{0} = \gamma\big(\mathcal{F}^0+\vec{\text{v}}\cdot\vec{\mathcal{F}}\big)\,,\hspace{15pt}\vec{f} = \vec{\mathcal{F}}+\gamma\mathcal{F}^{0}\vec{\text{v}}+(\gamma-1)\frac{\vec{\text{v}}\cdot\vec{\mathcal{F}}}{\text{v}^2}\vec{\text{v}}\,.
\end{equation}

We substitute Eqs.~\eqref{FT}-\eqref{FN} into the above equation to find
\begin{itemize}
    \item $T$-field 
    \begin{equation} \label{fT}
        f_T^0 = \beta\gamma Mr_+\,,\hspace{15pt}\vec{f}_T = \frac{1}{3}\beta\gamma^2\vec{J}\times\vec{\text{v}}\,.
    \end{equation}

    \item $S$-field 
    \begin{align} \label{fS}
        &f_S^0 = -\frac{1}{3}\beta\gamma^2(\vec{\text{v}}\cdot\vec{N})\,\vec{J}\cdot\big(\vec{\text{v}}\times\vec{N}\big)\,,\\\nonumber
        &\vec{f}_S = -\beta\gamma(\vec{\text{v}}\cdot\vec{N})\Bigg[Mr_+\vec{N}-\frac{1}{3}\bigg(\vec{J}\times\vec{N}\\\nonumber
        &\hspace{26pt}+\frac{\gamma-1}{\text{v}^2}\Big[(\vec{\text{v}}\cdot\vec{N})\big(\vec{J}\times\vec{\text{v}}\big)+\big(\vec{J}\cdot(\vec{N}\times\vec{\text{v}})\big)\vec{\text{v}}\Big]\bigg)\Bigg]\,.
    \end{align}

    \item $N$-field 
    \begin{equation} \label{fN}
        f_N^{\mu} = (1+\vec{N}\cdot\vec{\text{v}})\bigg(f_T^{\mu}+\frac{1}{\vec{N}\cdot\vec{\text{v}}}f_S^{\mu}\bigg)\,.
    \end{equation}
\end{itemize}

We now look at some special cases for the black hole moving in each possible field configuration, starting with the $T$-field.

\subsection{Black Hole Moving in $T$-field} \label{s333}

If we denote
\begin{equation}
    \vec{P} = M\gamma\vec{\text{v}}\,,
\end{equation}
then from Eqs.~\eqref{dMJ}, \eqref{fF}, and \eqref{fT}, we assess that
\begin{align} \label{MJPT}
    &\frac{dM}{d\tau} = \beta\gamma^2Mr_+\,,\\\nonumber
    &\frac{d\vec{J}}{d\tau} = \frac{\beta}{6M}\bigg[\vec{P}\times\big(\vec{P}\times\vec{J}\big)+\frac{8}{5M^2}(\vec{P}\cdot\vec{J})\big(\vec{P}\times\vec{J}\big)\bigg]\,,\\\nonumber
    &\frac{d\vec{P}}{d\tau} = -\frac{\beta}{3M}\gamma \big(\vec{P}\times\vec{J}\big)\,.
\end{align}
Using some vector algebra identities, it is easy to verify that this set of equations implies
\begin{align}
    &\vec{P}^2 = const.\,,\hspace{15pt} \big(\vec{J}\cdot\vec{P}\big) = const.\,,\\\nonumber
    &\frac{1}{2}\frac{d\vec{J}^2}{d\tau} = -\frac{\beta}{6M}\Big[\vec{P}^2\vec{J}^2-\big(\vec{P}\cdot\vec{J}\big)^2\Big]\,.
\end{align}

Noting that $\gamma\geq 1$ and $r_+\geq M$, if we look at the first equation in Eq.~\eqref{MJPT}, we conclude that
\begin{equation}
    \frac{1}{M^2}\frac{dM}{d\tau} \geq \beta\,.
\end{equation}
Taking the initial condition $M(\tau=0)=M_0$, we can integrate the above inequality to arrive at
\begin{equation} \label{Mineq}
    M(\tau) \geq \frac{M_0}{1-\tau\beta M_0}\,.
\end{equation}
Let's denote a time $\tau_1$ as
\begin{equation}
    \tau_1 = \frac{1}{\beta M_0}\,.
\end{equation}
Then, via Eq.~\eqref{Mineq}, we can deduce that for some time $\tau_0\leq\tau_1$, the mass of the black hole formally approaches infinity.  This property distinguishes the $T$-field case from the $S$-field and $N$-field cases.  For more specific solutions for the $T$-field case, see \cite{Frolov:2024sf}.

We would also like to briefly note the similarities between a black hole immersed in this scalar $T$-field, and a black hole immersed in a bath of thermal radiation. In the latter case, it was also shown that the mass of a black hole can also formally approach infinity in finite time under specific circumstances \cite{Zeldovich:1967lct}.  In an expanding universe, one can expect the accretion of mass by the black hole from the surrounding environment to be slowed, but the way in which this slowing happens is unique to each specific case.  

For a black hole immersed in a bath of thermal radiation, it will only increase in mass if its internal temperature is less than that of the surrounding environment.  Therefore, if a black hole is immersed in a thermal bath in an expanding universe, two mechanisms are at play: (1) the thermal bath will decrease in temperature due to the expansion of the universe, and (2) the black hole will decrease in temperature due to its accretion of mass.  Depending on how fast the universe is expanding, the black hole may cease to grow in mass, and instead start radiating.  For a black hole surrounded by a scalar field with an initially constant field gradient in an expanding universe, the magnitude of the field gradient will also decrease (see e.g. \cite{Turner1983}), slowing the black hole's mass accretion rate.

\subsection{Black Hole Moving in $S$-field} \label{s334}

We will look at two specific cases for the motion of the rotating black hole in the $S$-field: when the black hole motion is transverse to the field gradient, and when it is parallel to the field gradient.

\subsubsection{Black Hole Motion Transverse to the Field Gradient} \label{s3341}

For this case, we will assume that $(\vec{\text{v}}\cdot\vec{N})=0$.  From Eq.~\eqref{fS}, this gives
\begin{equation}
    f_S^0 = 0\,,\hspace{15pt}\vec{f}_S = 0\,.
\end{equation}

This tells us that both the black hole's mass and velocity are constant.  Therefore, it is only left to study how the black hole's angular momentum changes in this scenario.  Eq.~\eqref{US} tells us that $\Upsilon_T=0$ and $\vec{\Upsilon}=\vec{N}$.  Inserting these into the torque equation in \eqref{ForTor} yields
\begin{equation} \label{TorS1}
    \vec{\mathcal{T}}=\frac{1}{6}\beta\bigg[M\vec{N}\times\big(\vec{N}\times\vec{J}\big)+\frac{8}{5M}(\vec{N}\cdot\vec{J})\big(\vec{N}\times\vec{J}\big)\bigg]\,.
\end{equation}
Using vector identities, one can see that
\begin{equation}
    \vec{\mathcal{T}}\cdot\vec{N} = 0\,.
\end{equation}
Therefore, it is sensible to split up the spin vector as
\begin{equation}
    \vec{J} = J_{\parallel}\vec{N} + \vec{J}_{\perp}\,,
\end{equation}
where $J_{\parallel}=const.$

Then Eq.~\eqref{TorS1} takes the form
\begin{equation} \label{JPeq}
    \frac{d\vec{J}_{\perp}}{d\tau} = \frac{1}{6}\beta\bigg[-M\vec{J}_{\perp}+\frac{8J_{\parallel}}{5M}\big(\vec{N}\times\vec{J}_{\perp}\big)\bigg]\,.
\end{equation}
This equation can be solved by defining a coordinate system $y^i=(y^1,y^2,y^3)$ such that
\begin{equation}
    \vec{N}=(0,\,0,\,1)\,,\hspace{15pt}\vec{J}_{\perp} = (J_1,\,J_2,\,0)\,.
\end{equation}
We can then use the trick of switching to complex coordinates, like one would do for a damped oscillator
\begin{equation}
    K = J_1 + iJ_2\,,\hspace{15pt}\Gamma = \frac{1}{6}\beta\bigg(M-i\frac{8J_{\parallel}}{5M}\bigg)\,,
\end{equation}
in which Eq.~\eqref{JPeq} takes the elegant form
\begin{equation}
    \frac{dK}{d\tau} = -\Gamma K\,.
\end{equation}

This is easily solved by
\begin{equation}
    K(\tau) = K_0\exp(-\Gamma\tau)\,.
\end{equation}
The constant $K_0=K(\tau=0)$ represents the initial values of $J_1$ and $J_2$.  Furthermore, we see that the spin of the black hole precesses around the field gradient vector with characteristic frequency $\omega=|\Im(\Gamma)|$, and its magnitude exponentially decreases according to the characteristic time $\tau_{ch}\sim1/\Re(\Gamma)$.

\subsubsection{Black Hole Motion Parallel to the Field Gradient} \label{s3342}

If we now consider black hole motion that is always parallel to the field gradient $\vec{\text{v}}=\text{v}\vec{N}$, then Eq.~\eqref{US} tells us that
\begin{equation}
    \Upsilon_T = \gamma\text{v}\,,\hspace{15pt}\vec{\Upsilon}=\gamma\vec{N}\,.
\end{equation}
Inserting these expressions into our four-force gives us
\begin{equation}
    f_S^0 = 0\,,\hspace{15pt}\vec{f}_S = -\beta\gamma\text{v}\bigg[Mr_+\vec{N}-\frac{1}{3}\gamma\big(\vec{J}\times\vec{N}\big)\bigg]\,.
\end{equation}
If we keep the spin direction $\vec{n}$ general, then it is possible to have a force component in the direction orthogonal to the field gradient.  Therefore, to maintain consistency with the assumption that the black hole motion stays parallel to the field gradient, we also require that $\vec{n}=\vec{N}$.  For this case, we have
\begin{equation}
    \Upsilon_T = \gamma\,,\hspace{15pt}\vec{\Upsilon} = \gamma\vec{N}\,.
\end{equation}
One can check that this implies $\vec{\mathcal{T}}=0$, so that $J=const.$, and so we are left with the following set of equations describing the system that we derive from Eqs.~\eqref{dMJ} and \eqref{fS}
\begin{equation} \label{JPeq2}
    \frac{dM}{d\tau} =  \beta\gamma^2\text{v}^2Mr_+\,,\hspace{15pt}\frac{d(\gamma M)}{d\tau} = 0\,,\hspace{15pt} \frac{d(\gamma M\text{v})}{d\tau} = -\beta\gamma\text{v}Mr_+\,.
\end{equation}

The second equation here tells us that the energy of the black hole is conserved, so if we denote $M\gamma=\mathcal{E}_0$, then we can derive the relation
\begin{equation} \label{MV}
    M = \mathcal{E}_0\sqrt{1-\text{v}^2}\,.
\end{equation}
We can also use this equation, in addition to the third equation in Eq.~\eqref{JPeq2} to derive the following relations
\begin{equation} \label{MVVT}
    \frac{dM}{d\text{v}} = -\frac{M\text{v}}{1-\text{v}^2}\,,\hspace{15pt}\frac{d\text{v}}{d\tau} = -\beta\text{v}r_+\,.
\end{equation}
The first of these two equations allows us to generate a plot in $(M,\text{v})$-space that tells us how mass and velocity of the black hole evolve given some initial point $(M_0,\text{v}_0)$ by tracing the flow lines of this map.  This is seen in Fig.~\ref{Fig_3_1} below.

\begin{figure}[!hbt]
    \centering
      \includegraphics[width=0.7\textwidth]{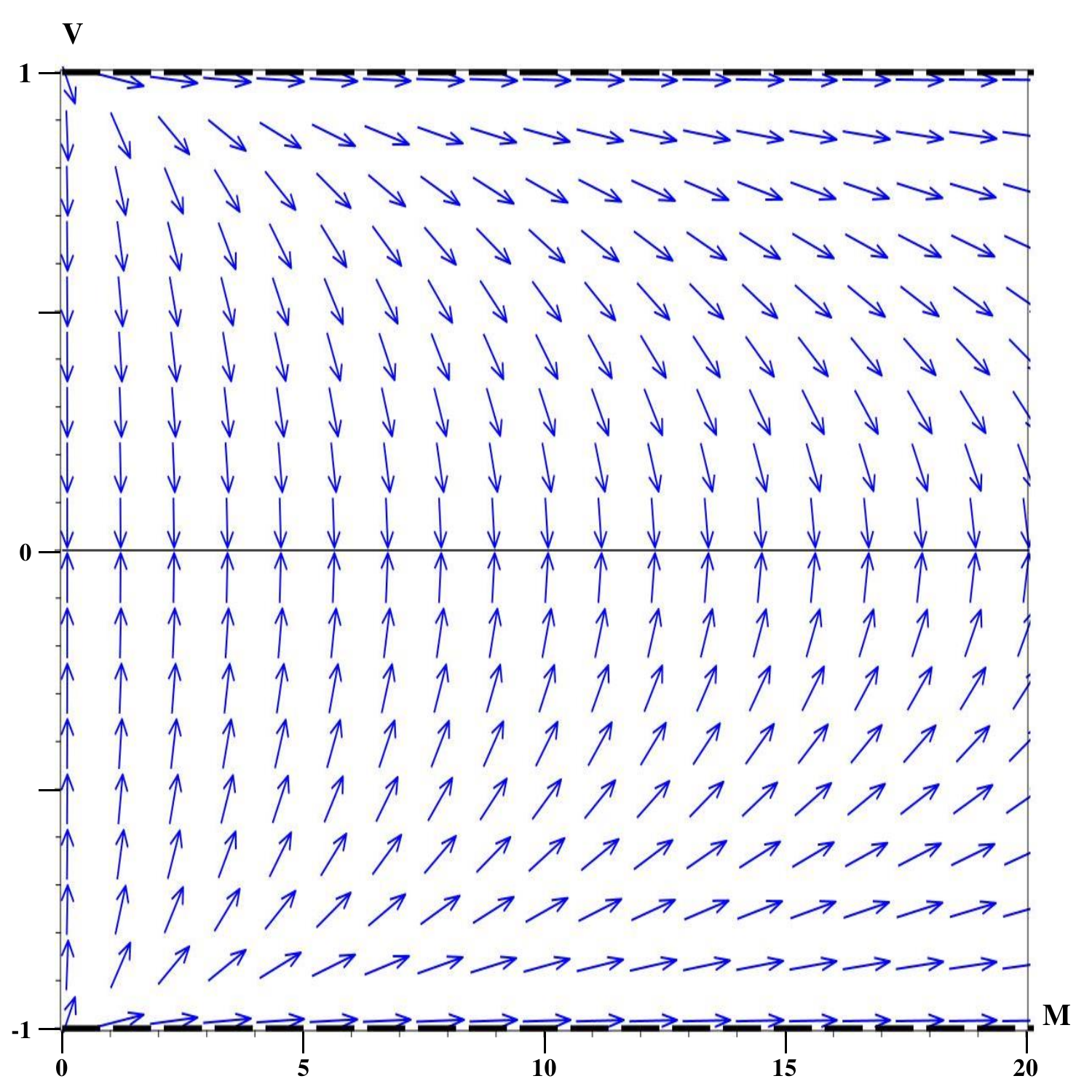}
    \caption{Phase space diagram of $M$ vs. $\text{v}$ for the $S$-solution
with $\vec{\text{v}}=\text{v}\vec{N}=\text{v}\vec{n}$.}
    \label{Fig_3_1}
\end{figure}

Furthermore, we can use Eq.~\eqref{MV} on the second equation in \eqref{MVVT} to fully decouple these variables and obtain a differential equation for velocity with respect to the proper time of the black hole
\begin{equation} \label{dvdtS}
    \frac{d\text{v}}{d\tau} = -\beta\mathcal{E}_0\text{v}\Bigg[\sqrt{1-\text{v}^2}+\sqrt{1-\text{v}^2-\frac{j^2}{1-\text{v}^2}}\;\Bigg]\,,
\end{equation}
where $j=J/\mathcal{E}_0^2=const.$

According to the flow lines of Fig.~\ref{Fig_3_1}, for any given initial mass, the system will tend towards a regime where velocity approaches zero.  With this in mind, it serves to analyze Eq.~\eqref{dvdtS} when $\text{v}\ll 1$.  Perturbing this equation to first order, we have
\begin{equation}
    \frac{d\text{v}}{d\tau} = -\lambda\text{v}+O\big(\text{v}^3\big)\,,\hspace{15pt}\lambda = \beta\mathcal{E}_0\sqrt{1-j^2}\,.
\end{equation}
This tells us that
\begin{equation}
    \text{v}\approx\text{v}_0\exp(-\lambda\tau)\,,
\end{equation}
and so $\text{v}\to 0$ asymptotically as $\tau\to\infty$.  If we include terms up to $O\big(\text{v}^3\big)$, then we obtain a Bernoulli equation, which is still easily solvable and will give us the same asymptotic motion.

\subsection{Black Hole Moving in $N$-field} \label{s335}

For this field configuration, we will consider the simplest case, where once more
\begin{equation}
    \vec{\text{v}} = \text{v}\vec{N} = \text{v}\vec{n}\,,
\end{equation}
so that the black hole is moving parallel to the field gradient, and the field gradient is aligned with the black hole's spin vector.  In this case,
\begin{align} \label{Neqs}
    &\Upsilon_T = \alpha\,,\hspace{15pt}\vec{\Upsilon}=\alpha\vec{N}\,,\hspace{15pt}\mathcal{F}_N^0 = \beta\alpha^2Mr_+\,,\\\nonumber
    &f_N^0 = \beta\alpha Mr_+\,,\hspace{15pt}\vec{f}_N = -\beta\alpha Mr_+\vec{N}\,,
\end{align}
where
\begin{equation}
    \alpha = \sqrt{\frac{1+\text{v}}{1-\text{v}}}\,.
\end{equation}
Under these conditions, $\vec{\mathcal{T}}=0$, and once more $J=const.$

\begin{figure}[!hbt]
    \centering
      \includegraphics[width=0.7\textwidth]{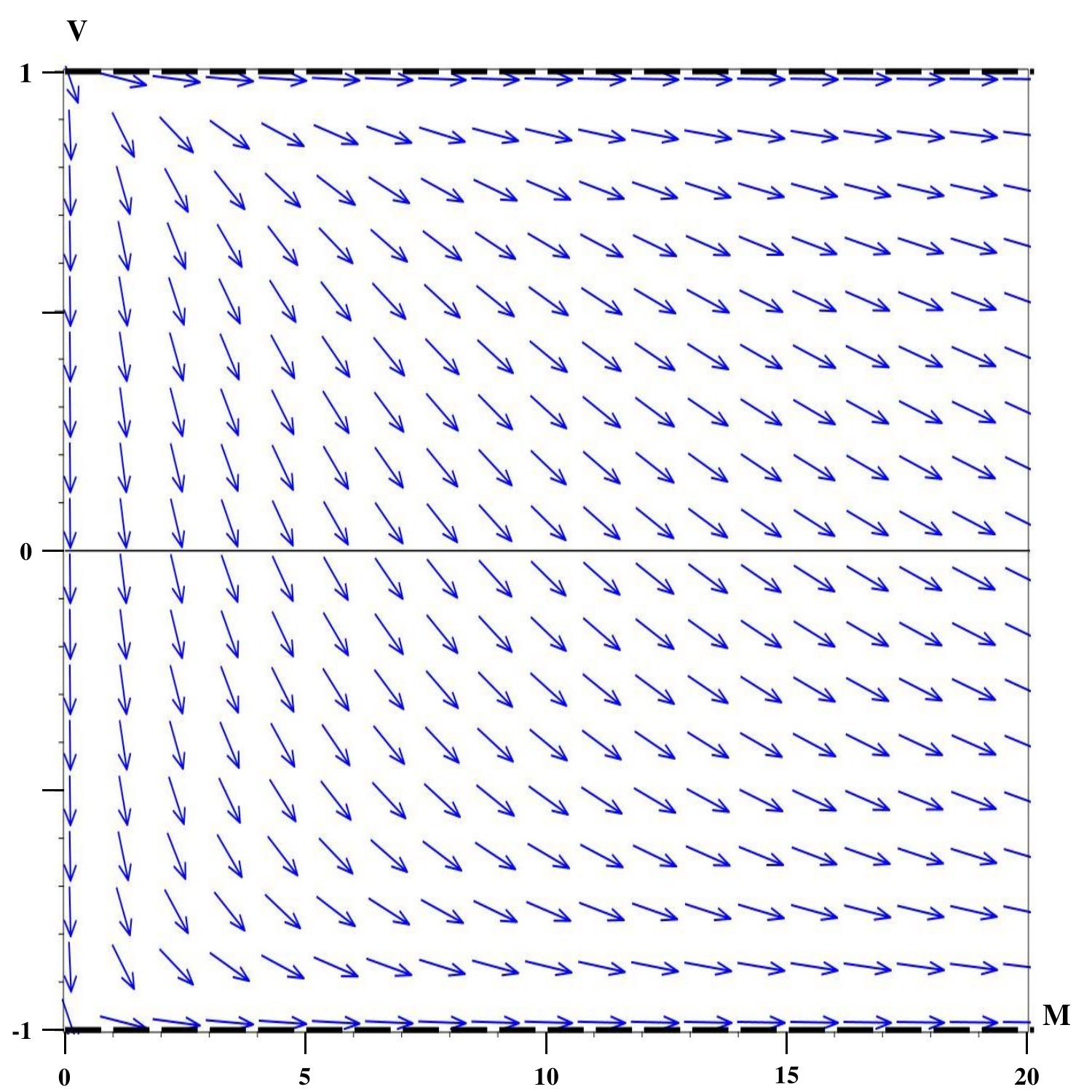}
    \caption{Phase space diagram of $M$ vs. $\text{v}$ for the $N$-solution
with $\vec{\text{v}}=\text{v}\vec{N}=\text{v}\vec{n}$.}
    \label{Fig_3_2}
\end{figure}

From Eq.~\eqref{Neqs}, we can derive
\begin{equation}
    \frac{dM}{d\tau} = \beta\alpha^2 Mr_+\,,\hspace{15pt}\frac{d(\gamma M)}{d\tau} = \beta\alpha Mr_+\,,\hspace{15pt}\frac{d(\gamma M\text{v})}{d\tau} = -\beta\alpha Mr_+\,.
\end{equation}
One can check that only two of these equations are linearly independent, as $\gamma=\gamma(\text{v})$, and so the third equation follows from the first two.  We can therefore search for a set of coupled differential equations between $M$ and $\text{v}$.  After a short search, we arrive at
\begin{equation} \label{MVN}
    \frac{d\text{v}}{d\tau} = -\beta r_+(1+\text{v})^2\,,\hspace{15pt}\frac{dM}{d\tau} = \beta Mr_+\frac{1+\text{v}}{1-\text{v}}\,.
\end{equation}
Once more we can combine these two equations to get a plot in $(M,\text{v})$-space
\begin{equation}
    \frac{dM}{d\text{v}} = -\frac{M}{1-\text{v}^2}\,,
\end{equation}
which is shown in Fig.~\ref{Fig_3_2}.  

This plot shows that for any initial mass-velocity pair $(M_0,\text{v}_0)$, after enough proper time passes $M\to\infty$ and $\text{v}\to -1$, meaning that the mass will asymptotically blow up and the velocity will asymptotically approach the speed of light in the direction opposite to the black hole's spin vector.  If we solve this differential equation for $M(\text{v})$, we find that
\begin{equation} \label{MofV}
    M = M_0\sqrt{\bigg(\frac{1-\text{v}}{1+\text{v}}\bigg)\bigg(\frac{1+\text{v}_0}{1-\text{v}_0}}\bigg)\,.
\end{equation}

Using this equation, along with Eq.~\eqref{MVN}, we can put some bounds on the growth rate of the black hole's mass with respect to its proper time.  Namely,
\begin{equation}
    \frac{dM}{d\tau} = \beta M_0^2\bigg(\frac{1+\text{v}_0}{1-\text{v}_0}\bigg)\frac{r_+}{M}\,.
\end{equation}
Recalling that from the definition of the outer horizon $M < r_+\leq 2M$, we see that
\begin{equation} \label{dMBounds}
    \beta M_0^2\bigg(\frac{1+\text{v}_0}{1-\text{v}_0}\bigg) < \frac{dM}{d\tau} \leq 2\beta M_0^2\bigg(\frac{1+\text{v}_0}{1-\text{v}_0}\bigg)\,.
\end{equation}

If we insert Eq.~\eqref{MofV} into the above equation, and consider the initial conditions $M(\tau=0)=M_0$ and $\text{v}(\tau=0)=0$, then we can put the following bounds on both $M(\tau)$ and $\text{v}(\tau)$
\begin{gather}
    M_0\sqrt{2e^{M_0\tau}-1} < M(\tau) \leq M_0\sqrt{2e^{2M_0\tau}-1}\,,\\[5pt]\nonumber
    e^{-2M_0\tau}-1\leq \text{v}(\tau) < e^{-M_0\tau}-1\,.
\end{gather}

\section{Concluding Remarks} \label{s34}

In this chapter, we discussed effects connected with the motion of a rotating black hole in a static homogeneous scalar field of general configuration. It extends the results of the earlier publication \cite{Frolov:2024sf} to the cases in which the vector of the field gradient is not only timelike, but can also be spacelike or null as well. Another difference from the previous publication is that a complete set of equations is obtained describing not only the change in magnitude of the black hole’s spin, but also its orientation. 

We demonstrated that the mass of a rotating black hole moving in a scalar field cannot decrease, while the magnitude of its spin can never increase. As a result, the surface area (and hence the entropy) of the black hole never decreases as well. This result is in complete agreement with the second law of black hole thermodynamics. Although these results are general and valid for any $T$-, $S$- and $N$-field configurations, there exist very important differences between these cases. In the $T$- and $N$-fields, the mass of the black hole always grows and cannot remain constant.

For motion within the $S$-field, the mass of the black
hole can remain constant under the right conditions. This happens when either the black hole is at rest with respect to the field frame, or when it moves with velocity orthogonal to the direction of the field gradient. In the case where the black hole’s velocity is parallel to the field
gradient, the growth of the black hole’s mass is accompanied by a decrease in its velocity. The final value of the black hole mass is finite and it is determined mainly by its initial energy. It takes infinite time to reach this asymptotic state in which $\text{v}$ = 0. 

For the black hole moving in the N-field, its mass and
the absolute value of the velocity $|\vec{\text{v}}|$ monotonically grows
with time. At late time the velocity $\text{v}$ is negative, meaning that the black hole moves in the negative $X$-direction. In the limit $\tau\to\infty$, the velocity goes as $\text{v}\to -1$. These results
have a simple and natural explanation. The stress-energy tensor of the scalar field in the $N$-state has the form of a left-moving null fluid. The black hole absorbs its energy and momentum, and this results in the growth of its mass $M$ and velocity $|\vec{\text{v}}|$.

The case of the $T$-field is singled out by the following
property: For a black hole with an arbitrary initial velocity (including the case when it is at rest), the mass of the black hole infinitely grows and (at least formally) it reaches an infinite value within a finite interval of time. This property was already discussed in the paper \cite{Frolov:2024sf}. A simple explanation of this phenomenon is given as follows: The rate of change of the black hole’s mass is proportional to its cross-section $\sim M^2$ (see e.g. \cite{Unruh:1976,Das:1997}),
\begin{equation}
    dM/d\tau\sim M^2\,.
\end{equation}
The integral
\begin{equation}
    \tau_f\sim\int^{M_f}\frac{dM}{M^2}\,,
\end{equation}
which shows that the time $\tau_f$ when the mass reaches the
value $M_f$ is finite for $M_f\to\infty$. This means that it takes
the black hole finite time $\tau$ to reach an infinite value
of mass. Let us note that for the $N$-field, the situation is
qualitatively different. Eq.~\eqref{dMBounds} shows that at a
late time, the rate of the change of the black hole mass
is practically constant. The explanation of the difference
with the $T$-field case is that for the $N$-field, when $\text{v}\to-1$, its null fluid energy flux as observed in the black
hole frame is highly suppressed by a Doppler shift (see
\eqref{Talpha}). This effect exactly compensates the effect of the black hole’s cross-sectional growth.

In many aspects, the motion of a black hole in the
$S$-field is similar to the motion of a rotating black hole
in a homogeneous electromagnetic field discussed in the following chapter (or also see \cite{A3}) with one important difference: In the latter case the mass of the black hole never changes.

In our discussion, we considered the simplest version of a scalar field. Namely, we assumed that it is a minimally coupled massless field obeying a linear equation. It would be interesting to use the developed approach and apply it to the study of the interaction of rotating moving black holes with more complicated fields, such as ghost condensate, khronon, and aether fields.

	\chapter{Black Holes in Homogeneous EM Fields}\label{ch4}

An interesting hypothesis explaining why magnetic fields are present in galaxies and galaxy clusters is that they have primordial origins stretching all the way back to the early universe \cite{Grasso:2000wj}.  While this hypothesis is yet to be confirmed, current and upcoming telescope array projects, such as the Hydrogen Epoch of Reionization Array (HERA), which is already operational, and the Square Kilometre Array (SKA), which is expected to start relaying data in 2027, will likely be able to prove or disprove the existence of such primordial magnetic fields (PMFs) \cite{Worku:2026sbp}.  If these PMFs were to exist, they would surely have an impact on the evolution of hypothesized primordial black holes.

Furthermore, the existence of primordial black holes (PBHs) could aid in resolving another long-standing problem of the early universe: the magnetic monopole problem.  It was first shown by t'Hooft that if the universe is described by some grand unified theory (GUT) that merges the three fundamental interaction forces within the standard model, then magnetic monopoles should arise as solutions to the field equations of these theories \cite{tHooft:1974kcl}.  Five years later, it was shown that the amount of monopoles that would be produced in the early universe by such a GUT is "unacceptably large," in the sense that it would be in stark contradiction with modern cosmological models \cite{Preskill:1979zi}.  The standard hypothesis which resolves the monopole problem is that rapid cosmic inflation heavily diluted the magnetic monopole density of the universe to values which make a detection event extremely improbable \cite{Guth:1980zm}.  However, another proposed solution to this problem is that, if PBHs did exist in the early universe, then they could absorb these magnetic monopoles and convert their energy into radiation via Hawking evaporation \cite{Stojkovic:2004hz}.

If one assumes that at some point in the early universe, primordial black holes could briefly retain magnetic charge, then a natural question to ask is: could they retain electric charge as well?  It is thought that the surrounding of macroscopic PBHs by plasmas in the early universe, as well as quantum particle emission \cite{Gibbons:1975pa}, will lead to rapid discharge, but there may exist cases in which this discharge is slow enough relative to the evolution of a PBH such that effects resulting from the interaction of electric charge of the black hole with its surrounding environment are still relevant. In addition to this, it was initially shown by Page \cite{Page:1977um} and recently confirmed in \cite{Vasquez2026} that, when one takes Hawking radiation into account, a non-rotating black hole retains a root mean square charge on the order of unity in terms of electron charge number.

The bottom line, taking into account all the above considerations, is that we are not really sure what the allowed charge profile of primordial black holes looks like.  Therefore, one can conclude that it is important to consider primordial black holes that contain both electric and magnetic charge, and see what happens when they interact with surrounding electromagnetic fields.  This is what we set out to do in the following chapter.

\section{Rotating Black Hole in a Homogeneous Electromagnetic Field} \label{s41}

As with our previous derivation for rotating black holes in homogeneous scalar fields, we will work with the Kerr metric \eqref{Kerr2} and make use of its asymptotic Killing vectors describing translations and rotations in oblate spheroidal Minkowski coordinates, given by Eqs.~\eqref{transKilling} and \eqref{rotKilling}.  

The solution to Maxwell's equations for a homogeneous magnetic field which is deformed by a Kerr black hole at its origin was found by Bi\v{c}\'{a}k and Dvo\v{r}\'{a}k by using Newman-Penrose formalism \cite{Bicak:1976ef}, and is described by the following four-potential
\begin{align} \label{AB}
    &A_t = B_Za\big(Mr(1+\cos^2\theta)/\Sigma-1\big)+\frac{aM\sin\theta\cos\theta}{\Sigma}\big[B_X(r\cos\psi-a\sin\psi)\\\nonumber
    &\hspace{27pt}+B_Y(r\sin\psi+a\cos\psi)\big]\,,\\\nonumber
    &A_r =-(r-M)\cos\theta\sin\theta(B_X\sin\psi-B_Y\cos\psi)\,,\\\nonumber
    &A_{\theta} = -(ar\sin^2\theta+aM\cos^2\theta)(B_X\cos\psi+B_Y\sin\psi)\\\nonumber
    &\hspace{27pt}+\big[r^2\cos^2\theta-(Mr-a^2)\cos(2\theta)\big](B_X\sin\psi-B_Y\cos\psi)\,,\\\nonumber
    &A_\phi = B_Z\sin^2\theta\big((r^2+a^2)/2-a^2Mr(1+\cos^2\theta)/\Sigma\big)\\\nonumber
    &\hspace{27pt}-\sin\theta\cos\theta\bigg[\Delta(B_X\cos\psi+B_Y\sin\psi)+\frac{M(r^2+a^2)}{\Sigma}\big(B_X(r\cos\psi-a\sin\psi)\\\nonumber
    &\hspace{27pt}+B_Y(r\sin\psi+a\cos\psi)\big)\bigg]\,.
\end{align}
Once again, we write $\phi$ in terms of the incoming Kerr coordinate $\psi$
\begin{equation}
    \psi = \phi + \frac{a}{2\sqrt{M^2-a^2}}\ln\bigg(\frac{r-r_+}{r-r_-}\bigg)\,.
\end{equation}
One should check that Eq.~\eqref{AB} has the following three properties:
\begin{enumerate}
    \item It solves the source-free Maxwell equations in the Kerr background.
    \item It is regular at the event horizon $r=r_+$.
    \item At asymptotically large $r$, it corresponds to the four-potential of a homogeneous magnetic field $\vec{B}=(B_X,\,B_Y,\,B_Z)$.
\end{enumerate}
The electromagnetic field tensor for this four-potential is given by
\begin{equation} \label{EMB}
    B_{\mu\nu} = \partial_{\mu}A_{\nu}-\partial_{\nu}A_{\mu}\,.
\end{equation}
The magnetic field vector $B^{\mu}$ described by this tensor as seen by some observer with four-velocity $\boldsymbol{U}$ is given by 
\begin{equation}
    B^{\mu} = -^\star\hspace{-1pt}B^{\mu\nu}U_{\nu}\,.
\end{equation}

What about the electromagnetic tensor corresponding to a homogeneous electric field?  For reasons we shall delve into shortly, we can guess that it has the form
\begin{equation} \label{EME}
    E_{\mu\nu} = \,^{\star}B_{\mu\nu}\big|_{\vec{B}\to-\vec{E}}\,,\hspace{15pt}^{\star}B_{\mu\nu} = \frac{1}{2}e_{\mu\nu\alpha\beta}B^{\alpha\beta}\,,
\end{equation}
where the star written above corresponds to the Hodge star operator, $e_{\mu\nu\alpha\beta}=\sqrt{-g}\,\epsilon_{\mu\nu\alpha\beta}$ is the Levi-Civita tensor, and $\epsilon_{\mu\nu\alpha\beta}$ is the Levi-Civita symbol with sign convention $\epsilon_{0123}=1$.  The notation $\{\dots\}|_{\vec{B}\to-\vec{E}}$ means that after taking the Hodge dual $^\star B_{\mu\nu}$, we should substitute
\begin{equation}
    B_X\to-E_X\,,\hspace{15pt}B_Y\to-E_Y\,,\hspace{15pt}B_Z\to-E_Z\,.
\end{equation}
This electric field should satisfy the same three properties that the magnetic field satisfies.  One can check that the electric field vector $E^{\mu}$ described by this tensor as seen by some observer with four-velocity $\boldsymbol{U}$ is given by
\begin{equation}
    E^{\mu}=E^{\mu\nu}U_{\nu}\,.
\end{equation}

We also note that in the case where the magnetic field vector aligns with the axial angular momentum vector of the black hole, Eq.~\eqref{AB} greatly simplifies.  Specifically, for $B_X=B_Y=0$,
\begin{align}
    &A_t = B_Za\big(Mr(1+\cos^2\theta)/\Sigma-1\big)\,,\\\nonumber
    &A_{\phi} = B_Z\sin^2\theta\big((r^2+a^2)/2-a^2Mr(1+\cos^2\theta)/\Sigma\big)\,,
\end{align}
and we can write
\begin{equation} \label{AZ}
    A_{\mu} = \frac{1}{2}B\zeta_{\mu}+aB\xi_{\mu}\,,
\end{equation}
which corresponds to the form of the homogeneous magnetic four-potential derived in \cite{Wald:1974mf}. We remind the reader that $\boldsymbol{\xi}=\partial_t$ and $\boldsymbol{\zeta}=\partial_\phi$ correspond to the two Killing vectors of the Kerr spacetime.

We should also note that we assume these electromagnetic fields are "weak," in the sense that we neglect the effects of their stress-energy on Einstein's equations.

\section{Weak Electric and Magnetic Monopole Charges} \label{s42}

Now that we know the form of the homogeneous electromagnetic field in the Kerr spacetime, we look to add charges to a rotating black hole.  This could be done by simply taking the Kerr-Newman form of the metric Eq.~\eqref{Kerr2}, but then complications would arise from backreaction effects.  To circumvent this, we assume that the electric and magnetic monopole charges of the black hole are also weak, so that they do not change the gravitational field of the spacetime.

The four-potential of a weak electric charge $q$ in the Kerr spacetime was derived by Wald \cite{Wald:1974mf}, and is given by
\begin{equation}
    \mathring{A}_{\mu} = -\frac{q}{2M}\xi_{\mu}\,.
\end{equation}
It therefore has the following electromagnetic tensor
\begin{equation} \label{EMQ}
    Q_{\mu\nu} = \partial_{\mu}\mathring{A}_{\nu}-\partial_{\nu}\mathring{A}_{\mu}\,.
\end{equation}
In correspondence with the previous section, we can derive the electromagnetic tensor for a monopole charge $p$ in the Kerr spacetime with the following trick
\begin{equation} \label{EMP}
    P_{\mu\nu} = \,^\star{Q}_{\mu\nu}\big|_{q\to p}\,,\hspace{15pt}^\star Q_{\mu\nu} = \frac{1}{2}e_{\mu\nu\alpha\beta}Q^{\alpha\beta}\,.
\end{equation}
As one would hope, both $Q_{\mu\nu}$ and $P_{\mu\nu}$ are regular at $r=r_+$, and they decrease as $\sim 1/r^2$ as $r\to\infty$.  

The total electromagnetic tensor from the weak homogeneous electromagnetic fields Eqs.~\eqref{EMB}-\eqref{EME} and weak electromagnetic monopole charges Eqs.~\eqref{EMQ}-\eqref{EMP} can be found by simple superposition
\begin{equation}
    F_{\mu\nu} = E_{\mu\nu}+B_{\mu\nu}+Q_{\mu\nu}+P_{\mu\nu}\,.
\end{equation}
A simple check to ensure that the field configurations we use really describe what we want them to describe is to calculate the field flux from Gauss' law for a two-dimensional surface that encapsulates the black hole, and check that only the monopole charges contribute to these flux integrals.  To this effect, we consider the two dimensional surface $S_0$ that surrounds the black hole, and is defined by the embedding map $x^{\mu}=x^{\mu}(y^a)$, where $y^a=(y^1,\,y^2)$.  The surface element induced by this coordinate reduction is
\begin{equation}
    d\sigma_{\mu\nu} = \frac{1}{2}e_{\mu\nu\alpha\beta}\frac{\partial x^{\alpha}}{\partial y^1}\frac{\partial x^\beta}{\partial y^2} dy^1dy^2\,.
\end{equation}
Then, the fluxes of $\boldsymbol{F}$ and $^{\star}\boldsymbol{F}$ through $S_0$ are
\begin{align}
    &\frac{1}{8\pi}\int F^{\mu\nu}\,d\sigma_{\mu\nu} = q\,,\\\nonumber
    &\frac{1}{8\pi}\int {^\star}F^{\mu\nu}\,d\sigma_{\mu\nu} = -p\,.
\end{align}

\section{Electromagnetic Duality} \label{s43}

One might ask themselves why the transformations in Eqs.~\eqref{EME} and \eqref{EMP} give us exactly what we need to produce an electric field corresponding to the magnetic field and a magnetic monopole corresponding to the electric monopole.  This is due to a deep symmetry of Maxwell's equations called "electromagnetic duality."  For any electromagnetic tensor $F_{\mu\nu}$, we can introduce another tensor $\hat{F}_{\mu\nu}$ via what we call a "duality transformation"
\begin{equation} \label{Fhat}
    \hat{F}_{\mu\nu} = F_{\mu\nu}\cos\varphi\,+\,^\star{F_{\mu\nu}}\sin\varphi\,.
\end{equation}
Via the following property of the Hodge star operator for two-forms in four-dimensional spacetime
\begin{equation}
    ^{\star\star}F_{\mu\nu} = -F_{\mu\nu}\,,
\end{equation}
we derive that
\begin{equation}
    ^\star\hat{F}_{\mu\nu} = -F_{\mu\nu}\sin\varphi\,+\,^\star F_{\mu\nu}\cos\varphi\,.
\end{equation}
It is a straightforward exercise to check that the stress-energy tensor for the electromagnetic tensor
\begin{align} \label{TFdual}
    T_{\mu\nu} &= \frac{1}{4\pi}\Big(F_{\mu\alpha}F_{\nu}^{\;\alpha}-\frac{1}{4}g_{\mu\nu}F_{\alpha\beta}F^{\alpha\beta}\Big)\\\nonumber
    &= \frac{1}{8\pi}\Big(F_{\mu\alpha}F_{\nu}^{\;\alpha}\,+\,^{\star}{F}_{\mu\alpha}\,^\star F_{\nu}^{\;\alpha}\Big)\,
\end{align}
is invariant under the transformation Eq.~\eqref{Fhat}.

Now, consider a local Lorentzian frame, where $\boldsymbol{U}$ is a unit timelike vector of an observer at rest in this frame.  For the electromagnetic tensor $\boldsymbol{F}$, its electric and magnetic fields $\boldsymbol{E}$ and $\boldsymbol{B}$ are given by
\begin{equation}
    E^{\mu} = F^{\mu\nu}U_{\nu}\,,\hspace{15pt}B^{\mu} = -^\star F^{\mu\nu}U_{\nu}
\end{equation}
in this rest frame, respectively.  If we make the special choice $\varphi=\pi/2$ for Eq.~\eqref{Fhat}, then the duality transformation becomes
\begin{equation}
    \hat{\boldsymbol{E}} = -\boldsymbol{B}\,,\hspace{15pt}\hat{\boldsymbol{B}} = \boldsymbol{E}\,,
\end{equation}
and similarly the electric and magnetic monopole charges transform as
\begin{equation}
    \hat{q} = -p\,,\hspace{15pt}\hat{p} = q\,.
\end{equation}
This explains our choices in Eqs.~\eqref{EME} and \eqref{EMP}.  We also note that due to the invariance of Eq.~\eqref{TFdual} under the transformation Eq.~\eqref{Fhat}, we can conclude that the stress-energy tensor will be invariant under the following transformation
\begin{equation} \label{dualTrans}
    \boldsymbol{E}\to\boldsymbol{B}\,,\hspace{15pt}\boldsymbol{B}\to-\boldsymbol{E}\,,\hspace{15pt}q\to p\,,\hspace{15pt}p\to -q\,.
\end{equation}

\section{Flux Redux} \label{s44}

We use the same formalism as that which we already developed in Sec.~\ref{s321} in order to discuss the energy-momentum fluxes. Rewriting Eqs.~\eqref{Eflux} and \eqref{PJ} for visual clarity,
\begin{align} \label{EPJ}
    &\mathcal{E} = \Delta\int_{S_0} T_{r\nu}\xi^{\nu} \,d\Omega^2\,,\\\nonumber
    &\mathcal{P}_{(i)} = -\lim_{r_0\to\infty}\bigg(\Delta\int_{S_0} T_{r\nu}\xi^{\nu}_{(i)} \,d\Omega^2\bigg)\,,\\\nonumber
    &\mathcal{J}_{(i)} = -\lim_{r_0\to\infty}\bigg(\Delta\int_{S_0} T_{r\nu}\zeta^{\nu}_{(i)} \,d\Omega^2\bigg)\,.
\end{align}

We can separate the three fluxes in Eq.~\eqref{EPJ} as
\begin{align}
   & \mathcal{E} = \mathcal{E}_{H} + \mathcal{E}_{I}\,,\\\nonumber
    &\mathcal{P}_{(i)} = \mathcal{P}_{H(i)} + \mathcal{P}_{I(i)}\,,\\\nonumber
    &\mathcal{J}_{(i)} = \mathcal{J}_{H(i)} + \mathcal{J}_{I(i)}\,,
\end{align}
where the $H$-subscript corresponds to flux contributions from the homogeneous fields, and the $I$-subscript corresponds to flux contributions from the interaction between the homogeneous fields and the monopole charges.

If we choose the axial angular momentum vector of the black hole to lie along the $Z$-axis of the asymptotic coordinates, so that $\vec{J}=(0,\,0,\,J)$, with $J=aM$, then the homogeneous energy-momentum fluxes are
\begin{align}
    &\mathcal{E}_H = 0\,,\\\nonumber
    &\mathcal{P}_{H(X)} = \frac{2}{3}Ma(B_XE_Z-B_ZE_X)\,,\\\nonumber
    &\mathcal{P}_{H(Y)} = \frac{2}{3}Ma(B_YE_Z-B_ZE_Y)\,,\\\nonumber
    &\mathcal{P}_{H(Z)} = 0 \,,
\end{align}
and the homogeneous angular momentum fluxes are
\begin{align}
    &\mathcal{J}_{H(X)} = \frac{2}{3}M^2a(B_XB_Z+E_XE_Z)+\frac{22}{15}Ma^2(B_YB_Z+E_YE_Z)\,,\\[4pt]\nonumber
    &\mathcal{J}_{H(Y)} = \frac{2}{3}M^2a(B_YB_Z+E_YE_Z)-\frac{22}{15}Ma^2(B_XB_Z+E_XE_Z)\,,\\[4pt]\nonumber
    &\mathcal{J}_{H(Z)} = -\frac{2}{3}M^2a(B_X^2+B_Y^2+E_X^2+E_Y^2)\,.
\end{align}
These were previously derived in \cite{Frolov:2024xyo}.

The new interaction fluxes are given by
\begin{align}
    &\mathcal{E}_{I} = 0\,,\\\nonumber
    &\mathcal{P}_{I(X)} = pB_X + qE_X\,,\\\nonumber
    &\mathcal{P}_{I(Y)} = pB_Y + qE_Y\,,\\\nonumber
    &\mathcal{P}_{I(Z)} = pB_Z + qE_Z\,,
\end{align}
and
\begin{align}
    &\mathcal{J}_{I(X)} = -a(qB_Y-pE_Y)\,,\\\nonumber
    &\mathcal{J}_{I(Y)} = a(qB_X-pE_X)\,,\\\nonumber
    &\mathcal{J}_{I(Z)} = 0\,.
\end{align}

\section{Evolution of Black Hole Parameters} \label{s45}

In accordance with the formalism developed in Ch.~\ref{ch3}, we define two different frames in which we study the interaction in question.  The first frame, $\mathcal{K}$, is the "BH frame" in which the black hole is stationary with respect to the metric.  The second frame, $\mathcal{K}'$ is the "field frame," which moves with respect to the BH frame at some velocity $\vec{\text{v}}$.  As previously mentioned, we take on the "zoomed-out" perspective, where an observer in the $\mathcal{K}'$ frame sees the black hole at such a far distance, that it can be treated as a point-particle.

\subsection{Black Hole Frame} \label{s451}

In the $\mathcal{K}$ frame, we know that we can treat the energy-momentum fluxes as a four-force $\mathcal{F}^{\mu}=(\mathcal{F}^0,\,\vec{\mathcal{F}})=(\mathcal{E},\,\mathcal{P}_{(X)},\,\mathcal{P}_{(Y)},\,\mathcal{P}_{(Z)})$ and a torque vector $\vec{\mathcal{T}} = (\mathcal{J}_{(X)},\,\mathcal{J}_{(Y)},\,\mathcal{J}_{(Z)})$.  Starting with the four-force $\boldsymbol{\mathcal{F}}$, we can break it up into homogeneous and interaction components
\begin{align}
    &\mathcal{F}^{\mu} = \mathcal{F}^{\mu}_H+\mathcal{F}^{\mu}_I\,,\hspace{15pt} \mathcal{F}^0_H=\mathcal{F}^0_I = 0\,,\\\nonumber
    &\vec{\mathcal{F}}_H = -\frac{2}{3}\vec{J}\times(\vec{E}\times\vec{B})\\\nonumber
    &\hspace{17pt}=-\frac{2}{3}\big[\vec{E}(\vec{J}\cdot\vec{B})-\vec{B}(\vec{J}\cdot\vec{E})\big]\,,\\\nonumber
    &\vec{\mathcal{F}}_I = q\vec{E}+p\vec{B}\,.
\end{align}

Let's take a moment to analyze these equations.  First and foremost, since $\mathcal{F}^0=0$, we see that the black hole mass is always constant. This is in agreement with the known result that the cross section of a black hole absorbing electromagnetic radiation goes to zero as the frequency of the radiation approaches zero, since the electromagnetic fields we are considering are stationary \cite{Starobinskil:1974nkd,Fabbri:1975sa,Page:1976ki}. Also, $\vec{\mathcal{F}}_I$ corresponds to the familiar Lorentz force law from classical electromagnetism.  As expected, it is also invariant under Eq.~\eqref{dualTrans}.

To better explain $\vec{\mathcal{F}}_H$, let's look at the special case where $\vec{J}$ points in the same direction as $\vec{B}$, but is orthogonal to $\vec{E}$.  This means that
\begin{equation} \label{FHZ}
    \vec{\mathcal{F}}_H = -\frac{2}{3}BJ\vec{E}\,,
\end{equation}
and so we see that this force acts on the black hole in the opposite direction of the electric field.  One can think of this force as arising from the following scenario:  We start with a neutral black hole rotating in a magnetic field given by Eq.~\eqref{AZ}.  This rotation in the presence of the magnetic field induces a charge on the black hole of magnitude $Q=-2BJ$.  The force then comes from the action of the electric field on this "fictitious" charge.

Next, we inspect the relative strength of the two force components $\mathcal{F}_H\equiv|\vec{\mathcal{F}}_H|$ and $\mathcal{F}_I\equiv|\vec{\mathcal{F}}_I|$.  For simplicity and familiarity, we will take $p=0$ in this comparison. Momentarily restoring Newton's gravitational constant $G$ and the speed of light $c$, Eq.~\eqref{FHZ} becomes
\begin{equation}
    \mathcal{F}_H\sim \alpha\frac{G^2M^2EB}{c^4}\,,
\end{equation}
where $\alpha=Jc/(GM^2)$ is the rotation parameter of the black hole, which can take on values $\alpha\in[0,\,1)$.  Similarly,
\begin{equation}
    \mathcal{F}_I\sim\beta\sqrt{G}ME\,,
\end{equation}
where $\beta=q/(\sqrt{G}M)$ is the charge parameter of the black hole, and also lies within the range $\beta\in[0,\,1)$.  The cases where $\alpha=1$ or $\beta=1$ are reserved for extremally rotating or extremally charged black holes, respectively.

Furthermore, we can write
\begin{equation}
    \epsilon_B = B^2\,,\hspace{15pt}\epsilon_{BH} = \frac{c^8}{G^3M^2}\,.
\end{equation}
The first parameter is the energy density associated with the magnetic field, and the second parameter is the effective energy density of the black hole, which can be derived by dividing its rest energy $Mc^2$ by its effective volume $R_g^3 = GM/c^2$.  If we denote
\begin{equation} \label{Z}
    \mathcal{Z} = \sqrt{\frac{\epsilon_B}{\epsilon_{BH}}}\,,
\end{equation}
then we can write the dimensionless ratio of the two forces at play as
\begin{equation}
    \mathcal{F}_H/\mathcal{F}_I \sim \frac{\alpha}{\beta}\mathcal{Z}\,.
\end{equation}
Since we require that the backreaction of the electromagnetic field on the spacetime geometry is small, we conclude that
\begin{equation}
    \mathcal{Z} \ll 1\,,
\end{equation}
meaning that as long as the charge parameter $\beta$ is not significantly smaller than the rotation parameter $\alpha$, then $\mathcal{F}_H\ll\mathcal{F}_I$.

Now that we have covered all aspects of the four-force $\mathcal{F}^{\mu}$ in the $\mathcal{K}$ frame, let's move on to a discussion about the torque $\vec{\mathcal{T}}$. We split it into homogeneous and interaction components, and further break up the homogeneous part into two more components to make it more digestible
\begin{align} \label{THI}
    &\vec{\mathcal{T}} = \vec{\mathcal{T}}_H + \vec{\mathcal{T}}_I\,,\hspace{15pt}\vec{\mathcal{T}}_H = \vec{\mathcal{T}}_1+\vec{\mathcal{T}}_2\,,\\\nonumber
    &\vec{\mathcal{T}}_1=-\frac{2M}{3}\big(\vec{B}\times(\vec{J}\times\vec{B})+\vec{E}\times(\vec{J}\times\vec{E})\big)\,,\\\nonumber
    &\vec{\mathcal{T}}_2 = -\frac{22}{15M}\big((\vec{J}\cdot\vec{B})(\vec{J}\times\vec{B})+(\vec{J}\cdot\vec{E})(\vec{J}\times\vec{E})\big)\,,\\\nonumber
    &\vec{\mathcal{T}}_I = \frac{1}{M}\vec{J}\times(q\vec{B}-p\vec{E})\,.
\end{align}
Note that every equation in Eq.~\eqref{THI} is invariant under Eq.~\eqref{dualTrans}, as required.

Like we did before in the previous chapter, we can see how the axial angular momentum vector of the black hole $\vec{J}$ changes with proper time $\tau$ via the equation
\begin{equation} \label{dJdt2}
    \frac{d\vec{J}}{d\tau} = \vec{\mathcal{T}}.
\end{equation}

\subsection{Field Frame} \label{s452}

There are precisely two scalar invariants that can be constructed from the electromagnetic tensor $\boldsymbol{F}$, which do not change under Lorentz transformations.  We denote them as
\begin{equation}
    \mathcal{I}_1 = F_{\mu\nu}F^{\mu\nu}\,,\hspace{15pt}\mathcal{I}_2 = F_{\mu\nu}\hspace{-2pt}\,^{\star}F^{\mu\nu}\,.
\end{equation}
If we suppose that $\mathcal{I}_1\neq 0$, then there must exist a reference frame in which the homogeneous electric and magnetic fields are parallel.  We designate this frame as the "field frame" $\mathcal{K}'$. If it so happens that in this frame, $\mathcal{I}_2=0$, then we know that one of the two fields also vanishes in this frame.  In the $\mathcal{K}'$ frame, we have that
\begin{equation}
    \vec{E}_0 = E_0\vec{N}\,,\hspace{15pt}\vec{B}_0=B_0\vec{N}\,,
\end{equation}
where $\vec{N}$ is a spacelike unit vector. This means that, switching back to the $\mathcal{K}$ frame which moves with respect to the $\mathcal{K}'$ frame with velocity $\vec{\text{v}}$, the electric and magnetic field vectors experience the following Lorentz transformations
\begin{align} \label{EBLorentz}
    &\vec{E} = \gamma(\vec{E}_0+\vec{\text{v}}\times\vec{B}_0)-(\gamma-1)\frac{\vec{\text{v}}\cdot\vec{E}_0}{\text{v}^2}\vec{\text{v}}\,,\\\nonumber
    &\vec{B} = \gamma(\vec{B}_0-\vec{\text{v}}\times\vec{E}_0)-(\gamma-1)\frac{\vec{\text{v}}\cdot\vec{B}_0}{\text{v}^2}\vec{\text{v}}\,.
\end{align}
One can see from Eq.~\eqref{EBLorentz} that our choice of "field frame" is not unique.  While the chosen frame no longer enjoys full Poincaré invariance, eight of the original ten boost generators survive, forming a subgroup for which the field frame remains invariant.  Namely, four spacetime translation generators, three rotation generators, and one of the original three boost generators survive.  One can check via Eq.~\eqref{EBLorentz} that this generator corresponds to boosts that are parallel to both $\vec{E}_0$ and $\vec{B}_0$, such that $\vec{\text{v}}=\text{v}\vec{N}$.

To discuss the forces acting on the black hole in the field frame, we must also perform a Lorentz transformation on the four-force $\boldsymbol{\mathcal{F}}$.  Since $\mathcal{F}^0=0$, we have
\begin{equation}
    f^0 = \gamma(\vec{\text{v}}\cdot\vec{\mathcal{F}})\,,\hspace{15pt}\vec{f} = \vec{\mathcal{F}}+(\gamma-1)\frac{\vec{\text{v}}\cdot\mathcal{F}}{\text{v}^2}\vec{\text{v}}\,.
\end{equation}
We can separate the four-force in the field frame as $\boldsymbol{f}=\boldsymbol{f}_H+\boldsymbol{f}_I$, and we have
\begin{align} \label{fEM}
    &f_H^0 = \frac{2}{3}\gamma^2\vec{J}\cdot\Big[(\vec{\text{v}}\cdot\vec{B}_0)(\vec{E}_0+\vec{\text{v}}\times\vec{B}_0)-(\vec{\text{v}}\cdot\vec{E}_0)(\vec{B}_0-\vec{\text{v}}\times\vec{E}_0)\Big]\,,\\\nonumber
    &f_I^0 = \gamma\vec{\text{v}}\cdot\big(q\vec{E}_0+p\vec{B}_0)\,,\\\nonumber
    &\vec{f}_H = \frac{2}{3}\gamma\bigg[\big(\vec{B}_0-\vec{\text{v}}\times\vec{E}_0\big)\Big(\vec{J}\cdot\big(\gamma(\vec{E}_0+\vec{\text{v}}\times\vec{B}_0)-(\gamma-1)\frac{\vec{\text{v}}\cdot\vec{E}_0}{\text{v}^2}\vec{\text{v}}\big)\Big)\\\nonumber
    &\hspace{27pt}-\big(\vec{E}_0+\vec{\text{v}}\times\vec{B}_0\big)\Big(\vec{J}\cdot\big(\gamma(\vec{B}_0-\vec{\text{v}}\times\vec{E}_0)-(\gamma-1)\frac{\vec{\text{v}}\cdot\vec{B}_0}{\text{v}^2}\vec{\text{v}}\big)\Big)\bigg]\,,\\\nonumber
    &\vec{f}_I = \gamma\big(q\vec{E}_0+p\vec{B}_0-\vec{\text{v}}\times(p\vec{E}_0-q\vec{B}_0)\big)\,.
\end{align}

The four-momentum associated with this vector in the $\mathcal{K}'$ frame is
\begin{equation}
    P^{\mu} = (\gamma M,\,\gamma M\vec{\text{v}})\,,
\end{equation}
which leads to the following evolution equations for the velocity of the black hole (since its mass remains constant):
\begin{equation} \label{ff0}
    M\frac{d\gamma}{d\tau} = f^0\,,\hspace{15pt}M\frac{d(\gamma\vec{\text{v}})}{d\tau} = \vec{f}\,.
\end{equation}

In the special case where the black hole does not rotate, $\vec{J}=0$, and our equations simplify greatly
\begin{equation}
    \vec{\mathcal{T}}=0\,,\hspace{15pt}f^0 = \gamma\vec{\text{v}}\cdot(q\vec{E}_0+p\vec{B}_0)\,,\hspace{15pt}\vec{f} = \gamma\big(q\vec{E}_0+p\vec{B}_0-\vec{\text{v}}\times(p\vec{E}_0-q\vec{B}_0)\big)\,.
\end{equation}
If we further simplify the last equation for the case where there is no magnetic monopole charge, $p=0$, we arrive at the familiar Lorentz force law
\begin{equation}
    \vec{f} = \gamma q\big(\vec{E}_0+\vec{\text{v}}\times\vec{B}_0\big)\,.
\end{equation}
One can then assess that for a non-rotating electrically-charged black hole, it moves like a point charge in a homogeneous electromagnetic field in flat Minkowski spacetime.

\subsection{Spin Evolution of the Black Hole} \label{s453}

One can derive from Eq.~\eqref{dJdt2} that
\begin{equation}
    \frac{dJ^2}{d\tau} = \frac{d\vec{J}^2}{d\tau} = 2\vec{J}\cdot\frac{d\vec{J}}{d\tau} = 2\vec{J}\cdot\vec{\mathcal{T}}\,.
\end{equation}
From Eq.~\eqref{THI}, we can easily see that $\vec{J}\cdot\vec{\mathcal{T}}_2=\vec{J}\cdot\vec{\mathcal{T}}_I=0$, and so the only term that survives is
\begin{equation}
    \vec{J}\cdot\vec{\mathcal{T}}_1 = -\frac{2M}{3}\big[J^2B^2-(\vec{J}\cdot\vec{B})^2+J^2E^2-(\vec{J}\cdot\vec{E})^2\big]\,.
\end{equation}
Taking $\vec{n}$, $\vec{e}$, and $\vec{b}$ to be the unit vectors along $\vec{J}$, $\vec{E}$, and $\vec{B}$, respectively
\begin{equation}
    \vec{J} = J\vec{n}\,,\hspace{15pt}\vec{E} = E\vec{e}\,,\hspace{15pt}\vec{B} = B\vec{b}\,,
\end{equation}
we arrive at
\begin{equation}
    \frac{dJ^2}{d\tau} = -\frac{4MJ}{3}\Big[B^2\big(1-(\vec{b}\cdot\vec{n})^2\big)+E^2\big(1-(\vec{e}\cdot\vec{n})^2\big)\Big]\,.
\end{equation}

The above equation shows that the magnitude of $\vec{J}$ cannot increase.  It can only decrease, or remain the same in the special case where $\vec{n}=\vec{e}=\vec{b}$ so that the spin of the black hole is aligned with the direction of both the electric and magnetic fields in the $\mathcal{K}$ frame.  Furthermore, via Eq.~\eqref{EBLorentz} we can conclude that the black hole frame is either at rest with respect to the field frame, or that it is moving at constant velocity in the direction parallel to $\vec{E}$ and $\vec{B}$.

Let's look at the case where $\vec{E}$ and $\vec{B}$ share a common direction in more detail.  We have the setup
\begin{equation}
    \vec{B} = B\vec{b}\,,\hspace{15pt}\vec{E} = E\vec{b}\,,\hspace{15pt}\vec{J} = J_{\parallel}\vec{b}+\vec{J}_{\perp}\,,
\end{equation}
where it is assumed that $\vec{J}_{\perp}\cdot\vec{b}=0$.  The torque components take on the following values
\begin{align}
    &\vec{\mathcal{T}}_1 = -\frac{2M}{3}(B^2+E^2)\vec{J}_{\perp}\,,\\\nonumber
    &\vec{\mathcal{T}}_2 =-\frac{22}{15M}(B^2+E^2)J_{\parallel}(\vec{J}_{\perp}\times\vec{b})\,,\\\nonumber
    &\vec{\mathcal{T}}_I = \frac{1}{M}(qB-pE)(\vec{J}_{\perp}\times\vec{b})\,.
\end{align}
This can be written in a more compact form as
\begin{align}
    &\vec{\mathcal{T}} = \mu\vec{J}_{\perp}+\nu(\vec{J}_{\perp}\times\vec{b})\,,\\\nonumber
    &\mu = -\frac{2M}{3}(B^2+E^2)\,,\\\nonumber
    &\nu = -\frac{22}{15M}(B^2+E^2)J_{\parallel}+\frac{1}{M}(qB-pE)\,.
\end{align}
From the above equation and Eq.~\eqref{dJdt2}, we conclude that $J_{\parallel}=const.$ and similarly that $\mu$ and $\nu$ are constant.

We can then easily solve Eq.~\eqref{dJdt2} for this scenario by recasting it in the complex plane.  First, we take $\vec{J}_{\perp}=(J_X,\,J_Y,\,0)$, which allows us to rewrite the equation
\begin{equation} \label{dJperp}
    \frac{d\vec{J}_{\perp}}{d\tau} = \mu\vec{J}_{\perp}+\nu(\vec{J}_{\perp}\times\vec{b})\,,
\end{equation}
as
\begin{equation}
    \frac{d\mathcal{J}}{d\tau} = \Gamma\mathcal{J}\,,\hspace{15pt}\mathcal{J}=J_X+iJ_Y\,,\hspace{15pt}\Gamma=\mu-i\nu.
\end{equation}
This equation has the simple solution
\begin{equation}
    \mathcal{J}(\tau) = \mathcal{J}_0\exp(\Gamma\tau)\,.
\end{equation}

Here, $\mathcal{J}_0$ corresponds to the initial values of $\vec{J}_{\perp}$ at $\tau=0$.  The parameter $\mu$ corresponds to the rate at which the magnitude of the black hole's angular momentum decays, and the parameter $\nu$ describes the rate of precession of the black hole's axial angular momentum vector around $\vec{b}$.  We note the striking similarity between this case and the case where a rotating black hole is moving transversely to the spacelike field gradient of a homogeneous scalar field in Sec.~\ref{s3341}.

\subsection{Rotating Black Hole Moving Transverse to Magnetic Field} \label{s454}

We now consider the case where the rotating black hole is moving transversely to the magnetic field in the field frame $\mathcal{K}'$, meaning that $\vec{\text{v}}\cdot\vec{B}_0=0$. Additionally, we assume that there is no electric field present in the field frame, so that $\vec{E}_0=0$.  This implies that in the black hole frame $\mathcal{K}$
\begin{equation}
    \vec{B} = \gamma\vec{B}_0\,,\hspace{15pt}\vec{E} = \gamma(\vec{\text{v}}\times\vec{B}_0)\,.
\end{equation}
We can simultaneously consider two separate cases.  In the first case, the spin of the black hole $\vec{J}$ is parallel to $\vec{B}_0$, such that $\vec{J}\times\vec{B}_0=0$ and $\vec{J}\cdot\vec{\text{v}}=0$.  In the second case, the spin of the black hole $\vec{J}$ is parallel to $\vec{\text{v}}$, so that $\vec{J}\times\vec{\text{v}}=0$ and $\vec{J}\cdot\vec{B}_0=0$.  These two cases can be compactified in the following form
\begin{equation}
    \vec{B}\cdot\vec{J} = \epsilon\gamma B_0J\,,\hspace{15pt}\vec{E}\cdot\vec{J}=0\,,
\end{equation}
where the former case corresponds to $\epsilon=1$, and the latter corresponds to $\epsilon=0$.

Eqs.~\eqref{THI} and \eqref{fEM} will then take on the following forms
\begin{equation} \label{TF1}
    \vec{\mathcal{T}} = -\frac{2}{3}MB_0\gamma^2\big[(1+\text{v}^2)B_0\vec{J}-\epsilon J\vec{B}_0\big]\,,\hspace{15pt}\vec{f} = \gamma\big(q-\frac{2}{3}\epsilon\gamma B_0J\big)(\vec{\text{v}}\times\vec{B}_0)\,.
\end{equation}
We also note that we have the identity
\begin{equation}
    \epsilon J\vec{B}_0 = \epsilon B_0\vec{J}\,,
\end{equation}
which allows us to write the first equation in Eq.~\eqref{TF1} as
\begin{equation}
    \vec{\mathcal{T}} = -\frac{2}{3}MB_0^2\gamma^2(1-\epsilon+\text{v}^2)\vec{J}\,.
\end{equation}
Since the second equation in Eq.~\eqref{TF1} implies that $\vec{\text{v}}\cdot\vec{f}=0$, we also have
\begin{equation}
    \frac{1}{2}M\frac{d(\gamma^2\text{v}^2)}{d\tau} = M\gamma\vec{\text{v}}\cdot\frac{d(\gamma\vec{\text{v}})}{d\tau} = \gamma\vec{\text{v}}\cdot\vec{f} = 0\,,
\end{equation}
via the definition in Eq.~\eqref{ff0}.  From this we can conclude that both $\gamma$ and $\text{v}$ are constant.

From this, we can write the spin evolution equation as
\begin{equation}
    \frac{d\vec{J}}{d\tau} = -\lambda\vec{J}\,,\hspace{15pt}\lambda=\frac{2}{3}MB_0^2\gamma^2(1-\epsilon+\text{v}^2)\,,
\end{equation}
which has the simple solution
\begin{equation} \label{Jsol}
    \vec{J}(\tau) = \vec{J}_0\exp(-\lambda\tau)\,.
\end{equation}
This tells us that the spin of the black hole is exponentially decreasing with characteristic time $\tau_{ch}=1/\lambda$.

Next we turn our attention to the motion of the black hole within the field frame $\mathcal{K}'$. Via Eq.~\eqref{ff0}, we know that this motion is governed by the equation
\begin{equation} \label{ff1}
    M\gamma\frac{d\vec{\text{v}}}{d\tau} = \vec{f}\,,
\end{equation}
where the magnitude of the velocity vector was found to be constant. We note that this equation has the same form as Eq.~\eqref{dJperp}, where $\mu=0$ in the case at hand. We proceed in a similar fashion.

Since $\vec{\text{v}}\cdot\vec{f}=0$, we know that if $\vec{\text{v}}\cdot\vec{B}_0=0$, then this will remain true for all future times as well. Then, we fix the direction $\vec{B}_0=(0,\,0,\,B_0)$, so that we can write $\vec{\text{v}}=(\text{v}_X,\,\text{v}_Y,\,0)$, or in the complex plane
\begin{equation}
    \mathcal{V} = \text{v}_X+i\text{v}_Y = \text{v}e^{-i\Phi(\tau)}\,.
\end{equation}
Then Eq.~\eqref{ff1} will take the form
\begin{equation} \label{dPhi}
    \frac{d\Phi}{d\tau} = \omega\,,\hspace{15pt}\omega = \frac{1}{M}\Big(qB_0 - \frac{2}{3}\epsilon\gamma B_0^2J\Big)\,.
\end{equation}
Substituting in the solution Eq.~\eqref{Jsol}, we can solve this equation to find that
\begin{equation}
    \Phi(\tau) = -\frac{\epsilon J_0}{\gamma M^2\text{v}^2}(1-e^{-\lambda\tau})+\frac{qB_0}{M}\tau\,.
\end{equation}

If we assume that the decay in the black hole's spin happens over large intervals of time, so that $\lambda\ll 1/M$, then we can assume the differential equation Eq.~\eqref{dPhi} reduces to
\begin{equation}
    \frac{d\Phi}{d\tau} = \omega\approx\frac{B_0}{M}\Big(q-\frac{2}{3}\epsilon\gamma B_0J_0\Big)\,.
\end{equation}
This tells us that the black hole moves on a near-circular trajectory with slowly-changing frequency given approximately by $\omega$ whose clock runs according to the proper time of the black hole $\tau$. If we wish to find the frequency in the field frame, which we will call $\Omega$, we use the time dilation relation $dt=\gamma d\tau$ to obtain
\begin{equation}
    \Omega\approx \frac{B_0}{M\gamma}\Big(q-\frac{2}{3}\epsilon\gamma B_0J_0\Big)\,.
\end{equation}
For a non-rotating black hole, the second term above vanishes, and we identify $\Omega$ with the synchrotron frequency of a relativistic point particle with mass $M$ and charge $q$ that is rotating around a homogeneous field $\vec{B}_0$ in flat Minkowski spacetime.  This gives us surmounting evidence that if we "zoom-out" far enough away from the black hole, then our problem will reduce to one of relativistic electrostatics.  Interestingly enough, for the case where $\vec{J}\times\vec{\text{v}}=0$, $\epsilon=0$ and the spin of the black hole is not "felt" by the magnetic field anyways, meaning that this setup still gives us a solution that corresponds to synchrotron motion.

If we look to the case where $\epsilon=1$, and denote $\hat{q}=(2/3)\gamma B_0J_0$, then we can write the ratio of $\hat{q}$ to $q$ in the same way that we did previously with Eq.~\eqref{Z}. Namely,
\begin{equation} \label{qhatq}
    \frac{\hat{q}}{q}\sim \gamma\frac{\alpha}{\beta}\mathcal{Z}\,,
\end{equation}
where $\alpha$ is the black hole's unit interval rotation parameter and $\beta$ is its unit interval charge parameter. Since we previously showed that $\mathcal{Z}\ll 1$, this means that the charge of the black hole $q$ will be the predominant factor in causing its rotation around the magnetic field, unless of course it is sufficiently small.

Since the angular frequency of the black hole's rotation $\Omega$ is slowly changing over time, we can also come to the same conclusion about its radius of rotation, which is roughly
\begin{equation}
    R = \frac{\text{v}}{\Omega} \approx \frac{\gamma\text{v}M}{B_0(q-\hat{q})}\,.
\end{equation}

\subsection{Charged Rotating Black Hole in an Electric Field} \label{s455}

Our last case of study in this chapter will be that of a charged rotating black hole that is immersed in a homogeneous electric field.  To simplify matters, we also assume that $\vec{\text{v}}\times\vec{E}_0=0$. In this case,
\begin{equation}
    \vec{B} = 0\,,\hspace{15pt}\vec{E}=\vec{E}_0=E_0\vec{e}\,,\hspace{15pt}\vec{\text{v}}=\text{v}\vec{e}\,.
\end{equation}

We have yet to fix the direction of the spin $\vec{J}$.  In the case where it is parallel to the electric field $\vec{E}$, the torque vanishes, and so the spin remains constant.  In the case where it is orthogonal to the electric field, \emph{i.e.} $\vec{J}\cdot\vec{E}=0$, we have
\begin{equation}
    \vec{\mathcal{T}} = -\frac{2M}{3}E_0^2\vec{J}\,.  
\end{equation}
Therefore, the component of spin that is along the direction of motion of the black hole remains constant, while the component of spin that is orthogonal to the black hole's motion decays with characteristic time $\tau_{ch}=3/(2ME_0^2)$.

In either of these two scenarios, the four-force is given by
\begin{equation}
    f^0=q\gamma\text{v}E_0\,,\hspace{15pt}\vec{f}=q\gamma \vec{E}_0\,.
\end{equation}
One can easily see that $\boldsymbol{f}^2=q^2E_0^2$ is constant.  This means that the black hole moves in the direction of the electric field with constant four-acceleration $w=qE_0/M$.  Once more, this is exactly what we would expect from a point-particle with mass $M$ and charge $q$ under the influence of a homogeneous electric field in relativistic electrostatics.  Let us choose $\vec{E}_0=(E_0,\,0,\,0)$, so that the black hole is moving along the $X$-axis.  If we choose $X(\tau=0)=0$, then we have the well-known solution
\begin{equation}
    X=\frac{1}{w}\big[\cosh(w\tau)-1\big]\,,\hspace{15pt}T=\frac{1}{w}\sinh(w\tau)\,,
\end{equation}
where $\tau$ is the proper time of the black hole in its rest frame.  Then, the magnitude of the three-velocity of the black hole is simply given by
\begin{equation}
    \text{v} = \frac{dX/d\tau}{dT/d\tau} = \tanh(w\tau)\,.
\end{equation}
Note that the acceleration of the rotating black hole has no dependence on its spin. 

\section{Order of Magnitude Estimate}

One might naturally ask about the scale of effects resulting from this phenomenon and its scope in relation to astrophysical problems.  In an attempt to help answer this question, we provide a simple "back of the envelope" estimate on the energy received from a black hole in the presence of a magnetic field.  Our construction is as follows:

Say that we have a weakly-charged rotating supermassive black hole with mass $M$, charge $Q$, and rotation parameter $a$.  Its rotation in the presence of a charge will induce a magnetic field on its surrounding environment.  We then assume that there is a smaller black hole with mass $m$ in a stable orbit around the larger black hole, and that its orbit lies at a distance $R$ from the center of the larger black hole.  We will also assume that this distance $R$ is sufficiently greater than $r_M$, the Schwarzschild radius of the larger black hole.  Its energy absorption per second from the magnetic field is then given by
\begin{equation}
    L = \epsilon_{B}\text{v}A\,,
\end{equation}
where $\epsilon_B$ is the energy density of the induced magnetic field, $\text{v}$ is the tangential velocity of the smaller black hole within its orbit, and $A$ is the gravitational capture area of the smaller black hole. In geometrized CGS units, these quantities are written as
\begin{equation}
    \epsilon_B = \frac{1}{2}B^2\,,\hspace{15pt}\text{v} = \frac{R\sqrt{MR}}{R^2+a\sqrt{MR}}\,,\hspace{15pt}A=27\pi m^2\,.
\end{equation}
Note that for the velocity $\text{v}$ we make the two following assumptions: (1) the smaller black hole takes a prograde orbit around the larger black hole, so that it is orbiting in the same direction as the larger black hole's rotation direction, and (2) the charge of the supermassive black hole in geometrized units is much smaller than its Schwarzschild radius, such that $Q \ll r_M$, and so we can use the formula for the tangential velocity of a test mass orbiting around a Kerr black hole.

If we utilize the assumption that the smaller black hole orbits at a far distance from the larger black hole $R\gg M > a$, we can simplify the velocity formula, and we have
\begin{equation}
    L \approx \frac{27\pi}{2}m^2B^2\sqrt{\frac{M}{R}}\,.
\end{equation}
Next, since $R\gg M$, we can use the standard Biot-Savart law to calculate the magnetic field induced by the supermassive black hole, which is
\begin{equation}
    B\approx \frac{Qa}{R^3}\,.
\end{equation}

An order of magnitude evaluation of this formula gives
\begin{equation}
    L\sim \Big(\frac{a}{R}\Big)^2\Big(\frac{Q}{R}\Big)^2\Big(\frac{m}{R}\Big)^2\Big(\frac{M}{R}\Big)^{1/2}\,,
\end{equation}
or restoring SI units,
\begin{equation} \label{lum}
    L\sim \Big(\frac{a}{R}\Big)^2\Big(\frac{r_Q}{R}\Big)^2\Big(\frac{r_m}{R}\Big)^2\Big(\frac{r_M}{R}\Big)^{1/2}L_{pl}\,,
\end{equation}
where in these units
\begin{equation}
    r_Q^2 = \frac{Q^2G}{4\pi\epsilon_0c^4}\,,\hspace{10pt}r_m = \frac{2Gm}{c^2}\,,\hspace{10pt}r_M=\frac{2GM}{c^2}\,,\hspace{10pt}L_{pl}=\frac{c^5}{G}\,.
\end{equation}
Here, $L_{pl}$ is the Planck luminosity, which is the conjectured upper limit on the power that any event in the universe, such as gravitational collapse, can produce \cite{Cardoso:2018nkg}.

Although this estimate is promising in that it appears to generate a large amount of energy, one should recall that black holes which are formed from gravitational collapse are believed to shed any charge they have quite rapidly, meaning that \eqref{lum} will typically be zero as well \cite{Gibbons:1975pa}. However, this process remains interesting for primordial black holes, which can be much smaller than any black hole formed from gravitational collapse, as these black holes may be able to retain some of their charge, albeit a small quantity \cite{Starobinskil:1974nkd,Fabbri:1975sa,Page:1976ki}.

\section{Concluding Remarks} \label{s46}

In this chapter, we explored the interaction between a weakly-charged rotating black hole and an external static weak homogeneous electromagnetic field. By weak, we mean that a back-reaction of the fields or the charges on
the metric is neglected. To calculate the force and torque acting on the moving black hole, we assumed that the change of the black hole’s parameters and the evolution of its velocity is slow. We first discussed a stationary solution of Maxwell’s equations in the background of the Kerr metric which is regular at the horizon and satisfies properly chosen asymptotic conditions at spatial infinity. Using this solution, we calculated the stress-energy tensor of the electromagnetic field arising from both the external homogeneous fields and from the electric and magnetic monopole charges. This allowed us to find the fluxes of the energy, momentum, and angular momentum into the black hole. We used these results to calculate the force and torque acting on the black hole. These quantities were obtained both in the black hole frame and in the frame of an observer at rest at asymptotic infinity.

Our three main results were the following:
\begin{itemize}
    \item For a moving charged rotating black hole in an external static homogeneous electromagnetic field, its mass remains constant.
    \item The spin of such a black hole remains constant only if (i) it is either at rest in the field frame or moves in it along the common directions of the fields, and (ii) the spin of the black hole is parallel to this common direction as well.
    \item  If the above conditions are not satisfied, then the black hole's spin decreases.
\end{itemize}

We also calculated three specific scenarios of black hole evolution, namely (i) a black hole at rest within a homogeneous magnetic field, (ii) a black hole moving perpendicular to a homogeneous magnetic field, and (iii) a black hole moving along the direction of a homogeneous electric field.  The dynamics of its motion and spin in these specific cases were discussed in depth.

Let us emphasize that for charged rotating black holes, the effects we discussed in this chapter usually contain two contributions depending on the black hole’s spin and charge, respectively. Usually, the ratio of these terms is controlled by a small parameter $\mathcal{Z}$ (see Eq.~\eqref{Z}). However, when the charge $q$ is extremely small or vanishes, the spin-dependent effects dominate. Let us note that in the case when a rotating black hole moves in a constant massless scalar field, considered in \cite{Frolov:2024sf,A2}, the equations of motion are quite similar to the equations discussed in the present paper, with one important difference. The mass of the black hole moving in a scalar field can increase, and in some cases it can infinitely grow in finite time.

In a real astrophysical set-up, the discussed effects connected with black hole spin are extremely small.  However, if during the evolution of the early universe there were phases with strong primordial magnetic fields (see e.g. \cite{Grasso:2000wj}), these effects might be of interest.

	\chapter{Spinoptics in the Kerr Spacetime}\label{ch5}

In this chapter we consider the following problem: we wish to describe the evolution of a high-frequency gravitational or electromagnetic wave as it passes through spacetime, specifically the spacetime of a rotating black hole.  In order to achieve this goal, one can use the well-established geometric optics approximation.  In this approximation, one considers a high-frequency electromagnetic wave which can be locally approximated by a monochromatic plane wave.  Solving Maxwell's equations for such an ansatz, one can recast the problem as the study of null rays with associated polarization vectors within a gravitational field.  One finds that these null rays are geodesics under this approximation, and that their polarization vectors are altered in the presence of a gravitational field.

We extend this analysis to the case where helicity-dependent corrections are considered.  In this prescription, a recasting of Maxwell's equations can once more break down the problem into the study of null rays, which now come with associated null tetrads.  We analyze the evolution of these null rays and their associated tetrads, and we find that the null rays are no longer geodesics, and that the deviation of these rays from their previously geodesic paths is helicity dependent.  This approach is known as the spinoptics approximation.  In this approximation we study propagation of polarized light as it is scattering by rotating black holes.

We study the spinoptics approximation in the case of spin waves moving at the speed of light which are scattered by rotating black holes, and analyze the resulting motion in detail. Lastly, we provide a few more examples of effects that arise from the spinoptics approximation, and present them for future study. We will now begin by describing the geometric optics approximation in greater detail.

\section{Geometric Optics Approximation} \label{s51}

For a more complete description of the geometric optics approximation and its first-order corrections in a curved spacetime, see \cite{Ehlers1967-lg,MTW,Dwivedi1972-df}.  In what follows, we will be following the calculations made in \cite{MTW}.  There are three main results we want to extract from the geometric optics approximation:
\begin{enumerate}
    \item Trajectories of light rays are null geodesics within a given spacetime.
    \item The polarization vector of the light ray is parallel propagated along the trajectory of the ray, and it always remains orthogonal to the tangent vector of the null ray.
    \item The photon number of the ray is conserved, at least under an adiabatic approximation of the system. More detail on this later.
\end{enumerate}

This approximation is possible as long as it is subject to two conditions:
\begin{enumerate}[label=\alph*)]
    \item The wavelength of the monochromatic wave $\lambda$ should be much smaller than the length scale on which its characteristic quantities (wavelength, amplitude, polarization, etc.) may vary, which we'll call $L_a$.
    \item The wavelength should also be much smaller than the characteristic length $L_b$ associated with the curvature of the spacetime.  In \cite{MTW}, a more exact definition of this scale is given.  In particular, for scattering of the light by a black hole this length parameter is on the order of its gravitational radius $L_b\sim M$.
\end{enumerate}
To meet both of the above conditions, we simply put $\lambda\ll\text{min}\{L_a,L_b\}$ and designate a small parameter to be $\lambdabar\equiv \lambda/(2\pi)=1/\omega$, where $\omega$ is the angular frequency of the electromagnetic wave. 

Assuming that $\lambdabar$ is sufficiently small, we are now able to formulate an ansatz for the waves in question under the geometric optics approximation.  Let's define a wave by the following four-vector
\begin{equation} \label{AmuA}
    A_{\mu} = \Re\big[\mathcal{A}_{\mu}\exp(i\theta/\lambdabar)\big]\,,
\end{equation}
where $\theta$ is proportional to the phase of the wave and $\mathcal{A}_{\mu}$ describes its complex amplitude.  To denote that $\theta$ is not exactly the phase function itself, we call it $\theta$ the "associated phase function." The key to calculating this approximation is to assume that the complex amplitude $\mathcal{A}_{\mu}$ and associated phase function $\theta$ change very slowly with respect to, say, a stationary observer's clock, depending on how small the wavelength is. To compensate for this we assume that the phase is changing very quickly with respect to said observer, at a rate inversely proportional to wavelength.  Mathematically, we would expect something like
\begin{equation} \label{AmuAnsatz}
    A_{\mu}\approx\Re\big[(a_\mu+\lambdabar b_\mu+\lambdabar^2 c_{\mu}+...)\exp(i\theta/\lambdabar)\big]\,.
\end{equation}
It turns out that this is all we will need to prove the three above assertions.  All that is left is to enforce that $\boldsymbol{A}$ satisfies the free wave equation in the Lorentz gauge.  Before doing this, we quickly define the wave's scalar amplitude, polarization vector, and wave vector to be $a=|\boldsymbol{a}|$, $\boldsymbol{f}=\boldsymbol{a}/a$, and $k_{\mu}=\lambdabar^{-1}\theta_{;\mu}$, respectively.

We will provide derivations for the above list of three assertions in order.  To achieve this, we start by applying the wave equation to the wave vector $\boldsymbol{A}$
\begin{align} \label{waveEq}
    0=&-\Box A^{\mu} +R^{\mu}_{\;\,\nu}A^{\nu}\\\nonumber
    =& -\Re\big[(ik^{\nu}_{\;\,;\nu}/\lambdabar)(a^{\mu}+\lambdabar b^{\mu}+...)e^{i\theta/\lambdabar}+(2ik^{\nu}/\lambdabar)(a^{\mu}+\lambdabar b^{\mu}+...)_{;\nu}\,e^{i\theta/\lambdabar}\\\nonumber
    &\hspace{27pt}-(k^{\nu}k_{\nu}/\lambdabar^2)(a^{\mu}+\lambdabar b^{\mu}+...)e^{i\theta/\lambdabar}+(a^{\mu}+\lambdabar b^{\mu}+...)^{;\nu}_{\;\,;\nu}\,e^{i\theta/\lambdabar}\big]\\\nonumber
    &+R^{\mu}_{\;\,\nu}\,\Re\big[(a^{\nu}+\lambdabar b^{\nu}+...)e^{i\theta/\lambdabar}\big]\,,
\end{align}
where $\Box\equiv\nabla^{\nu}\nabla_{\nu}$\footnote{This equation can be reformulated in terms of the wave operator $\Delta_{dR}\boldsymbol{A}=0$. A brief aside on where this wave operator, known as the "Laplace-de Rham wave operator" comes from.  In the language of physics, this equation follows from satisfying one of the source-free Maxwell equations in tensor form: $F^{\mu\nu}_{\;\;\;;\nu}=0$ where $F_{\mu\nu}=A_{\mu;\nu}-A_{\nu;\mu}$.  In the language of differential geometry, we are requiring that the $U(1)$ connection one-form $\boldsymbol{A}$, which is related to the curvature two-form $\boldsymbol{F}$ by $\boldsymbol{F}=d\boldsymbol{A}$, must also be a harmonic form, \emph{i.e.} $\Delta_{dR}\boldsymbol{A}=(d\delta+\delta d)\boldsymbol{A}=0$.  This has very deep physical and geometric implications that we will not delve further into here, but the interested reader can read more about this in e.g. \cite{Nakahara1990-dr}.}. 

Now, returning to equation \eqref{waveEq}, we can solve it perturbatively by matching terms of equal order in $\lambdabar$.  We will only need the lowest two orders of the expansion
\begin{align} \label{orders}
    O(\lambdabar^{-2}):&\hspace{20pt} k^{\mu}k_{\mu}=0\,,\\\nonumber
    O(\lambdabar^{-1}):&\hspace{20pt} k^{\nu}_{\,\;;\nu}a^{\mu}+2k^{\nu}a^{\mu}_{\;\,;\nu}=0\,.
\end{align}

The second of these equations will turn out to be what we need to prove the second and third geometric optics assumptions, so we will return back to that later.  For the first of the two equations, we will use it to derive the geodesic equation.  With this in mind, we take its covariant derivative, and play around with it to get
\begin{equation}
    0=(k^{\nu}k_{\nu})_{;\mu} = 2k^{\nu}k_{\nu;\mu} = 2k^{\nu}k_{\mu;\nu} \implies \nabla_{\boldsymbol{k}}\boldsymbol{k}=0\,.
\end{equation}
Thus, we have proved the first assertion.  The last equality before the arrow comes from the fact that $\boldsymbol{k}$ is the gradient of a scalar, so $k_{\nu;\mu}=\theta_{;\nu\mu}=\theta_{;\mu\nu}=k_{\mu;\nu}$.

Going back to the second equation in \eqref{orders}, with some clever work one can show that
\begin{equation}
    \nabla_{\boldsymbol{k}}\boldsymbol{f}=0\,.
\end{equation}
In other words, the polarization vector is parallel propagated along the null geodesic ray.

To obtain the final part of the second assumption, we enforce the Lorentz gauge on $\boldsymbol{A}$.  We have
\begin{equation}
    A^{\mu}_{\,\;;\mu} = 0\,. 
\end{equation}
At the lowest order, this gives us $k_{\mu}a^{\mu}=0$, which of course implies that $\boldsymbol{k}\cdot\boldsymbol{f}=0$.

Lastly, we can also further manipulate the second equation of \eqref{orders} to show that
\begin{equation}
    \nabla_{\mu}(a^2k^{\mu})=0\,.
\end{equation}
In a classical context, this corresponds to $a^2\boldsymbol{k}$ being a conserved current, but on the quantum level this would correspond to a conservation of photon number.  Once more, we stress this to be only approximately true, as is the case with photon number in general, since it is an adiabatic invariant.

We note that since both $\boldsymbol{k}$ and $\boldsymbol{f}$ are parallel propagated along the null geodesic ray, if $\boldsymbol{k}\cdot\boldsymbol{k}=\boldsymbol{k}\cdot\boldsymbol{f}=0$ and $\boldsymbol{f}\cdot\boldsymbol{\bar{f}}=1$ at one point along the null ray, it is true along the entire trajectory of the ray.

It should also be noted that in order for the polarization vector $\boldsymbol{f}$ to be parallel transported along the geodesic, it will necessarily experience a shift in orientation.  In fact, in the presence of either bending of the ray or a gravitational field of a rotating massive object in the spacetime, the polarization vector will accordingly rotate around the null ray.  This effect is known as the gravitational Faraday effect, or the "Skrotskii effect," which was first studied by Georgii Skrotskii in 1957 \cite{Skrotskii1957}.  As one would imagine, this effect is analogous to the electromagnetic Faraday effect.  The similarities between the propagation of electromagnetic waves in a gravitational field and the propagation of electromagnetic waves in an optical medium begs the question: what other analogous effects might be present that are currently missing from the geometric optics approximation detailed above?

It is a well-known result that one can describe the propagation of electromagnetic waves in the presence of a gravitational field as if it was instead propagating through some anisotropic optical medium (see e.g. \cite{Plebanksi:1960,Volkov:1971} and references therein).  This result can be corroborated by the presence of the gravitational Faraday effect and its analogous nature to the electromagnetic Faraday effect.  

In brief, the Faraday effect is a physical magneto-optical effect which causes a rotation in the polarization vectors of electromagnetic waves that is proportional to the projection of the magnetic field along the direction of light propagation.

Another equally important optical effect, which specifically occurs in inhomogeneous media, is the optical Magnus effect, which is also known as the spin Hall effect of light.  This effect can be summarized as a helicity-dependent altering of light wave trajectories due to the spin-orbit coupling between its intrinsic spin and its orbital motion within a specific medium.

Both of the above examples lead us to believe that the helicity of electromagnetic radiation as it propagates through a gravitational background is a crucial feature in understanding its motion and evolution.  For this reason, we set out to define helicity-dependent corrections to the geometric optics approximation and analyze how they affect the motion of the electromagnetic waves in question.

\section{The Spinoptics Equations} \label{s52}

As stated in \cite{Frolov2011-rw}, we can conclude that for some gravitational Faraday effect, \emph{i.e.} the path that a photon takes within a gravitational field affecting the direction of its spin, we should also expect an inverse effect.  If the spin of a photon is altered within a gravitational field, its path should be altered from that of a null geodesic ray.  This is what is known as the "spinoptics effect."  

Rough arguments and estimates on this phenomenon were given by Bahram Mashhoon in the early-mid 70's \cite{Mashhoon1973-ec,Mashhoon1974-ek,Mashhoon1974-vj,Mashhoon1975-ey}.  More specifically, it was shown that both rotating and non-rotating gravitational objects should scatter light in a manner that is helicity-dependent.  Left-circular and right-circular polarized light rays should be scattered differently, meaning that we should expect incoming unpolarized light passing by a black hole to end up polarized on its outgoing trajectory.

The first concrete analytical approximation of the spinoptics approximation was given by Frolov and Shoom in \cite{Frolov2011-rw}, and a covariant description of this was given shortly after by Yoo \cite{Yoo2012-vf}.  A more general covariant approach derived from a covariant action was discovered in \cite{Oancea:2020khc}, and was rediscovered and recast in a different formalism shortly after by \cite{Frolov:2020uhn}.  A few years later, this covariant approach was reformulated in the formalism of studying null rays and complete null tetrads associated with them \cite{Frolov2024-ms}. This is the approach that we will be using below to study the spinoptics effect, and it allows for the study of both electromagnetic and gravitational waves. 

Lastly, it should be noted that the spinoptics equations were recently shown to be integrable for a wide class of type D spacetimes via the discovery of a generalized Carter constant, but the result has yet to be extended to all type D spacetimes \cite{Andersson:2025bhq}. For a full history and list of papers on spinoptics, see \cite{A4} and references therein.

\subsection{Effective Action Formulation} \label{s521}

The starting point for the mathematical formulation of the spinoptics effect comes from the idea that we can describe the evolution of monochromatic waves of light by attaching null rays to them with associated null tetrads and studying the evolution of both the rays and the tetrads simultaneously \cite{Pineault:1977on,Pineault:1977tw}.  To see where these null tetrad vectors come from, we must study the Maxwell action in an arbitrary curved spacetime with metric $g_{\mu\nu}$, as was done in \cite{Frolov2024-ms}. For this subsection, we will be following the effective action formalism for the spinoptics equations developed in \cite{Frolov2024-ms}. The complexified Maxwell action in curved spacetime is given by
\begin{equation} \label{Maxwell}
    W = \frac{1}{8}\int\mathcal{F}^2\sqrt{-g}\,d^4x\,,\hspace{15pt}\mathcal{F}^2=F_{\mu\nu}\bar{F}^{\mu\nu}\,,
\end{equation}
where $F_{\mu\nu}$ is the complex electromagnetic tensor, $\bar{F}_{\mu\nu}$ is its complex conjugate, and it is assumed that $F^{\mu\nu}_{\;\;\;;\nu}=0$.  We use the complex form of Maxwell's equations because it is the most natural setting to talk about circularly-polarized light (see e.g. \cite{Landau1980Classical}).  To this effect, we omit the symbol $\Re$ in the previous ansatz in Eq.~\eqref{AmuA}, and use a complex electromagnetic four-potential for a circularly-polarized electromagnetic wave of the form
\begin{equation} \label{Amu}
    A_{\mu} = aM_{\mu}\exp{(i\omega S)}\,,
\end{equation}
where $\boldsymbol{M}$ is a complex null unit polarization vector, $a$ is the scalar magnitude of the four-potential, $\omega$ is the angular frequency of the wave, and $S$ is a scalar function that, when constant, describes the surfaces of the wavefronts.  

We enforce the following constraint conditions
\begin{equation} \label{Lagrange}
    M_{\mu}M^{\mu}=\bar{M}_{\mu}\bar{M}^{\mu}=0\,,\hspace{15pt}M_{\mu}\bar{M}^{\mu}=1\,,\hspace{15pt}M_{\mu}S^{,\mu}=\bar{M}_{\mu}S^{,\mu}=0\,.
\end{equation}
The first two conditions are exactly what are required to describe null unit polarization vectors for circularly polarized light.  The last condition corresponds to the enforcement of the Lorentz gauge for the light wave's four-potential \eqref{Amu} at lowest order in a $1/\omega$ expansion.

An explicit calculation of the electromagnetic tensor given the form of Eq.~\eqref{Amu} yields
\begin{equation}
    F_{\mu\nu} = A_{\nu;\mu}-A_{\mu;\nu} = \Big[i\omega a\big(S_{,\mu}M_{\nu}-S_{,\nu}M_{\mu}\big)+\big[(aM_{\nu})_{;\mu}-(aM_{\mu})_{;\nu}\big]\Big]e^{i\omega S}\,.
\end{equation}
Therefore, calculating $\mathcal{F}^2$ up to second order in a high-frequency expansion gives
\begin{equation} \label{Fapprox}
    \mathcal{F}^2\approx 4\omega^2 a^2\bigg[\frac{1}{2}(\nabla S)^2-\frac{1}{\omega}B^{\mu}S_{,\mu}+O(1/\omega^2)\bigg]\,,\hspace{15pt}B_{\mu}=\frac{i}{2}\big(\bar{M}^{\nu}M_{\nu;\mu}-M^{\nu}\bar{M}_{\nu;\mu}\big)\,.
\end{equation}
We are getting a bit ahead of ourselves with this comment, but here $B_{\mu}$ is the Berry's connection associated with the Berry's phase of some null ray trajectory.  For more information on this deep relationship, see Oancea, et. al. \cite{Oancea:2020khc} and references therein.  We also note that we are following the same validity assumptions that were used in the derivation of the geometric optics equations.  Therefore, the $O(1/\omega^2)$ term that is not present in the brackets of the left-hand equation in \eqref{Fapprox} is of the order $(\lambda/L_a)^2$ where, as we'll recall, $\lambda$ is the wavelength of the monochromatic wave and $L_a$ is the length scale at which the characteristics of the wave change.  Hence, it is reasonable to keep this omitted.

There are two facets to developing an effective action from Eq.~\eqref{Maxwell}.  The first, which we have already accomplished, is enforcing a high-frequency approximation so that we cut off the action after a first-order expansion. The second is to naturally add the physical constraints given in Eq.~\eqref{Lagrange} as Lagrange multipliers.  This results in the following effective action
\begin{align} \label{eff}
    &I = \int d^4{x}\sqrt{-g}\bigg[a^2\Big[\frac{1}{2}(\nabla S)^2-\frac{1}{\omega}B^{\mu}S_{,\mu}\Big]+\Lambda\bigg]\,,\\[5pt]\nonumber
    &\Lambda = \frac{1}{2}\bar{\lambda}_1 M_{\mu}M^{\mu}+\frac{1}{2}\lambda_1\bar{M}_{\mu}\bar{M}^{\mu}+\lambda_2(M_{\mu}\bar{M}^{\mu}-1)+\bar{\lambda}_3 M^{\mu}S_{,\mu}+\lambda_3\bar{M}^{\mu}S_{,\mu}\,.
\end{align}

From this action, we count four possible variations with respect to the previously-defined field variables $a$, $S$, $\boldsymbol{M}$, and $\bar{\boldsymbol{M}}$, as well as five additional variations that come from the Lagrange multipliers.  We will consider the variations with respect to the Lagrange multipliers done and satisfied, and so we turn our attention to the other possible variations.  They give the following equations:
\begin{itemize}
    \item Variation with respect to $a$
    \begin{equation} \label{v1}
        H \equiv \frac{1}{2}(\nabla S)^2 - \frac{1}{\omega}B^{\mu}S_{,\mu}=0\,.
    \end{equation}
    \item Variation with respect to $S$
    \begin{equation} \label{v2}
        J^{\mu}_{\;;\mu}=\Big[a^2\big(S^{,\mu}-\frac{1}{\omega}B^{\mu}\big)+\bar{\lambda}_3 M^{\mu}+\lambda_3\bar{M}^{\mu}\Big]_{;\mu} = 0\,.
    \end{equation}
    \item Variation with respect to $M^{\mu}$
    \begin{equation} \label{v3}
        \frac{a^2i}{\omega}S^{,\nu}\bar{M}_{\mu;\nu}+\bar{\lambda}_1 M_{\mu}+\bar{\lambda}_2\bar{M}_{\mu}+\bar{\lambda}_3 S_{,\mu}=0\,.
    \end{equation}
    \item Variation with respect to $\bar{M}^{\mu}$
    \begin{equation} \label{v4}
        \frac{a^2i}{\omega}S^{,\nu}{M}_{\mu;\nu}-{\lambda}_1 \bar{M}_{\mu}-{\lambda}_2{M}_{\mu}-{\lambda}_3 S_{,\mu}=0\,.
    \end{equation}
\end{itemize}

Looking at Eq.~\eqref{v2}, it clearly describes a current conservation equation.  Turning to Eqs.~\eqref{v3} and \eqref{v4}, we can multiply them by $\bar{M}^{\mu}$ and $M^{\mu}$, respectively, to obtain $\lambda_1=\bar{\lambda}_1=0$.  To eliminate $\lambda_2$ and $\bar{\lambda_2}$ from the equations, we must use one of the gauge freedoms that we have at hand.  Specifically, we will choose to fix the freedom to rotate the polarization vectors in their shared plane by taking
\begin{equation} \label{mtrans}
    M^{\mu}\to\check{M}^{\mu}=e^{i\psi}M^{\mu}\,,\hspace{15pt} \bar{M}^{\mu}\to\bar{\check{M}}^{\mu}=e^{-i\psi}\bar{M}^{\mu}\,.
\end{equation}
Accordingly,
\begin{equation}
    B^{\mu}S_{,\mu}\to\check{B}^{\mu}S_{,\mu}=(B^{\mu}-\psi^{,\mu})S_{,\mu}\,.
\end{equation}
Therefore, we have the freedom to fix $\psi$ such that $B^{\mu}S_{,\mu}=0$. Multiplying Eq.~\eqref{v3} by $M^{\mu}$ and Eq.~\eqref{v4} by $\bar{M}^{\mu}$, we find that $\lambda_2=\bar{\lambda}_2=0$.  We are left with
\begin{equation} \label{lambda3}
    \lambda_3 S_{,\mu} = \frac{i}{\omega}a^2S^{,\nu}M_{\mu;\nu}\,,\hspace{15pt}\bar\lambda_3 S_{,\mu} = -\frac{i}{\omega}a^2S^{,\nu}\bar{M}_{\mu;\nu}\,.
\end{equation}

Turning our attention to Eq.~\eqref{v1}, we can use the guidance of \cite{Stachel:1977cm} to identify it with the Hamilton-Jacobi equation for a wave with wavevector $S_{,\mu}$.  Furthermore, we can imagine that this wavevector corresponds to the generalized momentum of some ray trajectory, and take $S_{,\mu}=p_{\mu}$.  This gives us the following Hamiltonian
\begin{equation} \label{Ham}
    \mathcal{H}(p,x) = \frac{1}{2}g^{\mu\nu}p_{\mu}p_{\nu}-\frac{1}{\omega}B^{\mu}p_{\mu}\,.
\end{equation}
This Hamiltonian produces a corresponding Lagrangian
\begin{equation}
    L(x,\dot{x})=p_{\mu}\dot{x}^{\mu}-\mathcal{H}\,,\hspace{15pt}\dot{x}^{\mu}=\frac{\partial\mathcal{H}}{\partial p_{\mu}\,,}
\end{equation}
which allows us to write the equations of motion within a worldline formalism
\begin{equation}
    \mathcal{S} = \int L\,d\lambda = \frac{1}{2}\int\eta\dot{x}_{\mu}\dot{x}^{\mu}\,d\lambda+\frac{1}{\omega}\int B_{\mu}\dot{x}^{\mu}\,d\lambda\,,
\end{equation}
where $\eta$ is defined such that the above action preserves reparameterization invariance.

Now, instead of studying the entire motion of the full monochromatic wave, which analytically would be extremely complex in the background of a rotating black hole, we instead treat every point on a given initial wavefront as the initial condition for a single null ray, and trace the evolution of these individual rays.  Analytically, this is now a very feasible problem, as we will see below.

Applying Hamilton's equations to Eq.~\eqref{Ham}, we have
\begin{equation}
    \frac{dx^{\mu}}{d\lambda} = \frac{\partial\mathcal{H}}{\partial p_{\mu}} = p^{\mu} - \frac{1}{\omega}B^{\mu}\,,\hspace{15pt}\frac{dp_{\mu}}{d\lambda} = -\frac{\partial\mathcal{H}}{\partial x^{\mu}}\,.
\end{equation}
We identify the first equation with the tangent vector of the null ray, and after some algebra we can recast the second equation as an acceleration equation for the null ray.  We have
\begin{equation} \label{l}
    \frac{d x^{\mu}}{d\lambda} \equiv l^{\mu} = p^{\mu}-\frac{1}{\omega}B^{\mu}\,,
\end{equation}
and
\begin{equation} \label{w}
    w^{\mu}\equiv l^{\nu}l^{\mu}_{\;;\nu} = \frac{1}{\omega}\mathcal{K}^{\mu}_{\;\;\nu}\,l^{\nu}\,,\hspace{15pt}\mathcal{K}_{\mu\nu}=B_{\nu;\mu}-B_{\mu;\nu}\,.
\end{equation} 
One can see that this equation implies
\begin{equation}
    w^{\mu}l_{\mu} = 0\,,\hspace{15pt} l^{\mu}(l^{\nu}l_{\nu})_{;\mu}=0\,,
\end{equation}
meaning that the acceleration of the ray is always perpendicular to its tangent vector, and that if the ray is initially null, it will remain null for all of its trajectory. We note that if $B_{\mu}$ is the Berry connection associated with a geometric phase of the polarization vectors, then $K_{\mu\nu}$ is its Berry curvature.  

We would also like to mention that there exist several different applicable formulations of the spinoptics approximation, including but not limited to the use of Fermat's principle in order to find the shortest path given by this effect \cite{Duval:2005ky,Duval:2005ry}.

\subsection{Null Tetrad Evolution} \label{s522}

We will now consider a special form of a null tetrad, which we will call its "canonical form," and attempt to match this tetrad up to the evolution equations in the spinoptics approximation.  We start by considering the following general null tetrad $(\hat{\boldsymbol{l}},\,\hat{\boldsymbol{n}},\,\hat{\boldsymbol{m}},\,\bar{\hat{\boldsymbol{m}}})$, which obeys the relations
\begin{equation} \label{lnmmh}
    -(\hat{\boldsymbol{l}}\cdot\hat{\boldsymbol{n}}) = (\hat{\boldsymbol{m}}\cdot\bar{\hat{\boldsymbol{m}}}) = 1\,,
\end{equation}
and all other inner products within the tetrad vanish. A priori, we assume that the components of these tetrads are dimensionless in Cartesian or Cartesian-like coordinates. There are four important gauge transformations that preserve this inner product structure within the tetrad
\begin{enumerate}
    \item \label{i1}$
        m^{\mu}\to m^{\mu}+\beta l^{\mu}\,,\hspace{15pt}\bar{m}^{\mu}\to \bar{m}^{\mu}+\bar{\beta} l^{\mu}\,,\hspace{15pt}n^{\mu}\to n^{\mu}+\beta\bar{m}^{\mu}+\bar{\beta}m^{\mu}+\beta\bar{\beta}l^{\mu}$\,.
    \item \label{i2}$
        m^{\mu}\to e^{i\varphi}m^{\mu}\,,\hspace{15pt}\bar{m}^{\mu}\to e^{-i\varphi}\bar{m}^{\mu}$\,.
    \item \label{i3}
        $l^{\mu}\to\alpha l^{\mu}\,,\hspace{15pt}n^{\mu}\to\alpha^{-1}n^{\mu}$\,.
    \item \label{i4}
        $m^{\mu}\to\bar{m}^{\mu}\,,\hspace{15pt}\bar{m}^{\mu}\to m^{\mu}$\,.
\end{enumerate}
These gauge transformations are related to Wigner's "little group" for massless particles, which is a subgroup of the Poincar\'{e} group \cite{Wigner1939}. Via the three of these four transformations which are continuous, we will mold the evolution of the null tetrad into its "canonical form," and the fourth discrete transformation equation will come into play later.  First, note that the evolution of the $\hat{\boldsymbol{l}}$ vector in its most general form looks like
\begin{equation}
    \hat{D}\hat{l}^{\mu} = \hat{w}_0\hat{l}^{\mu}+\bar{\hat{\kappa}}\hat{m}^{\mu}+\hat{\kappa}\bar{\hat{m}}^{\mu}\,,
\end{equation}
where $\hat{w}_0$ is a real scalar and $\hat{\kappa}$ is a complex scalar. From here on out, we will adopt the shorthand
\begin{equation}
    D\equiv l^{\mu}\nabla_{\mu}\,,
\end{equation}
for any tangent vector $\boldsymbol{l}$.  Similarly, the evolution of $\hat{\boldsymbol{m}}$ should have the following general form
\begin{equation} \label{Dmh}
    \hat{D}\hat{m}^{\mu} = \hat{\mu}_0\hat{l}^{\mu}+i\hat{\mu}\hat{m}^{\mu}+\hat{\kappa}\hat{n}^{\mu}\,,
\end{equation}
where $\hat{\mu}$ is a real scalar and $\hat{\mu}_0$ is a complex scalar.

Utilizing Transformation \ref{i1}, we can write
\begin{align}
    &\hat m^{\mu}= \check m^{\mu}+\beta \check l^{\mu}\,,\hspace{15pt}\bar{\hat m}^{\mu}= \bar{\check m}^{\mu}+\bar{\beta} \check l^{\mu}\,,\\\nonumber
    &\hat n^{\mu}= \check n^{\mu}+\beta\bar{\check m}^{\mu}+\bar{\beta}\check m^{\mu}+\beta\bar{\beta}\check l^{\mu}\,,\hspace{15pt} \hat{l}^{\mu}=\check{l}^{\mu}\,.
\end{align}\
If we choose $\beta$ to obey the following differential equation
\begin{equation}
    \hat{D}\beta = \hat{\mu}_0+i\hat{\mu}\beta-\hat{w}_0\beta-\bar{\hat{\kappa}}\beta^2\,,
\end{equation}
then Eq.~\eqref{Dmh} becomes
\begin{equation} \label{Dmc}
    \check{D}\check{m}^{\mu} = i\check \mu \check m^{\mu} + \check\kappa\check n^{\mu}\,,
\end{equation}
where $\check\mu$ and $\check\kappa$ are defined in terms of the hatted scalars via Transformation \ref{i1}.

Now, using Transformation \ref{i2} on Eq.~\eqref{Dmc}, we write
\begin{equation}
    \check m^{\mu} =  e^{i\varphi} \mathring m^{\mu}\,,\hspace{15pt} \bar{\check m}^{\mu} =  e^{-i\varphi} \bar{\mathring m}^{\mu}\,,\hspace{15pt}\check l^{\mu} = \mathring l^{\mu}\,,\hspace{15pt}\check n^{\mu} = \mathring n^{\mu}\,.
\end{equation}
If we fix $\varphi$ such that
\begin{equation}
    \check D\varphi = \check\mu\,,
\end{equation}
then Eq.~\eqref{Dmc} becomes
\begin{equation}
    \mathring D\mathring m^{\mu} = \mathring\kappa\mathring n^{\mu}\,.
\end{equation}

One can check that after Transformations \ref{i1} and \ref{i2}, the evolution of the tetrad is given by
\begin{equation} \label{ring}
    \mathring D \mathring l^{\mu} = \mathring{w}_0\mathring l^{\mu} + \bar{\mathring{\kappa}}\mathring m^{\mu}+\mathring\kappa\bar{\mathring m}^{\mu}\,,\hspace{15pt}\mathring D \mathring m^{\mu} = \mathring\kappa\mathring n^{\mu}\,,\hspace{15pt}\mathring D \mathring n^{\mu} = -\mathring{w}_0\mathring n^{\mu}\,.
\end{equation}
Again, the ringed scalars are defined in terms of the hatted scalars from Transformation \ref{i2}.

We apply one last transformation, namely Transformation \ref{i3}, on this set of equations.  Taking
\begin{equation}
    \mathring l^{\mu} = \alpha l^{\mu}\,,\hspace{15pt}\mathring n^{\mu} = \alpha^{-1} n^{\mu}\,,\hspace{15pt} \mathring{m}^{\mu} = m^{\mu}\,,\hspace{15pt}\bar{\mathring{m}}^{\mu} = \bar{m}^{\mu}\,,
\end{equation}
and requiring that
\begin{equation}
    D\alpha = \mathring{w}_0\,,
\end{equation}
we finally arrive at the canonical tetrad equations for the tetrad $(\boldsymbol{l},\,\boldsymbol{n},\,\boldsymbol{m},\,\hat{\boldsymbol{m}})$:
\begin{equation} \label{canon0}
    Dl^{\mu} = \bar\kappa m^{\mu} + \kappa \bar m^{\mu}\,,\hspace{15pt}Dm^{\mu} = \kappa n^{\mu}\,,\hspace{15pt}Dn^{\mu}=0\,,
\end{equation}
where once more the scalars are defined in terms of the ringed scalars from previous transformations.  Let us summarize what we have done. We started with a general null tetrad that obeys the orthogonality relations \eqref{lnmmh}.  The evolution of these null tetrad vectors as they are transported along a null ray $\hat{\boldsymbol{l}}$ have a general form. Using gauge transformations that preserve these orthogonality relations, and which are related to Wigner's little subgroup of the Poincar\'{e} group for massless particles, we have simplified these evolution equations to a simpler form given by \eqref{canon0}, which we call the "canonical" form of the tetrad.

A couple of notes on Transformations \ref{i1}-\ref{i4}:
\begin{itemize}
    \item These "little group" gauge transformations are \emph{not} the same as the gauge transformations for electromagnetic fields that obey Maxwell's equations in curved spacetime.  They are two separate gauge schemes for two separate objects of study that may coincide in very specific instances.
    \item With the above statement in mind, if Transformation \ref{i1} is applied to the polarization vector $M^{\mu}$, it constitutes a restricted type of $U(1)$ gauge transformation, and therefore can be thought of a subgroup of the $U(1)$ gauge group in this specific scenario.
    \item When looking at Transformation \ref{i2} in the context of the polarization vectors, it corresponds to rotations of these vectors within their shared plane.  We already used this transformation in Eq.~\eqref{mtrans}, but we have yet to identify $\boldsymbol{M}$ with $\boldsymbol{m}$, so for now we think of them as unrelated objects.
    \item Transformation \ref{i3} in the context of $\boldsymbol{l}$ being a tangent vector to a null ray simply corresponds to a transformation of the affine parameter of the ray.
    \item Lastly, we expect that Transformation \ref{i4} is closely related to the helicity of electromagnetic waves.  This will be a crucial result in our analysis of the spinoptics approximation, and what sets it apart from geometric optics. More on this later...
\end{itemize}

We will use Eq.~\eqref{canon0} for future spinoptics calculations, but its form will slightly differ in that the right hand side of each equation will be accompanied by a factor of $1/\omega$.  This is because we are perturbatively expanding our solutions from zeroth-order solutions which are parallel transported, meaning that the perturbations will show up as $1/\omega$ for first order high-frequency corrections.  Therefore, the actual form of the "canonical" equations that we will use is
\begin{equation} \label{canon}
    Dl^{\mu} = \frac{1}{\omega}(\bar\kappa m^{\mu} + \kappa \bar m^{\mu})\,,\hspace{15pt}Dm^{\mu} = \frac{1}{\omega}\kappa n^{\mu}\,,\hspace{15pt}Dn^{\mu}=0\,.
\end{equation}
It is easy to confirm that if the frequency of a monochromatic wave approaches an infinite value $\omega\to\infty$, then the tetrad in Eq.~\eqref{canon} will be parallel transported.

We would now like to match the above tetrad with the previous spinoptics equations.  The first order of business is to identify the tangent vector $\boldsymbol{l}$ in Eq.~\eqref{l} with the first vector of the null tetrad.  It is easy to check via Eq.~\eqref{w} that this vector is indeed parallel transported if $\omega\to\infty$.  

Next, we assume that if the conditions of slow evolution of the system under this effect are correct, then all of the tetrad vectors should be able to be expanded perturbatively as $1/\omega$ expansions.  Therefore, at first order in the approximation, we should be able to write the polarization vector (and its complex conjugate) as
\begin{equation}
    M^{\mu} = m^{\mu} + \frac{1}{\omega}\mu^{\mu}\,.
\end{equation}
Going further with this approximation, the Berry connection becomes
\begin{equation}
    B^{\mu} = b^{\mu} + \frac{1}{\omega}\beta^{\mu}\,,\hspace{15pt}b_{\mu} = \frac{i}{2}(\bar{m}^{\nu}m_{\nu;\mu}-m^{\nu}\bar{m}_{\nu;\mu})\,,\hspace{15pt}\beta_{\mu} = i(\bar{\mu}^{\nu}m_{\nu;\mu}+\bar{m}^{\nu}\mu_{\nu;\mu})\,.
\end{equation}
Therefore, at first order in the expansion, the evolution equation \eqref{w} reduces to
\begin{equation}
    Dl^{\mu} = \frac{1}{\omega}\mathcal{K}^{\mu\nu}_{0}l_{\nu}\,,\hspace{15pt}\mathcal{K}_0^{\mu\nu} = b^{\nu;\mu}-b^{\mu;\nu}\,,\hspace{15pt}l_{\mu} = p_{\mu}-\frac{1}{\omega}b_{\mu}\,.
\end{equation}

Simplifying even further, if we assume that we can find a canonical tetrad given by \eqref{canon} that describes the system, then
\begin{align}
    Dl_{\mu} = &\frac{i}{\omega}\Big[(m_{\alpha;\nu\mu}-m_{\alpha;\mu\nu})\bar{m}^{\alpha}l^{\nu}+(\bar{m}_{\alpha;\mu}Dm^{\alpha}-m_{\alpha;\mu}D\bar{m}^{\alpha})\Big]\\
    =&\frac{i}{\omega}(m_{\alpha;\nu\mu}-m_{\alpha;\mu\nu})\bar{m}^{\alpha}l^{\nu}\,,
\end{align}
where in the second line, we have employed the fact that $Dm^{\mu}=O(1/\omega)$ and $D\bar{m}^{\mu}=O(1/\omega)$ and cut off the equation at first order.  Lastly, using identities of the Riemann tensor, we arrive at the spinoptics evolution equation
\begin{equation} \label{spin1}
    Dl^{\mu} = \frac{i}{\omega}R^{\mu}_{\;\nu\alpha\beta}\,l^{\nu}\bar{m}^{\alpha}m^{\beta}\,.
\end{equation}
A couple of notes about this equation before we put it to use:

At first glance, Eq.~\eqref{spin1} is not necessarily in canonical form, although it is very close.  One can see that it is possible that an $n_{\mu}Dl^{\mu}$ term is still present here, but in order to get rid of it we would simply need to use Transformation \ref{i3} above to make the proper change in affine parameter.

Also, we never solved the equation \eqref{lambda3} for $\lambda_3$ for the action from which this equation was derived.  However, if we stick to the assumption that $Dm^{\mu}=O(1/\omega)$ and $D\bar{m}^{\mu}=O(1/\omega)$, then $\lambda_3=O(1/\omega^2)$ and $\bar{\lambda}_3=O(1/\omega^2)$, and therefore they can be thought of as having no contribution to the motion of the rays.

Lastly, we previously suggested that our equations may be helicity-dependent.  If we perform Transformation \ref{i4} on Eq.~\eqref{spin1}, we can see that due to properties of the Riemann tensor, it is helicity sensitive.  This is also true for the equations \eqref{canon} of the canonical tetrad.  For this reason, if we want to include helicity dependence in the spinoptics equations, we must include the variable $\sigma=\{-1,\,1\}$, so that the full form of the spinoptic evolution equations becomes\footnote{ Let us note that in our presentation, we restricted ourselves by considering only electromagnetic waves. However, it was shown that a similar spinoptics formalism can be developed for high frequency gravitational waves \cite{Frolov:2024qow}. The spinoptics equation for this case is identical to \eqref{spin} with only one change: $\sigma$ takes on the values $\pm 2$. }
\begin{equation} \label{spin}
    Dl^{\mu} = \frac{i\sigma}{\omega}R^{\mu}_{\;\nu\alpha\beta}\,l^{\nu}\bar{m}^{\alpha}m^{\beta}\,,\hspace{15pt}Dm^{\mu} = \frac{\sigma}{\omega}\kappa n^{\mu}\,,\hspace{15pt}Dn^{\mu} = 0\,.
\end{equation}

\section{Spinoptics Equations in the Kerr Spacetime} \label{s53}

We will now study the spinoptics effect on the scattering of circularly polarized light waves in the Kerr spacetime. To simplify our formulas, we use dimensionless quantities. Namely, we denote
\begin{equation}
    \tau = t/M\,,\hspace{15pt}\rho = r/M\,,\hspace{15pt}\alpha = a/M\,,
\end{equation}
so that the Kerr metric in its dimensionless form is
\begin{align} \label{KerrDim}
   &ds^2 = -\Big(1-\frac{2\rho}{\Sigma}\Big)d\tau^2+\frac{\Sigma}{\Delta}d\rho^2 - \frac{4\rho\alpha\sin^2\theta}{\Sigma}d\tau d\phi\\\nonumber
  &\hspace{30pt} +\Sigma d\theta^2+\Big(\rho^2+\alpha^2+\frac{2\rho\alpha^2}{\Sigma}\sin^2\theta\Big)\sin^2\theta d\phi^2\,,\\[5pt]\nonumber
   &\Sigma = \rho^2+\alpha^2\sin^2\theta\,,\hspace{15pt}\Delta=\rho^2-2\rho+\alpha^2\,,\hspace{15pt}\sqrt{-g}=\Sigma\sin\theta\,.
\end{align}
From here on out, we will use these redefinitions of $\Sigma$, $\Delta$, and $\sqrt{-g}$ in their dimensionless forms.  The outer horizon and ergosurface of the black hole are then defined by
\begin{equation}
    \rho_h = 1 + \sqrt{1-\alpha^2}\,,\hspace{15pt}\rho_e = 1+\sqrt{1-\alpha^2\cos^2\theta}\,.
\end{equation}

In the general case, solving ray and polarization transport equations is a rather complicated problem. To simplify calculations, we use the assumption that helicity dependent corrections are small (of order $1/\omega$). Thus, one can first solve the geodesic equations for a null ray and construct its associated parallel transported null tetrad, and then substitute a curvature-dependent "force" term into the expressions for the parallel propagated ray along this geodesic complex null tetrad in order to perturb the tetrad at first order in $1/\omega$. Fortunately, for the Kerr spacetime both null geodesics and the parallel transport equations for them and their associated tetrads allow analytical solutions due to explicit and hidden symmetries of the Kerr metric.  For definitions of these explicit and hidden symmetries, both in general and for the Kerr spacetime, see Appendix~\ref{A2}.  

\subsection{Construction of the Null Tetrad} \label{s531}

According to Liouville's theorem \cite{Liouville1855}, to integrate dynamical equations for particle and light motion it is sufficient to have four independent and commuting integrals of motion. In the Kerr spacetime, for the geodesic motion of a null ray with tangent vector $\boldsymbol{l}$, these integrals are: the energy $E$ conserved by the timelike Killing vector $\boldsymbol{\xi}$, the axial angular momentum $L_z$ conserved by the azimuthal Killing vector $\boldsymbol{\zeta}$, the mass of the particle $m$ conserved by the Killing tensor (and metric) $\boldsymbol{g}$, and the Carter constant $Q$ conserved by the Killing tensor $\boldsymbol{K}$
\begin{equation} \label{integrals0}
    E = -\xi_{\mu}l^{\mu}\,,\hspace{15pt}L_z = \zeta_{\mu}l^{\mu}\,,\hspace{15pt}m=g_{\mu\nu}l^{\mu}l^{\nu}\,,\hspace{15pt}Q=K_{\mu\nu}l^{\mu}l^{\nu}\,,
\end{equation}
where a definition for $\boldsymbol{K}$ in the Kerr metric is given by Eq.~\eqref{KTKerr}. For a null geodesic, $m=0$. It should be noted that these quantities are only conserved at the lowest order of the spinoptics approximation, and do not correspond to the fully-conserved quantities derived in \cite{Andersson:2025bhq}. Studying the spinoptics tetrad while including the full higher-order behavior of the scalar conserved quantities, one can expect that the evolution of null rays will exhibit a behavior that is akin to that of helically phased beams, as given in \cite{Yao:2011epa}.  This is a problem for future study.

The above quantities in \eqref{integrals0} are dimensional, and so we use an affine parameterization $\lambda\to E\lambda$, so that everything is multiplied by the right power of the energy to make it dimensionless, and
\begin{equation} \label{integrals}
     1 = -\xi_{\mu}l^{\mu}\,,\hspace{15pt}\ell_z = \zeta_{\mu}l^{\mu}\,,\hspace{15pt}0=g_{\mu\nu}l^{\mu}l^{\nu}\,,\hspace{15pt}\mathcal{Q}=K_{\mu\nu}l^{\mu}l^{\nu}\,,
\end{equation}
where $\ell_z=L_z/E$ and $\mathcal{Q}=Q/E^2$.

For some tangent vector $\boldsymbol{l}$, the hidden symmetries of the Kerr spacetime also give us a prescription for how to construct the other vectors in the null tetrad.  We write the following
\begin{equation} \label{een}
    e_2^{\mu} = \frac{1}{\sqrt{\mathcal{Q}}}k^{\mu\nu}l_{\nu}\,,\hspace{15pt}\tilde{e}_3^{\mu}=\frac{1}{\sqrt{\mathcal{Q}}}h^{\mu\nu}l_{\nu}\,,\hspace{15pt}\tilde{n}^{\mu} = \frac{1}{\sqrt{\mathcal{Q}}}(h^{\mu\nu}\tilde{e}_{3\nu}+\gamma l^{\mu})\,,
\end{equation}
where $\gamma$ will be determined later. Here, $\boldsymbol{h}$ and $\boldsymbol{k}$ are the principal tensor and Killing-Yano tensor in the Kerr spacetime, given by Eqs.~\eqref{principal} and \eqref{kYano}, respectively. There is a lot to discuss about this tetrad construction and the magic it beholds, so let's get into it.  

We first assume that the ray's tangent vector is a null geodesic, so that it obeys
\begin{equation}
    Dl^{\mu} = l^{\nu}\nabla_{\nu}l^{\mu} = 0\,.
\end{equation}
Then
\begin{equation}
    De_{2\mu} = \frac{1}{\sqrt{\mathcal{Q}}}\big(l^{\alpha}l^{\nu}k_{\mu\nu;\alpha}+l^{\alpha}l^{\nu}_{\;;\alpha}k_{\mu\nu}\big)=0\,.
\end{equation}
The first term in the above equality disappears due to the definition of a Killing-Yano tensor, given by Eq.~\eqref{KYTdef}, while the second term disappears due to the fact that $\boldsymbol{l}$ is parallel transported along the ray.  Therefore, from our construction of $\boldsymbol{e}_2$, it automatically is parallel transported as well.  Furthermore, since $\boldsymbol{k}$ is an antisymmetric tensor, we have
\begin{equation}
    \boldsymbol{e}_2\cdot\boldsymbol{l} = \frac{1}{\sqrt{\mathcal{Q}}}k_{\mu\nu}l^{\mu}l^{\nu} = 0\,.
\end{equation}

For our construction of $\tilde{\boldsymbol{e}}_3$, we are not so fortunate as to immediately extract a parallel transported vector, but we are close.  Namely, it is still orthogonal to the first two vectors.  Orthogonality to $\boldsymbol{l}$ is given by the fact that $\boldsymbol{h}$ is also antisymmetric
\begin{equation}
    \boldsymbol{\tilde e}_3\cdot\boldsymbol{l} = \frac{1}{\sqrt{\mathcal{Q}}}h_{\mu\nu}l^{\mu}l^{\nu} = 0\,.
\end{equation}
To prove orthogonality with $\boldsymbol{e}_2$, we need a more direct calculation
\begin{equation}
    \boldsymbol{e}_2\cdot\boldsymbol{\tilde e}_3 = \frac{\alpha\rho\cos\theta}{\mathcal{Q}}g_{\mu\nu}l^{\mu}l^{\nu} = 0\,.
\end{equation}
Next, by the definition of $\boldsymbol{h}$ given in Eq.~\eqref{hdef}, we see that
\begin{equation}
    D\tilde{e}_{3\mu} = \frac{1}{\sqrt{\mathcal{Q}}}(\boldsymbol{\xi}\cdot\boldsymbol{l})l_{\mu} = -\frac{1}{\sqrt{\mathcal{Q}}}l_{\mu}\,. 
\end{equation}
One can imagine that in order to make this vector parallel transported, we simply need to perform a transformation akin to Transformation \ref{i1} in Sec.~\ref{s522}
\begin{equation} \label{ee33}
    e_3^{\mu} = \tilde{e}_3^{\mu}+\Phi l^{\mu}\,,\hspace{15pt}\Phi = \frac{1}{\sqrt{\mathcal{Q}}}(\lambda+\lambda_0)\,,\hspace{15pt}De_3^{\mu}=0\,,
\end{equation}
where $\lambda_0$ is some constant that is yet to be fixed.  Then, to create two null vectors out of this, we choose
\begin{equation}
    Dm^{\mu} = \frac{1}{\sqrt{2}}(e_2^{\mu}+ie_3^{\mu})\,,\hspace{15pt}D\bar{m}^{\mu} = \frac{1}{\sqrt{2}}(e_2^{\mu}-ie_3^{\mu})\,,
\end{equation}
which obey
\begin{equation}
    Dm^{\mu}=D\bar{m}^{\mu} = 0\,.
\end{equation}
Although $\boldsymbol{m}$ and $\bar{\boldsymbol{m}}$ are the true null tetrad vectors, we will bounce back and forth between using these and using $\boldsymbol{e}_2$ and $\tilde{\boldsymbol{e}}_3$ for convenience and brevity.

Before we move on to our discussion of $\tilde{\boldsymbol{n}}$, we should discuss the magnitudes of $\boldsymbol{e}_2$ and $\tilde{\boldsymbol{e}}_3$.  By the property \eqref{KYTdef} that $\boldsymbol{k}$ is the "square root" of the Killing tensor, we have
\begin{equation}
    \boldsymbol{e}_2\cdot\boldsymbol{e}_2 = \frac{1}{{\mathcal{Q}}}K_{\mu\nu}l^{\mu}l^{\nu} = 1\,.
\end{equation}
Similarly, via definitions \eqref{Hdef} and \eqref{HK}, we have
\begin{equation}
    \tilde{\boldsymbol{e}}_3\cdot\tilde{\boldsymbol{e}}_3 = \frac{1}{{\mathcal{Q}}}K_{\mu\nu}l^{\mu}l^{\nu} = 1\,.
\end{equation}
From previous assertions, this also means that $\boldsymbol{e}_3\cdot\boldsymbol{e}_3=1$.

What about the magnitude of $\tilde{\boldsymbol{n}}$?  If we wish it to be null, then we simply have to fix $\gamma$ as
\begin{equation} \label{gamma}
    \gamma = \frac{1}{2\mathcal{}Q^{3/2}}H_{\mu\alpha}H^{\alpha}_{\;\nu}l^{\mu}l^{\nu}\,,
\end{equation}
where $\boldsymbol{H}$ in coordinate form is defined in \eqref{Hdef}.

All that is left is to check if $\tilde{\boldsymbol{n}}$ is parallel transported.  A quick calculation shows that it is not, but given the prescription by Transformation \ref{i1}, we can verify that the transformed vector
\begin{equation} \label{nnPP}
    n^{\mu} = \tilde{n}^{\mu}+\Phi\tilde{e}_3^{\mu}+\frac{1}{2}\Phi^2l^{\mu}\,,
\end{equation}
should be.  One can check that
\begin{equation}
    \boldsymbol{n}\cdot\boldsymbol{l}=-1\,,\hspace{15pt}\boldsymbol{n}\cdot\boldsymbol{e}_2=0\,,\hspace{15pt}\boldsymbol{n}\cdot\boldsymbol{e}_3=0\,,
\end{equation}
as desired.

In summary, we have constructed a null tetrad $(\boldsymbol{l},\,\boldsymbol{n},\,\boldsymbol{m},\,\bar{\boldsymbol{m}})$ in the Kerr spacetime that obeys the parallel transport equations 
\begin{equation}
    Dl^{\mu}=Dn^{\mu}=Dm^{\mu}=D\bar{m}^{\mu}=0\,.
\end{equation}

\subsection{Exact Equations for the Null Tetrad} \label{s532}

If we want to find precise formulas for the null tetrad defined in the previous section, we simply need to apply the above definitions to a null ray in the Kerr spacetime.  To this effect, we can use a dimensionless form of the tangent vector defined in Eq.~\eqref{leq}, which is given by
\begin{align} \label{leqD} 
    &\Sigma\dot{\tau} = -
    \alpha\big(\alpha\sin^2\theta-\ell_{z}\big)+\frac{\rho^2+\alpha^2}{\Delta}\big(\rho^2+\alpha^2-\ell_{z}\alpha\big)\, , \\\nonumber
    &\Sigma\dot{\rho} = \epsilon_r \sqrt{\mathcal{R}}\,,\\\nonumber
    &\Sigma\dot{\theta} = \epsilon_{\theta}\sqrt{\Theta}\,,\\\nonumber
    &\Sigma\dot{\phi} = -\Big(\alpha-\frac{\ell_z}{\sin^2\theta}\Big)+\frac{\alpha}{\Delta}\big(\rho^2+\alpha^2-\ell_{z}a\big)\, ,
\end{align}
where now
\begin{equation}
    \mathcal{R}= \Big(\rho^2+\alpha^2-\ell_z a\Big)^2 - \mathcal{Q}\Delta
    \,,\hspace{15pt}\Theta = \mathcal{Q}-\Big(\alpha\sin\theta-\frac{\ell_z}{\sin\theta}\Big)^2\,. 
\end{equation}

A dot over a variable signifies its derivative with respect to the affine parameter $\lambda$.
Since $\sqrt{\Theta}$ enters in the equations of motion, $\Theta$ should be non-negative, so that $\mathcal{Q}\ge 0$.
This set of equations determines a congruence of null geodesics with fixed parameters $\ell_z$ and $\mathcal{Q}$.
For completeness, we have also included the covariant form of the vectors contained in this tetrad in Appendix~\ref{apxC}.
The affine parameter $\lambda$ as a function of $\rho$ and $\theta$ is defined as follows
\begin{equation} \label{LLL1}
\lambda =\epsilon_\rho \int^{\rho}\frac{\epsilon_\rho\rho^2d\rho}{\sqrt{\mathcal{R}}} +\int^{\theta}\frac{\epsilon_{\theta}\alpha^2\cos^2\theta d\theta}{\sqrt{\Theta}}\,.
\end{equation}

In what follows, we focus on null geodesics which describe a photon being scattered by the black hole. Any such geodesic has one radial turning point at $\mathcal{R}=0$. The sign parameter $\epsilon_\rho$ takes the value $\epsilon_\rho=-1$ for the incoming branch of the null ray before it reaches the turning point. After this point, for the outgoing branch $\epsilon_\rho=+1$.
Similarly, $\epsilon_{\theta}=-1$ for motion with a decrease in $\theta$, and $\epsilon_{\theta}=1$ for the part of the trajectory where $\theta$ increases.
Null geodesics in the Kerr geometry and their properties have been discussed in many publications\footnote{See e.g. \cite{Carter:68,Bardeen,MTW,Chandrasekhar}. An early review of the theory of geodesics in black hole spacetimes and
a rather complete set of the corresponding references can be found in \cite{Sharp}. Additional information and references to more recent publications can be found, e.g., in \cite{Cieslik}.}.

Now, to define the other tetrad vectors, we simply use the equations in \eqref{een} along with definitions \eqref{principal}, \eqref{kYano}, and \eqref{gamma} to obtain
\begin{align} \label{eeee}
e_2^{\mu} = &\frac{1}{\Sigma\Delta\sqrt{\mathcal{Q}}}\bigg(\alpha\cos\theta(\rho^2+\alpha^2)\epsilon_{\rho}\sqrt{{\mathcal{R}}}-\alpha\rho\Delta\sin\theta\epsilon_{\theta}\sqrt{{\Theta}},\\\nonumber
&\hspace{42pt}\alpha\Delta\cos\theta\big((\rho^2+\alpha^2)-\alpha\ell_z\big),\\\nonumber
&\hspace{42pt}-\rho\Delta\sin\theta\big(\alpha-\frac{\ell_z}{\sin^2\theta}\big),\\\nonumber
&\hspace{42pt}\alpha^2\cos\theta\epsilon_{\rho}\sqrt{{\mathcal{R}}}-\rho\Delta\csc\theta\epsilon_{\theta}\sqrt{{\Theta}}\bigg)\, ,\\\nonumber
\tilde{e}_{3}^{\mu} = &\frac{-1}{\Sigma\Delta\sqrt{\mathcal{Q}}}\bigg(\rho(\rho^2+\alpha^2)\epsilon_{\rho}\sqrt{{\mathcal{R}}}+\alpha^2\sin\theta\cos\theta\Delta\epsilon_{\theta}\sqrt{{\Theta}},\\\nonumber
&\hspace{42pt}\rho\Delta\big((\rho^2+\alpha^2)-\alpha\ell_z\big),\\\nonumber
&\hspace{42pt}\alpha\Delta\sin\theta\cos\theta\big(\alpha-\frac{\ell_z}{\sin^2\theta}\big),\\\nonumber
&\hspace{42pt}\alpha\rho\epsilon_{\rho}\sqrt{{\mathcal{R}}}+\alpha\Delta\cot\theta\epsilon_{\theta}\sqrt{{\Theta}}\bigg)\,, \\\nonumber
\tilde{n}^{\mu}=&\frac{1}{2}\bigg(
\alpha(\alpha\sin^2\theta-\ell_z)+\frac{\rho^2+\alpha^2}{\Delta}(\rho^2+\alpha^2-\ell_{z}\alpha),\\\nonumber
&\hspace{42pt} \epsilon_\rho \mathcal{R}, -\epsilon_{\theta}\Theta,\\\nonumber
&\hspace{42pt} \big(\alpha-\frac{\ell_z}{\sin^2\theta}\big)+\frac{\alpha}{\Delta}(\rho^2+\alpha^2-\ell_{z}\alpha)\bigg)\, .
\end{align}
From this, we can also construct the parallel transported null tetrad vectors $\boldsymbol{n}$, $\boldsymbol{m}$, and $\bar{\boldsymbol{m}}$.  One can also check that this tetrad is right-handed
\begin{equation}
    e_{\mu\nu\alpha\beta}l^{\mu}n^{\nu}e_2^{\alpha}e_3^{\beta}=-1\,,
\end{equation}
and hence
\begin{equation}
   i e_{\mu\nu\alpha\beta}l^{\mu}n^{\nu}m^{\alpha}\bar{m}^{\beta} = -1\,,
\end{equation}
where $e_{\mu\nu\alpha\beta}$ is the Levi-Civita tensor with $e_{\tau\rho\theta\phi}=\sqrt{-g}$.

Let us consider a null ray being scattered by the black hole. Its orbit has one radial turning point. We denote its coordinate by $(\rho_m,\theta_m)$, where $\theta_m$ is the value of $\theta$ at the radial turning point, not to be confused with its angular turning point where $\Theta=0$. We have a procedure for fixing the ambiguity in the location of the
zero of the affine parameter $\lambda$, which in the special case of an
equatorial geodesic puts the zero at a radial turning point.  To this effect, we write the expression \eqref{LLL1} in the form
\begin{align} \label{LLL}
&\lambda =\epsilon_\rho \Lambda(\rho)+F(\theta)\, ,\hspace{15pt}\Lambda(\rho)=\int_{\rho_m}^{\rho}\dfrac{d\rho}{\sqrt{\mathcal{B}}}\, ,\\\nonumber
&F(\theta)=\int_{\theta_m}^{\theta} \dfrac{\epsilon_{\theta} \alpha^2 \cos^2\theta d\theta}{\sqrt{\Theta}}\, ,\hspace{15pt} \mathcal{B}=\dfrac{\mathcal{R}}{\rho^4}=\Big(1-\dfrac{\alpha(\ell_z-\alpha)}{\rho^2}\Big)^2 -\dfrac{\mathcal{Q}}{\rho^4}\Delta\, .
\end{align}

At this point, it is also convenient to define another specific frame of reference in the Kerr spacetime. This frame is an orthonormal tetrad tied to the coordinates $\rho$ and $\theta$. It has the form
\begin{align} \label{EEEE}
E_{(\tau)}&=E_{(\tau)}^{\mu}\partial_{\mu}=\sqrt{\dfrac{A}{\Delta\Sigma}}\big(
\dfrac{\partial}{\partial \tau}+\Omega \dfrac{\partial}{\partial \phi}
\big)\, ,\\\nonumber
E_{(\phi)}&=E_{(\phi)}^{\mu}\partial_{\mu}=\sqrt{\dfrac{\Sigma}{A}}\dfrac{1}{\sin\theta}
\dfrac{\partial}{\partial \phi}\, ,\\\nonumber
E_{(\rho)}&=E_{(\rho)}^{\mu}\partial_{\mu}=
\sqrt{\dfrac{\Delta}{\Sigma}}\dfrac{\partial}{\partial \rho}\, ,\\\nonumber
E_{(\theta)}&=E_{(\theta)}^{\mu}\partial_{\mu}=
\sqrt{\dfrac{1}{\Sigma}}\dfrac{\partial}{\partial \theta}\, ,
\end{align}
where
\begin{equation}
A = (\rho^2+\alpha^2)^2-\Delta \alpha^2\sin^2\theta\,.
\end{equation}
The world line of an observer that is at rest in this frame is orthogonal to a $\tau$=constant hypersurface. This local observer rotates with respect to infinity with angular velocity
\begin{equation}
\Omega=\dfrac{2\alpha\rho}{A}\, ,
\end{equation}
and has zero angular momentum \cite{Bardeen_1,Bardeen_2,Bardeen}.
The frame \eqref{EEEE} associated with this zero angular momentum observer (ZAMO) is called a ZAMO frame.  For a derivation of this ZAMO frame, see Appendix~\ref{B42}.  For the exact expansion of the null tetrad in the ZAMO basis, see Appendix~\ref{C1} 

\subsection{Geodesic Null Rays in the Equatorial Plane} \label{s533}

In our study of the spinoptics approximation, we assume that this effect is small, allowing us to utilize a perturbative analysis.  To this effect, we first study null geodesic rays and their associated complex null tetrads within the Kerr geometry in the equatorial plane, and then use them in our perturbative expansion to find higher-order helicity-dependent spinoptics corrections to the evolution of the ray and its tetrad.  Therefore, our first step is to consider null geodesics and the null tetrads associated with them in the equatorial plane of the Kerr metric.  For this case, many relations are greatly simplified. First, let us write down the dimensionless Kerr metric in the equatorial plane
\begin{equation}
ds^2 = -\Big(1-\frac{2}{\rho}\Big)d\tau^2+\frac{\rho^2}{\Delta}d\rho^2-\frac{4\alpha}{\rho}d\tau d\phi+\rho^2 d\theta^2+\Big(\rho^2+\alpha^2+\frac{2\alpha^2}{\rho}\Big)d\phi^2\,.
\end{equation}
It is easy to show that the black hole's outer ergosurface crosses the equatorial plane at $\rho=2$.

We also have the following nonzero components of the Riemann tensor in the equatorial plane
\begin{align}
R_{\tau\rho\tau\rho} = &-\frac{2\Delta+\alpha^2}{\Delta\rho^3}\,,\hspace{10pt} R_{\tau\theta\tau\theta}=\frac{\Delta+2\alpha^2}{\rho^3}\,,\hspace{10pt}  R_{\tau\phi\tau\phi}=\frac{\Delta}{\rho^3}\,,\\\nonumber
R_{\tau\rho\rho\phi}=&-\frac{\alpha(3\Delta+2\rho)}{\Delta\rho^3}\,,\hspace{10pt} R_{\tau\theta\theta\phi}=\frac{\alpha(3\Delta+4\rho)}{\rho^3}\,,\\\nonumber
R_{\rho\theta\rho\theta}=&-\frac{\rho}{\Delta}\,,\hspace{10pt} R_{\rho\phi\rho\phi}=-\frac{(\rho^2+\alpha^2)^2+2\Delta\alpha^2}{\Delta\rho^3}\,,\\\nonumber
R_{\theta\phi\theta\phi}=&\frac{2(\rho^2+\alpha^2)^2+\Delta\alpha^2}{\rho^3}\,.
\end{align}

Let us now look at the null geodesics\footnote{
Useful information about null geodesics in the Kerr spacetime can be found e.g. in \cite{FroZel2011}.
}. For null geodesics in the equatorial plane, $\Theta=0$, implying that
\begin{equation} \label{QQQQ}
\mathcal{Q}=(\ell_z-\alpha)^2\, ,
\end{equation}
where $\mathcal{Q}$ and $\ell_z$ are previously defined in \eqref{integrals}. This implies that the "Carter constant" $\mathcal{C}$ defined as $\mathcal{C}=\mathcal{Q}-(\ell_z-\alpha)^2$ vanishes for null geodesics in the equatorial plane, as it should.

Let us denote
\begin{equation}\label{BBDD}
\mathcal{B}=1-\dfrac{\ell_z^2-\alpha^2}{\rho^2}+\dfrac{2(\ell_z-\alpha)^2}{\rho^3}\, .
\end{equation}
Then the tangent vector to a null geodesic in the equatorial plane can be written as follows
\begin{align}  \label{tetradEQ}
& l^{\mu}=(\dot{\tau},\dot{\rho},0,\dot{\phi})\, ,\\\nonumber
&\dot{\tau}=\dfrac{\rho(\rho^2+\alpha^2)-2\alpha(\ell_z-\alpha)}{\rho \Delta}\, ,\\\nonumber
&\dot{\rho}=\epsilon_{\rho} \sqrt{\mathcal{B}}\, ,\\\nonumber
&\dot{\phi}=\dfrac{
(\rho-2)\ell_z +2\alpha
}{\rho \Delta}\, .
\end{align}

Every null geodesic has no more than one turning point. Such a turning point $\rho=\rho_m$, where $\mathcal{B}=0$, exists for null geodesics that come from infinity and then bounce back to infinity.
One can solve the equation $\mathcal{B}=0$ and get
\begin{equation}\label{LLRR}
\ell^{\pm}_z=\dfrac{\pm \rho_m\sqrt{\Delta_m}-2\alpha}{\rho_m-2}\,,\hspace{15pt} \Delta_m=\rho_m^2-2\rho_m+\alpha^2
\, .
\end{equation}
We choose the direction of the rotation of the black hole so that the dimensionless rotation parameter $\alpha$ is non-negative, $0\le \alpha < 1$.
It is possible to check that for a $+$ sign
in \eqref{LLRR},
the dimensionless impact parameter $\ell^+_z$ is positive. This case corresponds to {\it{prograde}} motion of the null ray. For a $-$ sign, $\ell^-_z$ is negative and exhibits {\it{retrograde}} motion.
For null rays which stay far away from the black hole, one has
\begin{equation} \label{rmlarge}
\ell_z^{\pm}\approx \pm(\rho_m +1)\, .
\end{equation}

\begin{figure}[!hbt]
    \centering
\includegraphics[width=0.7\textwidth]{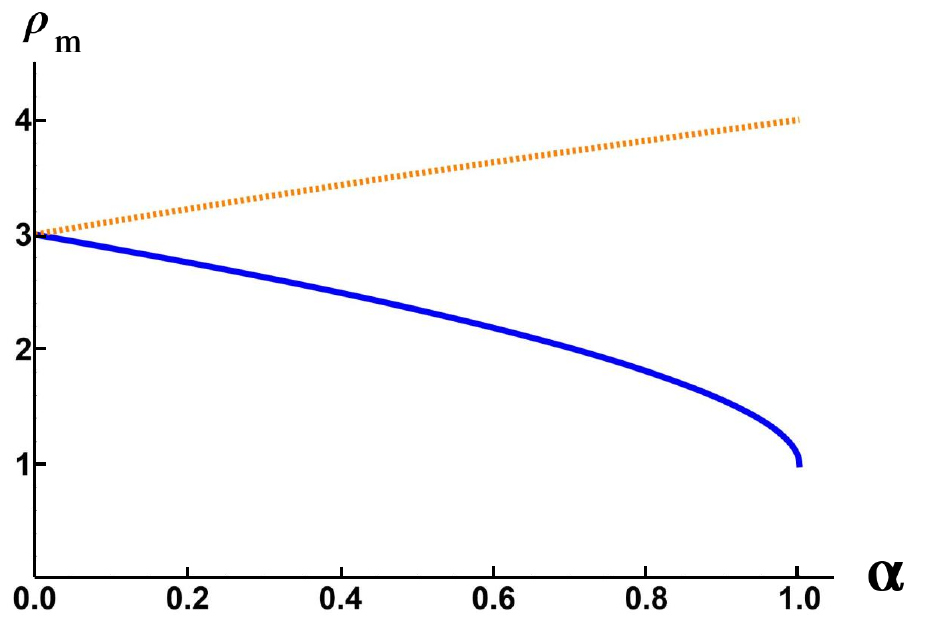}
\vspace{0.25cm}
\caption{\label{Fig_5_1}A plot of $\rho^{\pm}_m$ as a function of the dimensionless rotation parameter $\alpha$. The lower branch of the curve (solid) gives $\rho^{+}_m$ for prograde null rays, while the upper branch of the curve (dashed) gives $\rho^{-}_m$ for retrograde null rays.}
\end{figure}

Simple analysis shows that for $\rho_m>2$, $\ell^+_z$ as a function of $\rho_m$ has a minimum at some critical value $\rho_m=\rho_m^+$, while $\ell_z^-$ has a maximum at $\rho_m=\rho_m^-$, where
\begin{equation} \label{rhopm}
\rho_m^{\pm}=2+\cos\chi \mp \sqrt{3}\sin\chi\,,\hspace{15pt} \chi =\dfrac{2}{3}\arcsin\alpha\, ,
\end{equation}
which is derived in \cite{FroZel2011}, and briefly discussed in Appendix~\ref{B3}. The relation for $\rho_m$ as a function of $\alpha$ given by Eq.~\eqref{rhopm} for both prograde and retrograde null rays is displayed graphically in Fig.~\ref{Fig_5_1}.

In the equatorial plane, the vectors $\boldsymbol{e}_2$, $\tilde{\boldsymbol{e}}_3$ and $\tilde{\boldsymbol{n}}$ defined by \eqref{eeee} take the form
\begin{align}\label{TETR}
e_2^{\mu}\partial_{\mu} =& (1/\rho)\partial_{\theta}\,,\\\nonumber
\tilde{e}_3^{\mu}\partial_{\mu}=&-
\dfrac{\epsilon_\rho \rho(\rho^2+\alpha^2)\sqrt{\mathcal{B}}}{(\ell_z-\alpha)\Delta}\partial_{\tau}-\dfrac{\rho^2-\alpha(\ell_z-\alpha)}{(\ell_z-\alpha)\rho}\partial_{\rho}-\dfrac{\epsilon_\rho \alpha \rho\sqrt{\mathcal{B}}}{(\ell_z-\alpha)\Delta}\partial_{\phi}\, ,\\\nonumber
\tilde{n}^{\mu}\partial_{\mu} = &\frac{1}{2(\ell_z-\alpha)^2}\bigg(\Big(\frac{(\rho^2+\alpha^2)(\rho^2-\alpha(\ell_z-\alpha))}{\Delta}-\alpha(\ell_z-\alpha)\Big)\partial_{\tau}\\\nonumber
&+\epsilon_{\rho}\Big(\rho^2\sqrt{\mathcal B}\Big)\partial_{\rho}+\Big(\frac{\alpha(\rho^2-\alpha(\ell_z-\alpha))}{\Delta}-(\ell_z-\alpha)\Big)\partial_{\phi}\bigg)\,.
\end{align}

\begin{figure}[!hbt]
    \centering
\includegraphics[width=0.7\textwidth]{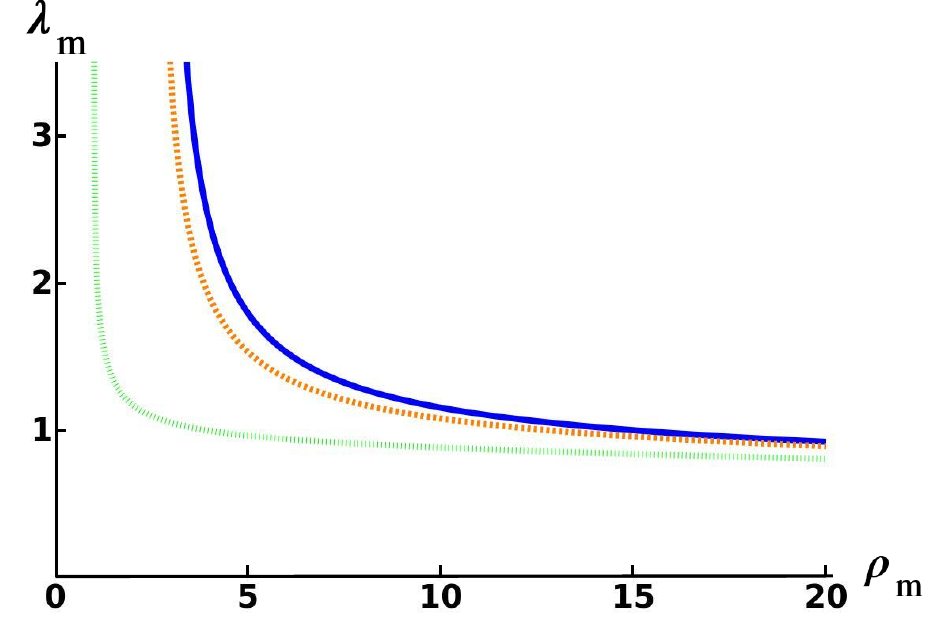}
\caption{\label{Fig_5_2}A plot of $\lambda_m=J_m-\rho_m$ as a function of $\rho_m$ for prograde photon scattering trajectories in the equatorial plane, where $\alpha =$ 0 (solid), 0.5 (dashed), 1 (dotted) enumerates each line going from right to left.}
\end{figure}

\begin{figure}[!hbt]
    \centering
\includegraphics[width=0.7\textwidth]{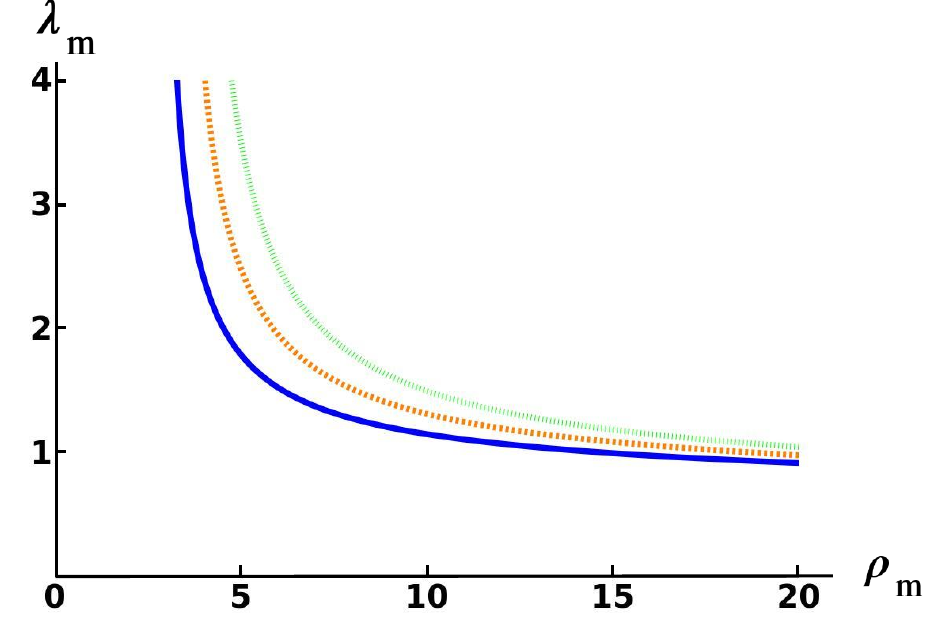}
\caption{\label{Fig_5_3}A plot of $\lambda_m=J_m-\rho_m$ as a function of $\rho_m$ for retrograde photon scattering trajectories in the equatorial plane, where $\alpha =$ 0 (solid), 0.5 (dashed), 1 (dotted) enumerates each line going from left to right.}
\end{figure}

In the equatorial plane, the function $F(\theta)$ in equation \eqref{LLL} for the affine parameter $\lambda$ vanishes, and for $\Lambda(\rho)$ defined in \eqref{LLL} one has
\begin{align} \label{LambdaEQ}
\Lambda&=\rho-\int_{\rho}^{\infty}\Big[
\dfrac{1}{\sqrt{\mathcal{B}}}-1
\Big] d\rho +\lambda_m\, ,\\\nonumber
\lambda_m&=J_m-\rho_m\,,\hspace{15pt}
J_m=\int_{\rho_m}^{\infty}\Big[
\dfrac{1}{\sqrt{\mathcal{B}}}-1\Big] d\rho\, .
\end{align}
One can see how the parameter $\lambda_m$ behaves for prograde and retrograde rays in Figs.~\ref{Fig_5_2} and \ref{Fig_5_3}, respectively. For large $\rho$, one has
\begin{equation}
\Lambda=\rho+\lambda_m+O(1/\rho)\, .
\end{equation}

For rays propagating far from the black hole where $\rho_m\gg 1$, one can use the approximate value of the impact parameter given in \eqref{BBDD}. Then
\begin{align} \label{Brm}
&\mathcal{B}\approx \dfrac{
\hat{\rho}^2-1}{\hat{\rho}^2}\big[
1-\dfrac{2}{\hat{\rho}(\hat{\rho}+1)}\dfrac{1}{\rho_m}
\big]\, ,\\\nonumber
& \dfrac{1}{\sqrt{\mathcal{B}}}-1\approx
\big(\dfrac{\hat{\rho}}{\sqrt{\hat{\rho}^2-1}}-1\big)+\dfrac{1}{(\hat{\rho}+1)^{3/2}(\hat{\rho}-1)^{1/2}}\dfrac{1}{\rho_m}\, .
\end{align}
Here we denote $\hat{\rho}=\rho/\rho_m$. Using \eqref{LambdaEQ}, one can write
\begin{equation}
\lambda_m/\rho_m=\int_{1}^{\infty}\Big[
\dfrac{1}{\sqrt{\mathcal{B}}}-1\Big] d\hat{\rho}-1\, .
\end{equation}
The integral in this expression can easily be taken approximately. Calculations give
\begin{equation}\label{LMR}
\lambda_m/\rho_m\approx 1/\rho_m\, .
\end{equation}
Thus, for large $\rho_m$, the parameter $\lambda_m$ has the limit $\lambda_m=1$. The plots for $\lambda_m$ shown in Figs.~\ref{Fig_5_2} and \ref{Fig_5_3} are in agreement with this asymptotic property of $\lambda_m$.

The parallel propagated versions of the vectors $\boldsymbol{\tilde{n}}$ and $\boldsymbol{\tilde{e}}_3$ can be found by using \eqref{ee33} and \eqref{nnPP}.
We write the function $\Phi$ which enters these relations in the following form
\begin{equation}\label{ZZZZ}
\Phi=\dfrac{\epsilon_\rho \Lambda(\rho)+\lambda_0}{\ell_z-\alpha}
\, .
\end{equation}
The parameter $\lambda_0$ is a constant that should be defined by the initial conditions of the corresponding problem.

Using  expressions \eqref{TET_ZAM} for the decomposition of the tetrad $(\boldsymbol{l},\boldsymbol{n},\boldsymbol{e}_2,\boldsymbol{e}_3)$ in the ZAMO basis, one finds that the following relations valid in the asymptotic domain $\rho\to\infty$
\begin{align}\label{ASSIM}
l^{\mu}=&\big(E_{(\tau)}^{\mu}+\epsilon_{\rho}E_{(\rho)}^{\mu}\big)\big(1+1/\rho\big)+\ell_{z}E_{(\phi)}+...\,,\\\nonumber
n^{\mu}=&\frac{1}{2}\big(E_{(\tau)}^{\mu}-\epsilon_{\rho}E_{(\rho)}^{\mu}\big)
+\dfrac{1}{2}\Delta\lambda^2 l^{\mu}+\epsilon_{\rho}\Delta\lambda E_{(\phi)}^{\mu}+ O(1/\rho)\,,\\\nonumber
e_{2}^{\mu}=&E_{(\theta)}^{\mu}\,,\\\nonumber
e_{3}^{\mu}=&\epsilon_{\rho}E_{(\phi)}^{\mu}+\Delta\lambda l^{\mu}+O(1/\rho)\,.
\end{align}
Here,
\begin{equation}
\Delta\lambda=\frac{\lambda_0+\epsilon_{\rho}\lambda_m}{\ell_z-\alpha}\,.
\end{equation}

We see that $\lambda_0=\lambda_m$ is a natural choice for scattering, as the null tetrad vectors $\boldsymbol{e}_2$ and $\boldsymbol{e}_3$ for incoming rays will "coincide" with the ZAMO basis vectors $\boldsymbol{E}_{\theta}$ and $\boldsymbol{E}_{\phi}$ as $\rho\to\infty$.  This choice will be adopted later on.

\subsection{Spinoptics Evolution in Kerr} \label{s534}

A ray describing photon motion in the geometric optics approximation is a null geodesic. 
In the spinoptics approximation, due to the interaction of the photon's spin with the curvature of spacetime, its worldline is still null but no longer a geodesic. As a result, the parallel propagated null tetrad associated with these rays will slightly differ from the null tetrads associated with null geodesics. In order to distinguish these two different tetrads we need to change our notation. We use the subscript $0$ to indicate that the corresponding tetrad is associated with null geodesics. Thus, from now on the tetrad discussed in the previous sections will be denoted by $(\boldsymbol{l}_0,\boldsymbol{n}_0,\boldsymbol{m}_{0},\bar{\boldsymbol{m}}_{0})$ , while the other (non-geodesic) tetrad
is $(\boldsymbol{l},\boldsymbol{n},\boldsymbol{m},\bar{\boldsymbol{m}})$.  We can also use the original vectors $\boldsymbol{e}_{20}$ and $\boldsymbol{e}_{30}$ via
\begin{equation}\label{MM0} 
m_0^{\mu}=\dfrac{1}{\sqrt{2}}(e_{20}^{\mu}+ie_{03}^{\mu})\,,\hspace{15pt}
\bar{m}_0^{\mu}=\dfrac{1}{\sqrt{2}}(e_{02}^{\mu}-ie_{03}^{\mu})\, .
\end{equation}

The parallel transport equations have the form
\begin{equation} \label{lnmm0}
D_0 l_0^{\mu}=D_0 n_0^{\mu}=D_0 m_0^{\mu}=
D_0 \bar{m}_0^{\mu}=0\, ,
\end{equation}
where $D_0=l_0^{\alpha}\nabla_{\alpha}$.

As we derived in \eqref{spin}, null rays in the spinoptics approximation obey \cite{Frolov:2020uhn}
\begin{align} \label{DDLLL}
D l^{\mu}&=\chi \Upsilon^{\mu}\,,\hspace{15pt}
D=l^{\alpha}\nabla_{\alpha}\, ,\\\nonumber
\Upsilon^{\mu}&=iR_{\mu\nu\alpha\beta}l^{\nu}\bar{m}^{\alpha}m^{\beta}\,,\hspace{15pt}
\chi=\dfrac{\sigma}{M\omega}
\,,
\end{align}
where we note that the above form is the dimensionless form of our equations, hence the factor of $1/M$ compared to the equations in \eqref{spin}. The term $\chi \Upsilon^{\mu}$ on the right-hand side of the equation for a null ray is nothing but its four-dimensional acceleration induced by the interaction of the spin with spacetime curvature. For $\chi=0$, it vanishes and the rays are geodesics. 

Eq.~\eqref{DDLLL} has the same form as Eq.~\eqref{ring} discussed in Sec.~\ref{s522}. As it is demonstrated in this subsection, using the freedom of choice of a complex null tetrad, one can choose it so that the following equations are satisfied \eqref{ring}
\begin{align} \label{EQLC}
& Dl^{\mu}=\chi \Upsilon^{\mu}=w_0 l^{\mu}+
\bar{\kappa} m^{\mu}+{\kappa}\bar{m}^{\mu}\, ,\\\nonumber
&Dn^{\mu} = -w_0 n^{\mu}\,,\hspace{15pt} Dm^{\mu}=\kappa n^{\mu}\,  .
\end{align}
As previously stated, the form of this tetrad is not in the "canonical form" given by Eq.~\eqref{canon0}, but a simple change in affine parameterization can get us there if we so desire.  We will find that, for the study of these equations for a ray starting out in the equatorial plane of the Kerr spacetime, this change in affine parameterization will not be necessary, as the equations will already appear in their canonical form for this special choice.

In the high-frequency approximation, which is adopted in this paper, the parameter $\chi$ is small, and therefore the deviation of the corresponding null ray from a geodesic ray is also small.
To find the corresponding approximate solution, we shall use the method of perturbations.
In this approximation, it is sufficient to use the complex null vectors that are parallel propagated along unperturbed null geodesics in the right-hand side of \eqref{DDLLL}. It is also convenient to use parallel propagated vectors $\boldsymbol{e}_{02}$ and $\boldsymbol{e}_{03}$ instead of $\boldsymbol{m}_0$ and $\bar{\boldsymbol{m}}_0$ (see \eqref{MM0}).
Under these assumptions, the ray equation \eqref{DDLLL} takes the form
\begin{equation} \label{psi}
Dl^{\mu}=\chi (\psi_0 l_0^{\mu} +
\psi_2 e_{02}^{\mu}+\psi_3 e_{03}^{\mu})\, ,
\end{equation}
where
\begin{align}
&\psi_0=-R_{\mu\nu\alpha\beta}n_0^{\mu}l_0^{\nu}e_{02}^{\alpha}e_{03}^{\beta}\,,\\\nonumber
&\psi_2=R_{\mu\nu\alpha\beta}e_{02}^{\mu}l_0^{\nu}e_{02}^{\alpha}e_{03}^{\beta}\,,\\\nonumber
&\psi_3=R_{\mu\nu\alpha\beta}e_{03}^{\mu}l_0^{\nu}e_{02}^{\alpha}e_{03}^{\beta}\, ,
\end{align}
where $R_{\mu\nu\alpha\beta}$ is the Riemann tensor for the Kerr metric. Calculations for $\psi_a$ give the following expressions
\begin{align} \nonumber
\psi_0&=-\frac{\alpha\cos\theta}{\Sigma^5}\Big(\Sigma^2\big(3\rho^2-\alpha^2\cos^2\theta\big)-3Q\Phi^2\big(5\rho^4-10\rho^2\alpha^2\cos^2\theta+\alpha^4\cos^4\theta\big)\Big)\,, \\
\psi_2&=-\frac{3\rho Q\Phi}{\Sigma^5}\big(\rho^4-10\rho^2\alpha^2\cos^2\theta+5\alpha^4\cos^4\theta\big) \,, \\\nonumber
\psi_3&=-\frac{3\alpha\cos\theta Q\Phi}{\Sigma^5}\big(5\rho^4-10\rho^2\alpha^2\cos^2\theta+\alpha^4\cos^4\theta\big)\, .
\end{align}
As we had previously hinted at, one can see that the tetrad will take on its canonical form in the equatorial plane, where $\theta=\pi/2$.

\section{Effects Associated With the Spinoptics Approximation} \label{s54}

In this section, we will derive two specific effects associated with the spinoptics approximation for null rays that initially start out in the equatorial plane at incoming radial infinity within the Kerr spacetime.

\subsection{Tilting Effect Near the Equatorial Plane} \label{s541}

\begin{figure}[!hbt]
    \centering
\includegraphics[width=0.7\textwidth]{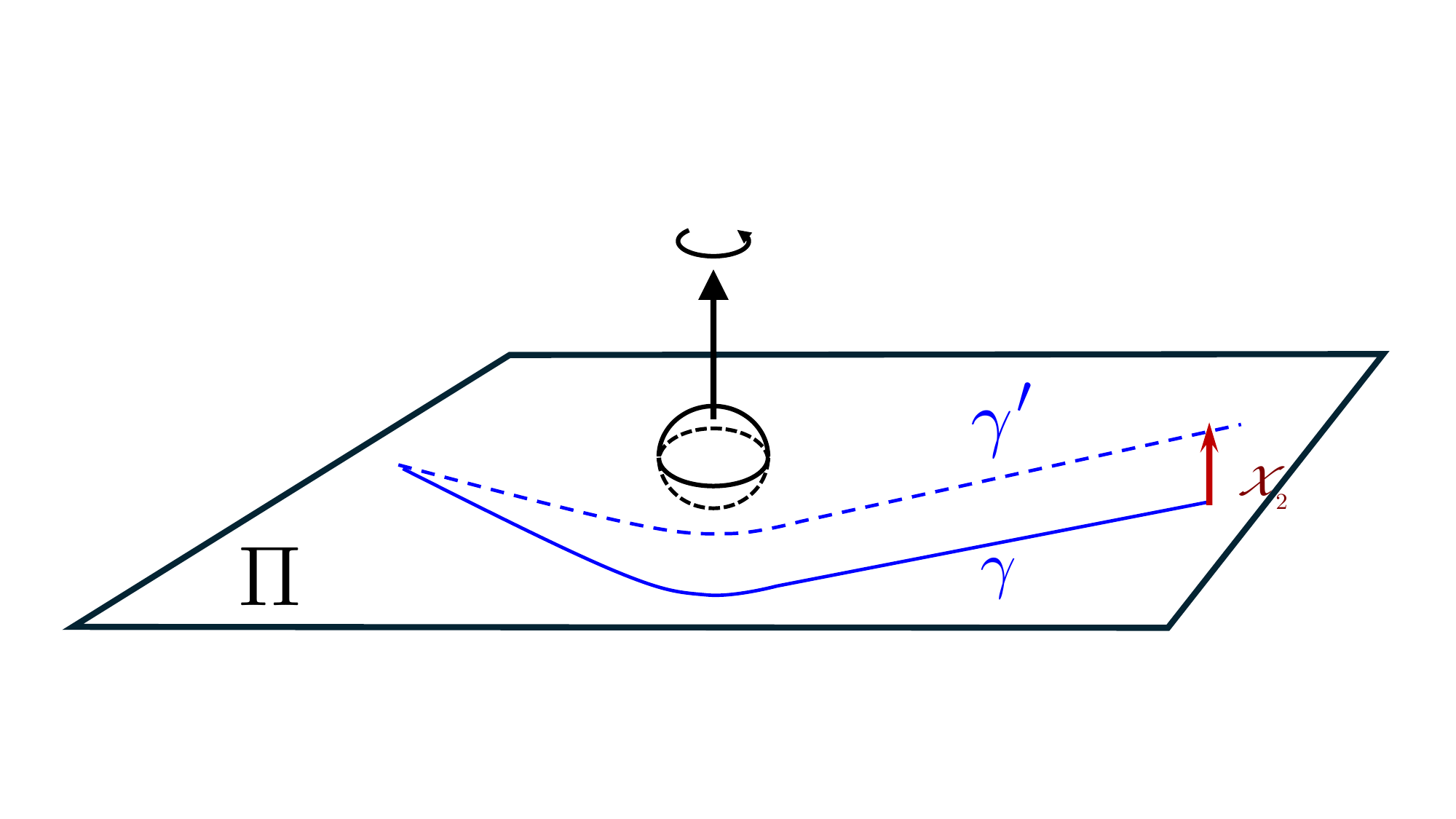}
\caption{\label{Fig_5_4}Displacement of the null ray near the equatorial plane.}
\end{figure}

We now apply the general equation \eqref{geoDev} derived in Appendix~\ref{B5} for the deflection of null rays in the spinoptics approximation to the case where these rays propagate close to the equatorial plane. Namely, we demonstrate that the spinoptics approximation results in the shift of the ray originally sent along the equatorial plane $\Pi$ away from it. This is schematically illustrated in Fig.~\ref{Fig_5_4}. The line $\gamma$ shows a geodesic null ray propagating along $\Pi$, while $\gamma'$ is a similar null ray but it follows the spinoptics approximation.

Let us denote
\begin{equation}
\mathcal{A}_{\mu\nu}=R_{\mu\alpha\beta\nu}l_0^{\alpha}l_0^{\beta}\,, \hspace{15pt}\Upsilon^{\mu}=\psi_0 l_0^{\mu} +
\psi_2 e_{02}^{\mu}+\psi_3 e_{03}^{\mu}\, .
\end{equation}
Then equation \eqref{geoDev}, written in dimensionless form, reads
\begin{equation} \label{EEDEF}
D_0^2 \delta x^{\mu} -\mathcal{A}^{\mu}_{\;\nu}\delta x^{\nu}=\chi\Upsilon^{\mu} \, .
\end{equation}

Calculations show that in the equatorial plane the only non-vanishing tetrad components of $\mathcal{A}_{\mu\nu}$ are
\begin{equation}
\mathcal{A}_{\mu\nu}n_0^{\mu}n_0^{\nu}=\frac{1}{\rho^5}\big(3Q\Phi^2-\rho^2\big)\,,\hspace{15pt}\mathcal{A}_{\mu\nu}e_{02}^{\mu}e_{02}^{\nu}=-\mathcal{A}_{\mu\nu}e_{03}^{\mu}e_{03}^{\nu}=-\frac{3Q}{\rho^5}\,.
\end{equation}

For null rays propagating near the equatorial plane $\theta=\pi/2$, the angle $\vartheta=\pi/2-\theta$ is small.
The vector $e_{02}^{\mu}$ is directed along $\theta$, while $-e_{02}^{\mu}$ is directed along $\vartheta$, and hence it points towards the upper side of the equatorial plane. We denote by $\mathcal{X}_2$ the following displacement of the ray
\begin{equation}\label{X2theta}
\mathcal{X}_2=-\delta x^{\mu}e_{02\mu}=\rho \vartheta\, .
\end{equation}

By contracting equation \eqref{EEDEF} with $e_{2\mu}$, one gets
\begin{equation} \label{XX2}
\dfrac{d^2 \mathcal{X}_2}{d\lambda^2}+\dfrac{3Q}{\rho^5} \mathcal{X}_2=-\chi \psi_2\, .
\end{equation}
In the equatorial plane, one has
\begin{equation} \label{psi22}
\psi_2=-\frac{3 Q\Phi}{\rho^5}\, .
\end{equation}
Relations \eqref{ASSIM} imply that for $\lambda_0=\lambda_m$, when $\Delta\lambda=0$, the polarization vector $\boldsymbol{m}_0$ for the incoming radiation has the standard form
\begin{equation}
\boldsymbol{m}_0\approx \dfrac{1}{\sqrt{2}\rho}(\partial_{\theta} +i\partial_{\phi})\, .
\end{equation}
We use this choice and write the function $\Phi$, which enters \eqref{psi22}, in the form
\begin{equation} \label{PPPP}
\Phi=\dfrac{1}{\ell_z-\alpha}(\epsilon_\rho\Lambda(\rho)+\lambda_m)\, .
\end{equation}

\begin{figure}[!hbt]
    \centering
\includegraphics[width=0.7\textwidth]{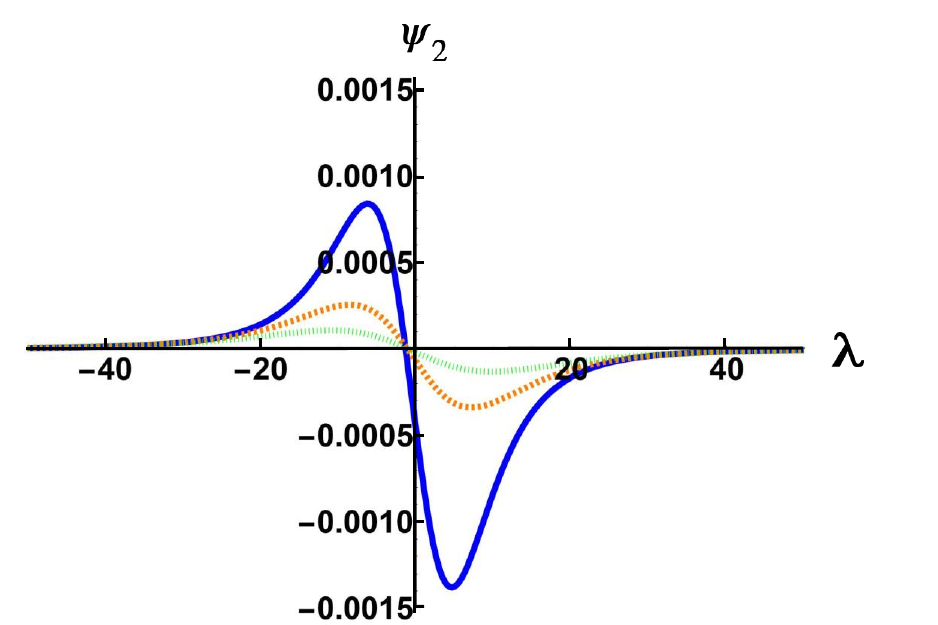}
\caption{\label{Fig_5_5}Driving function $\psi_2$ for prograde rays as a function of $\lambda$ for $\alpha=0.5$ and $\rho_m$=10 (solid), 15 (dashed), and 20 (dotted).}
\end{figure}

\begin{figure}[!hbt]
    \centering
\includegraphics[width=0.7\textwidth]{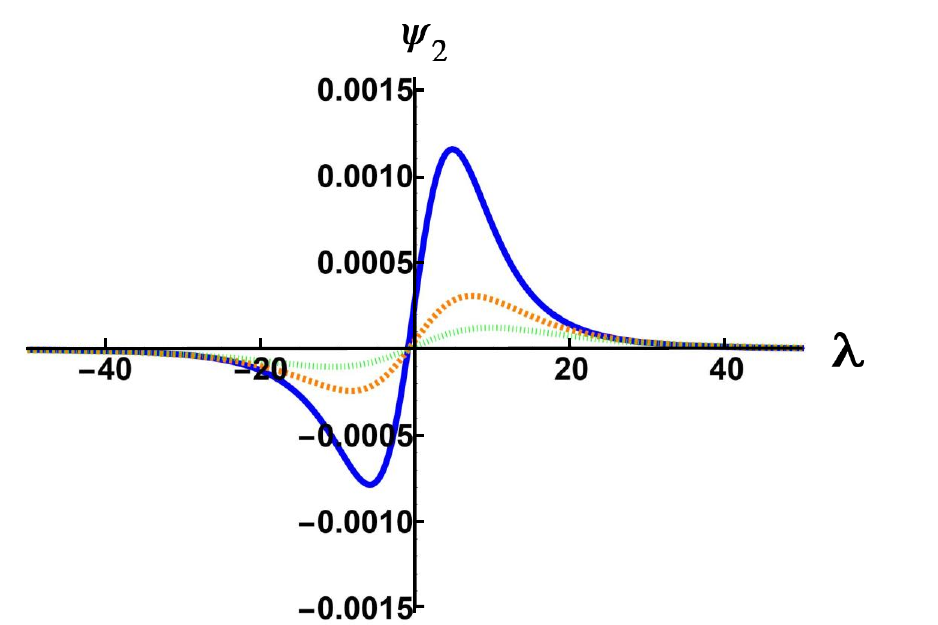}
\caption{\label{Fig_5_6}Driving function $\psi_2$ for retrograde rays as a function of $\lambda$ for $\alpha=0.5$ and $\rho_m$=10 (solid), 15 (dashed), and 20 (dotted).}
\end{figure}

Using equations \eqref{psi22} and \eqref{PPPP}, one gets
\begin{equation} \label{PPSI}
\psi_2=-\dfrac{3(\ell_z-\alpha)(\lambda+\lambda_m)}{\rho^5}\, .
\end{equation}
Figs.~\ref{Fig_5_5} and \ref{Fig_5_6} show $\psi_2$ for $\alpha=0.5$ and several values of $\rho_m$ for both prograde and retrograde rays, respectively.

\begin{figure}[!hbt]
    \centering
\includegraphics[width=0.7\textwidth]{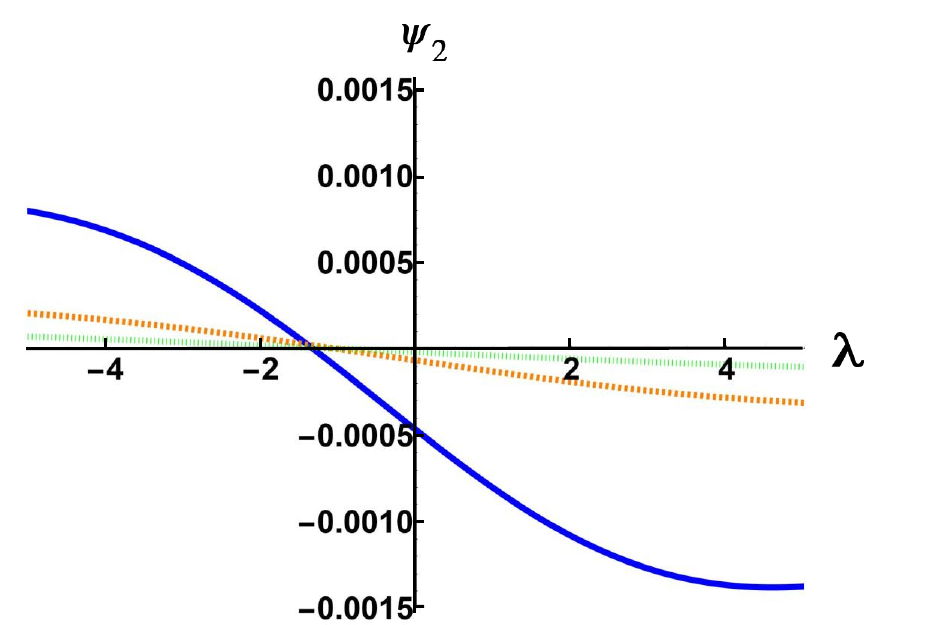}
\caption{\label{Fig_5_7}Close-up view of driving function $\psi_2$ for prograde rays as a function of $\lambda$ for $\alpha=0.5$ and $\rho_m$=10 (solid), 15 (dashed), and 20 (dotted).}
\end{figure}

\begin{figure}[!hbt]
    \centering
\includegraphics[width=0.7\textwidth]{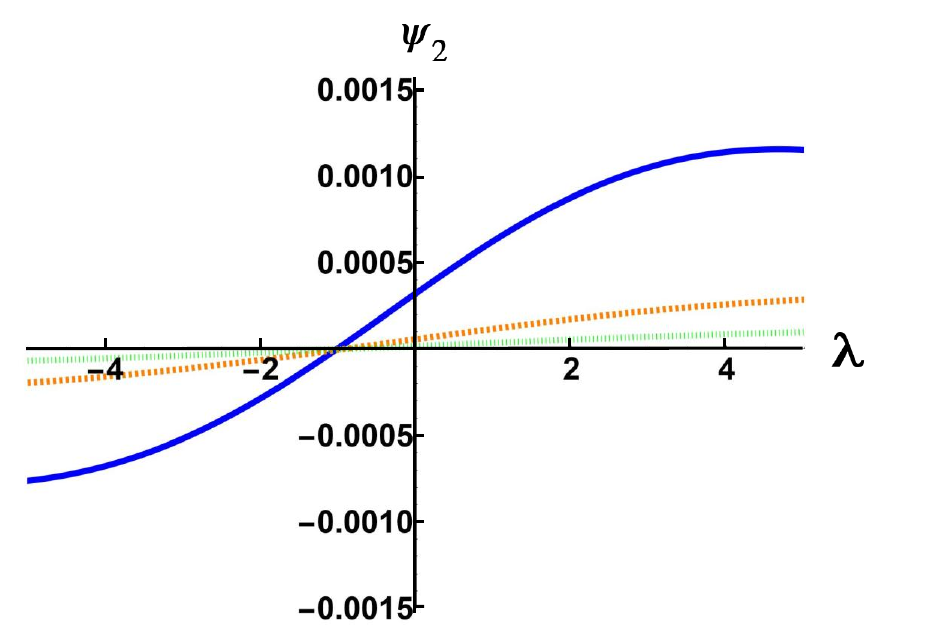}
\caption{\label{Fig_5_8}Close-up view of driving function $\psi_2$ for retrograde rays as a function of $\lambda$ for $\alpha=0.5$ and $\rho_m$=10 (solid), 15 (dashed), and 20 (dotted).}
\end{figure}

Let us note that if one puts $\lambda_m=0$ in \eqref{PPSI}, then $\psi_2=\psi_2(\lambda)$ would be an odd function of $\lambda$ and its plot would cross the $\lambda$ axis at $\lambda=0$. Due to the presence of the factor $\lambda_m$, the point where $\psi_2=0$ is slightly shifted to the point $\lambda=-\lambda_m$. To make this more visible, we present plots similar to \ref{Fig_5_5} and \ref{Fig_5_6} with magnified resolution of the region close to the origin, which are given by Figs.~\ref{Fig_5_7} and \ref{Fig_5_8}, respectively.

The equation \eqref{XX2} should be accompanied by the equation for $\rho=\rho(\lambda)$
\begin{equation} \label{dotr}
\dot{\rho}=\mbox{sgn}(\lambda) \sqrt{\mathcal{B}}\,.
\end{equation}
Since this equation is singular at the radial turning point $\rho=\rho_m$, we shall use the second order equation obtained from \eqref{dotr} by differentiating it with respect to $\lambda$
\begin{equation} \label{ddotr}
\ddot{\rho}=
\dfrac{\ell_z-\alpha}{\rho^4}\Big[
(\rho-3)\ell_z+(\rho+3)\alpha\Big]\, .
\end{equation}
Using \eqref{X2theta}, \eqref{PPSI}, \eqref{dotr}, and \eqref{ddotr} one obtains the following equation for the tilting angle $\vartheta$
\begin{equation} \label{ddotl}
\ddot{\vartheta}= -\frac{1}{\rho}\Big[2\dot{\rho}\dot{\vartheta}+\frac{\ell_z^2-\alpha^2}{\rho^3}\vartheta+\chi\psi_2\Big] \, .
\end{equation}

Let $\lambda_{in}$ be a large negative parameter. For the adopted value $\lambda_0=\lambda_m$, one has the following initial conditions for the incoming rays
\begin{equation} \label{TPics}
\rho_0=-\lambda_{in}\,,\hspace{15pt}
\dot{\rho}_0=-1\,,\hspace{15pt} \vartheta_0=0\,,\hspace{15pt}
\dot{\vartheta}_0=0\, .
\end{equation}
Solving \eqref{ddotr} and \eqref{ddotl} gives us the function $\vartheta=\vartheta(\lambda)$, which in the limit $\lambda\to \infty$ determines the asymptotic tilting angle for the outgoing trajectory of the null rays.

\begin{figure}[!hbt]
    \centering
\includegraphics[width=0.7\textwidth]{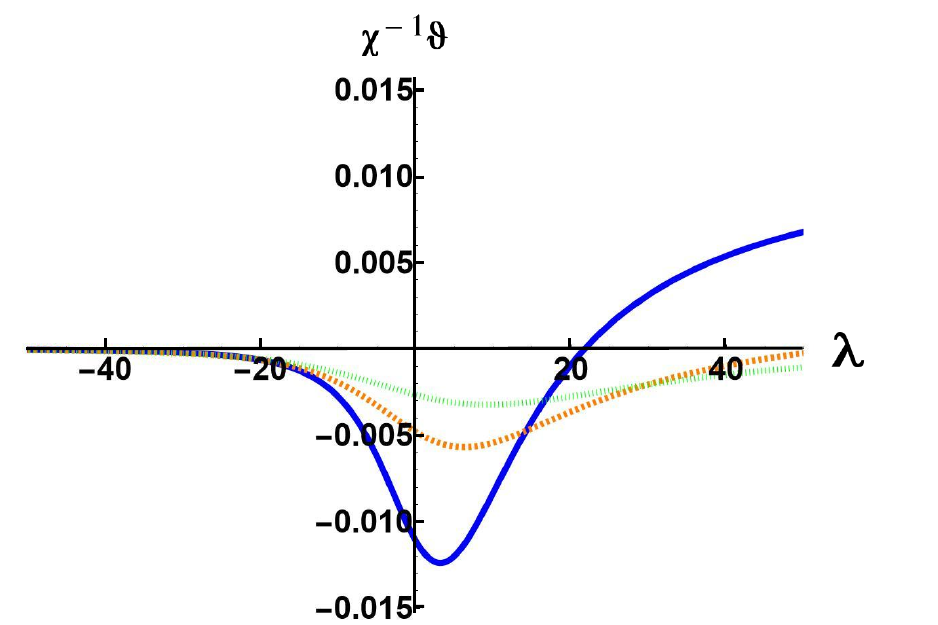}
\caption{\label{Fig_5_9}A plot of $\chi^{-1}\vartheta(\lambda)$ for prograde rays with fixed rotation parameter $\alpha=0.5$ and varying $\rho_m$=10 (solid), 15 (dashed), and 20 (dotted).}
\end{figure}

\begin{figure}[!hbt]
    \centering
\includegraphics[width=0.7\textwidth]{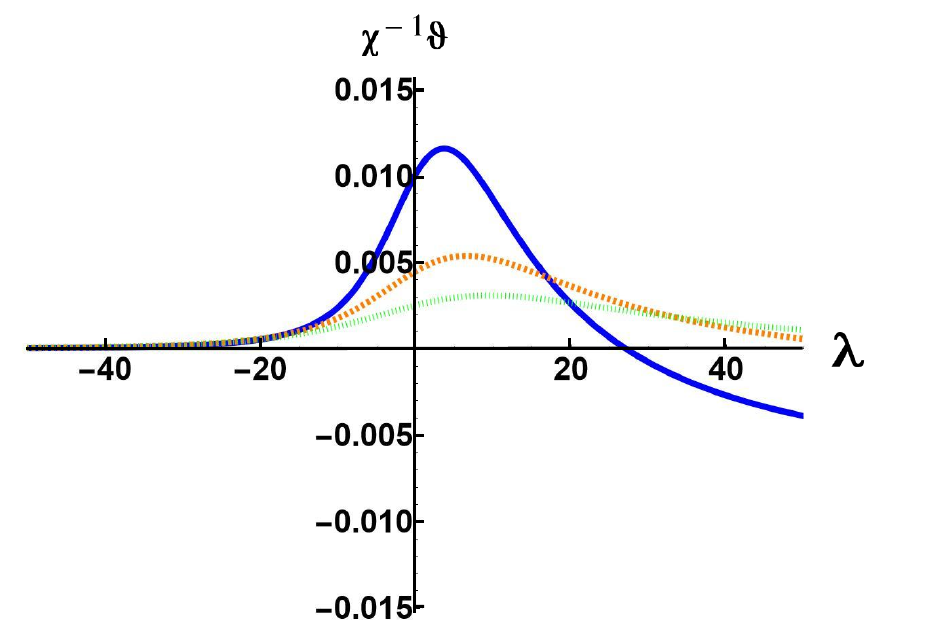}
\caption{\label{Fig_5_10}A plot of $\chi^{-1}\vartheta(\lambda)$ for retrograde rays with fixed rotation parameter $\alpha=0.5$ and varying $\rho_m$=10 (solid), 15 (dashed), and 20 (dotted).}
\end{figure}

In Figs.~\ref{Fig_5_9} and \ref{Fig_5_10}, we see how $\vartheta(\lambda)$ behaves for some fixed value of $\alpha$ and varying values of $\rho_m$, for both prograde and retrograde rays, respectively.

In the asymptotic region where $\lambda\to \infty$, $\dot{\rho}\approx 1$, and equation \eqref{ddotl} reduces to
\begin{equation}
\ddot{\vartheta}\approx -\dfrac{2\dot{\vartheta}}{\rho}\, .
\end{equation}
This gives $\dot{\vartheta}\approx C/\rho^2$, and
\begin{equation}
\vartheta\approx \vartheta^{\infty}-\dfrac{C}{\rho}\, .
\end{equation}

\begin{figure}[!hbt]
    \centering
\includegraphics[width=0.7\textwidth]{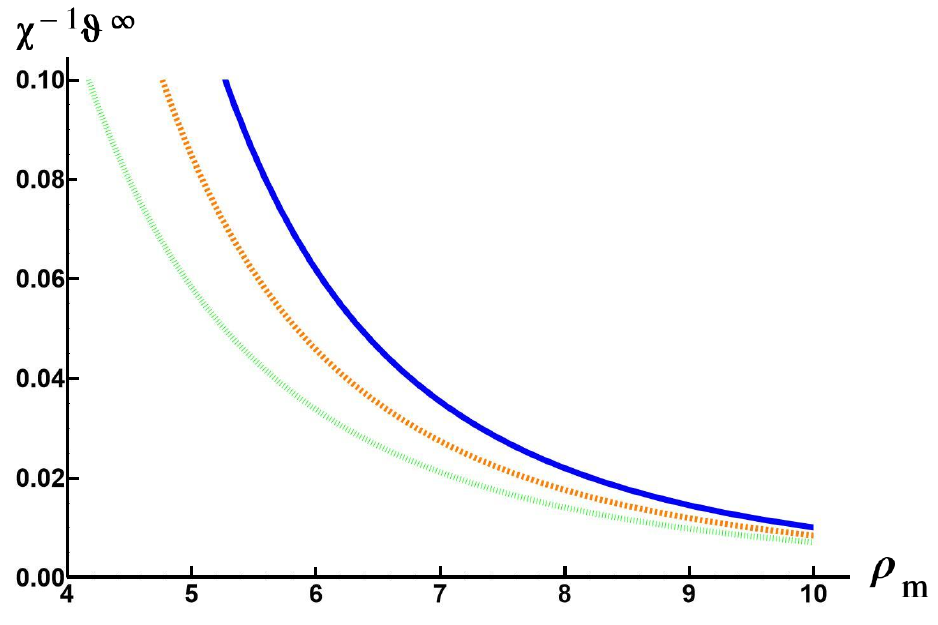}
\vspace{-6pt}
\caption{\label{Fig_5_11}A plot of $\chi^{-1}\,\vartheta^{\infty}$ as a function of $\rho_m$ for prograde photon scattering in the equatorial plane, with $\alpha =$ 0 (solid), 0.5 (dashed), 1 (dotted).}
\end{figure}

\begin{figure}[!hbt]
    \centering
\includegraphics[width=0.7\textwidth]{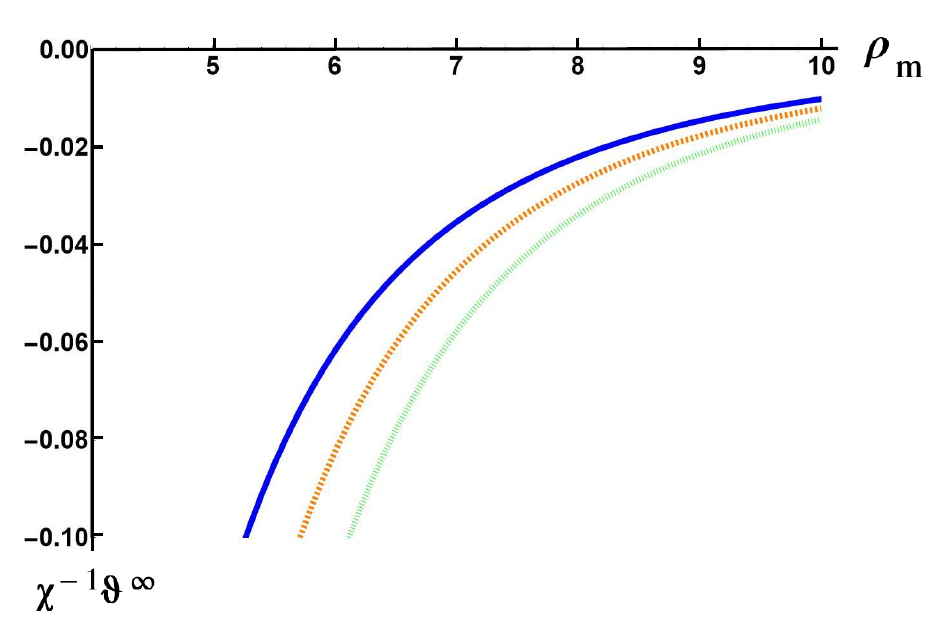}
\vspace{-6pt}
\caption{\label{Fig_5_12}A plot of $\chi^{-1}\,\vartheta^{\infty}$ as a function of $\rho_m$ for retrograde photon scattering in the equatorial plane, with $\alpha =$ 0 (solid), 0.5 (dashed), 1 (dotted).}
\end{figure}

Here, $C$ is a constant of integration, and $\vartheta^{\infty}$ is the asymptotic value of the tilting angle. By integrating equations \eqref{ddotr} and \eqref{ddotl}, one can find this asymptotic value of the tilting angle.

We perform these calculations for the values $\alpha=0$, $\alpha=0.5$, and $\alpha=1$ of the dimensionless rotation parameter $\alpha$. For each of these values, we find $\vartheta^{\infty}$ as a function of the parameter $\rho_m$, which is the radial coordinate of the radial turning point of the ray trajectory. These plots for prograde and retrograde rays are presented in Figs.~\ref{Fig_5_11} and \ref{Fig_5_12}, respectively.

For $\chi>0$, the asymptotic tilting angles $\vartheta^{\infty}$ for prograde and retrograde rays are correspondingly positive and negative. This means that after scattering, the prograde rays with positive helicity are reflected from the equatorial plane to the upper "half-space", that is, to the "north pole" direction of the black hole. The deflection of the negative helicity retrograde rays is in the opposite direction.

The plots \ref{Fig_5_11} and \ref{Fig_5_12} show that the absolute value of the tilting angle $\vartheta^{\infty}$ decreases with the growth of the parameter $\rho_m$. This is quite natural since rays with larger $\rho_m$ values pass at larger distance from the black hole, where the curvature is small. In Appendix~\ref{C2}, the asymptotic values of the tilting angle $\vartheta^{\infty}$ for prograde and retrograde rays with large parameter $\rho_m$ are obtained. It is shown that 
\begin{equation} \label{asvar}
\vartheta^{\infty}\approx \pm \dfrac{8\chi}{\rho^3_m}\,.
\end{equation}
This expression does not depend on the rotation parameter $\alpha$. The dependence on $\alpha$ is present only in the higher-order terms of a $\vartheta^{\infty}$ expansion over $1/\rho_m$. 

As we already mentioned, the spinoptics scattering problem in the Schwarzschild spacetime was considered in a recent paper \cite{Murk:2024qgj}. One can check that the acceleration vector $w^{\mu}$ of this paper coincides with the vector $\Upsilon^{\mu}$ evaluated for $\alpha=0$. The asymptotic expression for the tilting angle for prograde rays obtained in 
\cite{Murk:2024qgj} for large $\rho_m$ (after writing it in the units and conventions adopted in the present paper) has the form
\begin{equation}
|\vartheta^{\infty}|\approx \dfrac{C\chi}{\rho_m^3}\, ,
\end{equation}
where $C\approx 7.12$. This expression is similar to \eqref{asvar}, the only difference being that $C=8$ in \eqref{asvar}.

One other check we can provide is to see how our derived tilting angle compares to Mashhoon's original estimates in \cite{Mashhoon1974-ek} for the spinoptics deflection angle attained by a ray passing by a weakly rotating massive object along its axis of rotation. He estimates that for 8 GHz radio waves that pass by the sun at 7.8 solar radii, their spinoptics differential deflection angle should be $\lesssim10^{-23}\,rad$.  Let's assume that \eqref{asvar} is valid in this regime, and compare. Restoring $c$ and $G$ for this equation and applying it to the sun in SI units, we have
\begin{equation}
    \vartheta^{\infty}_\odot\sim \frac{G^{2}M^{2}_{\odot}}{fc^{3}r_m^3}\,,
\end{equation}
where $M_\odot\approx 2\times10^{30}\,kg$ is the solar mass, and $r_m=7.8\,R_\odot$, where $R_\odot=7\times10^8\,m$ is the solar radius.  Therefore, we find
\begin{equation}
    \vartheta^\infty_{\odot} \sim 10^{-24}\,rad\,.
\end{equation}
This is in very good agreement with Mashhoon's original estimate, although it is yet to be seen what deflection angles the spinoptics equations yield when applied to rays initially traveling parallel to the axis of rotation of a Kerr black hole.  This remains a question for future study.

Comparing the plots for prograde rays in Fig.~\ref{Fig_5_11}, one can see that for a given parameter $\rho_m$, the absolute value of the asymptotic tilting angle decreases with an increase in the rotation parameter $\alpha$. For retrograde rays the situation is different: the absolute value of the asymptotic tilting angle increases with an increase in the rotation parameter $\alpha$. This dependence of $\vartheta^{\infty}$ on the rotation can be explained as follows. The rotation dragging effect forces the prograde rays to "move faster" near the black hole, so that they spend less time in the domain of high curvature. For retrograde rays, the effect is the opposite. For both types of rays, the asymptotic tilting angle formally grows infinitely when the parameter $\rho_m$ reaches its limiting value $\rho_m^{\pm}$, signaling a breakdown of the perturbative expansion used in the problem at hand. This occurs because the spinoptics tetrad formalism that we use in this problem is not well-defined for rays which take on circular-orbits at lowest perturbative order (see \eqref{LambdaEQ} and \eqref{ZZZZ}).

\subsection{Rainbow Effect} \label{s542}

Without delving too deep into it, we would now like to briefly mention another effect associated with the spinoptics approximation that requires further study.  To do this, we need to develop the notion of a screen projection of the scattered rays as they reach a distant observer.

To this effect, we consider a ZAMO located at a far distance $\rho_0$ from the black hole along an angle $\theta_0$. Assume that this observer measures the directions of the photons relative to the center of symmetry of the asymptotic spacetime in correspondence with the black hole's center of mass. Let us denote the following components of $\boldsymbol{l}_0$ (the unperturbed tangent vector of the ray) in the ZAMO frame \eqref{EEEE}
\begin{equation}
l_{0(\theta)}=E_{(\theta)}^{\mu}l_{0\mu}\, \  \mbox{\ and\ } l_{0(\phi)}=E_{(\phi)}^{\mu}l_{0\mu}\,.
\end{equation}
Then it is possible to show that in the limit $\rho_0\to \infty$, the two impact parameters have the following limits
\begin{align} \label{imParam}
&\alpha_{0(\theta)}=\lim_{\rho_0\to\infty}\rho_0 l_{0(\theta)}=\pm\Big(\mathcal{Q}-\big(\alpha\sin\theta-\ell_z/\sin\theta \big)^2\Big)^{1/2}\,,\\\nonumber
&\alpha_{0(\phi)}=\lim_{\rho_0\to\infty}-\rho_0 l_{0(\phi)}=-\dfrac{\ell_z}{\sin\theta_0}\, .
\end{align}

A photon with the parameters $(\alpha_{0(\theta)},\alpha_{0(\phi)})$ is represented by a point in the image plane \cite{Bardeen}, at least for null rays in the zeroth-order geometric optics approximation.
In particular, the photons with impact parameters passing at the wedge of the black hole shadow surface are represented by an image of the black hole's shadow.

If we wished to calculate the first-order corrections to these images, we would simply apply Eq.~\eqref{imParam} to the first-order corrections of the tangent vector $\boldsymbol{l}$, such that
\begin{equation} \label{imParam1}
\alpha_{1(\theta)}=\lim_{\rho_0\to\infty}\chi\rho_0 l_{1(\theta)}\,,\hspace{15pt}
\alpha_{1(\phi)}=\lim_{\rho_0\to\infty}-\chi\rho_0 l_{1(\phi)}\, .
\end{equation}
As an illustrative example, we will do this for the polar impact parameter $\alpha_{1(\theta)}$ in the case where the null ray starts out on a trajectory with a very large initial impact parameter.  This calculation is worked out in Appendix~\ref{C3}, and we find that the first-order contribution to the polar impact parameter is given by Eq.~\eqref{a1sol}, which we repeat here for clarity
\begin{equation}
    \alpha_{1(\theta)} = \mp\frac{2\chi}{\rho_m^3}\,.
\end{equation}
This equation tells us that if a \emph{polychromatic} wave is scattered by a Kerr black hole, an observer will see a sort of "rainbow effect" due to spinoptics.  In more direct terms, one sees that the final location of the rays on the observer's screen is affected by their frequencies.

\section{Concluding Remarks} \label{s55}

In this chapter, we discussed how the interaction of the spin of a massless field with spacetime curvature modifies how it propagates in curved spacetime\footnote{Let us mention that this effect is quite similar to the effect of the spin-curvature interaction of a spinning massive particle moving in a gravitational field (see e.g. \cite{SPIN_KERR_0,SPIN_KERR_1,SPIN_KERR_2} and references therein).}. To describe this effect, we used the spinoptics approach \cite{Oancea:2019pgm,Oancea:2020khc,
Frolov:2020uhn,Dahal:2022gop,Frolov2024-ms,Frolov:2024qow}, which is a modification of the well-known geometric optics approach. In this approach, the problem of finding an approximate high-frequency solution of the wave equation is reduced to the study of dynamical equations for null rays and the null tetrads associated with them. In the leading zeroth order of a $1/\omega$ expansion, these equations describe geodesic null rays, while the associated tetrads are parallel propagated along them. In the spinoptics approximation, these equations are modified by the inclusion of first-order $1/\omega$ helicity-dependent corrections.

We focused on the study of the spinoptics approximation in the Kerr spacetime describing the gravitational field of a rotating black hole. In the Kerr metric, the geodesic equations and the parallel transport equations for the tetrad vectors can be solved in an explicit analytic form \cite{MARCK_1}. In the present paper, we used the method of solving the parallel transport equations developed in \cite{Kubiznak:2008zs}. (See also \cite{Dahal:2023ncl}). The spinoptics approximation is characterized by a small dimensionless parameter $\chi=\sigma/(M\omega)$. To describe the interaction of the spin of the massless field with the spacetime curvature, we used the method of perturbations. Since the "driving force" that describes this interaction is proportional to $\chi$, one can use unperturbed null tetrads associated with a corresponding null geodesic for the calculation of the driving force.

To describe a deviation of the "accelerated" null ray from a geodesic one, we derived a ray deviation equation, which is a natural generalization of the deviation equation for null geodesics. We used this equation to study the propagation of null rays in the vicinity of the equatorial plane. We demonstrated that for the scattering of circularly polarized waves, the asymptotic planes of the orbits of incoming and outgoing null rays are different. We calculated the corresponding tilting angles and studied their dependence on the ray impact parameter and on the spin of the black hole.
For a non-rotating black hole, the result for the calculated tilting angle agrees with the results presented in \cite{Dahal:2021qel,Murk:2024qgj,Frolov:2024olb}.

A recent paper by Dahal \cite{Dahal:2023ncl} discussed some spinoptics effects in the Kerr metric. Its main result was the calculation of the tilting angle for scattering of null rays passing at a large distance from the black hole. Our results are a far-reaching generalization of the results presented in \cite{Dahal:2023ncl}. In particular, we discussed and compared the asymptotic tilting angles for the scattering of polarized light near the equatorial plane for prograde and retrograde rays. This allowed us to illustrate the dependence of the tilting angle on the rotation parameter of the black hole.

Let us mention that there exist other interesting spinoptics-related effects for light and gravitational wave propagation near black holes. If a small-size object emits polarized radiation, then its image on an observer's screen evaluated in the spinoptics approximation would be slightly shifted. This shift depends on the helicity, and for a given helicity, it also depends on the frequency of the radiation. This is the "rainbow effect" detailed in the previous section. Another spin-optical effect is an additional helicity- and frequency-dependent time delay for waves propagating near black holes, which may be observable by the next generation of gravitational wave detectors \cite{Kubota:2023dlz}. Let us also mention that an interesting question is how the spinoptics approximation can modify the shadow of a black hole. For example, it is expected that in the Schwarzschild spacetime, we will see no altering of the shadow, however in the Kerr spacetime we expect some helicity-dependent deviations from the shadow which is predicted using classical general relativity \cite{oancea2021spin,Almeida:2026ewq}. Two more recent studies also show that spinoptics approximation may be used as a signature to prove or disprove the no-hair theorem for spherically-symmetric black holes \cite{Alves:2026jyc}, and that this effect may be carried into the quantum regime \cite{Ito:2026byb}. Lastly, we mention the question of gauge invariance.  The spinoptics equations are derived by initially considering Maxwell's equations in a curved spacetime.  These equations exhibit local $U(1)$ gauge invariance, as well as conformal invariance. After all of the perturbative approximations and assumptions made in our calculations, the question remains: how much of this gauge invariance is preserved in our spinoptics evolution equations? With all of the above studies in mind, we see that there are many problems yet to be solved regarding the spinoptics approximation, and we hope to address and continue the study of some of these questions in future research.

	\chapter{Conclusions}\label{ch6}

In this thesis, we set out to show by example that there are a wealth of effects in classical general relativity that are yet to be studied, some of which could hopefully lead to observational signatures.  

In Chapter~\ref{ch2}, we derived a method to boost compact masses with arbitrary mass distributions and angular momentum distributions to an object that lies in a two-dimensional null plane. We then outlined a nifty trick for which calculations of null ray scattering trajectories in the background of these massive compact rotating objects reduce to deflections and deviations from the two-dimensional impulse plane.  We showed that this procedure is similar to taking the Penrose limit of a black hole spacetime.  Within this ultrarelativistic limit, we were able to apply this trick to recover three main results for the scattering of null rays by black holes: (i) the azimuthal scattering angle of null rays that are scattered in the equatorial plane of a Kerr black hole, (ii) the frame-dragging shift effect that comes from a null ray initially traveling parallel to the rotation axis of a Kerr black hole, and (iii) the Shapiro time delay effect for a null ray scattered and deflected by a Schwarzschild black hole.  We also related our procedure to the gravitational memory effect.  Although this ultrarelativistic limit was only used for a handful of scenarios, one can imagine that it could be useful in the study of pp-waves in any gravitational background, including cosmological backgrounds or even backgrounds of black holes that are constructed from modified theories of gravity.

In Chapters~\ref{ch3} and \ref{ch4}, we described the evolution of rotating black holes traveling in homogeneous scalar and electromagnetic fields, showing that their masses, angular momenta, and velocities will all be affected by such setups.  In the case of homogeneous scalar fields, we considered three specific field configurations, which we called $T$-, $S$-, and $N$-fields.  These field configurations correspond to homogeneous scalar fields with constant timelike, spacelike, and null gradient vectors, respectively.  Although we were able to show that these black holes obey the second law of black hole thermodynamics within any of the three field configurations, they still exhibit other strange properties.  Specifically, in the case of the $T$-field and the $N$-field, the black hole may approach infinite mass in finite time for the $T$-field, and may approach infinite mass and the speed of light asymptotically for the $N$-field.  For each field configuration, we considered specific cases of motion depending on how the black hole's spin axis aligned with the field gradient.  In the case of the $S$-field, we analyzed two of these separate cases; one where the black hole's angular momentum vector precesses around the field-gradient vector and slowly decreases with time, and another where the black hole will slow down its speed to a halt in an infinite amount of time.  We note that these scalar fields were analyzed in a "weak" field approximation, in the sense that the backreaction of their energy-momentum on the black hole spacetime was ignored.  Furthermore, the relative rate of change of the black hole's characteristic quantities was considered to be small, so that we could analyze its evolution in a stationary Kerr background.

For the case of charged rotating black holes evolving in homogeneous electromagnetic fields, we also considered the electromagnetic fields and the electric and magnetic monopole charges to be "weak," in the sense that their backreaction on the background spacetime was small enough to be ignored.  Here, we looked at four specific alignments of black holes with respect to electric and magnetic fields, and analyzed the black holes' motion and evolution in these cases.  It should be noted that some of these cases have a striking similarity to the rotating black hole moving in a homogeneous $S$-field described above, with the main difference being that in the case of these charged rotating black holes moving in homogeneous electromagnetic fields, the mass of the black holes always remains constant, and can never increase.  Another takeaway from this study was that in most cases, these black holes act exactly like charged point particles in electromagnetic backgrounds.

Lastly, in Chapter~\ref{ch5} we delved into the spinoptics approximation, and calculated its impact on null ray scattering in the Kerr spacetime.  We showed that for null rays that are initially traveling in the equatorial plane of a Kerr black hole, they will experience some upwards or downwards acceleration due to the spinoptics approximation.  The direction of this spinoptics acceleration is dependent on whether the ray is being scattered in the prograde or retrograde direction, and if its polarization is left-handed (counterclockwise) or right-handed (clockwise).  The result of this acceleration is that the rays will achieve a final tilting angle with respect to the equatorial plane solely due to spinoptics effects.  The magnitude of this tilting angle is weakly dependent on the strength of the black hole's spin, but more importantly dependent on the frequency and initial impact parameter of the null ray, as well as the mass of the black hole.  We finished off our analysis of the spinoptics approximation by discussing other effects from this phenomenon that warrant future study, such as a rainbow effect, a time delay effect, and the preservation of gauge and diffeomorphism invariance.

Within all of the above studies there are some common themes: they all were studied within the framework of classical general relativity and classical field theory, they all can be related to observational effects, and they all lend themselves to future investigation, with the possibility of being extended to modified theories of gravity to note the difference between these effects in classical relativity and theories that go beyond relativity.  Although some of these effects are currently too small or unfeasible to be observed by the arsenal of telescopes and observatories that we as physicists have at our disposal, the universe often lends itself to be a magnificent laboratory, and when the future generation of gravitational wave detectors come online, we should be armed and ready with testable theories and effects that can either confirm or deny the robustness of general relativity in the vicinity of a black hole, or in the background of the universe.








	\printbibliography[heading=bibintoc]

@article{A1,
    author = "Frolov, Valeri P. and Koek, Alex",
    title = "{Gravitational lensing, memory, and the Penrose limit}",
    journal = "Phys. Rev. D",
    volume = "106",
    number = "6",
    pages = "064026",
    year = "2022"
}

@article{A2,
    author = "Frolov, Valeri P. and Koek, Alex",
    title = "{Motion of rotating black holes in homogeneous scalar fields: A general case}",
    journal = "Phys. Rev. D",
    volume = "109",
    number = "8",
    pages = "084067",
    year = "2024"
}

@article{A3,
    author = "Frolov, Valeri P. and Koek, Alex",
    title = "{Motion of a weakly charged rotating black hole in a homogeneous electromagnetic field}",
    journal = "Phys. Rev. D",
    volume = "110",
    number = "6",
    pages = "064052",
    year = "2024"
}

@article{A4,
    author = "Frolov, Valeri P. and Koek, Alex",
    title = "{Spinoptics in the Kerr spacetime: Polarized wave scattering}",
    journal = "Phys. Rev. D",
    volume = "111",
    number = "10",
    pages = "104081",
    year = "2025"
}

@ARTICLE{EHT,
  title     = "First {Sagittarius A*} {Event Horizon Telescope} results. {I}. {T}he
               shadow of the supermassive black hole in the center of the {Milky
               Way}",
  author    = "{Event Horizon Telescope Collaboration} and Akiyama, Kazunori
               and Alberdi, Antxon and Alef, Walter and Algaba, Juan Carlos and
               Anantua, Richard and Asada, Keiichi and Azulay, Rebecca and
               Bach, Uwe and Baczko, Anne-Kathrin and Ball, David and
               Balokovi{\'c}, Mislav and Barrett, John and Baub{\"o}ck, Michi
               and Benson, Bradford A and Bintley, Dan and Blackburn, Lindy and
               Blundell, Raymond and Bouman, Katherine L and Bower, Geoffrey C
               and Boyce, Hope and Bremer, Michael and Brinkerink, Christiaan D
               and Brissenden, Roger and Britzen, Silke and Broderick, Avery E
               and Broguiere, Dominique and Bronzwaer, Thomas and Bustamante,
               Sandra and Byun, Do-Young and Carlstrom, John E and Ceccobello,
               Chiara and Chael, Andrew and Chan, Chi-Kwan and Chatterjee,
               Koushik and Chatterjee, Shami and Chen, Ming-Tang and Chen,
               Yongjun and Cheng, Xiaopeng and Cho, Ilje and Christian, Pierre
               and Conroy, Nicholas S and Conway, John E and Cordes, James M
               and Crawford, Thomas M and Crew, Geoffrey B and Cruz-Osorio,
               Alejandro and Cui, Yuzhu and Davelaar, Jordy and Laurentis,
               Mariafelicia De and Deane, Roger and Dempsey, Jessica and
               Desvignes, Gregory and Dexter, Jason and Dhruv, Vedant and
               Doeleman, Sheperd S and Dougal, Sean and Dzib, Sergio A and
               Eatough, Ralph P and Emami, Razieh and Falcke, Heino and Farah,
               Joseph and Fish, Vincent L and Fomalont, Ed and Ford, H Alyson
               and Fraga-Encinas, Raquel and Freeman, William T and Friberg,
               Per and Fromm, Christian M and Fuentes, Antonio and Galison,
               Peter and Gammie, Charles F and Garc{\'\i}a, Roberto and Gentaz,
               Olivier and Georgiev, Boris and Goddi, Ciriaco and Gold, Roman
               and G{\'o}mez-Ruiz, Arturo I and G{\'o}mez, Jos{\'e} L and Gu,
               Minfeng and Gurwell, Mark and Hada, Kazuhiro and Haggard, Daryl
               and Haworth, Kari and Hecht, Michael H and Hesper, Ronald and
               Heumann, Dirk and Ho, Luis C and Ho, Paul and Honma, Mareki and
               Huang, Chih-Wei L and Huang, Lei and Hughes, David H and Ikeda,
               Shiro and Impellizzeri, C M Violette and Inoue, Makoto and
               Issaoun, Sara and James, David J and Jannuzi, Buell T and
               Janssen, Michael and Jeter, Britton and Jiang, Wu and
               Jim{\'e}nez-Rosales, Alejandra and Johnson, Michael D and
               Jorstad, Svetlana and Joshi, Abhishek V and Jung, Taehyun and
               Karami, Mansour and Karuppusamy, Ramesh and Kawashima, Tomohisa
               and Keating, Garrett K and Kettenis, Mark and Kim, Dong-Jin and
               Kim, Jae-Young and Kim, Jongsoo and Kim, Junhan and Kino, Motoki
               and Koay, Jun Yi and Kocherlakota, Prashant and Kofuji, Yutaro
               and Koch, Patrick M and Koyama, Shoko and Kramer, Carsten and
               Kramer, Michael and Krichbaum, Thomas P and Kuo, Cheng-Yu and
               Bella, Noemi La and Lauer, Tod R and Lee, Daeyoung and Lee,
               Sang-Sung and Leung, Po Kin and Levis, Aviad and Li, Zhiyuan and
               Lico, Rocco and Lindahl, Greg and Lindqvist, Michael and
               Lisakov, Mikhail and Liu, Jun and Liu, Kuo and Liuzzo,
               Elisabetta and Lo, Wen-Ping and Lobanov, Andrei P and Loinard,
               Laurent and Lonsdale, Colin J and Lu, Ru-Sen and Mao, Jirong and
               Marchili, Nicola and Markoff, Sera and Marrone, Daniel P and
               Marscher, Alan P and Mart{\'\i}-Vidal, Iv{\'a}n and Matsushita,
               Satoki and Matthews, Lynn D and Medeiros, Lia and Menten, Karl M
               and Michalik, Daniel and Mizuno, Izumi and Mizuno, Yosuke and
               Moran, James M and Moriyama, Kotaro and Moscibrodzka, Monika and
               M{\"u}ller, Cornelia and Mus, Alejandro and Musoke, Gibwa and
               Myserlis, Ioannis and Nadolski, Andrew and Nagai, Hiroshi and
               Nagar, Neil M and Nakamura, Masanori and Narayan, Ramesh and
               Narayanan, Gopal and Natarajan, Iniyan and Nathanail, Antonios
               and Fuentes, Santiago Navarro and Neilsen, Joey and Neri,
               Roberto and Ni, Chunchong and Noutsos, Aristeidis and Nowak,
               Michael A and Oh, Junghwan and Okino, Hiroki and Olivares,
               H{\'e}ctor and Ortiz-Le{\'o}n, Gisela N and Oyama, Tomoaki and
               {\"O}zel, Feryal and Palumbo, Daniel C M and Paraschos, Georgios
               Filippos and Park, Jongho and Parsons, Harriet and Patel, Nimesh
               and Pen, Ue-Li and Pesce, Dominic W and Pi{\'e}tu, Vincent and
               Plambeck, Richard and PopStefanija, Aleksandar and Porth, Oliver
               and P{\"o}tzl, Felix M and Prather, Ben and Preciado-L{\'o}pez,
               Jorge A and Psaltis, Dimitrios and Pu, Hung-Yi and Ramakrishnan,
               Venkatessh and Rao, Ramprasad and Rawlings, Mark G and Raymond,
               Alexander W and Rezzolla, Luciano and Ricarte, Angelo and
               Ripperda, Bart and Roelofs, Freek and Rogers, Alan and Ros,
               Eduardo and Romero-Ca{\~n}izales, Cristina and Roshanineshat,
               Arash and Rottmann, Helge and Roy, Alan L and Ruiz, Ignacio and
               Ruszczyk, Chet and Rygl, Kazi L J and S{\'a}nchez, Salvador and
               S{\'a}nchez-Arg{\"u}elles, David and S{\'a}nchez-Portal, Miguel
               and Sasada, Mahito and Satapathy, Kaushik and Savolainen, Tuomas
               and Schloerb, F Peter and Schonfeld, Jonathan and Schuster,
               Karl-Friedrich and Shao, Lijing and Shen, Zhiqiang and Small,
               Des and Sohn, Bong Won and SooHoo, Jason and Souccar, Kamal and
               Sun, He and Tazaki, Fumie and Tetarenko, Alexandra J and Tiede,
               Paul and Tilanus, Remo P J and Titus, Michael and Torne, Pablo
               and Traianou, Efthalia and Trent, Tyler and Trippe, Sascha and
               Turk, Matthew and van Bemmel, Ilse and van Langevelde, Huib Jan
               and van Rossum, Daniel R and Vos, Jesse and Wagner, Jan and
               Ward-Thompson, Derek and Wardle, John and Weintroub, Jonathan
               and Wex, Norbert and Wharton, Robert and Wielgus, Maciek and
               Wiik, Kaj and Witzel, Gunther and Wondrak, Michael F and Wong,
               George N and Wu, Qingwen and Yamaguchi, Paul and Yoon, Doosoo
               and Young, Andr{\'e} and Young, Ken and Younsi, Ziri and Yuan,
               Feng and Yuan, Ye-Fei and Zensus, J Anton and Zhang, Shuo and
               Zhao, Guang-Yao and Zhao, Shan-Shan and Agurto, Claudio and
               Allardi, Alexander and Amestica, Rodrigo and Araneda, Juan Pablo
               and Arriagada, Oriel and Berghuis, Jennie L and Bertarini,
               Alessandra and Berthold, Ryan and Blanchard, Jay and Brown, Ken
               and C{\'a}rdenas, Mauricio and Cantzler, Michael and Caro,
               Patricio and Castillo-Dom{\'\i}nguez, Edgar and Chan, Tin Lok
               and Chang, Chih-Cheng and Chang, Dominic O and Chang, Shu-Hao
               and Chang, Song-Chu and Chen, Chung-Chen and Chilson, Ryan and
               Chuter, Tim C and Ciechanowicz, Miroslaw and Colin-Beltran,
               Edgar and Coulson, Iain M and Crowley, Joseph and Degenaar,
               Nathalie and Dornbusch, Sven and Dur{\'a}n, Carlos A and
               Everett, Wendeline B and Faber, Aaron and Forster, Karl and
               Fuchs, Miriam M and Gale, David M and Geertsema, Gertie and
               Gonz{\'a}lez, Edouard and Graham, Dave and Gueth,
               Fr{\'e}d{\'e}ric and Halverson, Nils W and Han, Chih-Chiang and
               Han, Kuo-Chang and Hasegawa, Yutaka and Hern{\'a}ndez-Rebollar,
               Jos{\'e} Luis and Herrera, Cristian and Herrero-Illana, Ruben
               and Heyminck, Stefan and Hirota, Akihiko and Hoge, James and
               Hostler Schimpf, Shelbi R and Howie, Ryan E and Huang, Yau-De
               and Jiang, Homin and Jinchi, Hao and John, David and Kimura,
               Kimihiro and Klein, Thomas and Kubo, Derek and Kuroda, John and
               Kwon, Caleb and Lacasse, Richard and Laing, Robert and Leitch,
               Erik M and Li, Chao-Te and Liu, Ching-Tang and Liu, Kuan-Yu and
               Lin, Lupin C-C and Lu, Li-Ming and Mac-Auliffe, Felipe and
               Martin-Cocher, Pierre and Matulonis, Callie and Maute, John K
               and Messias, Hugo and Meyer-Zhao, Zheng and Monta{\~n}a, Alfredo
               and Montenegro-Montes, Francisco and Montgomerie, William and
               Moreno Nolasco, Marcos Emir and Muders, Dirk and Nishioka,
               Hiroaki and Norton, Timothy J and Nystrom, George and Ogawa,
               Hideo and Olivares, Rodrigo and Oshiro, Peter and
               P{\'e}rez-Beaupuits, Juan Pablo and Parra, Rodrigo and Phillips,
               Neil M and Poirier, Michael and Pradel, Nicolas and Qiu, Richard
               and Raffin, Philippe A and Rahlin, Alexandra S and Ram{\'\i}rez,
               Jorge and Ressler, Sean and Reynolds, Mark and
               Rodr{\'\i}guez-Montoya, Iv{\'a}n and Saez-Madain, Alejandro F
               and Santana, Jorge and Shaw, Paul and Shirkey, Jr, Leslie E and
               Silva, Kevin M and Snow, William and Sousa, Don and Sridharan, T
               K and Stahm, William and Stark, Anthony A and Test, John and
               Torstensson, Karl and Venegas, Paulina and Walther, Craig and
               Wei, Ta-Shun and White, Chris and Wieching, Gundolf and
               Wijnands, Rudy and Wouterloot, Jan G A and Yu, Chen-Yu and Yu
               (于威), Wei and Zeballos, Milagros",
  journal   = "Astrophys. J. Lett.",
  publisher = "American Astronomical Society",
  volume    =  930,
  number    =  2,
  pages     = "L12",
  year      =  2022,
  copyright = "http://creativecommons.org/licenses/by/4.0/"
}

@article{LIGO,
  title = {Observation of Gravitational Waves from a Binary Black Hole Merger},
  author = {Abbott, B. P. and Abbott, R. and Abbott, T. D. and Abernathy, M. R. and Acernese, F. and Ackley, K. and Adams, C. and Adams, T. and Addesso, P. and Adhikari, R. X. and Adya, V. B. and Affeldt, C. and Agathos, M. and Agatsuma, K. and Aggarwal, N. and Aguiar, O. D. and Aiello, L. and Ain, A. and Ajith, P. and Allen, B. and Allocca, A. and Altin, P. A. and Anderson, S. B. and Anderson, W. G. and Arai, K. and Arain, M. A. and Araya, M. C. and Arceneaux, C. C. and Areeda, J. S. and Arnaud, N. and Arun, K. G. and Ascenzi, S. and Ashton, G. and Ast, M. and Aston, S. M. and Astone, P. and Aufmuth, P. and Aulbert, C. and Babak, S. and Bacon, P. and Bader, M. K. M. and Baker, P. T. and Baldaccini, F. and Ballardin, G. and Ballmer, S. W. and Barayoga, J. C. and Barclay, S. E. and Barish, B. C. and Barker, D. and Barone, F. and Barr, B. and Barsotti, L. and Barsuglia, M. and Barta, D. and Bartlett, J. and Barton, M. A. and Bartos, I. and Bassiri, R. and Basti, A. and Batch, J. C. and Baune, C. and Bavigadda, V. and Bazzan, M. and Behnke, B. and Bejger, M. and Belczynski, C. and Bell, A. S. and Bell, C. J. and Berger, B. K. and Bergman, J. and Bergmann, G. and Berry, C. P. L. and Bersanetti, D. and Bertolini, A. and Betzwieser, J. and Bhagwat, S. and Bhandare, R. and Bilenko, I. A. and Billingsley, G. and Birch, J. and Birney, I. A. and Birnholtz, O. and Biscans, S. and Bisht, A. and Bitossi, M. and Biwer, C. and Bizouard, M. A. and Blackburn, J. K. and Blair, C. D. and Blair, D. G. and Blair, R. M. and Bloemen, S. and Bock, O. and Bodiya, T. P. and Boer, M. and Bogaert, G. and Bogan, C. and Bohe, A. and Bojtos, P. and Bond, C. and Bondu, F. and Bonnand, R. and Boom, B. A. and Bork, R. and Boschi, V. and Bose, S. and Bouffanais, Y. and Bozzi, A. and Bradaschia, C. and Brady, P. R. and Braginsky, V. B. and Branchesi, M. and Brau, J. E. and Briant, T. and Brillet, A. and Brinkmann, M. and Brisson, V. and Brockill, P. and Brooks, A. F. and Brown, D. A. and Brown, D. D. and Brown, N. M. and Buchanan, C. C. and Buikema, A. and Bulik, T. and Bulten, H. J. and Buonanno, A. and Buskulic, D. and Buy, C. and Byer, R. L. and Cabero, M. and Cadonati, L. and Cagnoli, G. and Cahillane, C. and Bustillo, J. Calder\'on and Callister, T. and Calloni, E. and Camp, J. B. and Cannon, K. C. and Cao, J. and Capano, C. D. and Capocasa, E. and Carbognani, F. and Caride, S. and Diaz, J. Casanueva and Casentini, C. and Caudill, S. and Cavagli\`a, M. and Cavalier, F. and Cavalieri, R. and Cella, G. and Cepeda, C. B. and Baiardi, L. Cerboni and Cerretani, G. and Cesarini, E. and Chakraborty, R. and Chalermsongsak, T. and Chamberlin, S. J. and Chan, M. and Chao, S. and Charlton, P. and Chassande-Mottin, E. and Chen, H. Y. and Chen, Y. and Cheng, C. and Chincarini, A. and Chiummo, A. and Cho, H. S. and Cho, M. and Chow, J. H. and Christensen, N. and Chu, Q. and Chua, S. and Chung, S. and Ciani, G. and Clara, F. and Clark, J. A. and Cleva, F. and Coccia, E. and Cohadon, P.-F. and Colla, A. and Collette, C. G. and Cominsky, L. and Constancio, M. and Conte, A. and Conti, L. and Cook, D. and Corbitt, T. R. and Cornish, N. and Corsi, A. and Cortese, S. and Costa, C. A. and Coughlin, M. W. and Coughlin, S. B. and Coulon, J.-P. and Countryman, S. T. and Couvares, P. and Cowan, E. E. and Coward, D. M. and Cowart, M. J. and Coyne, D. C. and Coyne, R. and Craig, K. and Creighton, J. D. E. and Creighton, T. D. and Cripe, J. and Crowder, S. G. and Cruise, A. M. and Cumming, A. and Cunningham, L. and Cuoco, E. and Canton, T. Dal and Danilishin, S. L. and D'Antonio, S. and Danzmann, K. and Darman, N. S. and Da Silva Costa, C. F. and Dattilo, V. and Dave, I. and Daveloza, H. P. and Davier, M. and Davies, G. S. and Daw, E. J. and Day, R. and De, S. and DeBra, D. and Debreczeni, G. and Degallaix, J. and De Laurentis, M. and Del\'eglise, S. and Del Pozzo, W. and Denker, T. and Dent, T. and Dereli, H. and Dergachev, V. and DeRosa, R. T. and De Rosa, R. and DeSalvo, R. and Dhurandhar, S. and D\'{\i}az, M. C. and Di Fiore, L. and Di Giovanni, M. and Di Lieto, A. and Di Pace, S. and Di Palma, I. and Di Virgilio, A. and Dojcinoski, G. and Dolique, V. and Donovan, F. and Dooley, K. L. and Doravari, S. and Douglas, R. and Downes, T. P. and Drago, M. and Drever, R. W. P. and Driggers, J. C. and Du, Z. and Ducrot, M. and Dwyer, S. E. and Edo, T. B. and Edwards, M. C. and Effler, A. and Eggenstein, H.-B. and Ehrens, P. and Eichholz, J. and Eikenberry, S. S. and Engels, W. and Essick, R. C. and Etzel, T. and Evans, M. and Evans, T. M. and Everett, R. and Factourovich, M. and Fafone, V. and Fair, H. and Fairhurst, S. and Fan, X. and Fang, Q. and Farinon, S. and Farr, B. and Farr, W. M. and Favata, M. and Fays, M. and Fehrmann, H. and Fejer, M. M. and Feldbaum, D. and Ferrante, I. and Ferreira, E. C. and Ferrini, F. and Fidecaro, F. and Finn, L. S. and Fiori, I. and Fiorucci, D. and Fisher, R. P. and Flaminio, R. and Fletcher, M. and Fong, H. and Fournier, J.-D. and Franco, S. and Frasca, S. and Frasconi, F. and Frede, M. and Frei, Z. and Freise, A. and Frey, R. and Frey, V. and Fricke, T. T. and Fritschel, P. and Frolov, V. V. and Fulda, P. and Fyffe, M. and Gabbard, H. A. G. and Gair, J. R. and Gammaitoni, L. and Gaonkar, S. G. and Garufi, F. and Gatto, A. and Gaur, G. and Gehrels, N. and Gemme, G. and Gendre, B. and Genin, E. and Gennai, A. and George, J. and Gergely, L. and Germain, V. and Ghosh, Abhirup and Ghosh, Archisman and Ghosh, S. and Giaime, J. A. and Giardina, K. D. and Giazotto, A. and Gill, K. and Glaefke, A. and Gleason, J. R. and Goetz, E. and Goetz, R. and Gondan, L. and Gonz\'alez, G. and Castro, J. M. Gonzalez and Gopakumar, A. and Gordon, N. A. and Gorodetsky, M. L. and Gossan, S. E. and Gosselin, M. and Gouaty, R. and Graef, C. and Graff, P. B. and Granata, M. and Grant, A. and Gras, S. and Gray, C. and Greco, G. and Green, A. C. and Greenhalgh, R. J. S. and Groot, P. and Grote, H. and Grunewald, S. and Guidi, G. M. and Guo, X. and Gupta, A. and Gupta, M. K. and Gushwa, K. E. and Gustafson, E. K. and Gustafson, R. and Hacker, J. J. and Hall, B. R. and Hall, E. D. and Hammond, G. and Haney, M. and Hanke, M. M. and Hanks, J. and Hanna, C. and Hannam, M. D. and Hanson, J. and Hardwick, T. and Harms, J. and Harry, G. M. and Harry, I. W. and Hart, M. J. and Hartman, M. T. and Haster, C.-J. and Haughian, K. and Healy, J. and Heefner, J. and Heidmann, A. and Heintze, M. C. and Heinzel, G. and Heitmann, H. and Hello, P. and Hemming, G. and Hendry, M. and Heng, I. S. and Hennig, J. and Heptonstall, A. W. and Heurs, M. and Hild, S. and Hoak, D. and Hodge, K. A. and Hofman, D. and Hollitt, S. E. and Holt, K. and Holz, D. E. and Hopkins, P. and Hosken, D. J. and Hough, J. and Houston, E. A. and Howell, E. J. and Hu, Y. M. and Huang, S. and Huerta, E. A. and Huet, D. and Hughey, B. and Husa, S. and Huttner, S. H. and Huynh-Dinh, T. and Idrisy, A. and Indik, N. and Ingram, D. R. and Inta, R. and Isa, H. N. and Isac, J.-M. and Isi, M. and Islas, G. and Isogai, T. and Iyer, B. R. and Izumi, K. and Jacobson, M. B. and Jacqmin, T. and Jang, H. and Jani, K. and Jaranowski, P. and Jawahar, S. and Jim\'enez-Forteza, F. and Johnson, W. W. and Johnson-McDaniel, N. K. and Jones, D. I. and Jones, R. and Jonker, R. J. G. and Ju, L. and Haris, K. and Kalaghatgi, C. V. and Kalogera, V. and Kandhasamy, S. and Kang, G. and Kanner, J. B. and Karki, S. and Kasprzack, M. and Katsavounidis, E. and Katzman, W. and Kaufer, S. and Kaur, T. and Kawabe, K. and Kawazoe, F. and K\'ef\'elian, F. and Kehl, M. S. and Keitel, D. and Kelley, D. B. and Kells, W. and Kennedy, R. and Keppel, D. G. and Key, J. S. and Khalaidovski, A. and Khalili, F. Y. and Khan, I. and Khan, S. and Khan, Z. and Khazanov, E. A. and Kijbunchoo, N. and Kim, C. and Kim, J. and Kim, K. and Kim, Nam-Gyu and Kim, Namjun and Kim, Y.-M. and King, E. J. and King, P. J. and Kinzel, D. L. and Kissel, J. S. and Kleybolte, L. and Klimenko, S. and Koehlenbeck, S. M. and Kokeyama, K. and Koley, S. and Kondrashov, V. and Kontos, A. and Koranda, S. and Korobko, M. and Korth, W. Z. and Kowalska, I. and Kozak, D. B. and Kringel, V. and Krishnan, B. and Kr\'olak, A. and Krueger, C. and Kuehn, G. and Kumar, P. and Kumar, R. and Kuo, L. and Kutynia, A. and Kwee, P. and Lackey, B. D. and Landry, M. and Lange, J. and Lantz, B. and Lasky, P. D. and Lazzarini, A. and Lazzaro, C. and Leaci, P. and Leavey, S. and Lebigot, E. O. and Lee, C. H. and Lee, H. K. and Lee, H. M. and Lee, K. and Lenon, A. and Leonardi, M. and Leong, J. R. and Leroy, N. and Letendre, N. and Levin, Y. and Levine, B. M. and Li, T. G. F. and Libson, A. and Littenberg, T. B. and Lockerbie, N. A. and Logue, J. and Lombardi, A. L. and London, L. T. and Lord, J. E. and Lorenzini, M. and Loriette, V. and Lormand, M. and Losurdo, G. and Lough, J. D. and Lousto, C. O. and Lovelace, G. and L\"uck, H. and Lundgren, A. P. and Luo, J. and Lynch, R. and Ma, Y. and MacDonald, T. and Machenschalk, B. and MacInnis, M. and Macleod, D. M. and Maga\~na-Sandoval, F. and Magee, R. M. and Mageswaran, M. and Majorana, E. and Maksimovic, I. and Malvezzi, V. and Man, N. and Mandel, I. and Mandic, V. and Mangano, V. and Mansell, G. L. and Manske, M. and Mantovani, M. and Marchesoni, F. and Marion, F. and M\'arka, S. and M\'arka, Z. and Markosyan, A. S. and Maros, E. and Martelli, F. and Martellini, L. and Martin, I. W. and Martin, R. M. and Martynov, D. V. and Marx, J. N. and Mason, K. and Masserot, A. and Massinger, T. J. and Masso-Reid, M. and Matichard, F. and Matone, L. and Mavalvala, N. and Mazumder, N. and Mazzolo, G. and McCarthy, R. and McClelland, D. E. and McCormick, S. and McGuire, S. C. and McIntyre, G. and McIver, J. and McManus, D. J. and McWilliams, S. T. and Meacher, D. and Meadors, G. D. and Meidam, J. and Melatos, A. and Mendell, G. and Mendoza-Gandara, D. and Mercer, R. A. and Merilh, E. and Merzougui, M. and Meshkov, S. and Messenger, C. and Messick, C. and Meyers, P. M. and Mezzani, F. and Miao, H. and Michel, C. and Middleton, H. and Mikhailov, E. E. and Milano, L. and Miller, J. and Millhouse, M. and Minenkov, Y. and Ming, J. and Mirshekari, S. and Mishra, C. and Mitra, S. and Mitrofanov, V. P. and Mitselmakher, G. and Mittleman, R. and Moggi, A. and Mohan, M. and Mohapatra, S. R. P. and Montani, M. and Moore, B. C. and Moore, C. J. and Moraru, D. and Moreno, G. and Morriss, S. R. and Mossavi, K. and Mours, B. and Mow-Lowry, C. M. and Mueller, C. L. and Mueller, G. and Muir, A. W. and Mukherjee, Arunava and Mukherjee, D. and Mukherjee, S. and Mukund, N. and Mullavey, A. and Munch, J. and Murphy, D. J. and Murray, P. G. and Mytidis, A. and Nardecchia, I. and Naticchioni, L. and Nayak, R. K. and Necula, V. and Nedkova, K. and Nelemans, G. and Neri, M. and Neunzert, A. and Newton, G. and Nguyen, T. T. and Nielsen, A. B. and Nissanke, S. and Nitz, A. and Nocera, F. and Nolting, D. and Normandin, M. E. N. and Nuttall, L. K. and Oberling, J. and Ochsner, E. and O'Dell, J. and Oelker, E. and Ogin, G. H. and Oh, J. J. and Oh, S. H. and Ohme, F. and Oliver, M. and Oppermann, P. and Oram, Richard J. and O'Reilly, B. and O'Shaughnessy, R. and Ott, C. D. and Ottaway, D. J. and Ottens, R. S. and Overmier, H. and Owen, B. J. and Pai, A. and Pai, S. A. and Palamos, J. R. and Palashov, O. and Palomba, C. and Pal-Singh, A. and Pan, H. and Pan, Y. and Pankow, C. and Pannarale, F. and Pant, B. C. and Paoletti, F. and Paoli, A. and Papa, M. A. and Paris, H. R. and Parker, W. and Pascucci, D. and Pasqualetti, A. and Passaquieti, R. and Passuello, D. and Patricelli, B. and Patrick, Z. and Pearlstone, B. L. and Pedraza, M. and Pedurand, R. and Pekowsky, L. and Pele, A. and Penn, S. and Perreca, A. and Pfeiffer, H. P. and Phelps, M. and Piccinni, O. and Pichot, M. and Pickenpack, M. and Piergiovanni, F. and Pierro, V. and Pillant, G. and Pinard, L. and Pinto, I. M. and Pitkin, M. and Poeld, J. H. and Poggiani, R. and Popolizio, P. and Post, A. and Powell, J. and Prasad, J. and Predoi, V. and Premachandra, S. S. and Prestegard, T. and Price, L. R. and Prijatelj, M. and Principe, M. and Privitera, S. and Prix, R. and Prodi, G. A. and Prokhorov, L. and Puncken, O. and Punturo, M. and Puppo, P. and P\"urrer, M. and Qi, H. and Qin, J. and Quetschke, V. and Quintero, E. A. and Quitzow-James, R. and Raab, F. J. and Rabeling, D. S. and Radkins, H. and Raffai, P. and Raja, S. and Rakhmanov, M. and Ramet, C. R. and Rapagnani, P. and Raymond, V. and Razzano, M. and Re, V. and Read, J. and Reed, C. M. and Regimbau, T. and Rei, L. and Reid, S. and Reitze, D. H. and Rew, H. and Reyes, S. D. and Ricci, F. and Riles, K. and Robertson, N. A. and Robie, R. and Robinet, F. and Rocchi, A. and Rolland, L. and Rollins, J. G. and Roma, V. J. and Romano, J. D. and Romano, R. and Romanov, G. and Romie, J. H. and Rosi\ifmmode \acute{n}\else \'{n}\fi{}ska, D. and Rowan, S. and R\"udiger, A. and Ruggi, P. and Ryan, K. and Sachdev, S. and Sadecki, T. and Sadeghian, L. and Salconi, L. and Saleem, M. and Salemi, F. and Samajdar, A. and Sammut, L. and Sampson, L. M. and Sanchez, E. J. and Sandberg, V. and Sandeen, B. and Sanders, G. H. and Sanders, J. R. and Sassolas, B. and Sathyaprakash, B. S. and Saulson, P. R. and Sauter, O. and Savage, R. L. and Sawadsky, A. and Schale, P. and Schilling, R. and Schmidt, J. and Schmidt, P. and Schnabel, R. and Schofield, R. M. S. and Sch\"onbeck, A. and Schreiber, E. and Schuette, D. and Schutz, B. F. and Scott, J. and Scott, S. M. and Sellers, D. and Sengupta, A. S. and Sentenac, D. and Sequino, V. and Sergeev, A. and Serna, G. and Setyawati, Y. and Sevigny, A. and Shaddock, D. A. and Shaffer, T. and Shah, S. and Shahriar, M. S. and Shaltev, M. and Shao, Z. and Shapiro, B. and Shawhan, P. and Sheperd, A. and Shoemaker, D. H. and Shoemaker, D. M. and Siellez, K. and Siemens, X. and Sigg, D. and Silva, A. D. and Simakov, D. and Singer, A. and Singer, L. P. and Singh, A. and Singh, R. and Singhal, A. and Sintes, A. M. and Slagmolen, B. J. J. and Smith, J. R. and Smith, M. R. and Smith, N. D. and Smith, R. J. E. and Son, E. J. and Sorazu, B. and Sorrentino, F. and Souradeep, T. and Srivastava, A. K. and Staley, A. and Steinke, M. and Steinlechner, J. and Steinlechner, S. and Steinmeyer, D. and Stephens, B. C. and Stevenson, S. P. and Stone, R. and Strain, K. A. and Straniero, N. and Stratta, G. and Strauss, N. A. and Strigin, S. and Sturani, R. and Stuver, A. L. and Summerscales, T. Z. and Sun, L. and Sutton, P. J. and Swinkels, B. L. and Szczepa\ifmmode \acute{n}\else \'{n}\fi{}czyk, M. J. and Tacca, M. and Talukder, D. and Tanner, D. B. and T\'apai, M. and Tarabrin, S. P. and Taracchini, A. and Taylor, R. and Theeg, T. and Thirugnanasambandam, M. P. and Thomas, E. G. and Thomas, M. and Thomas, P. and Thorne, K. A. and Thorne, K. S. and Thrane, E. and Tiwari, S. and Tiwari, V. and Tokmakov, K. V. and Tomlinson, C. and Tonelli, M. and Torres, C. V. and Torrie, C. I. and T\"oyr\"a, D. and Travasso, F. and Traylor, G. and Trifir\`o, D. and Tringali, M. C. and Trozzo, L. and Tse, M. and Turconi, M. and Tuyenbayev, D. and Ugolini, D. and Unnikrishnan, C. S. and Urban, A. L. and Usman, S. A. and Vahlbruch, H. and Vajente, G. and Valdes, G. and Vallisneri, M. and van Bakel, N. and van Beuzekom, M. and van den Brand, J. F. J. and Van Den Broeck, C. and Vander-Hyde, D. C. and van der Schaaf, L. and van Heijningen, J. V. and van Veggel, A. A. and Vardaro, M. and Vass, S. and Vas\'uth, M. and Vaulin, R. and Vecchio, A. and Vedovato, G. and Veitch, J. and Veitch, P. J. and Venkateswara, K. and Verkindt, D. and Vetrano, F. and Vicer\'e, A. and Vinciguerra, S. and Vine, D. J. and Vinet, J.-Y. and Vitale, S. and Vo, T. and Vocca, H. and Vorvick, C. and Voss, D. and Vousden, W. D. and Vyatchanin, S. P. and Wade, A. R. and Wade, L. E. and Wade, M. and Waldman, S. J. and Walker, M. and Wallace, L. and Walsh, S. and Wang, G. and Wang, H. and Wang, M. and Wang, X. and Wang, Y. and Ward, H. and Ward, R. L. and Warner, J. and Was, M. and Weaver, B. and Wei, L.-W. and Weinert, M. and Weinstein, A. J. and Weiss, R. and Welborn, T. and Wen, L. and We\ss{}els, P. and Westphal, T. and Wette, K. and Whelan, J. T. and Whitcomb, S. E. and White, D. J. and Whiting, B. F. and Wiesner, K. and Wilkinson, C. and Willems, P. A. and Williams, L. and Williams, R. D. and Williamson, A. R. and Willis, J. L. and Willke, B. and Wimmer, M. H. and Winkelmann, L. and Winkler, W. and Wipf, C. C. and Wiseman, A. G. and Wittel, H. and Woan, G. and Worden, J. and Wright, J. L. and Wu, G. and Yablon, J. and Yakushin, I. and Yam, W. and Yamamoto, H. and Yancey, C. C. and Yap, M. J. and Yu, H. and Yvert, M. and Zadro\ifmmode \dot{z}\else \.{z}\fi{}ny, A. and Zangrando, L. and Zanolin, M. and Zendri, J.-P. and Zevin, M. and Zhang, F. and Zhang, L. and Zhang, M. and Zhang, Y. and Zhao, C. and Zhou, M. and Zhou, Z. and Zhu, X. J. and Zucker, M. E. and Zuraw, S. E. and Zweizig, J.},
  collaboration = {LIGO Scientific Collaboration and Virgo Collaboration},
  journal = {Phys. Rev. Lett.},
  volume = {116},
  issue = {6},
  pages = {061102},
  numpages = {16},
  year = {2016},
  month = {Feb},
  publisher = {American Physical Society},
}

@book{MTW,
    author = "Misner, Charles W. and Thorne, K. S. and Wheeler, J. A.",
    title = "{Gravitation}",
    publisher = "W. H. Freeman",
    address = "San Francisco",
    year = 1973,
}

@ARTICLE{RicciLC1900,
  title     = "M\'{e}thodes de calcul diff\'{e}rentiel absolu et leurs applications",
  author    = "Ricci, M M G and Levi-Civita, T",
  journal   = "Math. Ann.",
  publisher = "Springer Science and Business Media LLC",
  volume    =  54,
  number    = "1-2",
  pages     = "125--201",
  year      =  1900,
}

@article{Slebodzinksky1931,
    author = "\'{S}lebodzi\'{n}ski, W." ,
    title = "Sur les \'{e}quations de {H}amilton",
    journal = "Bull. Acad. Roy. D. Belg.",
    year = 1931,
    volume = 17,
    number = "5",
    pages = "864--870",
}

@ARTICLE{HawkingPenrose1970,
  title     = "The singularities of gravitational collapse and cosmology",
  author    = "Hawking, S W and Penrose, R",
  journal   = "Proc. R. Soc. Lond.",
  publisher = "The Royal Society",
  volume    =  314,
  number    =  1519,
  pages     = "529--548",
  year      =  1970,
}

@ARTICLE{Penrose1965,
  title     = "Gravitational collapse and space-time singularities",
  author    = "Penrose, Roger",
  journal   = "Phys. Rev. Lett.",
  publisher = "American Physical Society (APS)",
  volume    =  14,
  number    =  3,
  pages     = "57--59",
  year      =  1965,
}

@ARTICLE{Kerr1963,
  title     = "Gravitational field of a spinning mass as an example of
               algebraically special metrics",
  author    = "Kerr, Roy P",
  journal   = "Phys. Rev. Lett.",
  publisher = "American Physical Society (APS)",
  volume    =  11,
  number    =  5,
  pages     = "237--238",
  year      =  1963,
}

@ARTICLE{Boyer1967,
  title     = "Maximal analytic extension of the {Kerr} metric",
  author    = "Boyer, Robert H and Lindquist, Richard W",
  journal   = "J. Math. Phys.",
  publisher = "AIP Publishing",
  volume    =  8,
  number    =  2,
  pages     = "265--281",
  year      =  1967,
}

@BOOK{FroZel2011,
  title     = {{Introduction to Black Hole Physics}},
  author    = "Frolov, Valeri P and Zelnikov, Andrei",
  publisher = "Oxford University Press",
  year      =  2011,
  address   = "London, England"
}

@ARTICLE{Bardeen_1,
       author = {{Bardeen}, James M.},
        title = {A Variational Principle for Rotating Stars in General Relativity},
      journal = {The Astrophysical Journal},
         year = 1970,
        month = oct,
       volume = {162},
        pages = {71},
}

@ARTICLE{Bardeen_2,
       author = {{Bardeen}, James M. and {Press}, William H. and {Teukolsky}, Saul A.},
        title = {Rotating Black Holes: Locally Nonrotating Frames, Energy Extraction, and Scalar Synchrotron Radiation},
      journal = {The Astrophysical Journal},
         year = 1972,
        month = dec,
       volume = {178},
        pages = {347-370},
}

@article{Bardeen,
  title={Timelike and null geodesics in the {K}err metric},
  author={Bardeen, James M and others},
  journal={Black holes},
  volume={215},
  year={1973},
  publisher={Gordon and Breach New York}
}

@article{Shapiro64,
  title = {Fourth Test of General Relativity},
  author = {Shapiro, Irwin I.},
  journal = {Phys. Rev. Lett.},
  volume = {13},
  issue = {26},
  pages = {789--791},
  numpages = {0},
  year = {1964},
  month = {Dec},
  publisher = {American Physical Society},
}

@BOOK{Carroll2019-di,
  title     = "Spacetime and {G}eometry",
  author    = "Carroll, Sean M",
  publisher = "Cambridge University Press (Virtual Publishing)",
  year      =  2019,
  address   = "Cambridge, England",
}

@article{Frolov2025-bn,
    author = "Frolov, Valeri P. and Koek, Alex",
    title = "{Spinoptics in the Kerr spacetime: Polarized wave scattering}",
    journal = "Phys. Rev. D",
    volume = "111",
    number = "10",
    pages = "104081",
    year = "2025"
}

@ARTICLE{LT1918,
       author = {{Lense}, Josef and {Thirring}, Hans},
        title = "{{\"U}ber den Einflu{\ss} der Eigenrotation der Zentralk{\"o}rper auf die Bewegung der Planeten und Monde nach der Einsteinschen Gravitationstheorie}",
      journal = {Physikalische Zeitschrift},
         year = 1918,
        month = jan,
       volume = {19},
        pages = {156},
}

@incollection{PenroseLimit,
    author = "Penrose, Roger",
    title = "Any Space-Time has a Plane Wave as a Limit",
    booktitle = "Differential Geometry and Relativity:
A Volume in Honour of André Lichnerowicz on His 60th Birthday",
    publisher = "Springer Dordrecht",
    year = {1976}
}

@article{AichelburgSexl,
    author = "Aichelburg, P. C. and Sexl, R. U.",
    title = "{On the gravitational field of a massless particle}",
    journal = "Gen. Rel. Grav.",
    volume = "2",
    pages = "303--312",
    year = "1971"
}

@article{Memory1,
  title = {Gravitational-wave bursts with memory: The {C}hristodoulou effect},
  author = {Thorne, Kip S.},
  journal = {Phys. Rev. D},
  volume = {45},
  issue = {2},
  pages = {520--524},
  numpages = {0},
  year = {1992},
  month = {Jan},
}

@article{Memory2,
year = {2010},
month = {apr},
publisher = {},
volume = {27},
number = {8},
pages = {084036},
author = {Favata, Marc},
title = {The gravitational-wave memory effect},
journal = {Classical and Quantum Gravity},
}

@article{Memory3,
    author = "Pasterski, Sabrina and Strominger, Andrew and Zhiboedov, Alexander",
    title = {New Gravitational Memories},
    journal = "JHEP",
    volume = "12",
    pages = "053",
    year = "2016"
}

@article{Memory4,
title = {The memory effect for plane gravitational waves},
author = "Zhang, P. -M and Duval, C. and Gibbons, G. W. and Horvathy, P. A.",
journal = {Physics Letters B},
volume = {772},
pages = {743-746},
year = {2017},
issn = {0370-2693},
}

@article{Memory5,
  title = {Soft gravitons and the memory effect for plane gravitational waves},
  author = {Zhang, P.-M. and Duval, C. and Gibbons, G. W. and Horvathy, P. A.},
  journal = {Phys. Rev. D},
  volume = {96},
  issue = {6},
  pages = {064013},
  numpages = {20},
  year = {2017},
  month = {Sep},
}

@book{Memory6,
    author = "Strominger, Andrew",
    title = "{Lectures on the Infrared Structure of Gravity and Gauge Theory}",
    publisher = "Princeton University Press",
    year = "2018"
}

@article{Memory7,
    author = "Hollands, Stefan and Ishibashi, Akihiro and Wald, Robert M.",
    title = {{BMS} Supertranslations and Memory in Four and Higher Dimensions},
    journal = "Class. Quant. Grav.",
    volume = "34",
    number = "15",
    pages = "155005",
    year = "2017"
}

@article{HiggsTh,
  title = {Broken Symmetries and the Masses of Gauge Bosons},
  author = {Higgs, Peter W.},
  journal = {Phys. Rev. Lett.},
  volume = {13},
  issue = {16},
  pages = {508--509},
  year = {1964},
  month = {Oct},
  publisher = {American Physical Society},
}

@article{HiggsEx,
    author = "Aad, Georges and others",
    collaboration = "ATLAS",
    title = "{Observation of a new particle in the search for the Standard Model Higgs boson with the ATLAS detector at the LHC}",
    journal = "Phys. Lett. B",
    volume = "716",
    pages = "1--29",
    year = "2012"
}

@article{ArkaniHamed1,
    author = "Arkani-Hamed, Nima and Cheng, Hsin-Chia and Luty, Markus A. and Mukohyama, Shinji",
    title = "{Ghost condensation and a consistent infrared modification of gravity}",
    journal = "JHEP",
    volume = "05",
    pages = "074",
    year = "2004"
}

@article{ArkaniHamed2,
    author = "Arkani-Hamed, Nima and Creminelli, Paolo and Mukohyama, Shinji and Zaldarriaga, Matias",
    title = "{Ghost inflation}",
    journal = "JCAP",
    volume = "04",
    pages = "001",
    year = "2004"
}

@article{Mattingly2005,
    author = "Mattingly, David",
    title = "{Modern tests of Lorentz invariance}",
    journal = "Living Rev. Rel.",
    volume = "8",
    pages = "5",
    year = "2005"
}

@article{Endlich2012,
    author = "Endlich, Solomon and Nicolis, Alberto and Wang, Junpu",
    title = "{Solid inflation}",
    journal = "JCAP",
    volume = "10",
    pages = "011",
    year = "2013"
}

@article{Moffat:1992bf,
    author = "Moffat, J. W.",
    title = "{Quantum gravity, the origin of time and time's arrow}",
    journal = "Found. Phys.",
    volume = "23",
    pages = "411--437",
    year = "1993"
}

@article{Babichev:2007dw,
    author = "Babichev, Eugeny and Mukhanov, Viatcheslav and Vikman, Alexander",
    title = "{K-essence, superluminal propagation, causality and emergent geometry}",
    journal = "JHEP",
    volume = "02",
    pages = "101",
    year = "2008"
}

@article{Mukohyama:2013ls,
  title = {From configuration to dynamics: Emergence of {L}orentz signature in classical field theory},
  author = {Mukohyama, Shinji and Uzan, Jean-Philippe},
  journal = {Phys. Rev. D},
  volume = {87},
  issue = {6},
  pages = {065020},
  numpages = {17},
  year = {2013},
  month = {Mar},
}

@article{Mukohyama:2013em,
  title = {Emergence of time in power-counting renormalizable {R}iemannian theory of gravity},
  author = {Mukohyama, Shinji},
  journal = {Phys. Rev. D},
  volume = {87},
  issue = {8},
  pages = {085030},
  numpages = {7},
  year = {2013},
  month = {Apr},
}

@article{Feng:2026em,
  title = {Emergent big bang scenario},
  author = {Feng, Justin C. and Mukohyama, Shinji and Uzan, Jean-Philippe},
  journal = {Phys. Rev. D},
  volume = {113},
  issue = {8},
  pages = {084025},
  numpages = {9},
  year = {2026},
  month = {Apr},
}

@article{Blas:2010hb,
    author = "Blas, Diego and Pujolas, Oriol and Sibiryakov, Sergey",
    title = "{Models of non-relativistic quantum gravity: The good, the bad and the healthy}",
    journal = "JHEP",
    volume = "04",
    pages = "018",
    year = "2011"
}

@article{Kovachik:2023fos,
    author = "Kovachik, Andrew and Sibiryakov, Sergey",
    title = "{Slowly moving black holes in khronometric model}",
    journal = "Phys. Rev. D",
    volume = "111",
    number = "4",
    pages = "044042",
    year = "2025"
}

@article{Jacobson:2000xp,
    author = "Jacobson, Ted and Mattingly, David",
    title = "{Gravity with a dynamical preferred frame}",
    journal = "Phys. Rev. D",
    volume = "64",
    pages = "024028",
    year = "2001"
}

@article{Jacobson:2007veq,
    author = "Jacobson, Ted",
    title = "{Einstein-aether gravity: A status report}",
    journal = "PoS",
    volume = "QG-PH",
    pages = "020",
    year = "2007"
}

@article{Oost:2018ea,
  title = {Constraints on {E}instein-aether theory after {GW170817}},
  author = {Oost, Jacob and Mukohyama, Shinji and Wang, Anzhong},
  journal = {Phys. Rev. D},
  volume = {97},
  issue = {12},
  pages = {124023},
  numpages = {8},
  year = {2018},
  month = {Jun},
}

@article{Israel:1967nh,
  title = {Event Horizons in Static Vacuum Space-Times},
  author = {Israel, Werner},
  journal = {Phys. Rev.},
  volume = {164},
  issue = {5},
  pages = {1776--1779},
  numpages = {0},
  year = {1967},
  month = {Dec},
}

@article{Israel:1967za,
    author = "Israel, Werner",
    title = "{Event horizons in static electrovac space-times}",
    journal = "Commun. Math. Phys.",
    volume = "8",
    pages = "245--260",
    year = "1968"
}

@article{Carter:1971bh,
  title = {Axisymmetric Black Hole Has Only Two Degrees of Freedom},
  author = {Carter, B.},
  journal = {Phys. Rev. Lett.},
  volume = {26},
  issue = {6},
  pages = {331--333},
  numpages = {0},
  year = {1971},
  month = {Feb},
}

@ARTICLE{Frolov2017-wa,
  title     = "Black holes, hidden symmetries, and complete integrability",
  author    = "Frolov, Valeri P and Krtou{\v s}, Pavel and Kubiz{\v n}{\'a}k,
               David",
  journal   = "Living Rev. Relativ.",
  publisher = "Springer Science and Business Media LLC",
  volume    =  20,
  number    =  1,
  pages     = "6",
  year      =  2017,
}

@BOOK{Nakahara1990-dr,
  title     = "Geometry, Topology and Physics",
  author    = "Nakahara, Mikio",
  publisher = "Institute of Physics Publishing",
  edition   =  2,
  year      =  1990,
  address   = "London, England",
}

@misc{grtensorii,
  title = "{{GRTensorII} Computer Algebra Package}",
  author = {Lake, Kayll and Pelavas, Nicos and Coley, Alan},
  howpublished = {Queen's University, Kingston, Ontario, Canada},
  year = {1994},
}

@article{Frolov:2024sf,
  title = {Motion of a rotating black hole in a homogeneous scalar field},
  author = {Frolov, Valeri P.},
  journal = {Phys. Rev. D},
  volume = {109},
  issue = {2},
  pages = {024055},
  numpages = {13},
  year = {2024},
  month = {Jan},
}

@article{Grasso:2000wj,
    author = "Grasso, Dario and Rubinstein, Hector R.",
    title = "{Magnetic fields in the early universe}",
    journal = "Phys. Rept.",
    volume = "348",
    pages = "163--266",
    year = "2001"
}

@article{Worku:2026sbp,
    author = "Worku, Keduse and Cruz, Hector Afonso G. and Kamionkowski, Marc",
    title = {Primordial Magnetic Fields at Cosmic Dawn: 21-cm Forecasts with {HERA} and {SKA}},
    eprint = "2605.05323",
    archivePrefix = "arXiv",
    primaryClass = "astro-ph.CO",
    month = "5",
    year = "2026"
}

@article{tHooft:1974kcl,
    author = "'t Hooft, Gerard",
    editor = "Taylor, J. C.",
    title = {Magnetic Monopoles in Unified Gauge Theories},
    journal = "Nucl. Phys. B",
    volume = "79",
    pages = "276--284",
    year = "1974"
}

@article{Preskill:1979zi,
    author = "Preskill, John",
    title = {Cosmological Production of Superheavy Magnetic Monopoles},
    journal = "Phys. Rev. Lett.",
    volume = "43",
    pages = "1365",
    year = "1979"
}

@article{Stojkovic:2004hz,
    author = "Stojkovic, Dejan and Freese, Katherine",
    title = "{A black hole solution to the cosmological monopole problem}",
    journal = "Phys. Lett. B",
    volume = "606",
    pages = "251--257",
    year = "2005"
}

@article{Gibbons:1975pa,
    author = "Gibbons, G. W.",
    title = "{Vacuum polarization and the spontaneous loss of charge by black holes}",
    journal = "Commun. Math. Phys.",
    volume = "44",
    pages = "245--264",
    year = "1975",
}

@article{Bicak:1976ef,
  title={Stationary electromagnetic fields around black holes. {II}. {G}eneral solutions and the fields of some special sources near a {K}err black hole},
  author={Jiř{\'i} Bi{\v{c}}{\'a}k and Leo{\vs} Dvoř{\'a}k},
  journal={General Relativity and Gravitation},
  year={1976},
  volume={7},
  pages={959-983},
}

@article{Wald:1974mf,
  title = {Black hole in a uniform magnetic field},
  author = {Wald, Robert M.},
  journal = {Phys. Rev. D},
  volume = {10},
  issue = {6},
  pages = {1680--1685},
  numpages = {0},
  year = {1974},
  month = {Sep},
}

@article{Frolov:2024xyo,
    author = "Frolov, Valeri P.",
    title = "{Motion of a rotating black hole in a homogeneous electromagnetic field}",
    journal = "Phys. Rev. D",
    volume = "109",
    number = "6",
    pages = "064045",
    year = "2024"
}

@ARTICLE{Ehlers1967-lg,
  title     = "{Zum {\"U}bergang von der Wellenoptik zur geometrischen Optik in
               der allgemeinen Relativit{\"a}tstheorie}",
  author    = "Ehlers, J{\"u}rgen",
  journal   = "Z. Naturforsch. A",
  publisher = "Walter de Gruyter GmbH",
  volume    =  22,
  number    =  9,
  pages     = "1328--1332",
  year      =  1967,
}

@article{Skrotskii1957,
  title     = {The Influence of Gravitation on the Propagation of Light},
  author    = {Skrotskii, G. V.},
  journal   = {Soviet Physics JETP},
  volume    = {5},
  number    = {2},
  pages     = {226--229},
  year      = {1957},
}

@article{Plebanksi:1960,
  title = {Electromagnetic Waves in Gravitational Fields},
  author = {Plebanski, Jerzy},
  journal = {Phys. Rev.},
  volume = {118},
  issue = {5},
  pages = {1396--1408},
  numpages = {0},
  year = {1960},
  month = {Jun},
}

@ARTICLE{Dwivedi1972-df,
  title     = "On the possibility of observing first-order corrections to
               geometrical optics in a curved space-time",
  author    = "Dwivedi, I and Kantowski, R",
  journal   = "J. Math. Phys.",
  publisher = "AIP Publishing",
  volume    =  13,
  number    =  12,
  pages     = "1941--1943",
  year      =  1972,
}

@ARTICLE{Frolov2011-rw,
  title     = "Spinoptics in a stationary spacetime",
  author    = "Frolov, Valeri P and Shoom, Andrey A",
  journal   = "Phys. Rev. D",
  publisher = "American Physical Society (APS)",
  volume    =  84,
  number    =  4,
pages = "051",
  year      =  2011,
}

@ARTICLE{Mashhoon1973-ec,
  title     = "Scattering of electromagnetic radiation from a black hole",
  author    = "Mashhoon, Bahram",
  journal   = "Phys. Rev. D Part. Fields",
  publisher = "American Physical Society (APS)",
  volume    =  7,
  number    =  10,
  pages     = "2807--2814",
  year      =  1973,
}

@ARTICLE{Mashhoon1974-ek,
  title     = "{Can Einstein's theory of gravitation be tested beyond the
               geometrical optics limit?}",
  author    = "Mashhoon, Bahram",
  journal   = "Nature",
  publisher = "Springer Science and Business Media LLC",
  volume    =  250,
  number    =  5464,
  pages     = "316--317",
  year      =  1974,
}

@ARTICLE{Mashhoon1974-vj,
  title     = "Electromagnetic scattering from a black hole and the glory
               effect",
  author    = "Mashhoon, Bahram",
  journal   = "Phys. Rev. D Part. Fields",
  publisher = "American Physical Society (APS)",
  volume    =  10,
  number    =  4,
  pages     = "1059--1063",
  year      =  1974,
}

@ARTICLE{Mashhoon1975-ey,
  title     = "Influence of gravitation on the propagation of electromagnetic
               radiation",
  author    = "Mashhoon, Bahram",
  journal   = "Phys. Rev. D Part. Fields",
  publisher = "American Physical Society (APS)",
  volume    =  11,
  number    =  10,
  pages     = "2679--2684",
  year      =  1975,
}

@ARTICLE{Yoo2012-vf,
  title     = "Notes on spinoptics in a stationary spacetime",
  author    = "Yoo, Chul-Moon",
  journal   = "Phys. Rev. D",
  publisher = "American Physical Society (APS)",
  volume    =  86,
  number    =  8,
  pages = "084005",
  year      =  2012,
}

@ARTICLE{Frolov2024-ms,
  title     = "Spinoptics in a curved spacetime",
  author    = "Frolov, Valeri P",
  journal   = "Phys. Rev. D",
  publisher = "American Physical Society (APS)",
  volume    =  110,
  number    =  6,
  pages = "064020",
  year      =  2024,
}

@article{Pineault:1977on,
  title = {Applications of Geometrical Optics to the {K}err Metric. {I}. {A}nalytical Results},
  author = {Pineault, S. and Roeder, R. C.},
  journal = {The Astrophysical Journal},
  volume = {212},
  pages = {541--548},
  year = {1977},
}

@article{Pineault:1977tw,
  title = {Applications of geometrical optics to the {Kerr} metric. {II}. {Analytical} results},
  author = {Pineault, S. and Roeder, R. C.},
  journal = {The Astrophysical Journal},
  volume = {212},
  pages = {548--558},
  year = {1977},
  month = {Mar},
}

@article{Oancea:2020khc,
    author = "Oancea, Marius A. and Joudioux, J{\'e}r{\'e}mie and Dodin, I. Y. and Ruiz, D. E. and Paganini, Claudio F. and Andersson, Lars",
    title = "{Gravitational spin Hall effect of light}",
    journal = "Phys. Rev. D",
    volume = "102",
    number = "2",
    pages = "024075",
    year = "2020"
}

@article{Stachel:1977cm,
    author = "Stachel, J. and Plebanski, J.",
    title = "{Classical particles with spin. {I}. The WKBJ approximation}",
    journal = "J. Math. Phys.",
    volume = "18",
    pages = "2368--2374",
    year = "1977"
}

@article{Frolov:2024qow,
    author = "Frolov, Valeri P. and Shoom, Andrey A.",
    title = "{Gravitational spinoptics in a curved space-time}",
    journal = "JCAP",
    volume = "10",
    pages = "039",
    year = "2024"
}

@article{Carter:68,
  title = {Global Structure of the {K}err Family of Gravitational Fields},
  author = {Carter, Brandon},
  journal = {Phys. Rev.},
  volume = {174},
  issue = {5},
  pages = {1559--1571},
  numpages = {0},
  year = {1968},
  month = {Oct},
}

@book{Chandrasekhar,
  title={The Mathematical Theory of Black Holes},
  author={Chandrasekhar, S.},
  series={International series of monographs on physics},
  year={1992},
  publisher={Oxford University Press}
}

@article{Sharp,
  title={Geodesics in black hole space-times},
  author={Nigel A. Sharp},
  journal={General Relativity and Gravitation},
  year={1979},
  volume={10},
  pages={659-670},
}

@article{Cieslik,
    author = "Cie\'slik, Adam and Hackmann, Eva and Mach, Patryk",
    title = "{Kerr geodesics in terms of Weierstrass elliptic functions}",
    journal = "Phys. Rev. D",
    volume = "108",
    number = "2",
    pages = "024056",
    year = "2023"
}

@article{Frolov:2020uhn,
    author = "Frolov, Valeri P.",
    title = "{Maxwell equations in a curved spacetime: Spin optics approximation}",
    journal = "Phys. Rev. D",
    volume = "102",
    number = "8",
    pages = "084013",
    year = "2020"
}

@article{Murk:2024qgj,
    author = "Murk, Sebastian and Terno, Daniel R. and Vadapalli, Rama",
    title = "{Gravity-induced birefringence in spherically symmetric spacetimes}",
    journal = "Phys. Rev. D",
    volume = "111",
    number = "4",
    pages = "044001",
    year = "2025"
}

@article{SPIN_KERR_0,
  title = {Motion of spinning test bodies in {K}err spacetime},
  author = {Hackmann, Eva and L\"ammerzahl, Claus and Obukhov, Yuri N. and Puetzfeld, Dirk and Schaffer, Isabell},
  journal = {Phys. Rev. D},
  volume = {90},
  issue = {6},
  pages = {064035},
  numpages = {12},
  year = {2014},
  month = {Sep},
}

@ARTICLE{SPIN_KERR_1,
      title={Extreme mass-ratio inspiral of a spinning body into a {K}err black hole {I}: Evolution along generic trajectories}, 
      author={Lisa V. Drummond and Alexandra G. Hanselman and Devin R. Becker and Scott A. Hughes},
      year={2023},
      eprint={2305.08919},
      archivePrefix={arXiv},
      primaryClass={gr-qc},
      journal = ""
}

@ARTICLE{SPIN_KERR_2,
      title={Extreme mass-ratio inspiral and waveforms for a spinning body into a {K}err black hole via osculating geodesics and near-identity transformations}, 
      author={Lisa V. Drummond and Philip Lynch and Alexandra G. Hanselman and Devin R. Becker and Scott A. Hughes},
      year={2023},
      eprint={2310.08438},
      archivePrefix={arXiv},
      primaryClass={gr-qc},
      journal = ""
}

@article{Oancea:2019pgm,
    author = "Oancea, Marius A. and Paganini, Claudio F. and Joudioux, Jeremie and Andersson, Lars",
    title = "{An overview of the gravitational spin Hall effect}",
    eprint = "1904.09963",
    archivePrefix = "arXiv",
    primaryClass = "gr-qc",
    month = "4",
    year = "2019",
    journal = ""
}

@article{Dahal:2022gop,
    author = "Dahal, Pravin Kumar",
    title = "{Covariant formulation of spin optics for electromagnetic waves}",
    journal = "Appl. Phys. B",
    volume = "129",
    number = "1",
    pages = "11",
    year = "2023"
}

@article{MARCK_1,
title = {Parallel-tetrad on null geodesics in {Kerr-Newman} space-time},
journal = {Physics Letters A},
volume = {97},
number = {4},
pages = {140-142},
year = {1983},
issn = {0375-9601},
author = {Jean-Alain Marck},
}

@article{Kubiznak:2008zs,
    author = "Kubiznak, David and Frolov, Valeri P. and Krtous, Pavel and Connell, Patrick",
    title = "{Parallel-propagated frame along null geodesics in higher-dimensional black hole spacetimes}",
    journal = "Phys. Rev. D",
    volume = "79",
    pages = "024018",
    year = "2009"
}

@article{Dahal:2023ncl,
    author = "Dahal, Pravin Kumar",
    title = "{Light propagation in Kerr spacetime}",
    journal = "Eur. Phys. J. Plus",
    volume = "138",
    number = "3",
    pages = "205",
    year = "2023"
}

@article{Dahal:2021qel,
    author = "Dahal, Pravin Kumar",
    title = {Spin Optics for Gravitational Waves},
    journal = "Astronomy",
    volume = "1",
    number = "3",
    pages = "271--287",
    year = "2022"
}

@article{Frolov:2024olb,
    author = "Frolov, Valeri P.",
    title = "{Spinoptics in the Schwarzschild spacetime}",
    journal = "JCAP",
    volume = "12",
    pages = "051",
    year = "2024"
}

@article{Yano1952,
    author = "Yano, Kentari",
    title = "On harmonic and {K}illing vector fields",
    journal = "Ann. of Math.",
    year = 1952,
    volume = 55,
    pages = "38--45"
}

@ARTICLE{Houri2007-hk,
  title     = "Closed conformal {Killing--Yano} tensor and {Kerr--NUT--de}
               {S}itter space--time uniqueness",
  author    = "Houri, Tsuyoshi and Oota, Takeshi and Yasui, Yukinori",
  journal   = "Phys. Lett. B",
  volume    =  656,
  number    = "4-5",
  pages     = "214--216",
  year      =  2007,
}

@ARTICLE{Houri2008-dp,
  title     = "Closed conformal {Killing--Yano} tensor and geodesic
               integrability",
  author    = "Houri, Tsuyoshi and Oota, Takeshi and Yasui, Yukinori",
  journal   = "J. Phys. A Math. Theor.",
  volume    =  41,
  number    =  2,
  pages     = "025204",
  year      =  2008
}

@ARTICLE{Krtous2008-gz,
  title     = "Hidden symmetries of higher-dimensional black holes and
               uniqueness of the {Kerr-NUT-(A)dS} spacetime",
  author    = "Krtou{\v s}, Pavel and Frolov, Valeri P and Kubiz{\v n}{\'a}k,
               David",
  journal   = "Phys. Rev. D",
  volume    =  78,
  number    =  6,
  pages = "064022",
  year      =  2008,
}

@article{Liouville1855,
    author = "Liouville, J",
    title = "Note sur l'int\'{e}gration des \'{e}quations diff\'{e}rentielles de la Dynamique",
    journal = "Journal de math\'{e}matiques pures et appliqu\'{e}es",
    volume = 20,
    pages = "137--138",
    year = 1855
}

@BOOK{Arnold1989-bc,
  title     = "{Mathematical Methods of Classical Mechanics}",
  author    = "Arnold, V I and Vogtmann, K and Weinstein, A",
  publisher = "Springer",
  series    = "Graduate Texts in Mathematics",
  edition   =  2,
  year      =  1989,
  address   = "New York, NY",
}

@ARTICLE{Page2007-ic,
  title     = "Complete integrability of geodesic motion in general
               higher-dimensional rotating black-hole spacetimes",
  author    = "Page, Don N and Kubizn{\'a}k, David and Vasudevan, Muraari and
               Krtous, Pavel",
  journal   = "Phys. Rev. Lett.",
  volume    =  98,
  number    =  6,
  pages     = "061102",
  year      =  2007,
}

@ARTICLE{Krtous2007-zv,
  title     = "{Killing-Yano} tensors, rank-2 {K}illing tensors, and conserved
               quantities in higher dimensions",
  author    = "Krtou{\v s}, Pavel and Kubizn{\'a}k, David and Page, Don N and
               Frolov, Valeri P",
  journal   = "J. High Energy Phys.",
  volume    =  2007,
  number    =  02,
  pages     = "004--004",
  year      =  2007
}

@ARTICLE{Krtous2007-fh,
  title     = "Constants of geodesic motion in higher-dimensional black-hole
               spacetimes",
  author    = "Krtou{\v s}, Pavel and Kubiz{\v n}{\'a}k, David and Page, Don N
               and Vasudevan, Muraari",
  journal   = "Phys. Rev. D",
  volume    =  76,
  number    =  8,
  pages = {084034},
  year      =  2007,
}

@article{Vaidya1943,
  title = {The external field of a radiating star in general relativity},
  author = {Vaidya, P. C.},
  journal = {Current Science},
  volume = {12},
  pages = {183},
  year = {1943},
}

@article{James:2015yla,
    author = "James, Oliver and von Tunzelmann, Eugenie and Franklin, Paul and Thorne, Kip S.",
    title = {Gravitational Lensing by Spinning Black Holes in Astrophysics, and in the Movie {I}nterstellar},
    journal = "Class. Quant. Grav.",
    volume = "32",
    number = "6",
    pages = "065001",
    year = "2015"
}

@article{Ciufolini:2004rq,
    author = "Ciufolini, I. and Pavlis, E. C.",
    title = "{A confirmation of the general relativistic prediction of the Lense-Thirring effect}",
    journal = "Nature",
    volume = "431",
    pages = "958--960",
    year = "2004"
}

@article{Blandford:1977ds,
    author = "Blandford, R. D. and Znajek, R. L.",
    title = "{Electromagnetic extractions of energy from Kerr black holes}",
    journal = "Mon. Not. Roy. Astron. Soc.",
    volume = "179",
    pages = "433--456",
    year = "1977"
}

@article{Blandford:1982xxl,
    author = "Blandford, R. D. and Payne, D. G.",
    title = "{Hydromagnetic flows from accretion discs and the production of radio jets}",
    journal = "Mon. Not. Roy. Astron. Soc.",
    volume = "199",
    number = "4",
    pages = "883--903",
    year = "1982"
}

@article{LeVerrier1859,
  title     = {Théorie du mouvement de {M}ercure},
  author    = {Le Verrier, Urbain Jean Joseph},
  journal   = {Annales de l'Observatoire Impérial de Paris},
  year      = {1859},
  volume    = {V},
  pages     = {1--195},
}

@article{Einstein1915,
  title = {{E}rklärung der {P}erihelbewegung des {M}erkur aus der allgemeinen {R}elativitätstheorie},
  author = {Albert Einstein},
  journal = {Sitzungsberichte der Königlich Preußischen Akademie der Wissenschaften},
  year = {1915},
  pages = {831--839},
  volume = {47}
}

@article{Einstein:1916vd,
    author = "Einstein, Albert",
    title = "{The foundation of the general theory of relativity.}",
    journal = "Annalen Phys.",
    volume = "49",
    number = "7",
    pages = "769--822",
    year = "1916"
}

@article{Hilbert1915,
  title     = {Die {G}rundlagen der {P}hysik},
  author    = {Hilbert, David},
  journal   = {Nachrichten von der K. Gesellschaft der Wissenschaften zu G{\"o}ttingen, Math.-phys. Klasse},
  year      = {1915},
  pages     = {395--407},
}

@article{Dyson1920,
  title = {A Determination of the Deflection of Light by the Sun's Gravitational Field, from Observations Made at the Total Eclipse of {M}ay 29, 1919},
  author = {Dyson, F. W. and Eddington, A. S. and Davidson, C.},
  journal = {Philosophical Transactions of the Royal Society of London. Series A, Containing Papers of a Mathematical or Physical Character},
  volume = {220},
  pages = {291--333},
  year = {1920},
  }

@article{Popper1954,
  title = {Red Shift in the Spectrum of 40 {E}ridani {B}},
  author = {Popper, Daniel M.},
  journal = {The Astrophysical Journal},
  volume = {120},
  pages = {316},
  year = {1954},
}

@article{PoundRebka1960,
  title = {Apparent Weight of Photons},
  author = {Pound, R. V. and Rebka, G. A.},
  journal = {Phys. Rev. Lett.},
  volume = {4},
  issue = {7},
  pages = {337--341},
  numpages = {0},
  year = {1960},
  month = {Apr},
}

@article{Penrose:1969pc,
    author = "Penrose, R.",
    title = "{Gravitational collapse: The role of general relativity}",
    journal = "Riv. Nuovo Cim.",
    volume = "1",
    pages = "252--276",
    year = "1969"
}

@article{Moffat:2006gz,
    author = "Moffat, J. W.",
    title = "{Time delay predictions in a modified gravity theory}",
    journal = "Class. Quant. Grav.",
    volume = "23",
    pages = "6767--6772",
    year = "2006"
}

@article{Kubota:2023dlz,
    author = "Kubota, Kei-ichiro and Arai, Shun and Mukohyama, Shinji",
    title = "{Spin optics for gravitational waves lensed by a rotating object}",
    journal = "Phys. Rev. D",
    volume = "109",
    number = "4",
    pages = "044027",
    year = "2024"
}

@article{Almeida:2026ewq,
    author = "Almeida, C. A. S.",
    title = "{Helicity-dependent corrections to black-hole shadows from the gravitational spin Hall effect}",
    eprint = "2605.02136",
    archivePrefix = "arXiv",
    primaryClass = "gr-qc",
    month = "5",
    year = "2026"
}

@article{Alves:2026jyc,
    author = "Alves, Kelvin S. and Cavalcanti, Rogerio T. and Perez Bergliaffa, Santiago E.",
    title = {Light Deflection due to Spinoptic Effects in Parametrized and Spherically Symmetric Hairy Black Holes},
    eprint = "2605.20524",
    archivePrefix = "arXiv",
    primaryClass = "gr-qc",
    month = "5",
    year = "2026"
}

@article{Ito:2026byb,
    author = "Ito, Ritsuki and Mameda, Kazuya and Yamamoto, Naoki",
    title = "{Spin Hall effect and Berry curvature of gravitons from quantum field theory}",
    eprint = "2605.19817",
    archivePrefix = "arXiv",
    primaryClass = "hep-th",
    reportNumber = "RIKEN-iTHEMS-Report-26",
    month = "5",
    year = "2026"
}

@article{Duval:2005ky,
    author = "Duval, C. and Horvath, Z. and Horvathy, P. A.",
    title = "{Fermat principle for spinning light}",
    journal = "Phys. Rev. D",
    volume = "74",
    pages = "021701",
    year = "2006"
}

@article{Duval:2005ry,
    author = "Duval, Christian and Horvath, Zalan and Horvathy, Peter",
    title = "{Geometrical spinoptics and the optical Hall effect}",
    journal = "J. Geom. Phys.",
    volume = "57",
    pages = "925--941",
    year = "2007"
}

@book{Schneider:1992bmb,
    author = {Schneider, Peter and Ehlers, J{\"u}rgen and Falco, Emilio E.},
    title = "{Gravitational Lenses}",
    publisher = "Springer",
    series = "Astronomy and Astrophysics Library",
    year = "1992"
}

@article{michell1784,
  title     = {On the means of discovering the distance, magnitude, \&c. of the fixed stars, in consequence of the diminution of the velocity of their light, in case such a diminution should be found to take place in any of them, and such other data should be procured from observations, as would be farther necessary for that purpose},
  author    = {Michell, John},
  journal   = {Philosophical Transactions of the Royal Society of London},
  volume    = {74},
  pages     = {35--57},
  year      = {1784},
}

@article{Schwarzschild:1916uq,
    author = "Schwarzschild, Karl",
    title = "{On the gravitational field of a mass point according to Einstein's theory}",
    journal = "Sitzungsber. Preuss. Akad. Wiss. Berlin (Math. Phys. )",
    volume = "1916",
    pages = "189--196",
    year = "1916"
}

@article{Eddington1924,
    title = {A Comparison of {Whitehead's and Einstein's} Formulae},author = {Arthur Stanley Eddington},
    journal = {Nature},
    volume = {113},
    number = {2832},
    pages = {192},
    year = {1924},
}

@article{Finkelstein1958,
  title = {Past-Future Asymmetry of the Gravitational Field of a Point Particle},
  author = {Finkelstein, David},
  journal = {Phys. Rev.},
  volume = {110},
  issue = {4},
  pages = {965--967},
  numpages = {0},
  year = {1958},
  month = {May},
}

@article{Kruskal:1959vx,
    author = "Kruskal, Martin D.",
    title = "{Maximal extension of {S}chwarzschild metric}",
    journal = "Phys. Rev.",
    volume = "119",
    pages = "1743--1745",
    year = "1960",
}

@article{Szekeres1960,
  author   = {Szekeres, George},
  title    = {On the singularities of a {R}iemannian manifold},
  journal  = {Publicationes Mathematicae Debrecen},
  volume   = {7},
  pages    = {285--301},
  year     = {1960}
}

@inproceedings{Penrose:1964dga,
    author      = "Penrose, Roger",
    title       = "{Conformal treatment of infinity}",
    booktitle   = "{Relativity, Groups and Topology}",
    editor      = "DeWitt, C. B. and DeWitt, B. S.",
    publisher   = "Gordon and Breach",
    address     = "London",
    pages       = "565--584",
    year        = "1964"
}

@article{Reissner1916,
  author  = {Reissner, Hans},
  title   = {{\"U}ber die Gravitation des elektrischen {F}eldes nach der {Einsteinschen} {T}heorie},
  journal = {Annalen der Physik},
  volume  = {355},
  number  = {9},
  pages   = {106--108},
  year    = {1916}
}

@article{Nordstrom1918,
  author  = {Nordstr{\"o}m, Gunnar},
  title   = {On the energy of the gravitation field in {Einstein's} theory},
  journal = {Proceedings of the Koninklijke Nederlandse Akademie van Wetenschappen},
  volume  = {20},
  pages   = {1238--1245},
  year    = {1918}
}

@article{Newman:1965my,
    author = "Newman, E. T. and Couch, E. and Chinnapared, K. and Exton, A. and Prakash, A. and Torrence, R.",
    title = "{Metric of a rotating, charged mass}",
    journal = "J. Math. Phys.",
    volume = "6",
    pages = "918--919",
    year = "1965",
}

@incollection{Riemann1861,
  author    = {Riemann, Bernhard},
  title     = {Commentatio mathematica, qua respondere tentatur quaestioni ab Illma Academia Parisiensi propositae},
  booktitle = {Bernhard Riemann's Gesammelte Mathematische Werke},
  pages     = {391--404},
  year      = {1892},
  publisher = {B.G. Teubner},
}

@article{Christoffel1869,
  author   = {Christoffel, Elwin Bruno},
  title    = {{Über die Transformation der homogenen Differentialausdrücke zweiten Grades}},
  journal  = {Journal für die reine und angewandte Mathematik},
  volume   = {70},
  pages    = {46--70},
  year     = {1869},
}

@article{Lemaitre:1933gd,
    author = "Lemaitre, G.",
    title = "{The expanding universe}",
    journal = "Annales Soc. Sci. Bruxelles A",
    volume = "53",
    pages = "51--85",
    year = "1933"
}

@article{Synge:1950gk,
    author = "Synge, J. L.",
    title = "{The gravitational field of a particle}",
    journal = "Proc. Roy. Irish Acad. A",
    volume = "53",
    pages = "83",
    year = "1950"
}

@article{Jebsen1921,
  author  = {Jebsen, J. T.},
  title   = {{\"U}ber die allgemeinen kugelsymmetrischen {L}{\"o}sungen der {Einsteinschen} {G}ravitationsgleichungen im {V}akuum},
  journal = {Arkiv f{\"o}r Matematik, Astronomi och Fysik},
  volume  = {15},
  number  = {18},
  pages   = {1--9},
  year    = {1921}
}

@book{Birkhoff1923,
  title={Relativity and Modern Physics},
  author={Birkhoff, George David and Langer, Rudolph Ernest},
  year={1923},
  publisher={Harvard University Press},
  address={Cambridge, MA}
}

@article{Carter:66,
  title = {Complete Analytic Extension of the Symmetry Axis of {K}err's Solution of {E}instein's Equations},
  author = {Carter, Brandon},
  journal = {Phys. Rev.},
  volume = {141},
  issue = {4},
  pages = {1242--1247},
  numpages = {0},
  year = {1966},
  month = {Jan},
}

@article{Zeldovich:1967lct,
    author = "Zel'dovich, Ya. B. and Novikov, I. D.",
    title = {The Hypothesis of Cores Retarded during Expansion and the Hot Cosmological Model},
    journal = "Sov. Astron.",
    volume = "10",
    pages = "602",
    year = "1967"
}

@article{Turner1983,
  title = {Coherent scalar-field oscillations in an expanding universe},
  author = {Turner, Michael S.},
  journal = {Phys. Rev. D},
  volume = {28},
  issue = {6},
  pages = {1243--1247},
  numpages = {0},
  year = {1983},
  month = {Sep},
}

@article{Cardoso:2018nkg,
    author = "Cardoso, Vitor and Ikeda, Taishi and Moore, Christopher J. and Yoo, Chul-Moon",
    title = "{Remarks on the maximum luminosity}",
    journal = "Phys. Rev. D",
    volume = "97",
    number = "8",
    pages = "084013",
    year = "2018"
}

@article{Balasin:1995tj,
    author = "Balasin, Herbert and Nachbagauer, Herbert",
    title = "{Boosting the Kerr geometry into an arbitrary direction}",
    journal = "Class. Quant. Grav.",
    volume = "13",
    pages = "731--738",
    year = "1996"
}

@article{Lousto:1992th,
    author = "Lousto, C. O. and Sanchez, Norma G.",
    title = "{The ultrarelativistic limit of the boosted Kerr-Newman geometry and the scattering of spin 1/2 particles}",
    journal = "Nucl. Phys. B",
    volume = "383",
    pages = "377--394",
    year = "1992"
}

@article{Frolov:2005gy,
  title = {Gravitational field of a spinning radiation beam pulse in higher dimensions},
  author = {Frolov, Valeri P. and Fursaev, Dmitri V.},
  journal = {Phys. Rev. D},
  volume = {71},
  issue = {10},
  pages = {104034},
  numpages = {16},
  year = {2005},
  month = {May},
}

@article{Frolov:2005zq,
    author = "Frolov, Valeri P. and Israel, Werner and Zelnikov, Andrei",
    title = "{Gravitational field of relativistic gyratons}",
    journal = "Phys. Rev. D",
    volume = "72",
    pages = "084031",
    year = "2005"
}

@article{Frolov:2005ja,
    author = "Frolov, Valeri P. and Zelnikov, Andrei",
    title = "{Gravitational field of charged gyratons}",
    journal = "Class. Quant. Grav.",
    volume = "23",
    pages = "2119--2128",
    year = "2006"
}

@article{Zeldovich:1974gvh,
    author = "Zel'dovich, Y. B. and Polnarev, A. G.",
    title = "{Radiation of gravitational waves by a cluster of superdense stars}",
    journal = "Sov. Astron.",
    volume = "18",
    pages = "17",
    year = "1974"
}

@article{Christodoulou:1991me,
  title = {Nonlinear nature of gravitation and gravitational-wave experiments},
  author = {Christodoulou, Demetrios},
  journal = {Phys. Rev. Lett.},
  volume = {67},
  issue = {12},
  pages = {1486--1489},
  numpages = {0},
  year = {1991},
  month = {Sep},
}

@article{Giddings:2001bu,
    author = "Giddings, Steven B. and Thomas, Scott D.",
    title = "{High-energy colliders as black hole factories: The end of short distance physics}",
    journal = "Phys. Rev. D",
    volume = "65",
    pages = "056010",
    year = "2002"
}

@article{Eardley:2002re,
    author = "Eardley, Douglas M. and Giddings, Steven B.",
    title = "{Classical black hole production in high-energy collisions}",
    journal = "Phys. Rev. D",
    volume = "66",
    pages = "044011",
    year = "2002"
}

@article{Kanti:2004nr,
    author = "Kanti, Panagiota",
    title = "{Black holes in theories with large extra dimensions: A Review}",
    journal = "Int. J. Mod. Phys. A",
    volume = "19",
    pages = "4899--4951",
    year = "2004"
}

@article{Robinson1975,
  title = {Uniqueness of the {K}err Black Hole},
  author = {Robinson, D. C.},
  journal = {Phys. Rev. Lett.},
  volume = {34},
  issue = {14},
  pages = {905--906},
  numpages = {0},
  year = {1975},
  month = {Apr},
}

@article{Mazur:1982,
    author = "Mazur, P. O.",
    title = "{Proof of uniqueness of the Kerr-Newman black hole solution}",
    journal = "J. Phys. A",
    volume = "15",
    pages = "3173--3180",
    year = "1982",
}

@phdthesis{Bunting1983,
    title     = {Proof of the Uniqueness Conjecture for Black Holes},
    author    = {Bunting, Gary L.},
    school    = {University of New England},
    address   = {Armidale, N.~S.~W.},
    year      = {1983},
    type      = {Ph.D. Thesis},
    note      = {Unpublished}
}

@book{Will1986,
    title = {Was Einstein Right?: Putting General Relativity to the Test},
    author = {Will, Clifford M.},
    year = {1986},
    publisher = {Basic Books},
    address = {New York, NY},
    isbn = {0-465-09086-9}
}

@article{Einstein1907,
    author = {Einstein, Albert},
    title = {{{\"U}ber das {R}elativit{\"a}tsprinzip und die aus demselben gezogenen {F}olgerungen}},
    journal = {Jahrbuch der Radioaktivit{"a}t und Elektronik},year = {1907},
    volume = {4},
    pages = {411--462}
}

@book{will1993theory,
  title     = {Theory and Experiment in Gravitational Physics},
  author    = {Will, Clifford M.},
  year      = {1993},
  publisher = {Cambridge University Press},
  address   = {Cambridge, England},
  isbn      = {978-0521375733}
}

@book{poisson2014gravity,
  title     = {Gravity: Newtonian, Post-Newtonian, Relativistic},
  author    = {Poisson, Eric and Will, Clifford M.},
  year      = {2014},
  publisher = {Cambridge University Press},
  address   = {Cambridge, United Kingdom},
  isbn      = {9781107032866}
}

@article{Einstein:1918btx,
    author = "Einstein, Albert",
    title = {{{\"U}ber Gravitationswellen}},
    journal = "Sitzungsber. Preuss. Akad. Wiss. Berlin (Math. Phys. )",
    volume = "1918",
    pages = "154--167",
    year = "1918"
}

@article{Hulse_Taylor_1975,
  title = {Discovery of a pulsar in a binary system},
  author = {Hulse, R. A. and Taylor, J. H.},
  journal = {The Astrophysical Journal},
  volume = {195},
  pages = {L51--L53},
  year = {1975},
  month = {Jan},
}

@article{Weisberg1981,
  title = {Gravitational radiation from an orbiting pulsar},
  author = {Weisberg, Joel M. and Taylor, Joseph H.},
  journal = {General Relativity and Gravitation},
  year = {1981},
  volume = {13},
  number = {1},
  pages = {1--6},
}

@misc{nobelphx2017,
  author       = {Nobel Foundation},
  title        = {The {N}obel Prize in Physics 2017},
  howpublished = {Nobel Foundation},
  year         = {2017},
}

@article{Dreyer:2003bv,
    author = "Dreyer, Olaf and Kelly, Bernard J. and Krishnan, Badri and Finn, Lee Samuel and Garrison, David and Lopez-Aleman, Ramon",
    title = "{Black hole spectroscopy: Testing general relativity through gravitational wave observations}",
    journal = "Class. Quant. Grav.",
    volume = "21",
    pages = "787--804",
    year = "2004"
}

@article{Isi:2019aib,
    author = "Isi, Maximiliano and Giesler, Matthew and Farr, Will M. and Scheel, Mark A. and Teukolsky, Saul A.",
    title = "{Testing the no-hair theorem with GW150914}",
    journal = "Phys. Rev. Lett.",
    volume = "123",
    number = "11",
    pages = "111102",
    year = "2019"
}

@article{Berti:2025hly,
    author = "Berti, Emanuele and others",
    editor = "Berti, Emanuele and Cardoso, Vitor and Carullo, Gregorio",
    title = "{Black hole spectroscopy: From theory to experiment}",
    journal = "Class. Quant. Grav.",
    volume = "43",
    number = "12",
    pages = "123001",
    year = "2026"
}

@article{LIGOScientific:2026sit,
    author = "Abac, A. G. and others",
    collaboration = "LIGO Scientific, VIRGO, KAGRA",
    title = {{GWTC}-5.0: {A}n Introduction to Version 5.0 of the Gravitational-Wave Transient Catalog},
    eprint = "2605.27223",
    archivePrefix = "arXiv",
    primaryClass = "gr-qc",
    reportNumber = "LIGO-P2500701",
    month = "5",
    year = "2026"
}

@article{Page:1977um,
    author = "Page, Don N.",
    title = "{Particle Emission Rates from a Black Hole. {III}. {C}harged Leptons from a Nonrotating Hole}",
    journal = "Phys. Rev. D",
    volume = "16",
    pages = "2402--2411",
    year = "1977"
}

@article{Vasquez2026,
      title={Corrections to {H}awking radiation from asteroid-mass primordial black holes: {A}nalytic and numerical evaluation of the stochastic charge effect}, 
      author={Gabriel Vasquez and Bowen Chen and Cara Nel and Emily Koivu and Makana Silva and Christopher M. Hirata},
      year={2026},
      eprint={2608.04346},
      archivePrefix={arXiv},
      primaryClass={gr-qc}
}

@article{Volkov:1971,
  title = {The propagation of electromagnetic waves in a {Riemannian} space},
  author = {Volkov, A. M. and Izmest'ev, A. A. and Skrotskii, G. V.},
  journal = {Journal of Experimental and Theoretical Physics},
  volume = {32},
  number = {4},
  pages = {686--691},
  year = {1971},
}

@book{Landau1980Classical,
  title     = {The Classical Theory of Fields},
  author    = {Landau, L. D. and Lifshitz, E. M.},
  year      = {1980},
  publisher = {Butterworth-Heinemann},
  edition   = {4th},
  series    = {Course of Theoretical Physics},
  volume    = {2},
  isbn      = {0750627689},
}

@article{Wigner1939,
  title = {On Unitary Representations of the Inhomogeneous {L}orentz Group},
  author = {Wigner, Eugene P.},
  journal = {Annals of Mathematics},
  volume = {40},
  number = {1},
  pages = {149--204},
  year = {1939},
}

@phdthesis{oancea2021spin,
  title={Spin {H}all effects in general relativity},
  author={Oancea, Marius-Adrian},
  year={2021},
  school={Universit{\"a}t Potsdam},
}

@article{Ricarte:2022wpd,
    author = "Ricarte, Angelo and Palumbo, Daniel C. M. and Narayan, Ramesh and Roelofs, Freek and Emami, Razieh",
    title = {Observational Signatures of Frame Dragging in Strong Gravity},
    journal = "Astrophys. J. Lett.",
    volume = "941",
    number = "1",
    pages = "L12",
    year = "2022"
}

@article{Guth:1980zm,
    author = "Guth, Alan H.",
    title = {The Inflationary Universe: {A} Possible Solution to the Horizon and Flatness Problems},
    journal = "Phys. Rev. D",
    volume = "23",
    pages = "347--356",
    year = "1981"
}

@article{Andersson:2025bhq,
    author = "Andersson, Lars and Gray, Finnian and Oancea, Marius A.",
    title = "{Conserved quantities and integrability for massless spinning particles in general relativity}",
    journal = "Phys. Rev. D",
    volume = "113",
    number = "6",
    pages = "064061",
    year = "2026"
}

@article{Yao:2011epa,
    author = "Yao, Alison M. and Padgett, Miles J.",
    title = "{Orbital angular momentum: {O}rigins, behavior and applications}",
    journal = "Adv. Opt. Photon.",
    volume = "3",
    number = "2",
    pages = "161--204",
    year = "2011"
}

@article{Unruh:1976,
  title = {Absorption cross section of small black holes},
  author = {Unruh, W. G.},
  journal = {Phys. Rev. D},
  volume = {14},
  issue = {12},
  pages = {3251--3259},
  year = {1976},
}

@article{Das:1997,
  title = {Universality of Low Energy Absorption Cross Sections for Black Holes},
  author = {Das, Sumit R. and Gibbons, Gary and Mathur, Samir D.},
  journal = {Phys. Rev. Lett.},
  volume = {78},
  issue = {3},
  pages = {417--419},
  year = {1997},
}

@article{Fabbri:1975sa,
    author = "Fabbri, R.",
    title = "{Scattering and absorption of electromagnetic waves by a {S}chwarzschild black hole}",
    journal = "Phys. Rev. D",
    volume = "12",
    pages = "933--942",
    year = "1975"
}

@article{Page:1976ki,
    author = "Page, Don N.",
    title = "{Particle Emission Rates from a Black Hole. {II}. {M}assless Particles from a Rotating Hole}",
    journal = "Phys. Rev. D",
    volume = "14",
    pages = "3260--3273",
    year = "1976"
}

@article{Starobinskil:1974nkd,
    author = "Starobinskil, Alexei A. and Churilov, S. M.",
    title = "{Amplification of electromagnetic and gravitational waves scattered by a rotating ''black hole''}",
    journal = "Sov. Phys. JETP",
    volume = "65",
    number = "1",
    pages = "1--5",
    year = "1974"
}

\appendix
	\truedoublespacing            

%

	\chapter{Geometric Considerations}\label{apxA}

In this appendix, we will collect geometric formulas related to general relativity, differential geometry, and hidden symmetries within black hole spacetimes.

\section{Differential Geometry} \label{A1}

In this section, we define geometrical objects from differential geometry that will be used throughout the thesis.

For a spacetime with the metric $g_{\mu\nu}$, the Riemann curvature tensor $R^{\mu}_{\;\nu\alpha\beta}$\footnote{This tensor was originally defined in a collection of essays by Riemann in 1861 \cite{Riemann1861}, but its coordinate description can be attributed to the work of Christoffel in 1869 \cite{Christoffel1869}.} is defined as
\begin{equation} \label{Rabcd}
R^{\mu}_{\;\nu\alpha\beta} = \partial_{\alpha}\Gamma^{\mu}_{\;\nu\beta}-\partial_{\beta}\Gamma^{\mu}_{\;\nu\alpha}+\Gamma^{\mu}_{\;\alpha\lambda}\Gamma^{\lambda}_{\;\nu\beta}-\Gamma^{\mu}_{\;\beta\lambda}\Gamma^{\lambda}_{\;\nu\alpha}\,,
\end{equation}
where the $\Gamma^{\mu}_{\;\nu\alpha}$ symbols are called Christoffel symbols.  They are written in terms of the metric as
\begin{equation}
\Gamma^{\mu}_{\;\nu\alpha} = \frac{1}{2}g^{\mu\lambda}(\partial_{\nu}g_{\lambda\alpha}+\partial_{\alpha}g_{\lambda\nu}-\partial_{\lambda}g_{\nu\alpha})\,.
\end{equation}
The Ricci tensor and Ricci scalar are simply obtained from the Riemann tensor via the index contractions $R_{\mu\nu}=R^{\lambda}_{\;\mu\lambda\nu}$ and $R=R^{\mu}_{\;\mu}$.

The Christoffel symbols are also important for defining a derivative on manifolds that transforms tensors into tensors. The covariant derivative of a vector on a manifold is defined as
\begin{equation}
\nabla_{\mu}V^{\nu} = \partial_{\mu}V^{\nu}+\Gamma_{\mu\lambda}^{\nu}V^{\lambda}\,,
\end{equation}
for any vector field $\boldsymbol{V}=V^{\mu}\partial_{\mu}$.  In the same vein, we define the covariant derivative for a generic mixed tensor field to be
\begin{align}
\nabla_{\lambda}T^{\mu_1...\mu_p}_{\hspace{30pt}\nu_1...\nu_q}=\partial_{\lambda}T^{\mu_1...\mu_p}_{\hspace{30pt}\nu_1...\nu_q}&+\Gamma^{\mu_1}_{\lambda\kappa}T^{\kappa...\mu_p}_{\hspace{30pt}\nu_1...\nu_q}+...+\Gamma^{\mu_p}_{\lambda\kappa}T^{\mu_1...\kappa}_{\hspace{30pt}\nu_1...\nu_q}\\\nonumber
&-\Gamma^{\kappa}_{\lambda\nu_1}T^{\mu_1...\mu_p}_{\hspace{30pt}\kappa...\nu_q}-...-\Gamma^{\kappa}_{\lambda\nu_q}T^{\mu_1...\mu_p}_{\hspace{30pt}\nu_1...\kappa}\,.
\end{align}
This derivative was introduced in 1901 by Gregorio Ricci-Curbastro and Tullio Levi-Civita, and was the first coordinate-independent description of a derivative on a manifold \cite{RicciLC1900}.

There is another notion of a derivative on a manifold that we will need to define, namely the Lie derivative.  This derivative was originally formulated by Sophus Lie for the cases of scalar and vector fields along the flow of a vector, but it was concretely formulated for any tensor field flowing along a vector field by W\l{}adys\l{}aw \'{S}lebodzi\'{n}ski in 1931 \cite{Slebodzinksky1931}.  For some mixed rank-$(r,s)$ tensor field $\boldsymbol{T}$, we say that its Lie derivative along a vector field $\boldsymbol{X}$ is given by
\begin{align}
    (\mathcal{L}_{\boldsymbol{X}}\boldsymbol{T})^{\mu_1...\mu_r}_{\hspace{28pt}\nu_1...\nu_s} = &X^{\lambda}\partial_{\lambda}T^{\mu_1...\mu_r}_{\hspace{28pt}\nu_1...\nu_s}\\\nonumber-&T^{\lambda...\mu_r}_{\hspace{28pt}\nu_1...\nu_s}\partial_{\lambda}X^{\mu_1}-...-T^{\mu_1...\lambda}_{\hspace{28pt}\nu_1...\nu_s}\partial_{\lambda}X^{\mu_r}\\\nonumber
    +&T^{\mu_1...\mu_r}_{\hspace{28pt}\lambda...\nu_s}\partial_{\nu_1}X^{\lambda}+...+T^{\mu_1...\mu_r}_{\hspace{28pt}\nu_1...\lambda}\partial_{\nu_s}X^{\lambda}\,.
\end{align}
A special and important case of this is when the Lie derivative of another vector field, say $\boldsymbol{Y}$, is taken.  This is called the Lie bracket of $\boldsymbol{Y}$ along $\boldsymbol{X}$, and is written as $\mathcal{L}_{\boldsymbol{X}}\boldsymbol{Y}=[\boldsymbol{X},\boldsymbol{Y}]$. An interesting and rather restrictive feature of the Lie derivative is that it is necessarily defined with respect to some vector field.

Now we will turn back to the Riemann tensor.  Another definition of this tensor is the measure by which covariant derivatives fail to commute.  To this effect, we first define the covariant derivative in the direction of any arbitrary vector $\boldsymbol{X}$ to be $\nabla_{\boldsymbol{X}}\equiv X^{\mu}\nabla_{\mu}$, and for two arbitrary vector fields $\boldsymbol{X}$  and $\boldsymbol{Y}$, the second covariant derivative is defined as $\nabla^2_{\boldsymbol{X},\boldsymbol{Y}}\equiv X^{\mu}Y^{\nu}\nabla_{\mu}\nabla_{\nu}$.  Then, we can define the Riemann tensor as 
\begin{equation}
    \boldsymbol{R}(\boldsymbol{X},\boldsymbol{Y},\;\boldsymbol{\cdot}\;) = \nabla_{\boldsymbol{X},\boldsymbol{Y}} - \nabla_{\boldsymbol{Y},\boldsymbol{X}}\,.
\end{equation}
A more common way to write this definition is 
\begin{equation} \label{RXY}
    \boldsymbol{R}(\boldsymbol{X},\boldsymbol{Y},\;\boldsymbol{\cdot}\;) = \big[\nabla_{\boldsymbol{X}},\nabla_{\boldsymbol{Y}}\big]-\nabla_{[\boldsymbol{X},\boldsymbol{Y}]}\,,
\end{equation}
where $\nabla_{[\boldsymbol{X},\boldsymbol{Y}]}\equiv\nabla_{\nabla_{\boldsymbol{X}}\boldsymbol{Y}}-\nabla_{\nabla_{\boldsymbol{Y}}\boldsymbol{X}}$.

In coordinate form, if we operate on a third vector $\boldsymbol{Z}$, we have
\begin{equation}
    \boldsymbol{R}(\boldsymbol{X},\boldsymbol{Y},\boldsymbol{Z}) = R^{\lambda}_{\;\kappa\mu\nu}X^{\mu}Y^{\nu}Z^{\kappa}\,.
\end{equation}

\section{Explicit and Hidden Symmetries} \label{A2}

In this section, we set out to describe explicit and hidden symmetries of spacetimes which generate conserved quantities for particles that are traveling in these spacetimes via Noether's theorem. 

\subsection{Killing Tensors} \label{A21}

We first start out with the definition of a Killing vector. A vector $\boldsymbol{\xi}$ is called a Killing vector of a spacetime with metric $\boldsymbol{g}$ if the Lie derivative of the metric along $\boldsymbol{\xi}$ vanishes, $\mathcal{L}_{\boldsymbol{\xi}}\hspace{1pt}g=0$.  In coordinate form, this equation can be written as
\begin{equation} \label{KillingV}
    \nabla_{\mu}\xi_{\nu}+\nabla_{\nu}\xi_{\mu} = 0 \iff \nabla_{(\mu}\xi_{\nu)}=0\,.
\end{equation}
This is called the Killing equation.  The fact that $\boldsymbol{\xi}$ is a Killing vector leads to a conservation law for some particle traveling within the spacetime defined by $\boldsymbol{g}$ as follows:  say that we have an affinely-parameterized geodesic representing the motion of a particle that follows the path $x(\lambda)$ with tangent vector $p^{\mu}=dx^{\mu}/d\lambda$.  This means that we can require the geodesic to obey the equation $Dp^{\mu}=0$, where the derivative operator $D$ is defined as $D\equiv p^{\nu}\nabla_{\nu}$.  Therefore, if we define a scalar quantity of a particle following the geodesic to be $Q=p_{\mu}\xi^{\mu}$, then it should be conserved along the path of the geodesic, since
\begin{equation}
DQ = D(p_{\mu}\xi^{\mu}) = \xi_{\mu}Dp^{\mu}+p_{\mu}D\xi^{\mu}=0.
\end{equation}
The first term in the product rule goes to zero due to the fact that $p^{\mu}$ is the tangent vector of a geodesic.  To see why the second term disappears, we write it out explicitly
\begin{equation}
    p_{\mu}D\xi^{\mu} = p^{\mu}p^{\nu}\nabla_{\nu}\xi_{\mu}=p^{\mu}p^{\nu}(\nabla_{[\nu}\xi_{\mu]}+\nabla_{(\nu}\xi_{\mu)})=0\,.
\end{equation}
In the last equality, the first term naturally is zero, and the second term is zero due to the Killing equation \eqref{KillingV}. Symmetries related to Killing vectors are called explicit symmetries.

We now describe hidden symmetries, which are generated by higher-rank tensors that are present in our spacetime.  A natural generalization of the Killing vector is the Killing tensor, which is fully symmetric, and obeys a similar symmetrized covariance equation to that of its vector counterpart
\begin{equation} \label{KillingT}
\nabla_{(\nu}K_{\mu_1...\mu_m)}=0\,,\hspace{15pt}K_{(\mu_1...\mu_m)}=K_{\mu_1...\mu_m}\,.
\end{equation}
We therefore say that $\boldsymbol{K}$ is a rank-$m$ Killing tensor within a given spacetime.  Just as before, we can expect conserved quantities to appear for particles in this spacetime due to the Killing vector if we contract each tensor index with the tangent vector of a geodesic corresponding to a freely moving particle.  Let $Q_m=K_{\mu_1...\mu_m}p^{\mu_1}...p^{\mu_m}$, then
\begin{equation}
    DQ_m = p^{\mu_1}...p^{\mu_m}p^{\nu}\nabla_{\nu}K_{\mu_1...\mu_m}\,,
\end{equation}
where all other components disappear due to the geodesic equation $Dp^{\mu}=0$.  Now, since $\boldsymbol{K}$ is both symmetric and a Killing tensor, we can write
\begin{equation}
p^{\mu_1}...p^{\mu_m}p^{\nu}\nabla_{\nu}K_{\mu_1...\mu_m}=-mp^{\mu_1}...p^{\mu_m}p^{\nu}\nabla_{\nu}K_{\mu_1...\mu_m}\,.
\end{equation}
This comes from expanding the first equation in \eqref{KillingT}, grouping all of the symmetrizations of $\boldsymbol{K}$ that exclude the index $\nu$, and then renaming the indices on the remaining terms so that all terms match.  Clearly this equation implies that $DQ_m=0$.  More detailed information on Killing vectors and tensors can be found in e.g. \cite{FroZel2011,Frolov2017-wa}

\subsection{Other Important Tensors} \label{A22}

The equations of this subsection can be found in much greater detail within \cite{Frolov2017-wa}.  In addition to Killing tensors, there are a myriad of other tensors which have deep geometric meaning and are extremely useful tools when investigating black hole metrics.  The first of these objects we will look at, which was discovered by Kentari Yano in 1952, is called a Killing-Yano tensor \cite{Yano1952}.  A rank-$m$ Killing-Yano tensor $\boldsymbol{k}$ obeys the following equations
\begin{equation} \label{KYTdef}
    \nabla_{(\nu}k_{\mu_1)...\mu_m} = 0\,,\hspace{15pt}k_{\mu_1...\mu_m}=k_{[\mu_1...\mu_m]}\,.
\end{equation}
The Killing-Yano tensor is often called the "square root" of the Killing tensor, because from any Killing-Yano tensor, we can obtain a rank-2 Killing tensor via the contraction $K_{\mu\nu}=k_{\mu\mu_2...\mu_m}k_{\nu}^{\;\,\mu_2...\mu_m}$.  

Even more importantly, rank-$m$ Killing-Yano tensors give rise to rank-$(m-1)$ tensors that are both parallel propagated along geodesics, and orthogonal to the geodesic's tangent vector in the sense that contracting the tangent vector with any of the conserved tensor's indices gives zero. 

Another very important tensor object for our study is defined by taking the Hodge dual of a rank-2 Killing-Yano tensor $\boldsymbol{k}$ in four-dimensional spacetime.  We define it as $\boldsymbol{h}=\star \boldsymbol{k}$, or in component-form
\begin{equation}
    h_{\mu\nu} = \frac{1}{2}e_{\mu\nu\alpha\beta}k^{\alpha\beta}\,,
\end{equation}
where $e_{\mu\nu\alpha\beta}$ is the Levi-Civita tensor with sign convention $e_{0123}=\sqrt{-g}$.  It turns out that $\boldsymbol{h}$ plays a pivotal role in the geometrical nature of spacetime, and can only exist for a very specific class of spacetimes, including the Kerr spacetime.  It is called the "principal tensor," and we will briefly return to the crucial role it plays at the end of this subsection.  

More generally, for a Killing-Yano tensor $\boldsymbol{k}$ of any rank, we can define a corresponding closed-conformal Killing-Yano tensor $\boldsymbol{h}$ by taking $\boldsymbol{h}=\star\boldsymbol{k}$.  If $\boldsymbol{h}$ is a rank-$m$ closed conformal Killing-Yano tensor in a $D$-dimensional spacetime, then it must satisfy the following equation
\begin{equation} \label{prin}
    \nabla_{\nu}h_{\mu_1...\mu_m}=\frac{m}{D-m+1}g_{\nu[\mu_1}\nabla^{\lambda}h_{|\lambda|\mu_2...\mu_m]}\,.
\end{equation}
As stated, $\boldsymbol{h}$ is closed in the sense that its exterior derivative is zero.  It is also completely antisymmetric in its indices.  Just as $\boldsymbol{k}$ corresponded to the "square root" of a rank-2 Killing tensor, $\boldsymbol{h}$ corresponds to the "square root" of some rank-2 conformal Killing tensor $\boldsymbol{H}$
\begin{equation}
    H_{\mu\nu} = h_{\mu\mu_2...\mu_m}h_{\nu}^{\;\,\mu_2...\mu_m}\,.
\end{equation}
If $\boldsymbol{H}$ were an arbitrary rank-$m$ conformal Killing tensor, it would obey the following equation
\begin{equation}
    \nabla_{(\nu}H_{\mu_1...\mu_m)}=g_{(\nu\mu_1}\alpha_{\mu_2...\mu_m)}\,,
\end{equation}
for some symmetric rank-$(m-1)$ tensor $\boldsymbol{\alpha}$.

Lastly, we are left to discuss the critical role that the principal tensor plays in black hole spacetimes.  It was shown separately by both Krotu\v{s}, Frolov, and Kubiz\v{n}\'{a}k as well as Houri, Oota, and Yasui that the most general geometry that solves Einstein's equations for four or more dimensions with a cosmological constant and admits a principal tensor turns out to be the Kerr-NUT-(A)dS spacetime \cite{Houri2007-hk,Houri2008-dp,Krtous2008-gz}.  This means that Kerr, Schwarzschild, Kerr-Newman, Reissner-Nordstr\"{o}m, (A)dS, and Minkowski spacetimes in any number of dimensions \emph{all} admit a principal tensor, since they are all just special cases of the Kerr-NUT-(A)dS spacetime.

It was also shown that {\bf{if a spacetime admits a principal tensor, then the geodesic equations corresponding to particle motion in such a spacetime are completely integrable, and the Hamilton-Jacobi, Klein-Gordon, and Dirac equations also allow separation of variables within said spacetime}}.  In fact, all massless fields allow for separation of variables in the four-dimensional Kerr spacetime, and the electromagnetic field allows for separation of variables in the Kerr-NUT-(A)dS spacetime in any number of dimensions.

We will give a brief summary of why the existence of this principal tensor leads to completely integrable geodesic equations.  Separability of the Hamilton-Jacobi, Klein-Gordon, and Dirac equations is more involved, so we will not touch on it, but for more on this subject, see \cite{Frolov2017-wa}.

The first thing to note with the principal tensor is that it generates a "Killing tower" of these rich geometric tensors that we have mentioned within this chapter.  Namely, in the language of differential forms, $j$ wedge products of the principal closed conformal Killing-Yano two-form give us a $2j$-form, which we can call $\boldsymbol{h}^{(j)}$, which is also a closed conformal Killing-Yano tensor.  

From there, if a spacetime is $D$-dimensional, then we can get $j$ rank-$(D-2j)$ Killing-Yano tensors by taking the Hodge dual $\boldsymbol{k}^{(j)}=(-1)^{2j(D-2j)+1}\star\boldsymbol{h}^{(j)}$.  We choose this sign as a prefactor so that, when employing Hodge duality, we have $\boldsymbol{h}^{(j)}=\star\boldsymbol{k}^{(j)}$. Each Killing-Yano tensor is the "square-root" of a Killing tensor, so now we have $j$ rank-2 Killing tensors
\begin{equation}
    K_{\mu\nu}^{(j)} = f^{(j)}_{\mu\mu_2...\mu_{D-2j-1}}f_{\nu}^{(j)\mu_2...\mu_{D-2j-1}}\,.
\end{equation}
Since $\boldsymbol{h}$ follows equation \eqref{prin} for a rank-2 tensor, we have that
\begin{equation} \label{hdef}
    \nabla_{\lambda}h_{\mu\nu}=g_{\lambda\mu}\xi_{\nu}-g_{\lambda\nu}\xi_{\mu}\,,\hspace{15pt}\xi^{\mu}=\frac{1}{D-1}\nabla^{\nu}h_{\nu\mu}\,.
\end{equation}
It turns out that $\boldsymbol{\xi}$ is a Killing vector.  We call it the "primary" Killing vector, because a list of $j$ more Killing vectors can be generated via the interior product $\boldsymbol{\zeta}^{(j)}=\boldsymbol{K}^{(j)}\cdot\boldsymbol{\xi}$.  

Since each Killing vector and Killing tensor gives one integral of motion for geodesic particle orbits, we get exactly $D$ integrals of motion for a $D$-dimensional metric.  To satisfy the Liouville-Arnold theorem, all that is left to show is that these tensors evolve in time independently of one another, and independently of the Hamiltonian \cite{Liouville1855,Arnold1989-bc}.  In other words, all of their Poisson brackets should commute, and all of their Poisson brackets should commute with the Hamiltonian.  This was shown throughout the following four papers \cite{Page2007-ic,Krtous2007-fh,Krtous2007-zv,Houri2008-dp}, thus proving that the existence of a principal tensor in a spacetime leads to completely integrable geodesics for particles in that spacetime.  Therefore, the Kerr-NUT-(A)dS spacetime has completely integrable geodesics for freely-moving particles in any dimension.

We should also quickly mention that since every closed form is locally exact, we can write $\boldsymbol{h}$ locally as the exterior derivative of some potential $\boldsymbol{h}=d\boldsymbol{b}$.

\subsection{Killing and Killing-Yano Tensors in the Kerr Metric} \label{A23}

We choose to outline the above procedure for the Kerr metric in 1+3 dimensions.  The goal here is to derive four integrals of motion for the four-dimensional spacetime.  This should come from two Killing vectors and two Killing tensors.

For the Kerr metric in Boyer-Lindquist coordinates \eqref{Kerr}, we have the following potential
\begin{equation}
    b_{\mu}dx^{\mu}=-\frac{1}{2}\big((r^2-a^2\cos^2\theta)dt-a(r^2\sin^2\theta-a^2\cos^2\theta)d\phi\big)\,.
\end{equation}
In coordinate form, this gives us the principal tensor $h_{\mu\nu}=\partial_{\mu}b_{\nu}-\partial_{\nu}b_{\mu}$,
\begin{align} \label{principal}
    h_{\mu\nu} = &r(\delta_{\mu}^{t}\delta_{\nu}^{r}-\delta_{\mu}^{r}\delta_{\nu}^{t}) + 
    a^2\sin\theta\cos\theta(\delta_{\mu}^{t}\delta_{\nu}^{\theta}-\delta_{\mu}^{\theta}\delta_{\nu}^{t}) \\\nonumber
    &+ a\sin\theta\cos\theta(r^2+a^2)(\delta_{\mu}^{\theta}\delta_{\nu}^{\phi}-\delta_{\mu}^{\phi}\delta_{\nu}^{\theta})\\\nonumber
    &+ ar\sin^2\theta(\delta_{\mu}^{r}\delta_{\nu}^{\phi}-\delta_{\mu}^{\phi}\delta_{\nu}^{r})\,.
\end{align}
Recall that the principal tensor also gives rise to the primary Killing vector
\begin{equation}
    \xi^{\mu}=\frac{1}{3}\nabla^{\nu}h_{\nu\mu}=\delta^{\mu}_{t}\,.
\end{equation}

Once more working in coordinate form, we can take the Hodge dual of the principal tensor (with a negative sign convention) to get the following Killing-Yano tensor
\begin{align} \label{kYano}
    k_{\mu\nu} = -\frac{1}{2}e_{\mu\nu\alpha\beta}h^{\alpha\beta}=&-a\cos\theta(\delta_{\mu}^{t}\delta_{\nu}^{r}-\delta_{\mu}^{r}\delta_{\nu}^{t}) + ar\sin\theta(\delta_{\mu}^{t}\delta_{\nu}^{\theta}-\delta_{\mu}^{\theta}\delta_{\nu}^{t}) \\\nonumber
    &- a^2\sin^2\theta\cos\theta(\delta_{\mu}^{r}\delta_{\nu}^{\phi}-\delta_{\mu}^{\phi}\delta_{\nu}^{r})\\\nonumber
    &+ r\sin\theta(r^2+a^2)(\delta_{\mu}^{\theta}\delta_{\nu}^{\phi}-\delta_{\mu}^{\phi}\delta_{\nu}^{\theta})\,.
\end{align}
As previously mentioned, "squaring" this tensor gives us a Killing tensor, which we write in matrix notation for convenience
\begin{equation} \label{KTKerr}
    K^{\mu}_{\hspace{5pt}\nu}=  k_{\mu}^{\;\,\lambda}k_{\nu\lambda} = \begin{bmatrix}
                    -a^2 & 0 & 0 & a\sin^2\theta(r^2+a^2)\\
                    0 & -a^2\cos^2\theta & 0 & 0\\
                    0 & 0 & r^2 & 0\\
                    -a & 0 & 0 & r^2+a^2\sin^2\theta
                \end{bmatrix}\,.
\end{equation}
For calculations, it will also be convenient to have the form of the rank-2 conformal Killing tensor that is obtained by contracting $\boldsymbol{h}$ with itself
\begin{equation} \label{Hdef}
    H^{\mu}_{\hspace{5pt}\nu}=  h_{\mu}^{\;\,\lambda}h_{\nu\lambda} = \begin{bmatrix}
                    -(r^2+a^2\sin^2\theta) & 0 & 0 & a\sin^2\theta(r^2+a^2)\\
                    0 & -r^2 & 0 & 0\\
                    0 & 0 & a^2\cos^2\theta & 0\\
                    -a & 0 & 0 & a^2
                \end{bmatrix}\,.
\end{equation}
These two tensors turn out to be related by the following equation, which can be verified by hand
\begin{equation} \label{HK}
    H_{\mu\nu} = K_{\mu\nu}-\frac{1}{2}g_{\mu\nu}K^{\alpha}_{\;\,\alpha}\,.
\end{equation}

We now have one Killing vector and one Killing tensor, equating to two integrals of motion.  For the second Killing vector, it can be obtained via the following equation
\begin{equation}
    \tilde{\zeta}^{\mu}=K^{\mu}_{\;\,\nu}\,\xi^{\nu} = -a^2\delta^{\mu}_{t}-a\delta^{\mu}_{\phi}\,.
\end{equation}
Due to linearity properties of the Lie derivative, the linear combination of two Killing vectors with constant coefficients is also a Killing vector.  With this in mind, it will be more convenient to define and use the following Killing vector in Boyer-Lindquist coordinates
\begin{equation}
    \zeta^{\mu} = -a\xi^{\mu}-(1/a)\tilde{\zeta}^{\mu}=\delta^{\mu}_{\phi}\,.
\end{equation}

What about the final Killing tensor?  By definition, if the metric is compatible, then it is automatically a Killing tensor as well.  Now that we have two desired Killing vectors and two desired Killing tensors, let's define their integrals of motion for particles that travel along geodesics with tangent vector $\boldsymbol{p}$. We choose them to be the following
\begin{equation} \label{IOM}
    E = -\xi_{\mu}p^{\mu}\,,\hspace{15pt} L_z = \zeta_{\mu}p^{\mu}\,,\hspace{15pt}Q=K_{\mu\nu}p^{\mu}p^{\nu}\,,\hspace{15pt}m^2=-g_{\mu\nu}p^{\mu}p^{\nu}\,.
\end{equation}
Here, $E$ and $L_z$ correspond to the energy of the geodesic particle and its angular momentum along the black hole's spin axis.  The quantity $m$ corresponds to the particle's mass.  Lastly, $Q$ is called "Carter's constant", which was originally discovered by Brandon Carter in 1968 and led to the complete integrability of geodesics in the Kerr spacetime \cite{Carter:68}.  We will see that a similar form of this constant can be derived in the Hamilton-Jacobi formulation of the geodesic equations for a free particle moving within a given spacetime.  A cross-check of the equations in the above section can be found in \cite{FroZel2011,Frolov2025-bn}.

    \chapter{Geodesic Equations and Geodesic Deviation} \label{apxB}

In this appendix, we derive the geodesic equations for a particle in the Kerr spacetime using the Hamilton-Jacobi approach.  We then use this geodesic equation to discuss important particle trajectories that one finds in the Kerr spacetime.  Lastly, we derive an equation of geodesic deviation, which describes how a non-geodesic particle experiencing acceleration deviates from a geodesic particle that initially starts on the same path as the non-geodesic particle.

\section{Hamilton-Jacobi Equation} \label{B1}

We first derive the geodesic equations for particles in the Kerr spacetime using the Hamilton-Jacobi equation, which can be found in \cite{MTW}.  We start off by assuming that any particle of mass $m$ in a specific spacetime should satisfy the mass-shell condition
\begin{equation} \label{massShell}
    p_{\mu}p^{\mu}=-m^2\,,
\end{equation}
where we have identified the momentum of the particle $\boldsymbol{p}$ with the tangent vector of its motion.

Now, to complete the derivation of the Hamilton-Jacobi equation, we look to the action functional $S$
\begin{equation}
    S[x(\tau)] = \int_{\tau_0}^{\tau}d\tau' L(x(\tau'),\dot{x}(\tau'),\tau')\,.
\end{equation}
In a typical derivation of the Euler-Lagrange equations, one assumes that the boundary points are fixed, and perturbs the action along the path between the boundary points.  In the case of the Hamilton-Jacobi equation, we assume that the action satisfies the Euler-Lagrange equations, and that the endpoints remain varied.  Therefore, we have the following
\begin{align}
    &\delta S = \Big[\frac{\partial L}{\partial \dot{x}^{\mu}}\delta x^{\mu}\Big]^{\tau}_{\tau_0} + \int_{\tau_0}^{\tau}d\tau'\Big(\frac{\partial L}{\partial x^{\mu}}-\frac{d}{d\tau}\frac{\partial L}{\partial \dot{x}^{\mu}}\Big)\delta x^{\mu}\\\nonumber
    &\hspace{14pt}=\frac{\partial L}{\partial \dot{x}^{\mu}}\delta x^{\mu}\\\nonumber
    &\hspace{14pt}=p_{\mu}\delta x^{\mu}\,.
\end{align}
This implies that
\begin{equation}
    \frac{\partial S}{\partial x^{\mu}} =  p_{\mu} \,.
\end{equation}
Inserting this into the mass-shell equation \eqref{massShell}, we arrive at the Hamilton-Jacobi equation
\begin{equation} \label{HJE}
    g_{\mu\nu}S_{,\mu}S_{,\nu} = -m^2\,.
\end{equation}

\section{Geodesic Equations in the Kerr Spacetime} \label{B2}
As stated in Appendix~\ref{apxA}, the Hamilton-Jacobi equations in the Kerr spacetime are separable.  This, along with the fact that $t$ and $\phi$ are cyclic variables in the Lagrangian, allows us to write the following ansatz for the action
\begin{equation}
    S = -Et + L_z\phi + S_r + S_{\theta}\,.
\end{equation}
Inserting this into the Hamilton-Jacobi equation \eqref{HJE} for a particle in the Kerr spacetime with Boyer-Lindquist coordinates \eqref{Kerr}, we have
\begin{align}
    -&\frac{1}{\Delta}\big[(r^2+a^2)^2E^2-4MarEL_z+a^2L_z^2\big]\\\nonumber
    +&\Delta S_{,r}^2+m^2r^2+2aL_zE+a^2\sin^2\theta E^2+\frac{L_z^2}{\sin^2\theta}\\\nonumber
    +&S_{,\theta}^{2}+m^2a^2\cos^2\theta-2aL_zE =0\,.
\end{align}
If we look at this equation for long enough, we notice that it can be completely separated into an $r$-dependent part and a $\theta$-dependent part.  This gives rise to a separation constant, which is exactly the same as the Carter's constant $Q$ defined in \eqref{IOM}
\begin{align}
    Q = &\Big(Ea\sin\theta-\frac{L_z}{\sin\theta}\Big)^2 + S_{,\theta}^2+m^2a^2\cos^2\theta\\\nonumber
    =&\frac{1}{\Delta}\big[(r^2+a^2)^2E^2-4MarEL_z+a^2L_z^2\big]
    -\Delta S_{,r}^2-2aL_zE-m^2r^2\,.
\end{align}
From here we can solve for $S_{,r}=p_r$ and $S_{,\theta}=p_{\theta}$ to obtain the equations of motion for a particle of mass $m$ in the Kerr spacetime, which of course are encoded in its tangent vector $p^{\mu}=(\dot{t},\dot{r},\dot{\theta},\dot{\phi})$
\begin{align} \label{leq} 
    &\Sigma\dot{t} = -a\big(Ea\sin^2\theta-L_{z}\big)+\frac{r^2+a^2}{\Delta}\big(E(r^2+a^2)-L_{z}a\big)\, , \\\nonumber
    &\Sigma\dot{r} = \epsilon_r \sqrt{\mathcal{R}}\,,\\\nonumber
    &\Sigma\dot{\theta} = \epsilon_{\theta}\sqrt{\Theta}\,,\\\nonumber
    &\Sigma\dot{\phi} = -\Big(Ea-\frac{L_z}{\sin^2\theta}\Big)+\frac{a}{\Delta}\big(E(r^2+a^2)-L_{z}a\big)\, ,
\end{align}
or in the more succinct covariant form
\begin{align}
    p_{t} =& -E\,,\\\nonumber
    p_{r} =& \epsilon_r\frac{\sqrt{\mathcal{R}}}{\Delta}\,,\\\nonumber
    p_{\theta} =& \epsilon_{\theta}\sqrt{\Theta}\,,\\\nonumber
    p_{\phi} =& L_z\,,
\end{align}
where
\begin{align} \label{RT}
   & \mathcal{R }= \Big(E(r^2+a^2)-L_z a\Big)^2 - \big(m^2r^2+Q\big)\Delta
    \,,\\\nonumber
    &\Theta = Q-m^2a^2\cos^2\theta-\Big(Ea\sin\theta-\frac{L_z}{\sin\theta}\Big)^2\,. 
\end{align}
Here, $\epsilon_r$ and $\epsilon_{\theta}$ define the radial and angular turning points of the geodesic ray.  Accordingly, we define $\epsilon_r=-1$ for incoming rays and $\epsilon_r=+1$ for outgoing rays, as well as $\epsilon_\theta=-1$ when $\theta$ is decreasing and $\epsilon_\theta=+1$ when $\theta$ is increasing.  This can be checked against \cite{FroZel2011}, where equation \eqref{leq} is calculated in the equatorial plane, or in \cite{MTW}, where it is calculated for the full Kerr spacetime.

Note that for the majority of this thesis, we will be working exclusively with null rays, in which case we can simply set $m=0$ in the above equations.

\section{ISCOs and the Photon Sphere} \label{B3}

We now look at some important trajectories that arise when considering particle orbits around Kerr black holes.  For this reason, we now restate Eq.~\eqref{leq}
\begin{align} \label{leq2} 
    &\Sigma\dot{t} = -a\big(Ea\sin^2\theta-L_{z}\big)+\frac{r^2+a^2}{\Delta}\big(E(r^2+a^2)-L_{z}a\big)\, , \\\nonumber
    &\Sigma\dot{r} = \epsilon_r \sqrt{\mathcal{R}}\,,\\\nonumber
    &\Sigma\dot{\theta} = \epsilon_{\theta}\sqrt{\Theta}\,,\\\nonumber
    &\Sigma\dot{\phi} = -\Big(Ea-\frac{L_z}{\sin^2\theta}\Big)+\frac{a}{\Delta}\big(E(r^2+a^2)-L_{z}a\big)\, ,
\end{align}
along with
\begin{align}
   & \mathcal{R }= \Big(E(r^2+a^2)-L_z a\Big)^2 - \big(m^2r^2+Q\big)\Delta
    \,,\\\nonumber
    &\Theta = Q-m^2a^2\cos^2\theta-\Big(Ea\sin\theta-\frac{L_z}{\sin\theta}\Big)^2\,. 
\end{align}

The equation that we use to determine orbits is of course the radial equation.  If we require that
\begin{equation} \label{circ1}
    \dot{r}=0\,,
\end{equation}
then we can find the turning points in these orbits where we go from decreasing radius to increasing radius.  This can be rewritten as
\begin{equation}
    \mathcal{V}_{eff} = E^2\,,
\end{equation}
where this equation is obtained by solving the quadratic equation $\mathcal{R}=0$ for $E$. In order for the orbits to remain circular, so that they do not have any radial acceleration, we also require that
\begin{equation} \label{circ2}
    \frac{d\mathcal{V}_{eff}}{dr} = 0\,.
\end{equation}
Therefore, a circular orbit at radius $r=r_0$ in the Kerr metric is determined by solving the system of equations given by \eqref{circ1} and \eqref{circ2}.

If we take $\mathcal{E}=E/m$ and $\mathcal{L}=L_z/m$ to be the energy and axial angular momentum of the particle per unit mass, then for any circular orbit $r=r_0$, we have
\begin{equation} \label{EL}
    \mathcal{E}_0 = \frac{r_0^2-2Mr_0\pm a\sqrt{Mr_0}}{r_0\big(r_0^2-3Mr_0\pm 2a\sqrt{Mr_0}\big)^{1/2}}\,,\hspace{15pt}\mathcal{L}_0 = \frac{\pm\sqrt{Mr_0}\big(r_0^2\mp 2a\sqrt{Mr_0}+a^2\big)}{{r_0\big(r_0^2-3Mr_0\pm 2a\sqrt{Mr_0}\big)^{1/2}}}\,.
\end{equation}
From here on out, we consider the $+$ sign to mean that the orbits are prograde (rotating with the black hole), and similarly orbits with a $-$ sign are retrograde (rotating against the black hole).

The stability of the circular orbits is given by investigating the sign of $\ddot{\mathcal{V}}_{eff}$.  Specifically, the orbits are considered to be stable if
\begin{equation}
    \ddot{\mathcal{V}}_{eff} \geq 0\,.
\end{equation}
Therefore, the orbits switch from stable to unstable when $\ddot{\mathcal{V}}_{eff}=0$.  For this reason, we call the case where \begin{equation} \label{ISCO}
    \ddot{\mathcal{V}}_{eff}=0
\end{equation}
the \emph{innermost stable circular orbit}, or ISCO for short.  Solving for radius $r$ via the system of equations given by \eqref{circ1}, \eqref{circ2}, and \eqref{ISCO} gives us
\begin{align}
    r_{ISCO} &= M\Big(3+Z_2\mp\big[(3-Z_1)(3+Z_1+2Z_2)\big]^{1/2}\Big)\,,\\\nonumber
    Z_1 &=1+(1-\alpha^2)^{1/3}\Big[(1+\alpha)^{1/3}+(1-\alpha)^{1/3}\Big]\,,\\\nonumber
    Z_2 &=\big(3\alpha^2+Z_1^2\big)^{1/2}\,,
\end{align}
where $\alpha=a/M$ is the dimensionless rotation parameter.

For photons, one can follow a very similar procedure by studying the radial equation in \eqref{leq2} when $m=0$.  When this equation equals zero, we once again have the turning points for photon trajectories.  These turning points are given by
\begin{equation}
    \ell_z = \frac{\pm\rho\sqrt{\rho^2-2\rho+\alpha^2}-2\alpha}{\rho-2}\,,
\end{equation}
where $\ell_z\equiv L_z/(EM)$ and $\rho=r/M$.  This function can be thought of as the impact parameter of scattering photon trajectories around the black hole.

For any given dimensionless rotation parameter $\alpha$, $\ell_z$ will have some minimum value that marks a divide between photons that are scattered by the black hole and go off to radial infinity, and photons which become trapped and fall into the black hole.  This limit corresponds to an unstable orbit for photons around the black hole, and for this reason it is called the \emph{photon sphere}.  This happens when $d\ell_z/d\rho=0$, and solving this equation gives us the radius of the photon sphere
\begin{equation} \label{pshere}
    \rho_{photon} = 2 + \cos\psi\mp\sqrt{3}\sin\psi\,,\hspace{15pt}\psi = \frac{2}{3}\arcsin\alpha\,.
\end{equation}
This photon sphere also gives us a direct definition of the black hole's "shadow," which is fixed by the boundary region between where we can see light around the black hole, and where incoming light is no longer emitted due to gravitational capture.

For a more in-depth derivation of these surfaces, see \cite{FroZel2011}.

\section{Scattering Problems} \label{B4}

The scattering of a null ray by a black hole leads to a wealth of observable and physically consequential effects that can be studied and exploited.  We will discuss a few of these effects below

\subsection{Deflection Angles and Shapiro Time Delay} \label{B41}

For simplicity, in the following subsection, we will set the rotation parameter $a=0$, so that we are working with Schwarzschild black holes.

A distinct difference in predictions between Newtonian gravity and general relativity is the value of the deflection angle of light scattered by a massive body.  In Newtonian gravity, this deflection angle is given by $\psi_N = 2M/b$, where $b$ is the impact parameter of the scattered light ray.  In the general relativistic case, we can calculate such a deflection angle via Eq.~\eqref{leq}.  This tells us that
\begin{equation}
    \frac{d\phi}{dr} = \epsilon_r \frac{L_z}{\sqrt{\mathcal{R}_S}\sin^2\theta}\,,
\end{equation}
where $\mathcal{R}_S=\mathcal{R}|_{a=0}$ is taken from Eq.~\eqref{RT}.
After integrating this equation via proper initial conditions and the use of elliptic functions, we come to the following expansion of the deflection angle by a Schwarzschild black hole for large impact parameter $b$:
\begin{equation}
    \psi_S = \frac{4M}{b} + \frac{15\pi}{4}\frac{M^2}{b^2}+O((M/b)^{3})\,.
\end{equation}
We call this the "weak-field limit," as we expect to recover Newtonian gravity at distance scales very far away from the black hole.  A more complete derivation for a rotating black hole can be found in \cite{FroZel2011}

Another important aspect of GR that is absent from Newtonian gravity is the time delay effect, discovered by Irwin Shapiro in 1964 \cite{Shapiro64}.  This effect is an important test of the consequences of GR, as it shows that under the influence of a massive object, the arrival time of light rays to a specific observer is longer than one would expect if the massive object was not present.  The reason for this time delay can be explained by gravitational time dilation.  Utilizing Eq.~\eqref{leq}, we have
\begin{equation}
    \frac{dt}{dr} = \epsilon_r\frac{Er^4}{\sqrt{\mathcal{R}_S}(r^2-2Mr)}\,.
\end{equation}
Once more, we integrate with the proper initial conditions.  For a light ray that is emitted from an observer $r_0$, is scattered by the black hole so that it reaches a mirror at $r_1$, and then is scattered back to the observer, the total proper time of its journey is given by
\begin{align} \label{shapiro}
    \Delta\tau = \sqrt{1-\frac{r_S}{r_0}}\Bigg[&2\sqrt{r_0^2-r_S^2}+2\sqrt{r_1^2-r_S^2}+2r_S\ln\bigg(\frac{r_0+\sqrt{r_0^2-r_{m}^2}}{r_{m}}\bigg)\\\nonumber
    &+2r_S\ln\bigg(\frac{r_1+\sqrt{r_1^2-r_{m}^2}}{r_{m}}\bigg)+r_S\sqrt{\frac{r_0-r_m}{r_0+r_m}}+r_S\sqrt{\frac{r_1-r_m}{r_1+r_m}}\Bigg]
\end{align}
Here, $r_{m}$ is the radius of closest approach for the ray passing by the black hole. For a more in-depth calculation of this phenomenon, see \cite{FroZel2011}.

\subsection{ZAMOs and the Black Hole Shadow} \label{B42}

The rotation of a Kerr black hole should also be thought of as a rotation of spacetime itself.  To see this, we introduce a concept known as the zero angular momentum observer (ZAMO), which was first derived by Bardeen, Press, and Teukolsky in 1972 \cite{Bardeen_2}.  We can derive an orthonormal tetrad attached to such an observer as follows:

First, its radial and polar components are not affected by rotation, and therefore should simply be of unit length and orthogonal to each other
\begin{equation}
    e_{(r)}^{\mu} = \delta_r^{\mu}/\sqrt{g_{rr}}\,,\hspace{10pt}e_{(\theta)}^{\mu} = \delta_{\theta}^{\mu}/\sqrt{g_{\theta\theta}}\,.
\end{equation}
Next, if we define the angular velocity of the black hole at any given radius as
\begin{equation}
    \Omega = -\frac{g_{t\phi}}{g_{\phi\phi}}\,,
\end{equation}
then for an observer with four-velocity $\boldsymbol{u}=u^{\mu}\partial_{\mu}$, their angular velocity should obey
\begin{equation}
    \omega \equiv \frac{u^{\phi}}{u^{t}} = \Omega\,.
\end{equation}
Assuming that the timelike four-velocity obeys $\boldsymbol{u}^2=u^{\mu}u_{\mu}=-1$, we get that
\begin{equation}
    u^{t} = 1/\sqrt{-(g_{tt}^2+2\Omega g_{t\phi}+\Omega^{2}g_{\phi\phi})}\,,
\end{equation}
which subsequently defines $u^{\phi}$.  From here, if we identify the timelike component of the tetrad with the observer's four-velocity $e^{\mu}_{(t)}=u^{\mu}$, then we can use orthonormality to find the remaining basis vector $e^{\mu}_{(\phi)}$.  

In summary, for the Kerr metric we find that
\begin{align}
    &\boldsymbol{e}_{(t)} = \sqrt{\frac{A}{\Sigma\Delta}}\partial_t + \frac{2Mar}{\sqrt{A\Sigma\Delta}}\partial_{\phi}\,,\\\nonumber
    &\boldsymbol{e}_{(r)} = \sqrt{\frac{\Delta}{\Sigma}}\partial_r\,,\\\nonumber
    &\boldsymbol{e}_{(\theta)} = \frac{1}{\sqrt{\Sigma}}\partial_{\theta}\,,\\\nonumber
    &\boldsymbol{e}_{(\phi)}=\sqrt{\frac{\Sigma}{A}}\frac{1}{\sin\theta}\partial_{\phi}\,,
\end{align}
where $A=(r^2+a^2)^2-\Delta a^2\sin^2\theta$.  One can easily check that the angular momentum of the observer along the axis of rotation is $L_z = \zeta_{\mu}u^{\mu}=0$, as desired.  We should note that a ZAMO is \emph{not} typically a geodesic, minus a few specific exceptions.

Lastly, one can show that the projected image of a null ray $\boldsymbol{l}$ traversing a black hole onto a screen that is orthogonal to the ray and whose origin is defined by ZAMO that is directly facing the center of the black hole at a far radial distance $r_0$ is simply given by the following arc lengths \cite{Bardeen}:
\begin{equation}
    \alpha_{(\theta)} = \lim_{r_0\to\infty} r_0 e^{\mu}_{(\theta)}l_{\mu}\,,\hspace{10pt}\alpha_{(\phi)} = \lim_{r_0\to\infty} -r_0 e^{\mu}_{(\phi)}l_{\mu}\,.
\end{equation}
When these null rays pass by the black hole at a radius of closest approach just above the radius at which they would be captured and absorbed by the black hole, this pair $(\alpha_{(\theta)},\alpha_{(\phi)})$ corresponds to a two-dimensional plane image of the black hole's shadow.

\section{Geodesic Deviation} \label{B5}

The principle of geodesic deviation describes the relative separation between two neighboring geodesics in a given spacetime due to the presence of curvature.  In this section, we derive a "modified" geodesic deviation equation that describes the same relative separation, but now considers the separation to be between a geodesic ray, and a non-geodesic ray that experiences a small acceleration in the direction orthogonal to the two rays' initial common path. In order to define this "modified" geodesic deviation, we follow the procedure outlined in \cite{Frolov2025-bn}, but one can find similar derivations in e.g. \cite{MTW}.  Normally, for a derivation of geodesic deviation between two rays, we would take the tangent vectors of two geodesic rays to be at the same starting point $O$, choose there to be an infinitesimal difference between them at said origin, and then see how they evolve with respect to each other as the difference in their paths traces out a two-dimensional worldsheet in spacetime.  

\begin{figure}[!hbt]
    \centering
\includegraphics[width=0.8\textwidth]{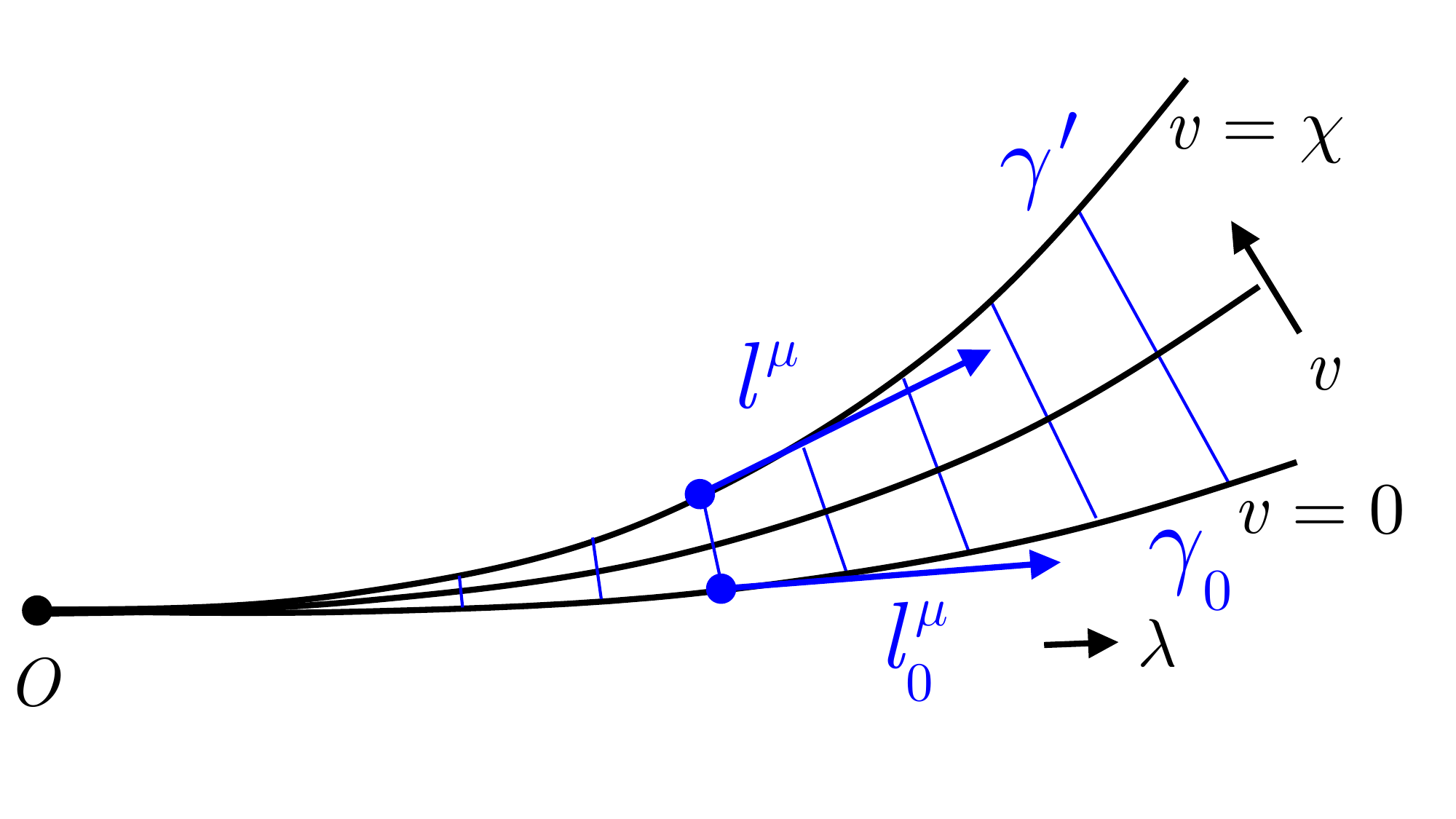}
\caption{\label{Fig_1_1}Congruence of null rays spanned by perturbing a geodesic null ray with a small acceleration $v\Upsilon^{\mu}$, governed by small parameter $v$.}
\end{figure}

For this derivation, the idea will be the same, except that now we choose the two tangent vectors to be exactly aligned at the origin $O$, and their growing deviation will be induced by the fact that one of them will no longer be a geodesic ray, \emph{i.e.} it will have some non-zero four-acceleration guiding it.  This scenario is described by Fig~\ref{Fig_1_1}, which was taken from \cite{Frolov2025-bn}

We define a congruence of null rays that is governed by two parameters, $x^{\mu}=x^{\mu}(\lambda,v)$, where $\lambda$ is the affine parameter and $v$ governs the separation of the non-geodesic ray from the geodesic one.  With this, we can assume that the non-geodesic ray has a very small acceleration $\nabla_{\boldsymbol{l}}l^{\mu}=v\Upsilon^{\mu}$.  We define the vectors
\begin{equation}
    l^{\mu}=\frac{\partial x^{\mu}}{\partial\lambda}\,,\hspace{15pt}p^{\mu}=\frac{\partial x^{\mu}}{\partial v}\,,
\end{equation}
where $l^{\mu}_0=l^{\mu}(\lambda,0)$.  The vector that measures the difference between $\boldsymbol{l}$ and $\boldsymbol{l_0}$ at the same value of $\lambda$ is then given by $\delta x^{\mu}(\lambda,v)=vp^{\mu}(\lambda,v)$.

By the commutativity of partial derivatives, we also have
\begin{align} \label{commute}
    &\frac{\partial^{2}x^{\mu}}{\partial\lambda\partial v} = \frac{\partial^{2}x^{\mu}}{\partial v\partial\lambda}\\[5pt]\nonumber
    \implies &l^{\nu}\partial_{\nu}p^{\mu} = p^{\nu}\partial_{\nu}l^{\mu}\\[5pt]\nonumber
    \implies &l^{\nu}\nabla_{\nu}p^{\mu} = p^{\nu}\nabla_{\nu}l^{\mu}\,,
\end{align}
where the last line comes from adding the term $\Gamma^{\mu}_{\alpha\beta}l^{\alpha}p^{\beta}$ to both sides of the equation.

Given that equation \eqref{commute} shows that $[\boldsymbol{l},\boldsymbol{p}]=0$, we utilize equation \eqref{RXY} to give us
\begin{equation}
    R^{\mu}_{\,\;\nu\alpha\beta}l^{\alpha}p^{\beta}l^{\nu} = \nabla_{\boldsymbol{l}}\nabla_{\boldsymbol{p}}l^{\mu}-\nabla_{\boldsymbol{p}}\nabla_{\boldsymbol{l}}l^{\mu}\,.
\end{equation}
With this, we are two steps away from deriving the full modified geodesic deviation equation.  All that is left is to use the commutation relation \eqref{commute} on the first term of the left hand side of the above equation, and then substitute the acceleration $\nabla_{\boldsymbol{l}}l^{\mu}=v\Upsilon^{\mu}$ into the second term.  Multiplying both sides by $v$, this leaves us with
\begin{equation} \label{geoDevFull}
    \nabla_{\boldsymbol{l}}\nabla_{\boldsymbol{l}}\delta x^{\mu} - R^{\mu}_{\,\;\nu\alpha\beta}l^{\nu}l^{\alpha}\delta x^{\beta} = v\big(\Upsilon^{\mu}+\nabla_{\boldsymbol{\delta x}}\Upsilon^{\mu}\big)\,.
\end{equation}
In theory, this equation exactly predicts the deviation of an accelerated ray from its geodesic counterpart.  In practice, this would not be possible to compute analytically, and so we look to employ a perturbative expansion in $\boldsymbol{l}$.  Namely, fix the small parameter $v=\chi$, assume that $\boldsymbol{l}$ has the form $\boldsymbol{l}=\boldsymbol{l}_0+\chi\boldsymbol{l}_1+...$.  Keeping the lowest order in $\chi$ and recalling that $\delta x^{\mu}$ is small, we have
\begin{equation} \label{geoDev}
    \nabla_{\boldsymbol{l}_0}\nabla_{\boldsymbol{l}_0}\delta x^{\mu} - R^{\mu}_{\,\;\nu\alpha\beta}l_0^{\nu}l_0^{\alpha}\delta x^{\beta} = \chi\Upsilon^{\mu}\,.
\end{equation}

	\chapter{Spinoptics Calculations and Formulas}\label{apxC}

This appendix contains a collection of formulas and relations which are used throughout Chapter~\ref{ch5} of this thesis.

\section{Covariant Components of the Null Tetrad Vectors} \label{C1}

For completeness, we give the covariant form of the vectors of the tetrad $(\boldsymbol{l},\tilde{\boldsymbol{n}},\boldsymbol{e}_2,\boldsymbol{\tilde{e}}_3)$ associated with a null geodesic in the Kerr metric, whose contravariant components are defined by Eqs.~\eqref{leqD} and \eqref{eeee}. Written in  dimensionless coordinates $(\tau,\rho,\theta,\phi)$ these vectors are
\begin{align}
l_{\mu}=&\bigg(-1,\,\epsilon_{\rho}\,\frac{\sqrt{\mathcal{R}}}{\Delta},\,\epsilon_{\theta}\sqrt{\Theta},\,\ell_z\bigg)\,,\\\nonumber
\tilde{n}_{\mu}=&\frac{1}{2Q}\bigg(\Sigma-2(\rho^2+\alpha^2-\ell_{z}\alpha),\,\Sigma\epsilon_{\rho}\sqrt{\mathcal{R}}/\Delta,\,-\Sigma\epsilon_{\theta}\sqrt{\Theta},\\\nonumber
&\hspace{35pt}(\rho^2+\alpha^2)(\alpha\sin^2\theta-\ell_{z})+\alpha\sin^2\theta(\rho^2+\alpha^2-\ell_{z}\alpha)\bigg)\,,\\\nonumber
e_{2\mu}=&\frac{1}{\Sigma\sqrt{Q}}\bigg(-\alpha\big(\cos\theta\epsilon_{\rho}\sqrt{\mathcal{R}}-\rho\sin\theta \epsilon_{\theta}\sqrt{\Theta}\big),\,\frac{\Sigma \alpha\cos\theta}{\Delta}(\rho^2+\alpha^2-\ell_{z}\alpha),\\\nonumber
&\hspace{35pt}-\Sigma \rho\sin\theta\big(\alpha-\ell_{z}/\sin^2\theta\big),\,\sin\theta\Big(\alpha^2\sin\theta\cos\theta\epsilon_{\rho}\sqrt{\mathcal{R}}\\\nonumber
&\hspace{35pt}-\rho(\rho^2+\alpha^2)\epsilon_{\theta}\sqrt{\Theta}\Big)\bigg)\,,\\\nonumber
\tilde{e}_{3\mu}=&\frac{-1}{\Sigma\sqrt{Q}}\bigg(-\Big(\rho\epsilon_{\rho}\sqrt{\mathcal{R}}+\alpha^2\sin\theta\cos\theta\epsilon_{\theta}\sqrt{\Theta}\Big),\,+\frac{\Sigma}{\Delta}\Big(\rho(\rho^2+\alpha^2-\ell_{z}\alpha)\Big),\\\nonumber
&\hspace{35pt}+\Sigma\Big(\alpha\sin\theta\cos\theta\big(\alpha-\ell_{z}/\sin^2\theta\big)\Big),\,+\alpha\sin\theta\Big(\rho\sin\theta\epsilon_{\rho}\sqrt{\mathcal{R}}\\\nonumber
&\hspace{35pt}+(\rho^2+\alpha^2)\cos\theta\epsilon_{\theta}\sqrt{\Theta}\Big)\bigg)\,.
\end{align}

The expressions for the vectors of the parallel propagated tetrad $(\boldsymbol{l},\boldsymbol{{n}},\boldsymbol{e_2},\boldsymbol{{e}_3})$ can easily be obtained using relations \eqref{ee33} and \eqref{nnPP}.

Note that
\begin{equation}
l_{[\mu,\nu]} = 0\,.
\end{equation}
One can easily show that
\begin{equation}
l_{\mu}=U_{,\mu}\,,\hspace{15pt}U=-\tau+\ell_z\phi +\int \dfrac{\epsilon_\rho\sqrt{\mathcal{R}}d\rho}{\Delta}+\int \epsilon_{\theta} \sqrt{\Theta}d\theta\, .
\end{equation}

Note that we can also write the vectors $(\boldsymbol{l},\boldsymbol{\tilde{n}},\boldsymbol{e_2},\boldsymbol{\tilde{e}_3})$ in terms of the ZAMO basis given in equation \eqref{EEEE}.  In the equatorial plane, this amounts to
\begin{align}\label{TET_ZAM}
l^{\mu}=&\frac{A-2\rho\alpha\ell_z}{\rho\sqrt{\Delta A}}E_{(\tau)}^{\mu}+\epsilon_{\rho}\rho\sqrt{\frac{\mathcal{B}}{\Delta}}E_{(\rho)}^{\mu}+\frac{\rho\ell_z}{\sqrt{A}}E_{(\phi)}^{\mu}\,,\\\nonumber
\tilde{n}^{\mu}=&\frac{\rho\big((\rho^2+\alpha^2-\alpha\ell_z)^2+\alpha^2\Delta-\alpha\ell_z(\alpha\ell_z-2\rho)\big)}{2\sqrt{A\Delta}(\ell_z-\alpha)^2}E_{(\tau)}^{\mu}\\\nonumber
&+\frac{\rho\big(2\alpha(\rho^2+\alpha^2-\alpha\ell_z)-\ell_z\rho^2\big)}{2(\ell_z-\alpha)^2\sqrt{A}}E_{(\phi)}^{\mu}+\frac{\epsilon_{\rho}\rho^3\sqrt{\mathcal{B}}}{2(\ell_z-\alpha)^2\sqrt{\Delta}}E_{(\rho)}^{\mu}\,,\\\nonumber
e_{2}^{\mu}=&E_{(\theta)}^{\mu}\,,\\\nonumber
\tilde{e}_{3}^{\mu}=&\frac{\epsilon_{\rho}\rho^2(
\rho^2+\alpha^2)\sqrt{\mathcal{B}}}{(\ell_z-\alpha)\sqrt{A\Delta}}E_{(\tau)}^{\mu}-\frac{\rho^2-\alpha(\ell_z-\alpha)}{(\ell_z-\alpha)\sqrt{\Delta}}E_{(\rho)}^{\mu}-\frac{\epsilon_{\rho}\rho^{2}\alpha\sqrt{\mathcal{B}}}{(\ell_z-\alpha)\sqrt{A}}E_{(\phi)}^{\mu}\,.
\end{align}

\section{Scattering With Large Impact Parameter} \label{C2}

Here we discuss the spinoptics tilting angle for the scattering of rays with large impact parameter $\ell_z$ in the vicinity of the equatorial plane of the Kerr black hole. For large $\ell_z$, the dimensionless radius of the radial turning point $\rho_m$ is also large. To find the solutions of \eqref{XX2} in this regime, we introduce the following new variables
\begin{equation} \label{rhom}
\hat{\rho}=\rho/\rho_m\, ,\hspace{10pt} \hat{\ell}_z=\ell_z/\rho_m\, ,\hspace{10pt}
\hat{\lambda}=\lambda/\rho_m\, ,\hspace{10pt}Y=\rho_m \mathcal{X}_2/(3\chi \hat{\ell}_z)\,.
\end{equation}
Then \eqref{XX2} takes the form
\begin{equation}\label{YYY}
\dfrac{d^2 Y}{d\hat{\lambda}^2} +\dfrac{3 \hat{\ell}_z^2}{\hat{\rho}^5}\frac{Y}{\rho_m}=  \dfrac{\hat{\lambda}+\hat{\lambda}_m}{\hat{\rho}^5}\, .
\end{equation}
Using equations \eqref{rmlarge} and \eqref{LMR}, one has
\begin{equation} \label{hatlz}
\hat{\ell}_z\approx \pm (1+1/\rho_m)\,,\hspace{15pt}
\hat{\lambda}_m\approx 1/\rho_m\, .
\end{equation}

Let us write  a solution of \eqref{YYY} in the form
\begin{equation} \label{Y01}
Y=Y_0+\dfrac{1}{\rho_m}Y_1+...\, .
\end{equation}
Then
\begin{equation}\label{YY0}
\dfrac{d^2 Y_0}{d\hat{\lambda}^2}= \dfrac{\hat{\lambda}}{\hat{\rho}^5}\, .
\end{equation}
For large $\rho_m$, one can find a relation between $\hat{\lambda}$ and $\hat{\rho}$ by using Eq.~\eqref{Brm}. We put $\hat{\rho}=\cosh\mu$, and therefore we find that
\begin{equation}
\hat{\lambda}\approx\sinh\mu+(1/\rho_m)\tanh(\mu/2)\, .
\end{equation}
In the leading order $\hat{\lambda}\approx\sinh\mu$
and so \eqref{YY0} takes the form
\begin{equation}
\dfrac{d^2 Y_0}{d\mu^2}
-\dfrac{\sinh\mu}{\cosh\mu}\dfrac{d Y_0}{d\mu}
= \dfrac{\sinh\mu}{\cosh^3\mu}\, .
\end{equation}
A solution of this equation is
\begin{equation}
Y_0=-\frac{1}{3}\tanh\mu+c_1\sinh\mu+c_2\, ,
\end{equation}
where $c_1$ and $c_2$ are integration constants. The boundary condition $Y_0(\mu=-\infty)=0$ implies that
\begin{equation} \label{Y0SOL}
Y_0=-\frac{1}{3}(\tanh\mu+1)\, .
\end{equation}
Then for $\mu\to \infty$ one has $Y_0=-2/3$. This according to Eq.~\eqref{rhom}, at zeroth order
\begin{equation} \label{tY0}
\vartheta=\dfrac{3\chi\hat{\ell}_z}{\rho_m} \dfrac{Y_0}{\cosh\mu}\, .
\end{equation}
This means that for $Y_0$ the tilting angle is $\vartheta^{\infty}=0$.

Substituting the obtained solution \eqref{Y0SOL} and now taking \eqref{YYY} to first order in the perturbation \eqref{Y01}, one obtains a differential equation for $Y_1$
\begin{equation}
\frac{(\sinh\mu+\cosh\mu+1)^2}{\cosh^{5}\mu(\cosh\mu+1)}=\frac{d}{d\mu}\Big[\frac{1}{\cosh{\mu}}\frac{dY_1}{d\mu}+\frac{1}{3\cosh^4{\mu}(\cosh\mu+1)}\Big]\,.
\end{equation}
One can solve this equation to obtain $Y_1$ explicitly. As before, the condition that $Y_1(\mu=-\infty)=0$ restricts the two constants of integration, and we have
\begin{align}
Y_1 = &\frac{4}{3}(\cosh\mu+\sinh\mu)-\frac{1}{4}\tan^{-1}(\sinh\mu)-\tan^{-1}(\tanh(\mu/2))\\\nonumber
&+\frac{1}{3}\tanh(\mu/2)+\frac{\tanh(\mu/2)}{\tanh^2(\mu/2)+1}-\frac{2\tanh(\mu/2)}{3\big(\tanh^2(\mu/2)+1\big)^2}\\\nonumber
&-\frac{1}{4}\frac{\sinh\mu}{\cosh^{2}\mu}-\frac{2}{3}\frac{1}{\cosh\mu}+\frac{2}{3}-\frac{5\pi}{8}\,.
\end{align}
One can show that only the first term in parentheses on the right-hand side of this expression after its division by $\cosh\mu$ remains finite as $\mu\to\infty$, and one has
${Y_1}/{\cosh\mu}\to 8/3$.
This allows us to find $\vartheta^{\infty}$, which has the form
\begin{equation}\label{TILT}
\vartheta^{\infty} = \pm \dfrac{8\chi}{\rho^3_m}+O(\rho_m^{-4})\,
\end{equation}

\section{Screen Projection Calculation} \label{C3}

In this section we aim to derive equations related to the "rainbow effect" which were used in Sec.~\ref{s542}.  We proceed as follows:

We note that, although the contribution of the term $Y_0$ defined in \eqref{Y0SOL} to the spinoptics tilting angle is zero, as seen by Eq.~\eqref{tY0}, it will still have an effect on the first-order approximation of the screen projection of an observer given by Eq.~\eqref{imParam1}. Using this formula, we have that
\begin{equation} \label{A1EQ}
    \alpha_{1(\theta)} = \lim_{\rho_0\to\infty}\rho_0\sqrt{\Sigma}l^{\theta}_1\,.
\end{equation}

Here, we can imagine that $l_1^{\theta}=l_1^{\mu}\delta^{\theta}_{\mu}=\chi\dot\theta_1$ (where $\dot{\theta}_1$ is the change in the polar angle due to the first-order spinoptics approximation) is given by
\begin{equation} \label{l1EQ}
    l_1^{\theta} = \frac{3\chi\hat{\ell}_z}{\rho_m^2}\frac{d}{d\lambda}\bigg[\frac{Y_0}{\rho}\bigg]+O(\rho_m^3)\,.
\end{equation}
This comes from Eq.~\eqref{tY0}, and the fact that $\vartheta=\pi/2-\theta$. From here on out we will drop higher order terms and treat the lowest-order terms in the expansion as strict equalities.

Using Eqs.~\eqref{rhom}, \eqref{hatlz}, and \eqref{l1EQ}, we can rewrite Eq.~\eqref{A1EQ} as
\begin{equation}
    \alpha_{1(\theta)} = \mp\frac{3\chi}{\rho_m^3}\lim_{\mu\to\infty} \bigg[\frac{dY_0}{d\mu}-Y_0\tanh\mu\bigg]\,.
\end{equation}
Substituting in Eq.~\eqref{Y0SOL} and taking the limit, we arrive at
\begin{equation} \label{a1sol}
    \alpha_{1(\theta)} = \mp\frac{2\chi}{\rho_m^3}\,.
\end{equation}

\end{document}